# Room-Temperature Superconductivity

## Room-Temperature Superconductivity

**Andrei Marouchkine**

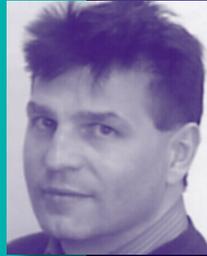

Andrei Mourachkine is a Senior Research Associate at the Nanoscience Center of the University of Cambridge. Andrei Mourachkine recieved the Master Degree in Chemical Physics from the Novosibirsk State University (Russia) in 1985, and the Ph.D. in Physics from The Free University of Brussels in 1996. Since 1992 Mourachkine has been working with high-temperature superconductors. He has authored numerous research articles on a variety of topics in high-temperature superconductivity. Andrei Mourachkine is the author of a book entitled "High-Temperature Superconductivity in Cuprates: The Nonlinear Mechanism and Tunneling Measurements" published by Kluwer Academic in August 2002. In 2003 he moved from Brussels to Cambridge.

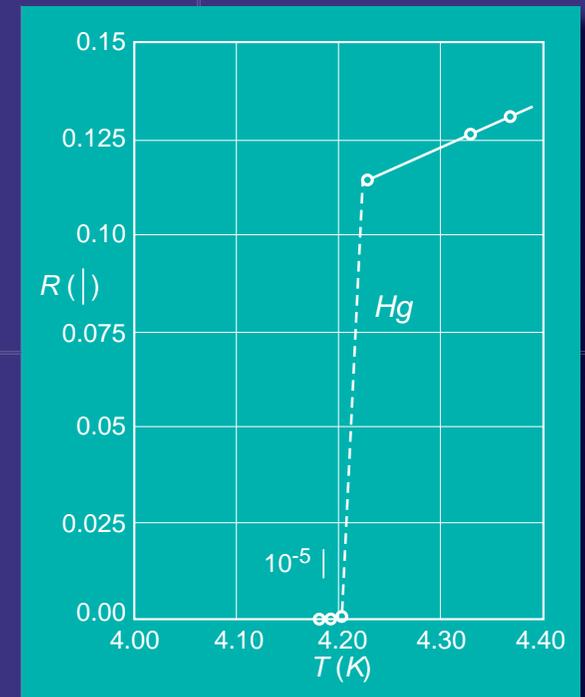

# Room-Temperature Superconductivity

# ROOM-TEMPERATURE SUPERCONDUCTIVITY

## Andrei Mourachkine

*University of Cambridge, Cambridge, United Kingdom*

*Dedicated to the Dutch
physicist H. Kamerlingh
Onnes and his assistant
Gilles Holst who made the
historic measurement*

*In order to create the living matter, Nature needed billions of years.*
*This experience is unique, and we must learn from it.*

# Contents

# Preface

In spite of the fact that it is Nature's "oversight," superconductivity is a remarkable phenomenon. Personally, I am fascinated by it. Superconductivity, indeed, was a major scientific mystery for a large part of the last century: discovered in 1911 by the Dutch physicist H. Kamerlingh Onnes and his assistant Gilles Holst, it was completely understood only in 1957. Generally speaking, superconductivity is a low-temperature phenomenon. As a result, it is commonly believed that it cannot occur at room temperature, $T \sim 300$ K.

The main purpose of the book is twofold. First, to show that, under suitable conditions, superconductivity can occur above room temperature. Second, to present general guidelines how to synthesize a room-temperature superconductor. The principal point of this book is that, in order to synthesize a room-temperature superconductor, we may use some of Nature's experience (see the prologue to the book).

The book is organized as follows. The first seven chapters of the book present an overview of the basic properties of the superconducting state and the mechanisms of superconductivity in various compounds. Chapter 1 is a historical review of major events related to the phenomenon of superconductivity. Chapter 2 gives an overview of the basic properties of the superconducting state. In all textbooks on superconductivity, the description of the superconducting state is based on the Bardeen-Cooper-Schrieffer (BCS) theory, assuming that the BCS model is the only possible mechanism of superconductivity (in fact, this is not the case). Contrary to this old tradition, Chapter 2 reviews the superconducting state independently of any specific mechanism. Chapter 3 gives an overview of superconducting materials. The next chapter, Chapter 4, presents the main principles of superconductivity as a phenomenon, valid for every superconductor independently of its characteristic properties. In various materials, the underlying mechanisms of superconductivity can be different, but these principles must be satisfied. The following three chapters, Chapters 5–7, describe the mechanisms of superconductivity in various com-pounds. The main purpose of Chapter 5 is to discuss the BCS mechanism of superconductivity characteristic for conventional superconductors. Chapter 6 reviews the mechanism of unconventional superconductivity. The mechanism of half-conventional superconductivity is discussed in Chapter 7. Thus, the first seven chapters make an introduction into the physics of the superconducting state and superconducting materials and, therefore, can be used by students.

The last three chapters of the book are mainly addressed to specialists in materials science and in the field of superconductivity. In Chapter 8, it is shown that the Cooper pairs exist above room temperature in organic materials. The main purpose of Chapter 9 is to discuss the onset of long-phase coherence in a room-temperature superconductor. The chief aim of Chapter 10 is to consider materials able to superconduct above room temperature. In the context of practical application, Chapter 10 is the most important in the book. The principal ideas of the last three chapters are based exclusively on experimental facts accumulated at the time of writing. Personally, I have no doubts that in 2011 superconductivity will celebrate its $100^{th}$ jubilee having a transition temperature above 300 K. I hope that the present book will make a valuable contribution to this event.

A few words about the history of this book: writing my first book entitled "High-Temperature Superconductivity in Cuprates: The Nonlinear Mechanism and Tunneling Measurements," I *could* add to the existing 12 chapters of the book one chapter more. The title of this additional chapter would be similar to the title of this book. However, I have decided to drop the additional chapter and to realize this project by myself, even if it will take a few years to synthesize a room-temperature superconductor. Unfortunately, I did not have a possibility to start the project locally (still in Brussels). Then, I have proposed this project to a few laboratories: I have sent a few e-mail messages. To my surprise, I did not receive even one reply. This is **how scientific relationships function in our society**: not even a simple "thank you for your proposal." In addition to a certain "culture" of relations among scientists, the disbelief mentioned above, namely, that superconductivity cannot occur at room temperature, was definitely the second reason why I did not get a reply. "Well," I said to myself, "Then I have no choice—if I cannot realize this project by myself, I will write a new book about how to synthesize a room-temperature superconductor." This is it; the story is very short. In fact, some readers can even thank these "nice" people to whom I have sent the proposal; otherwise, this book would not exist.

I thank three professors of physics J. W. Turner, J. Wickens and D. Johnson, and the publisher V. Riecansky for correcting English.

ANDREI MOURACHKINE
*Cambridge/Brussels, August 2003*

Chapter 1

# INTRODUCTION

*What Nature created at the Big Bang—the spin of the electron—she later tried to "get rid" of in the living matter.*

—From Ref. [19]

## 1. What is the superconducting state?

The exact definition of the superconducting state will be given in the following chapter. Here we discuss a bird's-eye view of the superconducting state.

First, one question: would you be able to notice the difference in taste between two glasses of your preferred drink—soft or hard, whatever—in one of which a $10^{-4}$ part, i.e. 0.01%, is replaced by another drink? I do not think so. However, it is not the case for a superconductor (not literally, of course).

In some metals for example, the superconducting state occurs due to the presence of a $10^{-4}$ fraction of "abnormal" electrons, while the other 99.99% free (conduction) electrons remain absolutely *normal*. The correlated behavior of the small fraction of these "abnormal" electrons overwhelms the rest. Amazing, is it not? Due to the presence of these "abnormal" electrons, the metal is no longer a metal but a superconductor, losing its ability to resist to a small-magnitude electrical current. The presence of normal (conduction) electrons is completely masked by that of the "abnormal" electrons, as if the normal electrons were not existing at all. (Of course, we talk only about electron transport properties of a metal; the crystal structure of a metal is almost unchanged below the critical temperature, i.e. when a metal becomes superconducting.)

What is even more interesting is that Nature had no intention at all to create the superconducting state. Superconductivity is rather Nature's oversight—it is an *instability*, an *anomaly*. What does the superconducting state literally mean? In the superconducting state, THERE IS NO FRICTION. In the real world, what does it mean? If friction were absent, Earth would be ideally round, no buildings, no clothes, and I am afraid that the living matter, including us, would not exist at all. Definitely, it was not Nature's intention. Humans however, after the discovery of the superconducting state, try to derive a good deal of benefit from use of its peculiar properties.

Nevertheless, the superconducting state is a *state of matter*, even if it is an instability, and in this book we shall discuss its characteristic properties. As any state of matter, superconductivity is not a property of isolated atoms, but is a collective effect determined by the structure of the whole sample.

The superconducting state is a *quantum* state occurring on a macroscopic scale. In a sense, the superconducting state is a "bridge" between the microworld and the macroworld. This "bridge" allows us to study the physics of the microworld directly. This is one of the reasons why superconductivity, driven only by a $10^{-4}$ fraction of "abnormal" electrons, has attracted the attention of so many scientists since its discovery in 1911 (thus, more than 90 years of *intensive* research!). Between 1911 and 1957, many best minds tried to unravel the mystery of this state caused only by 0.01 % of conduction electrons.

How do normal electrons in a superconductor become "abnormal"? At the Big Bang, Nature has created two types of elementary particles: *bosons* and *fermions*. Bosons have an integral spin, while fermions a half-integral spin. As a consequence, bosons and fermions conform to different statistics. Electrons are fermions with a spin of 1/2 and obey the *Fermi-Dirac* statistics. In a superconductor, two electrons can form a pair which is already a boson with zero spin (or a spin equal to 1). These electron pairs conform to the *Bose-Einstein* statistics and, being in a phase, can move in a crystal without friction. This is how, in a classical superconductor, a tiny fraction 0.01 % of all conduction electrons becomes "abnormal." Simple, is it not?

## 2.    A brief historical introduction

The history of superconductivity as a phenomenon is very rich, consisting of many events and discoveries. Therefore, it is not possible to describe all of them in one section. There are a few books devoted to the history of superconductivity—the reader who is interested to know more on this issue, is referred to these books (see, for example, [1]). The goal of this introductory section is primarily to give some historical perspective to the evolution of the subject.

In most textbooks on superconductivity, the subject is presented chronologically. The presentation in this book does not follow this tradition: in this section we consider the most important events and discoveries, and in the subse-quent chapters, we discuss the physics of the superconducting state in different compounds without emphasizing the historical order.

## 2.1 Phenomenon of superconductivity: its discovery and evolution

The phenomenon of superconductivity was discovered in 1911 by the Dutch physicist H. Kamerlingh Onnes and his assistant Gilles Holst in Leiden. They found that *dc* resistivity of mercury suddenly drops to zero below 4.2 K, as shown in Fig. 1.1. Gilles Holst actually made this measurement [1]. However, his name has become lost in the recesses of history, as is often the case with junior researchers working under a famous scientist. A year later, Kamerlingh Onnes and Holst discovered that a sufficiently strong magnetic field restores the resistivity in the sample as does a sufficiently strong electric current.

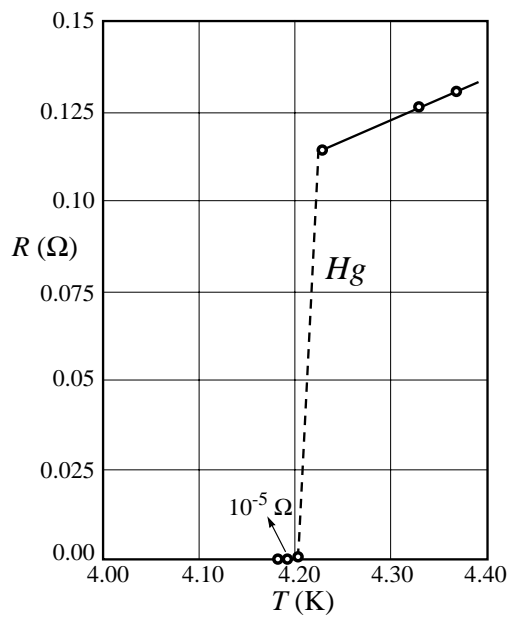

*Figure 1.1.* Experimental data obtained in mercury by Gilles Holst and H. Kamerlingh Onnes in 1911, showing for the first time the transition from the resistive state to the superconducting state.

In two years after the discovery of superconductivity in mercury, lead was found to superconduct at 7.2 K. In 1930, superconductivity was discovered in niobium, occurring at 9.2 K. This is the highest transition temperature among all elemental metals.

In 1933, W. Meissner and R. Ochsenfeld discovered in Berlin one of the most fundamental properties of superconductors: *perfect diamagnetism*. They found that the magnetic flux is expelled from the interior of the sample that is cooled below its critical temperature in weak external magnetic fields (see Fig. 2.2). Thus, they found that no applied magnetic field is allowed inside a metal when it becomes superconducting. This phenomenon is known today as the *Meissner effect*.

Dutch physicists C. J. Gorter and H. B. G. Casimir introduced in 1934 a phenomenological theory of superconductivity based on the assumption that, in the superconducting state, there are two components of the conducting electron "fluid": "normal" and "superconducting" (hence the name given this theory, the *two-fluid model*). The properties of "normal" electrons are identical to those of the electron system in a normal metal, and the "superconducting" electrons are responsible for the anomalous properties. In the superconducting state, these two components exist side by side as interpenetrating liquids. The two-fluid model proved a useful concept for analyzing, for example, the thermal and acoustic properties of superconductors.

Following the discovery of the expulsion of magnetic flux by a superconductor—the Meissner effect—the brothers F. and H. London together proposed in 1935 two equations to govern the microscopic (local) electric and magnetic fields. These two equations provided a description of the anomalous diamagnetism of superconductors in a weak external field. In the framework of the two-fluid model, the London equations, together with the Maxwell equations, describe the behavior of superconducting electrons, while the normal electrons behave according only to the Maxwell equations. The London equations explained not only the Meissner effect, but also provided an expression for the first characteristic length of superconductivity, namely what became known as the *London penetration depth* $\lambda_L$.

*Vortices* in superconductors were discovered by L. V. Shubnikov and co-workers in 1937. They found an unusual behavior for some superconductors in external magnetic fields. Actually, they discovered the existence of two critical magnetic fields for *type-II* superconductors and the new state of superconductors, known as the *mixed state* or the *Shubnikov phase*.

In 1950, H. Fröhlich proposed that vibrating atoms of a material must play an important role causing it to superconduct. He suggested that searching for an isotope effect in superconductors would establish whether or not lattice vibrations play some role in the interaction responsible for the onset of superconductivity. Following this proposal, the *isotope effect* was indeed found in the same year 1950 by E. Maxwell and C. A. Reynolds. The study of different superconducting isotopes of mercury established a relationship between the critical temperature $T_c$ and the isotope mass $M$: $T_c M^{1/2}$ = constant. Undoubtedly, this effect played the decisive role in showing the way to the correct theory of superconductivity.

Also in 1950, V. Ginzburg and L. Landau proposed an intuitive, phenomenological theory of superconductivity. The theory uses the general theory of the second-order phase transition, developed by L. Landau. The equations derived from the theory are highly non-trivial, and their validity was proven later on the basis of the microscopic theory. The *Ginzburg-Landau theory* played an important role in understanding the physics of the superconducting state. This theory is able to describe the behavior of superconductors (both conventional and unconventional) in strong magnetic fields. The Ginzburg-Landau theory provided the same expression for the penetration depth as the London equations and also an expression for the second characteristic length $\xi_{GL}$, called the *coherence length*.

By using the Ginzburg-Landau theory, A. A. Abrikosov theoretically found vortices and thus explained Shubnikov's experiments, suggesting that the Shubnikov phase is a state with vortices that actually form a periodic lattice. This result seemed so strange that he could not publish his work during five years; and even after 1957, when it was published, this idea was only accepted after experimental proof of several predicted effects.

In 1956 Leon Cooper showed that, in the presence of a very weak electron-phonon (lattice) interaction, two conducting electrons are capable of forming a stable paired state. After the discovery of the isotope effect, this was the second and the last breakthrough leading to the correct theory of superconductivity. This paired state is now referred to as the *Cooper pair*.

The first microscopic theory of superconductivity in metals was formulated by J. Bardeen, L. Cooper and R. Schrieffer in 1957, which is now known as the *BCS theory*. The central concept of the BCS theory is a weak electron-phonon interaction which leads to the appearance of an attractive potential between two electrons. As a consequence, they form the Cooper pairs. We shall discuss the BCS model in Chapter 5.

Quantum-mechanical tunneling of Cooper pairs through a thin insulating barrier (of the order of a few nanometers thick) between two superconductors was theoretically predicted by B. D. Josephson in 1962. After reading his paper, John Bardeen publicly dismissed young Josephson's tunneling-supercurrent assertion: "... pairing does not extend into the barrier, so that there can be no such superfluid flow." Josephson's predictions were confirmed within a year and the effects are known today as the *Josephson effects*. They play a special role in superconducting applications.

## 2.2 Era of high-temperature superconductivity

The issue of *room-temperature superconductivity* was for the first time *seriously* addressed in a paper written by W. A. Little in 1964 [2]. He proposed a model in which a high $T_c$ is obtained due to a non-phonon mediated mechanism of electron attraction, namely, an *exciton* model for Cooper-pair forma-tion in long organic molecules. Little's work revived the old dream of high $T_c$, and can be considered as the beginning of the search for high-temperature superconductivity.

Many new superconductors were discovered in the 1970s and 1980s. For example, the first representatives of two new classes of superconductors—*heavy fermions* and *organic superconductors*—were discovered in 1979. The experimental data obtained in organic superconductors and heavy fermions indicated that superconductivity in these compounds was unconventional. Before the discovery of superconductivity in *copper oxides* (*cuprates*), the highest critical temperature 23.2 K was observed in 1973 in $Nb_3Ge$. This type of superconductor is called an *A-15 compound*.

The *bisoliton* model of superconductivity was proposed in 1984 by L. S. Brizhik and A. S. Davydov [3] in order to explain superconductivity in quasi-one-dimensional organic conductors. In the framework of this model, the Cooper pairs are quasi-one-dimensional excitations coupled due to a moderately strong, nonlinear electron-phonon interaction.

In 1986, trying to explain the superconductivity in heavy fermions, K. Miyake, S. Schmitt-Rink and C. M. Varma considered the mechanism of superconductivity based on the exchange of antiferromagnetic spin fluctuations [4]. The calculations showed that the anisotropic even-parity pairings are assisted, and the odd-parity as well as the isotropic even-parity are impeded by antiferromagnetic spin fluctuations.

The real history of high-$T_c$ superconductivity began in 1986 when Bednorz and Müller found evidence for superconductivity at $\sim 30$ K in La-Ba-Cu-O ceramics [5]. This remarkable discovery has renewed the interest in superconductive research. In 1987, the groups at the Universities of Alabama and Houston under the direction of M. K. Wu and P. W. Chu, respectively, jointly announced the discovery of the 93 K superconductor Y-Ba-Cu-O. Just a year later—early in 1988—Bi- and Tl-based superconducting cuprates were discovered, having $T_c = 110$ and 125 K, respectively. Finally, Hg-based cuprates with the highest critical temperature $T_c = 135$ K were discovered in 1993 (at high pressure, $T_c$ increases up to 164 K). Figure 1.2 shows the superconducting critical temperature of several cuprates as a function of the year of discovery, as well as $T_c$ of some metallic superconductors. All these cuprates are hole-doped. One family of cuprates which was discovered in 1989 is electron-doped: $(Nd, Pr, Sm)$-Ce-Cu-O. Their maximum critical temperature is comparatively low, $T_{c,max} = 24$ K.

In 1986 the scientific world was astonished by the discovery of high-$T_c$ superconductivity in copper oxides because oxides are very bad conductors. The first reaction of most scientists working in the field of superconductivity was to think that there must be a new mechanism, since phonon-mediated superconductivity is impossible at so high a temperature. The discovery of super-conducting cuprates was followed by research growth at a rate unprecedented in the history of science: during 1987 the number of scientists working in the field of superconductivity increased, at least, by one order of magnitude. Data obtained in the cuprates, within a year after the discovery of their ability to superconduct, indeed showed that the characteristics of high-$T_c$ superconductors deviate from the predictions of the BCS theory as do those of organic superconductors and heavy fermions. For example, the BCS isotope effect is almost absent in optimally doped cuprates. As a consequence, this has prompted the exploration of non-phonon electronic coupling mechanisms. Ph. Anderson was probably the first to suggest a theoretical model which did not incorporate the phonon-electron interaction.

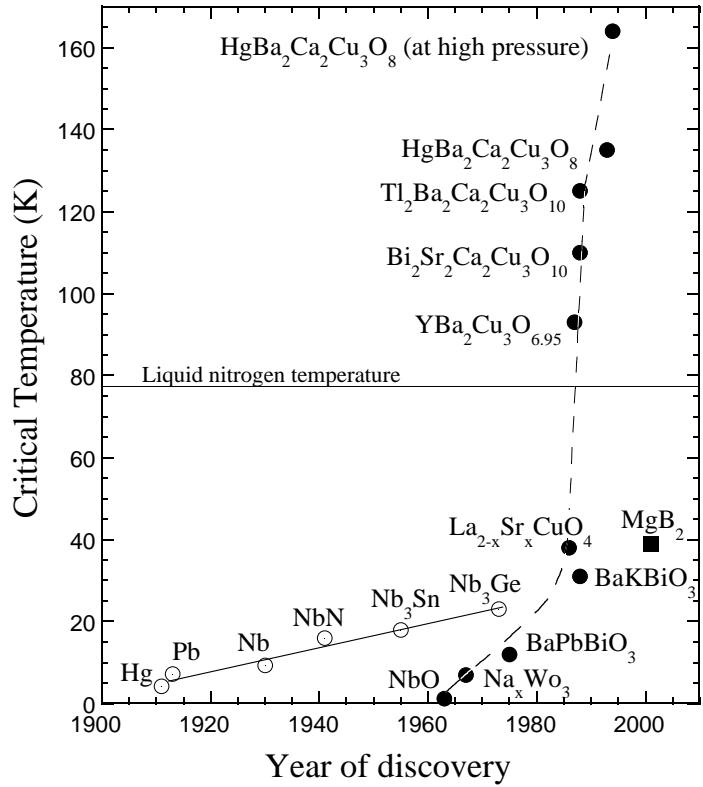

*Figure 1.2.* The time evolution of the superconducting critical temperature since the discovery of superconductivity in 1911. The solid line shows the $T_c$ evolution of metallic superconductors, and the dashed line marks the $T_c$ evolution of superconducting oxides.

The events presented below are important since they have led to our understanding of the mechanism of high-$T_c$ superconductivity. In 1987, L. P. Gor'kov and A. V. Sokol proposed the presence of two components of itiner-ant and more localized features in cuprates [6]. This kind of microscopic and dynamical phase separation was later rediscovered in other theoretical models. In 1988, A. S. Davydov suggested that high-$T_c$ superconductivity occurs due to the formation of bisolitons [7], as well as superconductivity in organic superconductors. The *pseudogap* above $T_c$ was observed for the first time in 1989 in nuclear magnetic resonance (NMR) measurements [8]. The pseudogap is a partial energy gap, a depletion of the density of states above the critical temperature.

In 1990, A. S. Davydov presented a theory of high-$T_c$ superconductivity based on the concept of a moderately strong electron-phonon coupling which results in perturbation theory being invalid [9, 10]. The theory utilizes the concept of *bisolitons*, or electron (or hole) pairs coupled in a singlet state due to local deformation of the -O-Cu-O-Cu- chain in the $CuO_2$ planes. We shall discuss the bisoliton model in Chapter 6. In the early 1990s, a few theorists autonomously proposed that, *independently* of the origin of the pairing mechanism, spin fluctuations mediate the long-range phase coherence in the cuprates. It turned out that this suggestion was correct. In 1994, A. S. Alexandrov and N. F. Mott pointed out that, in the cuprates, it is necessary to distinguish the "internal" wavefunction of a Cooper pair and the order parameter of a Bose-Einstein condensate, which may have different symmetries [11].

In 1995, V. J. Emery and S. A. Kivelson emphasized that superconductivity requires pairing and long-range phase coherence [12]. They demonstrated that, in the cuprates, the pairing may occur above $T_c$ without the onset of long-range phase coherence. In the same year 1995, J. M. Tranquada and co-workers found the presence of coupled, dynamical modulations of charges (holes) and spins in Nd-doped $La_{2-x}Sr_xCuO_4$ (LSCO) from neutron diffraction [13]. In LSCO, antiferromagnetic stripes of copper spins are separated by periodically spaced quasi-one-dimensional domain walls to which the holes segregate. The spin direction in antiferromagnetic domains rotates by $180°$ on crossing a domain wall. In 1997, V. J. Emery, S. A. Kivelson and O. Zachar presented a theoretical model of high-$T_c$ superconductivity based on the presence of charge stripes. It turned out that the model is not applicable to the cuprates (there is no charge-spin separation in the cuprates); however, it was the first model of high-$T_c$ superconductivity based on the presence of charge stripes in the $CuO_2$ planes.

In 1999, analysis of tunneling and neutron scattering measurements showed that, in $Bi_2Sr_2CaCu_2O_{8+x}$ (Bi2212) and $YBa_2Cu_3O_{6+x}$ (YBCO), the phase coherence is established due to spin fluctuations [14, 15]. We shall consider the mechanism of the onset of phase coherence in the cuprates in Chapter 6. In 2001, tunneling measurements provided evidence that the Cooper pairs in Bi2212 are pairs of quasi-one-dimensional solitonlike excitations [16–19]. We shall briefly discuss these data in Chapter 6.

## 2.3    History of the soliton

One may wonder how do *solitons* (or *solitary waves*) relate to the phenomenon of superconductivity? Simply because the Cooper pairs in superconducting cuprates and some other unconventional superconductors are pairs of soliton-like excitations, not electrons as in superconducting metals.

For a long time, linear equations have been used for describing different phenomena. However, the majority of real systems are *nonlinear*. For example, the fate of a wave travelling in a medium is determined by properties of the medium. *Nonlinearity* results in the distortion of the shape of large amplitude waves, for instance, in turbulence. The other source of distortion of a wave is the *dispersion*. Nonlinearity tends to make the hill of the wave steeper, while dispersion flattens it. The solitary wave lives "between" these two dangerous, destructive "forces." Thus, the *balance* between nonlinearity and dispersion is responsible for the existence of the solitary waves. As a consequence, the solitary waves are extremely robust.

The history of solitary waves or solitons is unique. The first *scientific* observation of the solitary wave was made by Russell in 1834 on the surface of water. One of the first mathematical equations describing solitary waves was formulated in 1895. But only in 1965 were solitary waves fully understood! Moreover, many phenomena which were well known before 1965 turned out to be solitons! Only after 1965 was it realized that solitary waves on the water surface, nerve pulses, vortices, tornados and many others belong to the same category: they are all solitons! That is not all, the most striking property of solitons is that they behave like particles!

In 1834 near Edinburgh (Scotland), John Scott Russell was observing a boat moving on a shallow channel and noticed that, when the boat suddenly stopped, the wave that it was pushing at its prow "rolled forward with great velocity, assuming the form of a large solitary elevation, a rounded, smooth and well defined heap of water which continued its course along the channel apparently without change of form or diminution of speed" [20]. He followed the wave along the channel for more than a mile.

In 1965 N. J. Zabusky and M. D. Kruskal performed computer simulations considering movements of a continuous nonlinear rubber string. They accounted for nonlinear forces by assuming that stretching the string by $\Delta\ell$ generates the force $k\Delta\ell + \alpha(\Delta\ell)^2$. The nonlinear correction to Hooke's law, $\alpha(\Delta\ell)^2$, was assumed to be small as compared to the linear force $k\Delta\ell$. After many attempts, they came to a striking conclusion: for small amplitudes, vibrations of the string are best described by a nonlinear equation formulated in 1895 by D. J. Korteweg and G. de Vries. It is only nowadays known that the Korteweg-de Vries equation describes a variety of nonlinear waves, and is suit-able for small amplitude waves in materials with weak dispersion. However, in 1965 it was a new finding.

In addition, Zabusky and Kruskal found that the solitary waves are not changed in collisions, like rigid bodies, and on passing through each other, two solitary waves accelerate. As a consequence, their trajectories deviate from straight lines, meaning that the solitons have particle-like properties. So they coined the term *soliton*, 131 years after its discovery.

The latest example of solitary waves is the so-called freak wave occurring in open ocean. The height of a freak wave can be as much as 30 meters. It suddenly appears from nowhere in weather conditions close to a storm. Only in 2001 it was scientifically shown that the freak wave is a solitary wave.

Mathematically, there is a difference between "solitons" and "solitary waves." *Solitons* are localized solutions of integrable equations, while *solitary waves* are localized solutions of non-integrable equations. Another characteristic feature of solitons is that they are solitary waves that are not deformed after collision with other solitons. Thus the variety of *solitary waves* is much wider than the variety of the "true" solitons. In fact, real systems do not carry exact soliton solutions in the strict mathematical sense (which implies an infinite life-time and an infinity of conservation laws) but *quasi*-solitons which have most of the features of true solitons. In particular, although they do not have an infinite life-time, quasi-solitons are generally so long-lived that their effect on the properties of the system is almost the same as that of true solitons. This is why physicists often use the word "soliton" in a relaxed way which does not agree with mathematical rigor.

## 3.    Room-temperature superconductivity

This issue is the main topic of this book and, in fact, an old dream. The dream of high-temperature superconductivity existed long before the development of the BCS theory. It had been expected that the future theory would not only explain the phenomenon of superconductivity, but also would show whether it is possible to create high-temperature superconductors and to predict the occurrence of superconductivity in different materials. The BCS theory, created in 1957, did explain the phenomenon of superconductivity in metals, however it did not provide a rule for predicting the occurrence of superconductivity in different compounds. In the framework of the phonon mechanism of superconductivity, the BCS formula showed that the *maximum* critical temperature $T_{c,max}$ should be approximately an order of magnitude less than the Debye temperature. Since the Debye temperature in many metals is around room temperature, this means that, in the framework of the BCS theory, superconductivity is a low-temperature phenomenon. Nevertheless, this estimation of maximum $T_c$ did not stop some dreamers to continue the search for high-temperature superconductors. The grand old man of superconductivity, Bernd Matthias, used to say "Never listen to theorists."

What is interesting is that this restriction imposed by the BCS theory on the maximum $T_c$ value has in its turn stimulated theorists to search for a new mechanism of superconductivity different from the phonon mechanism. As is mentioned above, Little proposed in 1964 the exciton model of superconductivity in long chainlike organic molecules [2]. In the framework of his model, the maximum critical temperature was estimated to be around 2200 K! Little's paper has encouraged the search for room-temperature superconductivity, especially in organic compounds. It is worth noting that in 1964 the idea of superconductivity in organic systems was not new. F. London already questioned in 1950 whether a superfluid-like state might occur in certain macromolecules which play an important role in biochemical reactions [21]. Such molecules usually have alternating single and double bonds, called *conjugate*, and contain molecular groups attached to certain carbons along the chain ("spine"). However, Little was the first to place the concept of high-temperature superconductivity in organic molecules on a serious theoretical footing.

Between 1964 and 1986, many new superconductors were discovered, including *organic* superconductors. However, none of them had a critical temperature going over the BCS limit, $\sim 30$ K. The increasing pessimism among experimentalists was crushed overnight in 1986 by Bednorz and Müller's discovery of superconductivity in cuprates. In 1993, a high-temperature superconductor having $T_c \simeq 135$ K became the reality. What about room-temperature superconductivity?

From the beginning, it is **important** to note that the issue of room-temperature superconductivity must be discussed without emotions. Everyone understands (otherwise see below) what technical marvels we can see if one day room-temperature superconductors become available. Therefore, it is sometimes very difficult to discuss this issue just as a physical phenomenon: the human brain in such situations tends not to function properly. However, we have to examine this question calmly. Only numbers obtained from estimations and experimental facts must be our guides to the "untouched" territory.

Secondly, it is necessary to note that the expression "a room-temperature superconductor" inherently contains an ambiguity. Some perceive this expression as a superconductor having a critical temperature $T_c \sim 300$ K, others as a superconductor functioning at 300 K. There is a huge difference between these two cases. From a technical point of view, superconductors only become useful when they are operated well below their critical temperature—one-half to two-third of that temperature provides a rule of thumb. Therefore, for the technologist, a room-temperature superconductor would be a substance whose resistance disappears somewhere above 450 K. Such a material could actually be used at room temperature for large-scale applications. At the same time, $T_c \sim 350$ K can already be useful for small-scale (low-power) applications. Consequently, unless specified, the expression "a room-temperature superconductor" will further be used to imply a superconductor having a critical temperature $T_c \simeq 350$ K. The case $T_c \simeq 450$ K will be discussed separately.

Consider the facts: a superconductor with $T_c = 135$ K is already available (since 1993). The first discovered superconductor—lead—has a critical temperature of 4.2 K. Taking into account that the ratio 135 K/4.2 K $\simeq 32$ is more than one order of magnitude larger than the ratio 350 K/135 K $\simeq 2.6$, one can conclude that the goal to have a room-temperature superconductor looks not only as a possibility but also a *near-future* possibility. In Fig. 1.2, if we assume that the rise of critical temperature will follow the same growth as that for copper oxides, then in 2010 we will have a room-temperature superconductor. This is one of the reasons why I believe that, in 2011, superconductivity will celebrate its 100-year anniversary having a critical temperature above 300 K. In Chapters 8 and 9, we shall see that, from the physics point of view, there is no formal limitation for superconductivity to occur above room temperature.

In the literature, one can find many papers (more than 20) reporting evidence of superconductivity near or above room temperature. Most researchers in superconductivity do not accept the validity of these results because they cannot be reproduced by others. Paul Chu, the discoverer of the 93 K superconductor Y-Ba-Cu-O (see above), calls these USOs—unidentified superconducting objects. The main problem with most of these results is that superconductivity is observed in samples containing many different conducting compounds, and the superconducting fraction (if such exists at all) of these samples is usually very small. Thus, it is possible that superconductivity *does* exist in these complex materials, but nobody knows what phase is responsible for its occurrence. In a few cases, however, the phase is known but superconductivity was observed exclusively on the surface. For any substance, the surface conditions differ from those inside the bulk, and the degree of this difference depends on many parameters, and some of them are extrinsic. In Chapter 8, we shall discuss the results of two works reporting superconductivity above room temperature.

In 1992, a diverse group of researchers gathered at a two-day workshop in Bodega Bay (California). They considered the issue of making much higher temperature superconductors. T. H. Geballe, who attended this workshop, summarized some guidelines in a two-page paper published in *Science* [22], that emerged from discussions:

- Materials should be multicomponent structures with more than two sites per unit cell, where one or more sites not involved in the conduction band can be used to introduce itinerant charge carriers.

- Compositions should be near the metal-insulator Mott transition.

- On the insulating side of the Mott transition, the localized states should have spin-1/2 ground states and antiferromagnetic ordering of the parent compound.

- The conduction band should be formed from antibonding tight-binding states that have a high degree of cation-anion hybridization near the Fermi level. There should be no extended metal-metal bonds.

- Structural features that are desirable include two-dimensional extended sheets or clusters with controllable linkage, or both.

All these hints are based on the working experience with cuprates. Personally, I have came across this paper when the main ideas of this book were already existing. *Basically*, these hints are correct but, however, not complete: I would add a few (remember that this paper was written in 1993). We shall discuss them in Chapter 10.

Finally, let us suppose that, one day, a room-temperature superconductor will be available, and suppose that in time, scientists and engineers figure out how to synthesize it in useful forms and build devices out of it. What technical marvels could we expect to see?

First of all, all devices made from the room-temperature superconductor will be reasonably cheap since its use would not involve cooling cost. The benefits would range from minor improvements in existing technology to revolutionary upheavals in the way we live our lives. Energy savings from many sources would add up to a reduced dependence on conventional power plants. Compact superconducting cables would replace unsightly power lines and revolutionize the electrical power industry. A world with room-temperature superconductivity would unquestionably be a cleaner world and a quieter world. Compact superconducting motors would replace many noisy, polluting engines. Advance transportation systems would lessen our demands on the automobile. Superconducting magnetic energy storage would become commonplace. Computers would be based on compact Josephson junctions. Thanks to the high-frequency, high-sensitivity operation of superconductive electronics, mobile phones would be so compact that could be made in the form of an earring. SQUID (Superconducting QUantum Interference Device) sensors would become ubiquitous in many areas of technology and medicine. Room-temperature superconductivity would undoubtedly trigger a revolution of scientific imagination. The effects of room-temperature superconductivity would be felt throughout society, including children who might well grow up playing with superconducting toys.

## 4.    Why the living matter is organic?

One may ask how does this question relate to the issue of room-temperature superconductivity? In fact, the superconducting state and *some* biological processes have, at least, one thing in common—they do not like the electron spin. The prologue to this chapter may look unusual, and to better understand its meaning here is a quote from the same book [19]: "From the physics point of view, the general understanding of many biological processes is still very limited. However, it is known that (i) in redox reactions occurring in living organisms, electrons are transferred from one molecule to another in pairs with opposite spins; and (ii) electron transport in the synthesis process of ATP (adenosine triphosphate) molecules in conjugate membranes of mitochondria and chloroplasts is realized by pairs, but not individually [9, 10]. Apparently, in living tissues, electron transfer is preferable in pairs in which two electrons are in a singlet state. ...

"At the Big Bang, spin was attached to what are now called the fermions in order to create diversity of possible forms of the existence of matter. Seemingly, in living tissues which appeared later, the spin of the electron became rather an obstacle in the evolution of the living matter. In many biological processes, in order to get rid of the electron spin, two electrons with opposite spins are coupled, forming a composite boson with $2e$ charge and zero spin. It happened that in inorganic solids two electrons, in some circumstances, can be paired too. This state of matter, which is in fact an instability in solids, is now called the superconducting state. Thus, the understanding of some biological processes can lead to better understanding of the phenomenon of superconductivity in solids. This is particularly true in the case of high-$T_c$ superconductivity. Superconductivity does not occur in living tissues because it requires not only the electron pairing but also the phase coherence among the pairs."

Nevertheless, the superconducting-like state exists *locally* in complex organic molecules with conjugate bonds [23]. Figure 1.3 shows a few examples of such molecules. Their main building blocks are carbon and hydrogen atoms. The characteristic feature of these conjugated hydrocarbons is the presence of a large number of $\pi$ electrons. These collectivized electrons are in the field of the so-called $\sigma$ electrons which are located close to the atomic nuclei and not much different from the ordinary atomic electrons. At the same time, the $\pi$ electrons are not localized near any particular atom, and they can travel throughout the entire molecular frame. This makes the molecule very similar to a metal. The framework of atoms plays the role of a crystal lattice, while the $\pi$ electrons that of the conduction electrons. As an example, Figure 1.4 schematically shows the formation of $\sigma$- and $\pi$-orbitals in ethene. It turns out, in fact, that the conjugated hydrocarbons with even number of carbon atoms are more than just similar to a metal, but are actually small superconductors [23]. Experimen-tally, conjugated hydrocarbons with even number of carbon atoms (thus, with even number of π electrons) exhibit properties similar to those of a superconductor: the Meissner-like effect, zero resistivity and the presence of an energy gap. The π electrons form bound pairs analogous to the Cooper pairs in an ordinary superconductor. The pair correlation mechanism is principally due to two effects: (i) the polarization of the σ electrons, and (ii) $\sigma - \pi$ virtual electron transitions. However, if the number of π electrons is odd, the properties of such conjugated hydrocarbons are different from those of a superconductor.

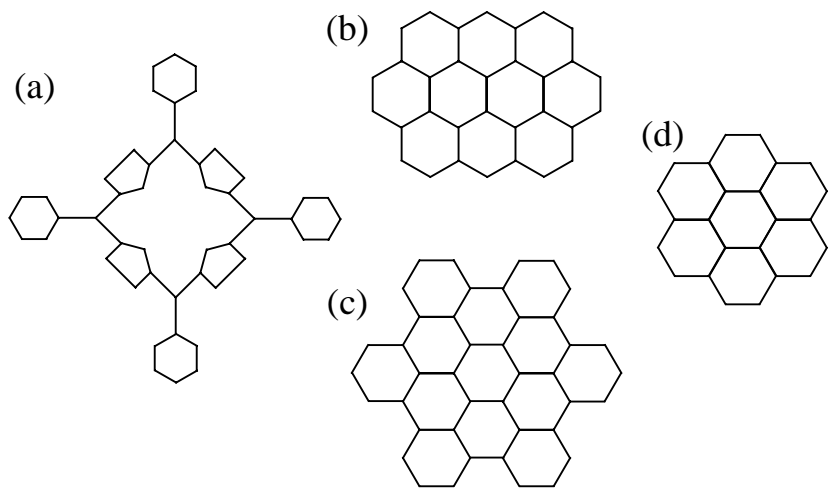

*Figure 1.3.* Organic molecules with delocalized π electrons: (a) tetraphenylporphin; (b) ovalene; (c) hexabenzocoronene, and (d) coronene [23].

Thus, "the essential fluidity of life agrees with the fluidity of the electronic cloud in conjugated molecules. Such systems may thus be considered as both the cradle and the main backbone of life" [24]. At the end, I would like to recall the prologue to this book: *In order to create the living matter, Nature*

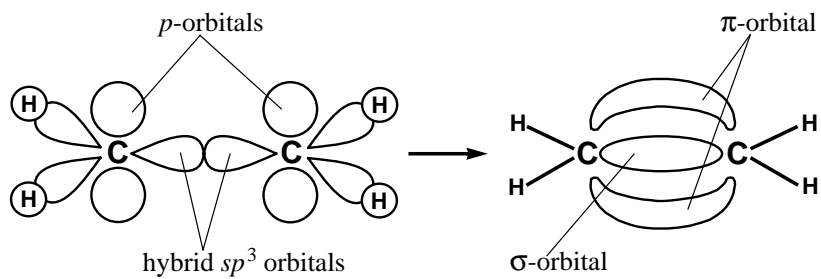

*Figure 1.4.* Formation of π- and σ-orbitals in ethene.

*needed billions of years. This experience is unique, and we must learn from it.*

Finally, aside from our problem of room-temperature superconductivity, why, indeed, is the living matter (including us) organic? Could we exist being made from, for example, B or N? The answer is no. Just by using common sense and the periodic table of chemical elements (the Mendeleev table) it is not difficult to show that the element which we could be made from can only be carbon. *Only C can perform this function.*

Chapter 2

# BASIC PROPERTIES OF THE
# SUPERCONDUCTING STATE

Any state of matter has its own specific characteristics and basic properties. For example, viscosity is a characteristic of a liquid, and a liquid takes the shape of a container which contains the liquid. The latter is one of the basic properties of a liquid. As remarked in Chapter 1, the superconducting state is a state of matter. Therefore, it has its own specific characteristics and basic properties, and we need to know them before we discuss room-temperature superconductivity. We also need to know why it occurs. What does cause superconductivity?

In earlier textbooks on the physics of superconductivity, the description of the superconducting state is based on the BCS theory, assuming that the BCS model is the only possible mechanism of superconductivity. In fact, as we shall see in the next chapter, all superconductors can be divided into three groups in accordance with the mechanism of superconductivity in each compound. Contrary to this old tradition, the purpose of this chapter is to characterize the superconducting state as whole, independently of any specific mechanism. Later, in Chapters 5, 6 and 7, we shall separately consider characteristic features of superconductivity in each group.

## 1.     What is the superconducting state?

Superconductivity was discovered by Kamerlingh Onnes and his assistant Gilles Holst in 1911: on measuring the electrical resistance of mercury at low temperatures, they found that, at 4.2 K, it dropped abruptly to zero (see Fig. 1.1). Subsequent investigations have shown that this sudden transition to perfect conductivity is characteristic of a number of metals and alloys. However, some metals never become superconducting. Bardeen, Cooper and Schrieffer reported in 1957 the first successful microscopic theory of superconductivity

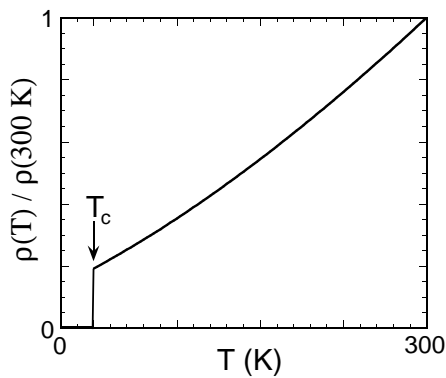

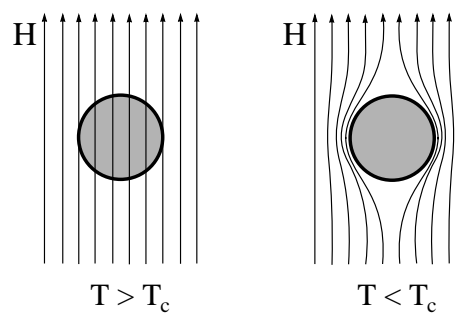

*Figure 2.1.* Temperature dependence of electrical resistivity of a superconductor. $T_c$ marks the transition to the superconducting state.

*Figure 2.2.* The Meissner effect: the expulsion of a weak, external magnetic field from the interior of a superconductor. The field is applied (a) at $T > T_c$, and (b) at $T < T_c$.

(BCS theory). Despite the existence of the BCS theory, there are no completely reliable rules for predicting whether a metal will superconduct at low temperature or not.

From a classical point of view, the superconducting state is characterized by two distinctive properties: *perfect electrical conductivity* ($\rho = 0$) and *perfect diamagnetism* ($\mathbf{B} = 0$ inside the superconductor), as shown in Figs 2.1 and 2.2, respectively. However, this definition of the superconducting state can soon be changed because, as recently found in an unconventional organic superconductor, the applied magnetic field *induces* superconductivity. Therefore, $B \neq 0$ inside this organic superconductor (see Section 4.10).

## 2.      Why does superconductivity occur?

At the Big Bang, Nature created two types of elementary particles: bosons and fermions. *Every elementary particle is either a boson or a fermion.* This is known as the quantum statistical postulate. Whether an elementary particle is a boson or fermion is related to the magnitude of its spin (in units of $\hbar$). Particles having an integer spin are bosons, while those with a half-integer spin are fermions. Electrons are fermions with a spin of 1/2. According to the *Pauli exclusion principle* no two electrons can occupy the same energy state. At the same time, bosons can occupy the same state multiply. This is the main difference between bosons and fermions. Thus, they conform to different quantum statistics. Fermions obey the Fermi-Dirac statistics: for a system of free fermions in equilibrium at temperature $T$, the probability of occupation of a level of energy $E$ is given by the *Fermi-Dirac distribution function*:

$$f_F(E) \equiv \frac{1}{\exp[(E - \mu)/k_B T] + 1}, \qquad (2.1)$$

where $\mu$ is the chemical potential (in metals, the chemical potential at low temperatures is very close to the Fermi level), and $k_B$ is the Boltzmann constant. Bosons obey the Bose-Einstein statistics: for a system of free bosons in equilibrium, the probability of occupation of a level of energy $E$ is given by the *Bose distribution function*:

$$f_B(E) \equiv \frac{1}{\exp[(E - \mu)/k_B T] - 1}. \qquad (2.2)$$

For fermions, Nature had however created a "loophole": under some circumstances, they can become bosons (but bosons can never be fermions). The moment fermions become bosons, they switch the statistics which they are obeying. As a consequence, the properties of the system are radically changed at this moment. This is exactly what happens in a superconductor at critical temperature: two electrons, *if there is a net attractive force acting between them*, form a pair which is already a boson with zero spin (or a spin equal to 1). These electron pairs *being in a phase* can move in a crystal without friction. In conventional superconductors, the pairing occurs in *momentum space* which is reciprocal to real space. In fact, it is not important whether the electron pairing occurs in momentum or, for example, real space. The most crucial circumstance for the electron pairing and, thus, for the onset of the superconducting state, is that the net force acting between two electrons must be attractive. In a superconductor, the electron pairs is usually called the *Cooper pairs*.

Are these electron pairs (composite bosons) really bosons? The answer is yes. For example, atomic nuclei are composed of protons and neutrons (which are fermions), and atoms are composed of nuclei and electrons. Thus, nuclei and atoms are composite objects. It has been experimentally demonstrated that they are indistinguishable quantum particles. Therefore, they are either bosons or fermions (this depends on the total number of elementary fermions in a composite particle). Two electrons, if there is an attraction between them, indeed, represent a boson. If the attractive force disappears, the two electrons will again behave as fermions do. As a matter of fact, all experimental and theoretical studies of superconductivity are in a first approximation reduced to finding the origin of this attractive force.

It is important to note that the phenomenon of fermion pairing gives rise not only to superconductivity, i.e. to the absence of electrical resistance in some solid conductors, but also to some other peculiar correlated states of matter. The latter ones can in a sense be considered as various manifestations of the superconducting state in Nature. For example, the fermion pairing gives rise to superfluidity. At 2.19 K, liquid $^4$He undergoes a superfluid transition. Below the transition temperature, liquid $^4$He exhibits frictionless (zero viscosity) flow remarkably similar to supercurrents in a superconductor. The $^4$He atoms consisting of 2 protons, 2 neutrons and 2 electrons, are composite bosons, and a finite fraction of them (about 7%) experience at 2.19 K (in fact, at 2.17 K) the Bose-Einstein condensation which we shall discuss in Chapter 4.

The fermion pairing gives also rise to the "superconducting" state in nuclei and neutron stars. The atomic nuclei are composed of protons and neutrons which have a spin of 1/2. If the total number of protons and neutrons in a nucleus is even, does the nucleus become superconducting? Yes and no. No, because, in a nucleus, there is no sense in discussing the absence of electrical resistivity—this concept has no meaning. Yes, because there are other indications of the "superconducting" state in nuclei having even number of protons and neutrons. For example, nuclei having *even* and *odd* number of protons and neutrons absorb radiation differently. In nuclei with even number of protons and neutrons, the fermions are paired. As a consequence, the energy of an incoming photon must be equal to or greater than the binding energy of a bound pair, otherwise, the radiation cannot be absorbed. Contrary to this, in nuclei having odd number of protons and neutrons, there is an unpaired fermion left over, which can absorb photons with much lower energy than that in the first case. Another indication of the fermion pairing in nuclei is provided by the fact that the measured nuclear moments of inertia are considerably smaller than the values calculated theoretically with the use of the noninteracting particle model. This effect is similar to that observed in superfluid helium. Thus, the paired fermions in a nucleus form a Bose condensate similar to that in liquid $^4$He. The development of the superfluid model of the atomic nucleus has predicted a large number of important results observed experimentally.

In neutron stars consisting almost entirely of neutrons, the neutron liquid is in a state analogous to that in an atomic nucleus. Thus, in neutrons stars, neutrons are paired. As we shall see below, superconductors have very low heat capacity. Due to this property, neutron stars cool very rapidly. Another indication of neutron pairing in neutron stars is the quantization of their angular momentum (every neutron star or *pulsar* rotates about its axis). This effect is similar to that in liquid helium. Generally speaking, the discreteness of any physical quantity is the fingerprint of the quantum world.

Undoubtedly, there are other manifestations of fermion pairing in Nature, which we are not yet aware of. The phenomenon of superconductivity is only one and, probably, the most spectacular exhibition of fermion pairing, occurring in some solids.

It is worth noting that in spite of the fact that this "loophole" for fermions was most likely created by Nature intentionally, one should however realize that the occurrence of the superconducting state on a *macroscale* is rather Na-ture's oversight, it is an anomaly (see the Introduction). The occurrence of the superconducting state on a macroscale requires *not only* the electron pairing *but also* the onset of long-range phase coherence. They are two different and independent phenomena.

## 2.1 What causes superconductivity?

Superconductivity is not a universal phenomenon. It shows up in materials in which the electron attraction overcomes the repulsion. What can cause the occurrence of this attractive force in solids? In all known cases at the moment of writing, *it is the interaction between electrons and the crystal lattice*. Thus, the *electron-phonon interaction* in solids is responsible for the electron attraction, leading to the electron pairing. It turns out that the electron-electron attraction provided by the lattice can overcome the electron-electron repulsion caused by the Coulomb force, so, the net force acting between them can be attractive.

## 3.    Characteristics of the superconducting state

Before we discuss the basic properties of the superconducting state, it is necessary, first, to know its specific characteristics. Such a sequence will simplify the understanding of this peculiar state of matter. For instance, in the aforementioned example, before studying a liquid, one must know what is the viscosity. Of course, some characteristics of the superconducting state are identical to the characteristics of the normal state. For example, the energy gap (see below) in a superconductor is tied to the *Fermi surface* which is the typical characteristic of a metal. In addition, a few characteristics such as the electron mass $m$, the electron charge $e$, the *Fermi velocity* $v_F$, the *electron mean free path* $\ell$ and so on, are simply indispensable for characterizing the superconducting state. Also, one should bear in mind that, in a superconductor at any $T > 0$, the vast majority of conduction electrons remain normal.

## 3.1    Critical temperature

The phase transition from normal into the superconducting state is a *second-order* transition, occurring at a temperature called the *critical temperature* $T_c$ shown in Fig. 2.2. The values of $T_c$ for some superconductors are given in Tables 2.1 and 2.2.

The superconducting state requires the electron pairing and the onset of long-range phase coherence, which in general occur at different temperatures. $T_c$ is the temperature controlled by the onset of long-range phase order.

For every superconducting material, the critical temperature is *exclusively* determined experimentally. At the moment of writing, there is no theoretical formula for predicting the value of critical temperature in a given compound.

*Table 2.1.*   Critical temperature $T_c$, the penetration depth $\lambda(0)$, the intrinsic coherence length $\xi_0$ and the critical magnetic field $H_c$ for some elemental superconductors

| Element | $T_c$ (K) | $\lambda(0)$ (Å) | $\xi_0$ (Å) | $H_c(T)$ |
|---------|-----------|------------------|-------------|----------|
| Al      | 1.1       | 500              | 16000       | 0.01     |
| Pb      | 7.2       | 390              | 830         | 0.08     |
| Sn      | 3.7       | 510              | 2300        | 0.03     |
| In      | 3.4       | 640              | 4400        | 0.03     |
| Tl      | 2.4       | 920              | -           | 0.02     |
| Cd      | 0.56      | 1300             | 7600        | 0.003    |

There is even no rule for predicting whether a certain substance will undergo the superconducting transition at low temperature or not. Actually, this is one of the main problems in the field of superconductivity—how to calculate the $T_c$ value in different materials. If we could know how to estimate the $T_c$ value for any specific material, there would be no need for this book. Stop reading and think this over for a while—this is an important point.

This book does not provide a formula for estimating the value of $T_c$ for any compound (this is in fact an impracticable task). Instead, this book presents an analysis of experimental facts, which is further used to show a way in achieving the goal, namely, $T_c \simeq 350$ K.

It is necessary to note that, in the framework of the BCS theory, there is in fact a formula for estimating the $T_c$ value (see Chapter 5); however, it is a *general* formula which does not take into account any specific features of a certain material.

Finally, one should remember that the critical temperature is a macroscopic quantity, while the Cooper-pair wavefunction and the order parameter are quantum ones.

## 3.2    Cooper-pair wavefunction

As discussed above, if in a solid, there is an attraction between two electrons, they become coupled, forming a composite boson. The electron pairing may occur in momentum or real space (see Chapter 4). Even if the lifetime of these paired electrons is very short, $\sim 10^{-12}$–$10^{-15}$ s, nevertheless, they live long enough, so that their effect on the properties of the system is almost the same as that of bosons with the infinite life-time.

In quantum mechanics, any particle is characterized by a *wavefunction*. So, a Cooper pair is also characterized by a wavefunction $\psi(\mathbf{r}_1 - \mathbf{r}_2)$, where $\mathbf{r}_1$ and $\mathbf{r}_2$ are the positions of each electron in real space, and the difference $\mathbf{r}_1 - \mathbf{r}_2$ is the relative coordinates. In Chapter 5, we shall see that in conventional superconductors, the Cooper-pair net spin is zero, $\mathbf{s}_1 + \mathbf{s}_2 = 0$, as well as the Cooper-pair net momentum, $\mathbf{k}_1 + \mathbf{k}_2 = 0$. Figure 2.3 schematically shows the Cooper-pair wavefunction. The wavefunction $\psi$ is a complex scalar having an amplitude and a phase. By definition, the probability to find a Cooper pair in real space is given by $\psi^*\psi$, where $\psi^*$ is the complex conjugate of $\psi$. The probability distribution of a Cooper pair in relative coordinates is schematically shown in Fig. 2.4.

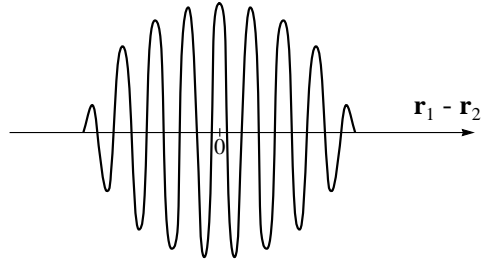

*Figure 2.3.* Schematic illustration of the Cooper-pair wavefunction in *conventional* superconductors. The diameter of a pair is around 100–1000 nm, and the wavelength is about 1 nm. So, the diameter of a pair is in fact equal to hundreds of wavelengths (this sketch shows just a few). The frequency of oscillations is of the order of $10^{15}$ Hz ($f = 2E_F/h$).

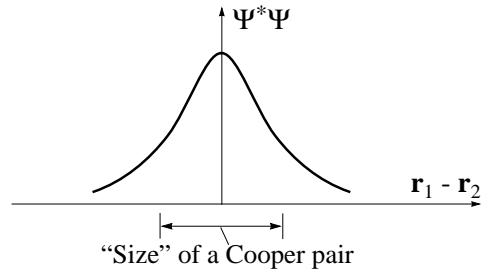

*Figure 2.4.* Schematic representation of the probability distribution of a Cooper pair in relative coordinates. The maximum probability is located between two electrons bound by a net attractive force.

## 3.3    Order parameter

The wavefunction of the superconducting condensate is called the *order parameter*. It is probably the most important parameter of the superconducting state. As mentioned in the Introduction, the superconducting state is a *quantum* state occurring on a macroscopic scale. This is the reason why the superconducting state is characterized by a single wavefunction $\Psi(\mathbf{r})$. Any wavefunction has an amplitude and a phase, therefore, it can be presented as

$$\Psi(\mathbf{r}) = |\Psi(\mathbf{r})|\, e^{i\theta(\mathbf{r})}, \tag{2.3}$$

where $\theta(\mathbf{r})$ is the phase. The order parameter has the following properties:

- It is a complex scalar which is continuous in real space.

- It is a single-valued function, that is, at any point, $\Psi^*(\mathbf{r})\Psi(\mathbf{r})$ can only have one value, where $\Psi^*(\mathbf{r})$ is the complex conjugate of $\Psi(\mathbf{r})$.

- In the absence of magnetic field, $\Psi \neq 0$ at $T < T_c$; and $\Psi = 0$ at $T \geq T_c$.

- $\Psi = 0$ outside a superconductor.

- The order parameter is usually normalized such that $|\Psi(\mathbf{r})|^2$ gives the number density of Cooper pairs at a point $\mathbf{r}$:

$$|\Psi(\mathbf{r})|^2 \equiv \Psi^*(\mathbf{r})\Psi(\mathbf{r}) = n_s/2, \qquad (2.4)$$

  where $n_s$ is the number of superconducting electrons and $n_s \equiv n - n_n$, where $n$ is the total number of free (conduction) electrons, and $n_n$ is the number of non-superconducting electrons. Then, in a conventional superconductor, $\Psi(\mathbf{r}) = (n_s/2)^{1/2}\mathrm{e}^{i\theta}$. Alternatively, the order parameter is sometimes normalized that $|\Psi(\mathbf{r})|^2 = n_s$, thus $|\Psi(\mathbf{r})|^2$ gives the number density of superconducting electrons.

- In momentum space, the variations of $|\Psi|$ are proportional to variations of the energy gap $\Delta$ (see below).

- In the absence of magnetic field, the phase is the same everywhere inside a superconductor at $T < T_c$, and $\theta(\mathbf{r}) = 0$ at $T \geq T_c$. In other words, below $T_c$ there is phase coherence in the whole sample.

- The phase is a periodic function in real space. Indeed, the addition of $2\pi\,n$ to $\theta(\mathbf{r})$, where $n = 0, \pm 1, \pm 2, \pm 3 \ldots$, does not change the function $\Psi(\mathbf{r}) = |\Psi(\mathbf{r})|\,\mathrm{e}^{i\theta(\mathbf{r})}$ because $\mathrm{e}^{2\pi\,i\,n} = 1$.

- Although absolute values of phase $\theta(\mathbf{r})$ cannot be measured, the gradient of the phase defines the supercurrent that flows between two superconducting regions (the *Josephson current*).

If the order parameter is known explicitly, then, *almost* complete information about the superconducting condensate is known too. As in quantum mechanics, any measurable value expected to be observed in the superconducting state can be obtained from the following expression

$$[\text{operator}]\,\Psi = (measured\ value)\,\Psi,$$

where [operator] is a quantum operator corresponding to a measurable quantity. This expression means that the measured value in quantum mechanics is the eigenvalue corresponding to the eigenfunction $\Psi$.

At the same time, knowledge of the order parameter does not provide automatically information about the attractive force that binds two electrons in a Cooper pair together. However, the symmetry of the order parameter gives a good hint. For example, if the order parameter has an s-wave symmetry, that is, $\Psi$ is positive (or negative) everywhere, it is very likely that the lattice is involved in the formation of Cooper pairs. If the order parameter has a p- or d-wave symmetry, that is, $\Psi$ has respectively two or four nodes where it changes sign, it is very likely that spin fluctuations mediate superconductivity.

Furthermore, knowledge of the order parameter does not provide automatically information about the $T_c$ value. In conventional superconductors, however, the $T_c$ value can be estimated from the maximum value of $|\Psi(\mathbf{r})|$ because $\Delta \propto |\Psi(\mathbf{r})|$. In unconventional superconductors, the situation is more complicated and, generally speaking, the ratio between $T_c$ and $\Delta$ is not fixed, i.e. depends on the material.

At a normal metal-superconductor interface, the order parameter does not change abruptly from a maximum value to zero. Instead, as we shall see below, it starts to diminish somewhat before the interface and even, going to zero, penetrates slightly into the normal metal. So, the order parameter never undergoes abrupt changes. This is a salient feature of the quantum world.

It is important to note that in conventional superconductors, that is, in most metallic superconductors, the order parameter can be considered as the wavefunction of a single Cooper pair. In unconventional superconductors, however, this is not the case. The order parameter of the superconducting condensate in unconventional superconductors does not coincide with the wavefunction of a single Cooper pair—they are different.

### 3.3.1 Symmetry of the order parameter

In conventional superconductors, each electron of a Cooper pair has opposite momentum and spin compared to the other: $\mathbf{k}_1 + \mathbf{k}_2 = 0$ and $\mathbf{s}_1 + \mathbf{s}_2 = 0$ (see Chapter 5). When the angular momentum of a pair is zero, $L = 0$, it is customary to say that the superconducting ground state has an s-wave symmetry (by analogy with the shape of atomic orbitals). When $L = 0$, the energy gap $\Delta$ has no nodes, and positive (negative) everywhere in momentum space. Since the momentum-space variations of $|\Psi|$ are proportional to variations of $\Delta$, it is also customary to say that the order parameter in conventional superconductors has an s-wave symmetry. This means that $|\Psi| \neq 0$ everywhere in real space. When $|\Psi|$ is constant, the s-wave symmetry of the order parameter is called *isotropic*. If $|\Psi|$ varies slightly in real space, the s-wave symmetry of the order parameter is called *anisotropic*.

In unconventional superconductors, the situation is slightly different. In most unconventional superconductors, each electron of a Cooper pair still has opposite momentum and spin compared to the other. However, the angular momentum of a pair is usually not zero. When $L = 2$, it is a custom to say that the superconducting ground state has a d-wave symmetry (by analogy with the shape of atomic orbitals). A key feature distinguishing a d-wave symmetry is that the energy gap has two positive and two negative lobes, and four nodes between the lobes. In this case, the order parameter also has a d-wave symmetry. It is necessary to underline that in *unconventional* superconductors, the symmetry of $|\Psi|$ coincides with the symmetry of **phase-coherence** energy gap $\Delta_c$ (unconventional superconductors have two energy gaps). The d-wave symmetry of the order parameter was first attributed to the superconducting ground state of heavy fermions just before the discovery of high-$T_c$ superconductors.

Theoretically, in some unconventional superconductors, electrons can be paired in a triplet state, so their spins are parallel, $\mathbf{s}_1 + \mathbf{s}_2 = 1$. In this case, it is customary to say that the order parameter has a p-wave symmetry (since the lowest value of the total angular moment is $L = 1$).

### 3.4    Penetration depth

The way in which a superconductor expels from its interior an applied magnetic field with the small magnitude (the Meissner effect) is by establishing a persistent supercurrent on its surface which exactly cancels the applied field inside the superconductor. This surface current flows in a very thin layer of thickness $\lambda$, which is called the *penetration depth*. The existence of a penetration depth was predicted by the London brothers (see the Introduction) and it was later confirmed by experiments.

Consider the two London equations to govern the *microscopic* electric and magnetic fields

$$\mathbf{E} = \frac{\mathrm{d}}{\mathrm{d}t}(\Lambda \mathbf{j}_s) \quad \text{and} \tag{2.5}$$

$$\mathbf{h} = -c\,\mathrm{curl}\,(\Lambda \mathbf{j}_s), \tag{2.6}$$

$$\text{where} \quad \Lambda = \frac{m^2}{n_s e^2} = \frac{4\pi\lambda^2}{c^2} \tag{2.7}$$

is a phenomenological parameter, and $\mathbf{j}_s$ is the supercurrent These two equations are derived in the framework of the two-fluid model which assumes that all free electrons are divided into two groups: superconducting and normal. The number density of superconducting electrons is $n_s$, and the number density of normal electrons is $n_n$. So, the total number of free (conduction) electrons is $n = n_s + n_n$. As the temperature increases from 0 to $T_c$, $n_s$ decreases from $n$ to 0. In addition, the number density $n_s$ is assumed to be the same everywhere, i.e. spatial variations of $n_s$ are disregarded.

The first London equation is simply Newton's second law for the superconducting electrons. It follows from this equation that in the stationary state, that is, when $\mathrm{d}\mathbf{j}_s/\mathrm{d}t = 0$, there is no electrical field inside the superconductor. In the second equation, **h** denotes the value of the flux density *locally* (in the first equation, we do not use **e** as a local value of **E** in the same way in order to avoid constant confusion with the charge $e$ of the electron). The second London equation, when combined with the Maxwell equation, curl $\mathbf{h} = 4\pi\mathbf{j}/c$, leads to

$$\nabla^2\mathbf{h} = \frac{\mathbf{h}}{\lambda^2}. \tag{2.8}$$

This implies that a magnetic field is exponentially screened from the interior of a sample with penetration depth $\lambda$, as shown in Fig. 2.5. This length clarifies the physical significance of the quantity $\lambda$ formally defined by Eq. (2.7), and is called the *London magnetic-field penetration depth*:

$$\lambda_L = \left(\frac{m^*c^2}{4\pi n_s e^2}\right)^{1/2}, \tag{2.9}$$

where $m^*$ is the effective mass of the charge carriers; $e$ is the electron charge, and $c$ is the speed of light in vacuum.

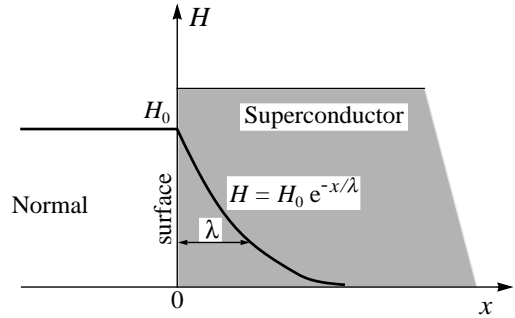

*Figure 2.5.* Penetration of the magnetic field into a superconducting sample. $\lambda$ is the penetration depth.

Equation (2.8) actually describes the Meissner effect. In a one-dimensional case, the solution of Eq. (2.8) is

$$h(x) = H_0\,\mathrm{e}^{-x/\lambda_L}, \tag{2.10}$$

where $H_0$ is the magnitude of magnetic field outside the superconductor, applied parallel to the surface. In Fig. 2.5, one can see that the external field actually penetrates the superconductor within $\lambda$.

It is important to underline that the magnitude of the penetration depth is directly related to the superfluid density $n_s$. As a consequence, it depends on temperature, since $n_s$ is temperature-dependent. In conventional superconductors, a good approximation for the temperature dependence of $\lambda$ is given by the

empirical formula

$$\lambda(T) = \frac{\lambda(0)}{[1 - (T/T_c)^4]^{1/2}}. \qquad (2.11)$$

This dependence $\lambda(T)$ is shown in Fig. 2.6. Let us estimate $\lambda(0)$. In a metal at $T = 0$, all conduction electrons are superconducting; then $n_s = n \approx 10^{22}$ cm$^{-3}$. Substituting this value into Eq. (2.9), together with $m \approx 9 \times 10^{-28}$ g, $c \approx 3 \times 10^{10}$ cm/s and $e = 4.8 \times 10^{-10}$ esu, we obtain that $\lambda_L \sim 530$ Å. It is worth noting that, in a metal, this length is considerably longer than the interatomic distance which is of the order of several Å. The values of $\lambda(0)$ for some metallic superconductors are listed in Table 2.1 and, for some unconventional superconductors, in Table 2.2.

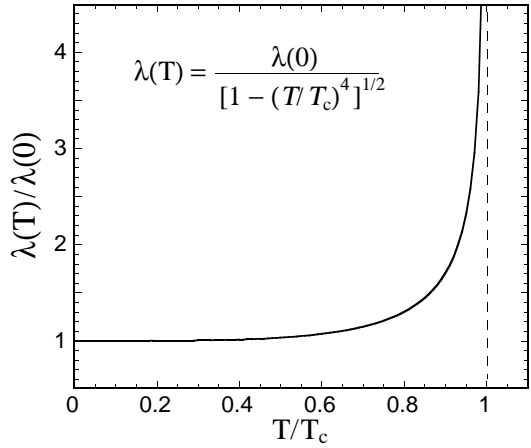

*Figure 2.6.*    Temperature dependence of penetration depth, $\lambda(T)$, given by Eq. (2.11).

Everything said so far about the electrodynamics of superconductors falls into the category of the so-called *local electrodynamics*. It means that the current at some point is given by the magnetic field at the same point. Therefore, strictly speaking, Equation (2.6) is applicable only if the size of the current carriers is much smaller than the characteristic length over which the magnetic field changes, that is, smaller than the penetration depth $\lambda_L$. We know that the superconducting current carriers are pairs of electrons. Let us denote the size of a Cooper pair by $\xi$ (see the following subsection). Then, the electrodynamics is local if $\xi \ll \lambda_L$. In pure metals, $\xi \sim 10^4$ Å and $\lambda_L \sim 10^2$–$10^3$ Å. Therefore, the local London electrodynamics is not applicable to pure metals because the magnetic field changes appreciably over the length $\xi$.

If the magnetic penetration depth is much smaller than the size of Cooper pairs, i.e. $\lambda_L \ll \xi$, the electrodynamics is *non-local*. In this case, the current at some point is given by the magnetic field at a different point. Such a non-local relation was considered for the first time by Pippard several years before the BCS theory of superconductivity appeared. Pippard calculated the magnetic penetration depth and found that the penetration distance of an applied magnetic field is in fact larger than $\lambda_L$. The zero-temperature value of the penetration depth in the case of non-local electrodynamics can be estimated from the following expression

$$\lambda_P \approx (\lambda_L^2 \xi)^{1/3}, \tag{2.12}$$

where the latter "P" denotes the name of Pippard. Indeed, if $\lambda_L \ll \xi$, then $\lambda_L \ll \lambda_P$. Of course, it is also assumed that $\lambda_P \ll \xi$, which is not always the case even for pure metals. For example, pure Al is described by non-local relations. At the same time, Pb, even of high purity, is a London superconductor. It is important to note that, on heating, when the temperature approaches $T_c$, all superconductors become local, i.e. London superconductors, because $\lambda(T)$ diverges at $T \to T_c$ while $\xi$ is independent of temperature.

Another important point which must be taken into account is how the mean free path of electrons $\ell$ is related to the Cooper-pair size $\xi$. Everything said so far applies to pure metals, that is, those characterized by a mean free path $\ell \gg \xi$. This case is also known as the *clean limit*. If a metal contains a large number of impurities, the mean electron free path can become smaller than the Cooper-pair size, i.e. $\ell \ll \xi$. This case is called the *dirty limit*. Alloys also fall into this category. For example, in Al, $\ell \simeq 1300$ Å and $\xi \simeq 16\,000$ Å; and $\ell \simeq 290$ Å and $\xi \simeq 380$ Å in Nb. In very dirty metals, the role of the coherence length (see below) is played by the mean electron free path $\ell$. In the framework of the microscopic theory of superconductivity (the BCS theory), the estimation of the magnetic-field penetration depth for "dirty" superconductors ($\ell \ll \xi$) is given by

$$\lambda_d \approx \lambda_L (\xi/\ell)^{1/2}. \tag{2.13}$$

Thus, $\lambda_d \gg \lambda_L$ always if $\ell \ll \xi$.

The value of the penetration depth can experimentally be obtained by different techniques such as microwave, infrared, muon-Spin-Rotation (Relaxation), ac-susceptibility, inductance measurements etc. In microwave measurements, for example, the penetration depth and its temperature dependence are inferred from the value and temperature dependence of surface reactance $X_s$ (the imaginary part of surface impedance). The relation between $X_s(T)$ and $\lambda(T)$ is given by

$$X_s(T) = \omega \mu_0 \lambda(T), \tag{2.14}$$

where $\mu_0$ is the permeability of the free space, and $\omega = 2\pi f$ is the microwave frequency.

*Table 2.2.* Critical temperature $T_c$, the penetration depth $\lambda(0)$, the Cooper-pair size $\xi(0)$ and the upper critical magnetic field $H_{c2}$ for type-II superconductors (for layered compounds, the in-plane values are given)

| Superconductor | $T_c$ (K) | $\lambda(0)$ (Å) | $\xi(0)$ (Å) | $H_{c2}$ (T) |
|---|---|---|---|---|
| Nb | 9.2 | 450 | 380 | 0.2 |
| NbTi | 9.5 | 1600 | 50 | 14 |
| NbN | 16 | 2000 | 50 | 16 |
| $Nb_3Sn$ | 18.4 | 800 | 35 | 24 |
| $Nb_3Ge$ | 23 | - | 35 | 38 |
| $Ba_{0.6}K_{0.4}BiO_3$ | 31 | 2200 | 35 | 32 |
| $MgB_2$ | 39 | 850 | 37 | 39 |
| $UPt_3$ | 0.5 | 7800 | 200 | 2.8 |
| $UBe_{13}$ | 0.9 | 3600 | 170 | 8 |
| $URu_2Si_2$ | 1.2 | - | 130 | 8 |
| $CeIrIn_5$ | 0.4 | 5300 | 250 | 1.0 |
| $CeCoIn_5$ | 2.3 | - | 80 | 11.9 |
| $TmNi_2B_2C$ | 11 | 800 | 150 | 10 |
| $LuNi_2B_2C$ | 16 | 760 | 70 | 7 |
| $K_3C_{60}$ | 19.5 | $\sim$4800 | 35 | $\sim$30 |
| $Rb_3C_{60}$ | 30 | $\sim$4200 | 30 | $\sim$55 |
| $YBa_2Cu_3O_7$ | 93 | 1450 | 13 | 150 |
| $HgBa_2Ca_2Cu_3O_{10}$ | 135 | 1770 | 13 | 190 |

## 3.5    Coherence length and the Cooper-pair size

In the framework of the Ginzburg-Landau theory (see below), the *coherence length* $\xi_{GL}$ is the characteristic scale over which variations of the order parameter $\Psi$ occur, for example, in a spatially-varying magnetic field or near a superconductor-normal metal boundary (see Figs. 2.7 and 2.10).

In many textbooks, one can find that the distance between two electrons in a Cooper pair (the Cooper-pair size), $\xi$, is also called the coherence length. However, in general, such a definition is incorrect: $\xi_{GL} \neq \xi$. Why? As we already know, superconductivity requires the electron pairing and the onset of long-range phase coherence. These two physical phenomena are different and independent of one another. The coherence length $\xi_{GL}$ defines variations of the order parameter of the superconducting condensate, whilst the pair size $\xi$ is related to the wavefunction of a Cooper pair (see Fig. 2.4). Thus, in general, the coherence length and the Cooper-pair size do not relate directly to one another. Secondly, the coherence length *depends* on temperature, $\xi_{GL}(T)$, while the Cooper-pair size is *temperature-independent*. The coherence length diverges at $T \to T_c$.

However, in conventional superconductors at zero temperature, $\xi_{GL}(0) = \xi$ because in conventional superconductors, the phase coherence is mediated by the overlap of the Cooper-pair wavefunctions—the process which does not give rise to a "new" order parameter. Instead, it simply "magnifies" the Cooper-pair wavefunctions to the level of the order parameter. In other words, in conventional superconductors, all the Cooper-pair wavefunctions below $T_c$ are *in phase*. Therefore, the electron pairing and the onset of phase coherence in conventional superconductors occur simultaneously at $T_c$. The overlap of Cooper-pair wavefunctions is also called the *Josephson coupling*. Thus, in conventional superconductors, the values of coherence length and Cooper-pair size coincide at $T = 0$. However, at $0 < T < T_c$, the value of the coherence length in "clean" conventional superconductors is always larger than the average size of Cooper pairs, $\xi < \xi_{GL}(T)$.

In unconventional superconductors, the long-range phase coherence is not mediated by the Josephson coupling; the phase-coherence mechanism is different (see Chapter 6). Therefore, in all unconventional superconductors, the order parameter has no relation with the Cooper-pair wavefunctions. As a consequence, in unconventional superconductors $\xi_{GL} \neq \xi$. Nevertheless, the values of $\xi_{GL}$ and $\xi$ in unconventional superconductors are of the same order of magnitude at $T \ll T_c$. In most unconventional superconductors, the electron pairing occurs above $T_c$, and the onset of long-range phase coherence appears at $T_c$.

In the framework of the BCS theory for *conventional* superconductors (see Chapter 5), the coherence length $\xi_0$ determined by the energy gap at zero temperature, $\Delta(T = 0)$ (see below), is called *intrinsic*:

$$\xi_0 = \frac{\hbar v_F}{\pi \Delta(0)}, \qquad (2.15)$$

where $v_F$ is the Fermi velocity (on the Fermi surface), and $\hbar = h/2\pi$ is the Planck constant. $\xi_0$ is also called the Pippard coherence length. Furthermore, in conventional superconductors, the values of the coherence length and Cooper-pair size coincide at $T = 0$, $\xi_0$ is also called the distance between electrons in a Cooper pair. Let us estimate $\xi_0$. In a metal superconductor, $\Delta(0) \sim 1$ meV. Substituting this value into Eq. (2.15), together with $v_F \approx 1.5 \times 10^8$ cm/s and $\hbar = h/2\pi \simeq 6.5 \times 10^{-13}$ meV s, we obtain $\xi_0 \simeq 3 \times 10^{-5}$ cm $= 3 \times 10^3$ Å.

In the framework of the Ginzburg-Landau theory, the temperature dependence of coherence length in "clean" superconductors ($\ell \gg \xi_0$) at temperatures close to $T_c$ is given by

$$\xi_{GL}^c(T) = 0.74 \, \xi_0 \, \left(1 - \frac{T}{T_c}\right)^{-1/2}. \qquad (2.16)$$

From this expression, one can see that the coherence length always exceeds the Cooper-pair size. For "dirty" superconductors ($\ell \ll \xi_0$), the Ginzburg-Landau temperature dependence of coherence length at temperatures close to $T_c$ is

$$\xi_{GL}^d(T) = 0.85 \, (\xi_0 \ell)^{1/2} \left(1 - \frac{T}{T_c}\right)^{-1/2}. \qquad (2.17)$$

From Eqs. (2.16) and (2.17), one can see that $\xi_{GL}^{c,d} \to \infty$ as $T \to T_c$. Such a temperature dependence is similar to that of $\lambda(T)$, shown in Fig. 2.6.

In the case of non-local electrodynamics, Pippard suggested an empirical relation for the coherence length

$$\frac{1}{\xi_P} = \frac{1}{\xi_0} + \frac{1}{\ell}. \qquad (2.18)$$

It follows from this expression that the Pippard coherence length $\xi_P$ is always smaller than $\xi_0$, and in very dirty metals ($\ell \ll \xi_0$), the role of the coherence length is played by the mean electron free path $\ell$.

In conventional superconductors, the intrinsic coherence length can be extremely large, $\sim 1000$ Å (see Table 2.1). In spite of the fact that two electrons in a Cooper pair in metallic superconductors are far apart from each other, the other Cooper pairs are only a few ten Å away (the period of a crystal lattice is several Å). In most unconventional superconductors, the values of coherence length and pair size at low temperature are very small: in cuprates, for example, $\xi$ is only a few periods of the crystal lattice (see Table 2.2).

## 3.6    Type-I and type-II superconductors

The ratio of the two characteristic lengths, defined above, is called the Ginzburg-Landau parameter $k$:

$$k = \frac{\lambda}{\xi_{GL}}. \qquad (2.19)$$

It is an important parameter that characterizes the superconducting material. Close to $T_c$, this dimensionless ratio is approximately independent of temperature, and allows one to distinguish between *type-I* and *type-II* superconductors. For example in Al, $\lambda = 500$ Å and $\xi_0 = 16\,000$ Å (see Table 2.1). Thus, in many conventional superconductors, $k \ll 1$.

As defined by Abrikosov, a superconductor is of type-I if $k < 1/\sqrt{2}$. If $k > 1/\sqrt{2}$, a superconductor is of type-II. Thus, the majority of metallic superconductors is of type-I. At the same time, in unconventional superconductors $k \gg 1$ (see Table 2.2). So, they are type-II superconductors. The main difference between these two types of superconductors is that they can show entirely different responses to an external magnetic field (the Meissner effect). While type-I superconductors expel magnetic flux completely from their interior, type-II superconductors do it completely only at small magnetic field magnitudes, but partially in higher external fields. The reason is that the surface energy of the interface between a normal and a superconducting region is positive for type-I superconductors and negative for those of type-II (see below).

As defined above, $\lambda$ measures the depth of penetration of the external magnetic field (see Fig. 2.5), and $\xi_{GL}$ is the characteristic scale over which variations of the order parameter $\Psi$ occur, for example, near a superconductor-normal metal interface. To visualize the difference between type-I and type-II superconductors, consider the two limiting cases: $k \ll 1$ and $k \gg 1$. Figure 2.7 illustrates these two cases. In Fig. 2.7a, $\lambda \ll \xi_{GL}$, and in Fig. 2.7b, $\lambda \gg \xi_{GL}$.

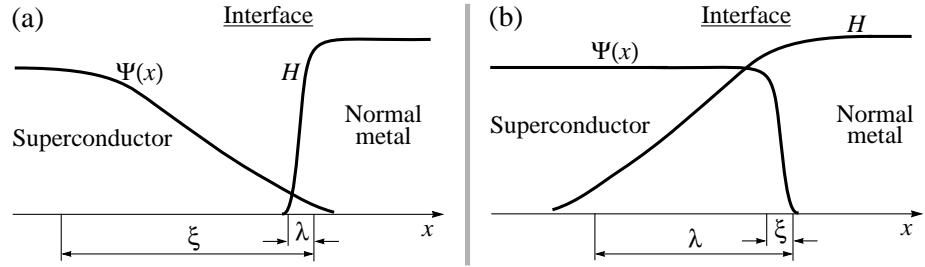

*Figure 2.7.* Spatial variations of the order parameter $\Psi$ and the magnetic field $H$ in the vicinity of a superconductor-normal metal interface for (a) $k \ll 1$ and (b) $k \gg 1$.

## 3.7 Critical magnetic fields

With the exception of Nb and V, all superconducting elements and some of their alloys are type-I superconductors. We already know that the superconducting state can be destroyed by a sufficiently strong magnetic field. The variation of the *thermodynamic critical field* $H_c$ with temperature for a type-I superconductor is approximately parabolic:

$$H_c(T) \simeq H_c(0)[1 - (T/T_c)^2],\tag{2.20}$$

where $H_c(0)$ is the value of the critical field at absolute zero. The dependence $H_c(T)$ is schematically shown in Fig. 2.8. For a type-II superconductor, there are two critical fields, the lower critical field $H_{c1}$ and the upper critical field $H_{c2}$, as shown in Fig. 2.9. In applied fields less than $H_{c1}$, the superconductor completely expels the field, just as a type-I superconductor does below $H_c$. At fields just above $H_{c1}$, flux, however, begins to penetrate the superconductor in microscopic filaments called *vortices* which form a regular (triangular) lattice. Each vortex consists of a normal core in which the magnetic field is large, surrounded by a superconducting region, and can be approximated by a long cylinder with its axis parallel to the external magnetic field. Inside the cylinder, the superconducting order parameter $\Psi$ is zero.

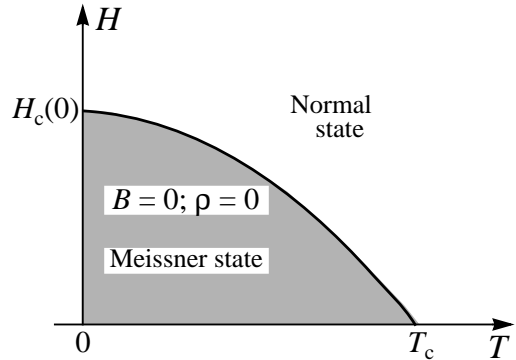

*Figure 2.8.* $H_c(T)$ dependence for a type-I superconductor, shown schematically.

The radius of the cylinder is of the order of the coherence length $\xi_{GL}$. The supercurrent circulates around the vortex within an area of radius $\sim \lambda$, the penetration depth (see Fig. 2.25). The spatial variations of the magnetic field and the order parameter inside and outside an isolated vortex are illustrated in Fig. 2.10. The vortex state of a superconductor, discovered experimentally by Shubnikov and theoretically by Abrikosov, is known as the *mixed state*. It exists for applied fields between $H_{c1}$ and $H_{c2}$. At $H_{c2}$, the superconductor becomes normal, and the field penetrates completely. Depending on the geometry of a superconducting sample and the direction of an applied field, the surface sheath of the superconductor may persist to even higher critical field $H_{c3}$, which is approximately $1.7H_{c2}$.

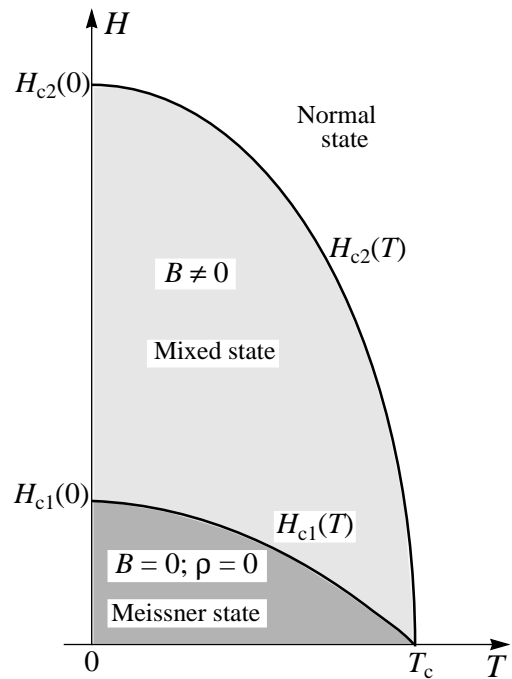

*Figure 2.9.* $H_{c1}(T)$ and $H_{c2}(T)$ dependences for a type-II superconductor, shown schematically.

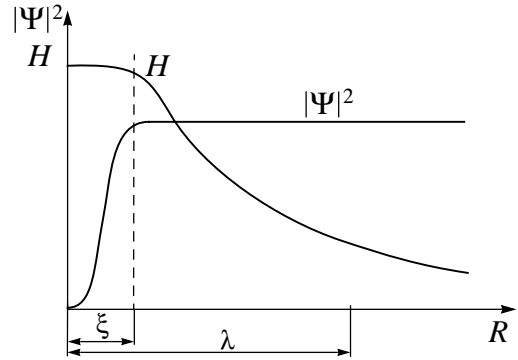

*Figure 2.10.* The spatial variations of the magnetic field $H$ and the order parameter $\Psi$ inside and outside an isolated vortex in an infinite superconductor. $R$ is the distance from the center of the vortex, and $\xi_{GL}$ and $\lambda$ are the coherence length and the penetration depth of the superconductor, respectively (in type-II superconductors, $\xi_{GL} < \lambda$).

As we shall see further below, the Ginzburg-Landau theory predicts that

$$H_c(T)\lambda(T)\xi_{GL}(T) = \frac{\hbar}{2\sqrt{2}e\mu_0} = \frac{\Phi_0}{2\sqrt{2}\pi\mu_0}, \qquad (2.21)$$

where $\quad \Phi_0 \equiv \dfrac{h}{2e} = 2.0679 \times 10^{-15} \quad \mathrm{T\,m^2}$ (or Weber) $\qquad (2.22)$

is the *magnetic flux quantum*. In the framework of the Ginzburg-Landau theory, $H_{c2} = \sqrt{2}kH_c$, where $k = \lambda/\xi_{GL}$ is the Ginzburg-Landau parameter. Then, substituting this expression into Eq. (2.21), we obtain

$$\Phi_0 = 2\pi\xi_{GL}^2 H_{c2}. \qquad (2.23)$$

This important relation is often used to obtain the values of the coherence length in type-II superconductors.

The magnitudes of the upper critical magnetic field of conventional type-II superconductors is very small, less than 1 T. However, in unconventional superconductors, these values can be extremely large (see Table 2.2). For example, in the Chevrel phase PbMo$_6$S$_8$, $H_{c2} = 60$ T. In three-layer high-$T_c$ superconductors, the critical magnetic field parallel to the $c$-axis can be $H_{c2}(0) \sim 50$ T and parallel to the $ab$-planes $H_{c2}(0) \sim 250$ T.

## 3.8    Critical current

The superconducting state can be destroyed not only by a magnetic field but by a *dc* electrical current as well. The *critical current* $J_c$ is the maximum current that a superconductor can support. Above $J_c$, the dc current breaks the Cooper pairs, and thus, destroys the superconducting state. In other words, $J_c$ is the minimum *pair-breaking* current. Thus, any superconductor is characterized by a critical *dc* current density $j_c$ (current divided by the cross-sectional area through which it flows).

The temperature dependence $J_c(T)$ (or $j_c(T)$) is similar to that of $H_c(T)$, shown in Fig. 2.8. At $T = 0$, the critical current density can be estimated by using the electron velocity on the Fermi surface, $v_F = \pi \Delta \, \xi_0 / \hbar$, the superfluid density $n_s$ given by Eq. (2.9), and the critical (maximum) velocity of a Cooper pair, $v_c \simeq \Delta / m v_F$, as

$$j_c = n_s e v_c \simeq n_s e \frac{\Delta}{m v_F} = \frac{\hbar c^2}{16 \pi e} \frac{1}{\lambda_L^2 \xi_0}, \qquad (2.24)$$

where $m$ is the electron mass. To estimate $j_c$, we take $\lambda_L \sim 10^3$ Å and $\xi_0 \sim 10^3$ Å. Substituting these values into Eq. (2.24), together with $c \simeq 3 \times 10^{10}$ cm/s, $\hbar = h/2\pi \simeq 10^{-27}$ erg s and $e = 4.8 \times 10^{-10}$ esu, we obtain $j_c \sim 4 \times 10^{16}$ CGS units. In Si units, it is equivalent to $j_s \sim 10^7$ A cm$^{-2}$.

The critical current density $j_c$ in Eq. (2.24) can be expressed in terms of critical magnetic field $H_c(0)$. Using the expressions for the condensation energy (see below) $\frac{H_c^2(0)}{8\pi} = \frac{1}{2} N(0) \Delta^2(0)$, where $N(0) = k_F m / (\pi \hbar)^2$ is the density of states near the Fermi surface, the electronic density $n_s(0) = k_F^3 / 3\pi^2$, and $\xi_0 = \hbar v_F / (\pi \Delta(0))$, we obtain in CGS units

$$j_s \simeq \frac{1}{4\pi\sqrt{3}} \frac{c \, H_c(0)}{\lambda_L(0)}. \qquad (2.25)$$

The Ginzburg-Landau and BCS theories give the same relation for $j_c$ with somewhat different numerical prefactors.

From Eq. (2.25), the maximum current density that can theoretically be sustained in a superconductor, is of the order of $H_c/\lambda_L$ (in SI units). Let us estimate $j_c$. Using $B_c \sim 0.1$ T, $\lambda_L \sim 10^3$ Å, and $\mu_0 = 4\pi \; 10^{-7}$ H/m, we obtain $j_c \approx 5 \times 10^6$ A cm$^{-2}$.

## 3.9    Energy scales

The superconducting state is characterized by a few energy scales. We already considered one energy scale given by the critical temperature, $k_B T_c$. The superconducting state is also characterized by an pairing energy gap, phase-coherence gap, phase stiffness and condensation energy. Let us consider the meaning of these energy scales.

### 3.9.1 Pairing energy gap

*The pairing energy gap* $2\Delta_p$ measures the strength of the binding of electrons (quasiparticles) into the Cooper pairs. In other words, the value of this gap corresponds to the binding energy that holds the electrons together. The magnitude of pairing energy gap is *temperature-dependent*.

As discussed above, the superconducting state requires the electron pairing *and* the onset of long-range phase coherence. They are two independent phenomena and, generally speaking, occur at different temperatures, $T_{pair}$ and $T_c$, respectively, and $T_c \leq T_{pair}$. In conventional superconductors, however, $T_{pair} = T_c$. At the same time, in most unconventional superconductors, $T_c < T_{pair}$. In a superconductor, the value of the *phase stiffness* (relative to $k_B T_c$) determines whether the electron pairing and the onset of long-range phase-coherence occur simultaneously or not.

The pairing gap is directly related to the $k_B T_{pair}$ energy scale, thus, $2\Delta_p \propto k_B T_{pair}$. At the same time, the magnitude of the *phase-coherence gap* is proportional to $k_B T_c$, i.e. $2\Delta_c \propto k_B T_c$. In general, the coefficients of proportionality in this two expressions are different and, as determined experimentally, varies between 3.2 and 6, depending on the case (in one heavy fermion, $\simeq 9$). The energy $2\Delta_p$ measures the strength of the binding of two electrons (quasiparticles) into a Cooper pair. At the same time, the energy $2\Delta_c$ is the condensation energy of a Cooper pair due to onset of phase coherence with other pairs. We shall discuss the phase-coherence gap in the following subsection.

Historically, conventional superconductors are the most studied. Furthermore, the physics of conventional superconductors is simpler than that of unconventional superconductors, because conventional superconductors have only one energy gap. Thus, let us discuss for the rest of this subsection the energy gap exclusively in conventional superconductors. The reason why the binding energy of two electrons is called the energy gap is because, when a metal undergoes a transition into the superconducting state, a small energy gap appears in the band at the Fermi level. As a result, the electronic system is unable to absorb arbitrary small amounts of energy.

The energy gap in a superconductor is quite different in its origin from that in a semiconductor. From the band theory, energy bands are a consequence of the static lattice structure. In a superconductor, the energy gap is far smaller, and results from an attractive force between electrons in the lattice which plays only an indirect role. In a superconductor, the gap occurs on either side of the Fermi level, as shown in Fig. 2.11. If, in a semiconductor, the energy gap is tied to the Brillouin zone (see Chapter 5), in a superconductor, the energy gap is carried by the Fermi surface. At $T = 0$ all electrons are accommodated in states below the energy gap, and a minimum energy $2\Delta(0)$ must be supplied to produce an excitation across the gap.

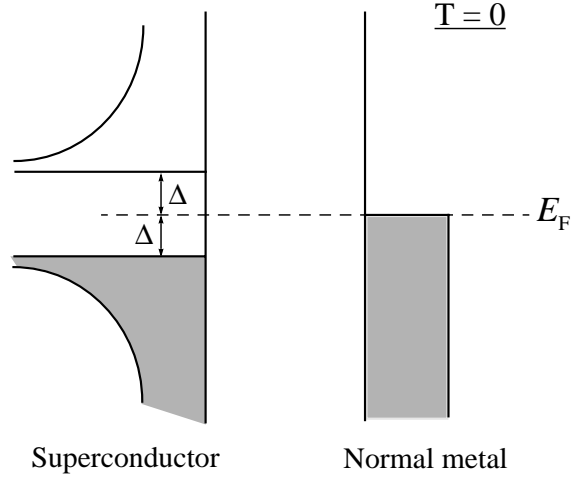

*Figure 2.11.* The density of states near the Fermi level $E_F$ in a superconductor, showing the energy gap $2\Delta$ at $T = 0$, and in a normal metal. All the states above the gap are assumed empty and those below, full.

As discussed above, the phase coherence in conventional superconductors is mediated by the overlap of the Cooper-pair wavefunctions—the process which does not give rise to a "new" order parameter. Instead, it simply "magnifies" the Cooper-pair wavefunctions to the level of the order parameter. Therefore, the electron pairing and the phase coherence in conventional superconductors occur simultaneously at $T_c$, so $T_{pair} = T_c$. As a consequence, the energy gap in the elementary excitation spectrum of conventional superconductors is exclusively determined by the pairing energy gap, $\Delta = \Delta_p$.

The BCS theory, developed for conventional superconductors, predicts that $2\Delta(0) = 3.52 k_B T_c$. Experimentally, the ratio $\frac{2\Delta}{k_B T_c}$ in conventional superconductors varies between 3.2 and 4.2. Figure 2.12 shows the temperature dependence of the energy gap in the framework of the BCS theory.

Since in unconventional superconductors, $\Delta_p$ and $k_B T_c$ do not relate with one another, the ratio $\frac{2\Delta_p}{k_B T_c}$ determined experimentally in unconventional superconductors, is usually larger than 4, and can be as large as 30.

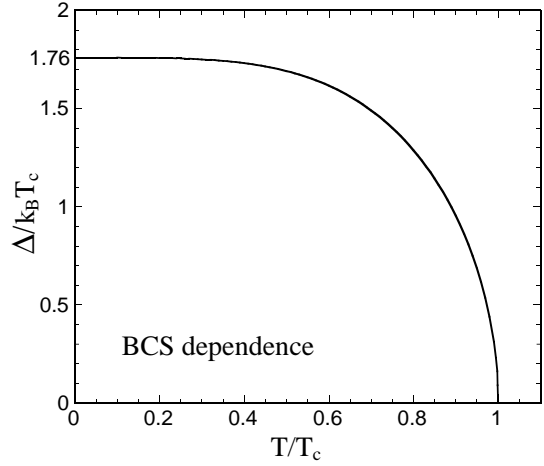

*Figure 2.12.* The BCS temperature dependence of the energy gap $\Delta(T)$.

### 3.9.2 Phase-coherence energy gap

The *phase-coherence energy gap* $2\Delta_c$ is the condensation energy of a Cooper pair when the long-range phase coherence appears. The condensation of Cooper pairs is *similar* to a Bose-Einstein condensation which occurs in momentum space (see Chapter 4). The magnitude of a phase-coherence gap is *temperature-dependent*, and the temperature dependence of $\Delta_c$ is similar to that in Fig. 2.12.

Generally speaking, the magnitudes of pairing and phase-coherence gaps depend on orientation: $\Delta_p(\mathbf{k})$ and $\Delta_c(\mathbf{k})$, where $\mathbf{k}$ is the vector in momentum space, and $\Delta_c$ and $\Delta_p$ have different symmetries. In many unconventional superconductors, the pairing and phase-coherence gaps are highly anisotropic, and often have nodes.

As was mentioned above, the variations of $|\Psi|$ in momentum space are *always* proportional to variations of **phase-coherence** gap $\Delta_c$, and not to those of $\Delta_p$. In conventional superconductors, however, the symmetry of order parameter coincides with the symmetry of the *pairing* gap $\Delta_p$, because conventional superconductors have only one energy gap: $\Delta = \Delta_p$.

### 3.9.3 Phase stiffness

The *phase stiffness* $\Omega_{ph}$ is the energy scale measuring the ability of the superconducting state to carry supercurrent. The magnitude of phase stiffness is mainly determined by zero-temperature superfluid density $n_s(0)$ and zero-temperature coherence length $\xi_{GL}(0)$. Thus, the phase stiffness is an energy scale defined at $T = 0$.

The phase stiffness is given by the following expression

$$\Omega_{ph} = \frac{A\, k_B\, \hbar^2 n_s(0)\, \xi_{GL}(0)}{4m^*} = \frac{A\, k_B (\hbar\, c)^2 \xi_{GL}(0)}{16e^2 \lambda_L^2(0)}, \qquad (2.26)$$

where $m^*$ is the effective mass of charge carriers; $\lambda_L(0)$ is the zero-temperature London penetration depth given by Eq. (2.9); $c$ is the speed of light; and $A$ is a dimensionless number of the order of 1 which depends on the details of the short distance physics [12]. For layered compounds, $\xi_{GL}(0) \rightarrow \xi_{GL,\perp}(0)$ in the above formula, where $\xi_{GL,\perp}(0)$ is the zero-temperature coherence length perpendicular to the layers.

In order to determine the importance of phase fluctuations, it is necessary to compare the values of the phase stiffness and energy scale $k_B T_c$. If $k_B T_c \ll \Omega_{ph}$, phase fluctuations are relatively unimportant. Then, the electron pairing and the onset of long-range phase coherence occur simultaneously at $T_c$. If $k_B T_c \approx \Omega_{ph}$, phase fluctuations are important. In this case, the electron pairing will most likely occur above $T_c$.

For example, the ratio $\Omega_{ph}/k_B T_c$ calculated for some conventional superconductors lies between $2 \times 10^2$ and $2 \times 10^5$ [12]. This means that the superfluid density in conventional superconductors is relatively high, and phase fluctuations in metal superconductors are practically absent. As a consequence, the pairing and the onset of long-range phase coherence in low-$T_c$ superconductors occur simultaneously at $T_c$. However, phase fluctuations play an important role in unconventional superconductors. The ratio $\Omega_{ph}/k_B T_c$ calculated for some superconductors with low superfluid density and small coherence length, such as organic and high-$T_c$ superconductors, is small and lies between 0.7 and 16 [12]. Thus, the pairing may occur well above $T_c$ which is controlled by the onset of a long-range phase order.

### 3.9.4    Condensation energy

The superconducting state is a more ordered state than the normal one. Therefore, the superconducting state is preferable to the normal state from the standpoint of free energy. However, the superconducting state can be destroyed by a critical magnetic field $H_c$ which is related thermodynamically to the free-energy difference between the normal and superconducting states in zero field. This difference is the *condensation energy* of the superconducting state. Thus, the thermodynamic critical field $H_c$ is determined by equating the energy $H_c^2/(8\pi)$ per unit volume (in CGS units), associated with holding the field out against the magnetic pressure, with the condensation energy:

$$F_n(T) - F_s(T) = \frac{H_c^2(T)}{8\pi}, \qquad (2.27)$$

where $F_n$ and $F_s$ are the *Helmholtz free energies* per unit volume in the respective phases in zero field. We shall discuss the condensation energy further below while considering the thermodynamic properties of superconductors.

Let us now estimate the condensation energy of a superconductor. As discussed above, the condensation energy of a single Cooper pair is $2\Delta_c$. Then, $n_s/2 \times 2\Delta_c$ is approximately the total condensation energy per unit volume, where $n_s$ is the density of Cooper pairs in a superconductor. The fraction of the electronic states directly involved in pairing approximately equals $\Delta_p(0)/E_F$, where $E_F$ is the Fermi energy. Recalling that a conventional superconductor has only one energy gap $\Delta = \Delta_p$, the condensation energy of a conventional superconductor is of the order of $\Delta^2(0)/E_F$.

In a conventional superconductor, $\Delta(0) \sim 0.5$–$1$ meV and $E_F \sim 5$–$10$ eV. Then, $\Delta(0)/E_F \approx 10^{-4}$, and the condensation energy is small as $\Delta^2(0)/E_F \approx 10^{-7}$–$10^{-8}$ eV per atom.

## 4. Basic properties of the superconducting state

We already considered the most important characteristics of the superconducting state; we know why the superconducting state occurs, and what causes superconductivity in solids. Now we are going to discuss basic properties of the superconducting state. The mechanisms of superconductivity occurring in different materials will be discussed in Chapters 5, 6 and 7. The Ginzburg-Landau theory will be considered at the end of this chapter.

The superconducting state, as any state of matter, has its own basic properties, so any superconductor, independently of the mechanism of superconductivity and the material, will exhibit these properties. Hence, a room-temperature superconductor will exhibit them too. The main basic properties of the superconducting state are the following: zero resistance, the Meissner effect, the magnetic flux quantization, the Josephson effects, the appearance of an energy gap in elementary excitation energy spectrum, and the proximity effect. Every superconducting transition is marked by a jump in specific heat. And lastly, in the mixed state, the behavior of type-II superconductors has the same pattern.

### 4.1 Zero resistance

*Every superconductor has zero resistivity*, i.e. infinite conductivity, for a small-amplitude *dc* current at any temperature below $T_c$. Is the resistivity of a superconductor really zero? Yes, its resistivity is zero as far as it can be measured. This property of the superconducting state was demonstrated by inducing a small-amplitude *dc* current around a closed ring of a conventional superconductor. The experiment continued over two and a half years—there was no measurable decay of the current. This means that the resistivity of a superconductor is smaller than $10^{-23}$ $\Omega$ m. This value is 18 orders of magni-tude smaller than the resistivity of copper at room temperature. Such a value of resistivity in a superconductor implies that the current lifetime in a superconducting ring in zero magnetic field is not less than $10^5$ years.

This intrinsic property of the superconducting state is probably the most fascinating one, and is widely used in different types of practical applications—from microchips to power lines.

It is worth to recall that the resistivity of a superconductor to an *ac* current is not zero. The *ac* current flows on the surface of a superconductor within a thin layer of thickness on the order of $\lambda_L$.

## 4.2    The Meissner effect

From a classical point of view, *every superconductor exhibits perfect diamagnetism*, i.e. $\mathbf{B} = 0$ inside the superconductor, as shown in Fig. 2.2b. In fact, as we already know, the magnetic field penetrates into the superconductor within a very thin surface layer having the thickness of the order of $\lambda_L$. To cancel $\mathbf{B}$, a superconductor creates a *dc* current on the surface, which gives rise to a magnetization $\mathbf{M}$, so that in the interior of the superconductor $4\pi\mathbf{M} + \mathbf{H} = 0$. Since the resistivity of the superconductor is zero, this surface current does not dissipate energy.

If the magnetic field was applied to a superconductor at $T > T_c$, and it is then cooled down to $T < T_c$, in this case, the field will remain inside the superconductor until it will be warmed up again through $T_c$. This "frozen" magnetic field will remain inside the superconductor independently of the presence of the external magnetic field.

Probably, the most spectacular demonstration of the Meissner effect is the *levitation effect*. A small magnet above $T_c$ simply rests on the surface of a superconductor having dimensions larger than those of the magnet. If the temperature is lowered below $T_c$, the magnet will float above the superconductor. The gravitational force exerted on the magnet is compensated by the magnetic pressure occurring due to supercurrent circulation on the surface of the superconductor.

## 4.3    Flux quantization

The quantum nature of the superconducting state manifests itself in *quantization of magnetic flux*. One of the characteristics of the quantum world is the quantization of a number of physical quantities, such as energy, spin, momentum etc. So, they can take on only a discrete set of values. Since the superconducting state is the quantum state occurring on a macroscopic scale, some physical quantities characterizing the superconducting state are quantized too.

Consider a bulk superconductor having a hole, as schematically shown in Fig. 2.13. Assume that the magnetic field $H_0$ was applied to the superconduc-tor at $T > T_c$, parallel to the hole walls. Then the superconductor was cooled down to $T < T_c$. In the non-superconducting hole, some magnetic flux will remain "frozen", produced by the supercurrent generated at the internal surface of the hole. Recalling that the order parameter of a conventional superconductor can be written in the form $\Psi(\mathbf{r}) = (n_s/2)^{1/2} e^{i\theta(\mathbf{r})}$, where $n_s$ is the density of superconducting electrons, and $\theta$ is the phase. In the superconductor with a hole, the order parameter has to go through an integral number of oscillations around the hole. The integral number of oscillations of $\Psi$ explains why magnetic flux inside the hole is quantized. Let us find the value of this "frozen" magnetic flux.

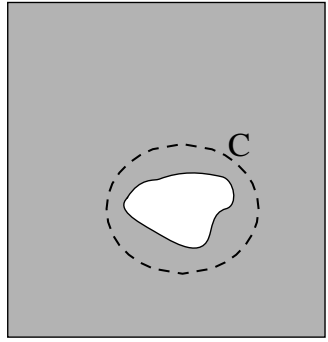

*Figure 2.13.* Superconductor with a hole. The contour of integration C goes around the hole through the interior of the superconductor.

First, we need to know the expression for a current-density vector operator in quantum mechanics. In classical mechanics, Hamilton's equations are expressed in terms of the canonical variables, $p_i, x_i$. In the absence of magnetic field and charge particles, the quantity $\mathbf{p}$ is the same as the ordinary or kinematic momentum, $\mathbf{p} = m\mathbf{v}$. However, when a particle carries charge $q$ and moves in a magnetic field $\mathbf{H}$ associated with the vector potential $\mathbf{A}$, the canonical and kinematic momenta are different, and related by

$$\mathbf{p} = m\mathbf{v} + q\mathbf{A}, \quad \text{or} \quad m\mathbf{v} = \mathbf{p} - q\mathbf{A}. \tag{2.28}$$

In electrodynamics, the vector potential $\mathbf{A}$ is defined as

$$\mathbf{H} = \nabla \times \mathbf{A} = \operatorname{curl} \mathbf{A}, \quad \text{with} \quad \nabla \cdot \mathbf{A} = \operatorname{grad} \mathbf{A} = 0. \tag{2.29}$$

In quantum mechanics, the relation between the canonical and kinematic momenta is maintained, but the canonical momentum is replaced by an operator: $\mathbf{p} \longrightarrow -i\hbar\nabla$. In quantum mechanics, the flow of particles characterized by a wavefunction (order parameter) $\Psi$ is described by a current density vector

$$\mathbf{J} = \frac{1}{2m}[(-i\hbar\nabla\Psi)^*\Psi + \Psi^*(-i\hbar\nabla\Psi)]. \tag{2.30}$$

If the particles are charged and moving in a vector potential $\mathbf{A}$, then the current density vector takes the following form

$$\mathbf{J} = \frac{1}{2m}\{[(-i\hbar\nabla - q\mathbf{A})\Psi]^*\Psi + \Psi^*(-i\hbar\nabla - q\mathbf{A})\Psi\}. \qquad (2.31)$$

To obtain the *electrical* current vector, $\mathbf{J}$ in the latter expression must be multiplied by a charge $q$.

We are now in a position find the value of the "frozen" magnetic flux. Substituting $\Psi(\mathbf{r}) = (n_s/2)^{1/2}e^{i\theta}$ into Eq. (2.31), and taking into account that each Cooper pair has a mass of $2m$ and a charge of $2e$, we obtain the expression for the supercurrent density

$$\mathbf{j}_s = \frac{1}{c\Lambda}\left(\frac{\Phi_0}{2\pi}\nabla\theta - \mathbf{A}\right), \qquad (2.32)$$

called the *generalized second London equation*. In this expression, $\Lambda$ is given by Eq. (2.7), and $\Phi_0 = \pi\hbar c/e$ (in CGS units) is the flux quantum in Eq. (2.22).

Consider the contour $C$ inside the superconductor, as shown in Fig. 2.13, enclosing the hole so that the distance between the contour and the internal surface of the hole is everywhere in excess of $\lambda_L$. Then at any point of the contour, the supercurrent is zero, $\mathbf{j}_s = 0$, and the path integral of supercurrent along the contour reduces to

$$\frac{\Phi_0}{2\pi}\oint_C \nabla\theta \cdot \mathrm{d}\mathbf{l} = \oint_C \mathbf{A} \cdot \mathrm{d}\mathbf{l}. \qquad (2.33)$$

Taking into account that in Eq. (2.33), the latter integral corresponds to the total flux through the contour $C$, i.e. $\oint_C \mathbf{A} \cdot \mathrm{d}\mathbf{l} = \Phi$, we have

$$\Phi = \frac{\Phi_0}{2\pi}\oint_C \nabla\theta \cdot \mathrm{d}\mathbf{l}. \qquad (2.34)$$

Since the order parameter $\Psi$ is single-valued, the change in $\theta$ after a full circle around the hole containing the magnetic flux must be an integral multiple of $2\pi$, i.e. $2\pi\,n$ ($n = 1, 2, 3, \ldots$), because the addition of $2\pi n$ to $\theta$ does not change the exponent: $e^{\theta+2\pi in} = e^{\theta}$. Therefore,

$$\Phi = \frac{\Phi_0}{2\pi}\cdot 2\pi n = n\,\Phi_0. \qquad (2.35)$$

Thus, the "frozen" magnetic flux through the contour $C$ is always an integral number of the flux quantum $\Phi_0$. It also follows from Eq. (2.35) that the minimum possible value of magnetic flux is $\Phi_0$.

If the magnetic flux enclosed in the hole is quantized, then the current circulating around the hole cannot be of an arbitrary magnitude, and cannot change continuously—it is also quantized.

## 4.4 The Josephson effects

In 1962, Josephson calculated the current that could be expected to flow during tunneling of Cooper pairs through a thin insulating barrier (the order of a few nanometers thick), and found that a current of paired electrons (supercurrent) would flow at zero bias in addition to the usual current that results from the tunneling of single electrons (single or unpaired electrons are present in a superconductor along with bound pairs). The zero-voltage current flow resulting from the tunneling of Cooper pairs is known as the *dc Josephson effect*, and was experimentally observed soon after its theoretical prediction. Josephson also predicted that if a constant nonzero voltage $V$ is maintained across the tunnel barrier, an alternating supercurrent will flow through the barrier in addition to the *dc* current produced by the tunneling of single electrons. The angular frequency of the *ac* supercurrent is $\omega = 2eV/\hbar$. The oscillating current of Cooper pairs that flows when a steady voltage is maintained across a tunnel barrier is known as the *ac Josephson effect*. These Josephson effects play a special role in superconducting applications.

In fact, the Josephson effects exist not only in tunneling junctions, but also in other kinds of the so-called weak links, that is, short sections of superconducting circuits where the critical currents is substantially suppressed. Some examples of weak links are shown in Fig. 2.14. Let us now discuss these effects in detail.

Consider a superconductor-insulator-superconductor junction in *thermodynamic equilibrium* at $T \ll T_c$. For simplicity, assume that the superconductors on both sides of the junction are conventional and identical. Then, the order parameters of the two superconductors can be presented as $\Psi_1(\mathbf{r}) = (n_s/2)^{1/2}$ $e^{i\theta_1(\mathbf{r})}$ and $\Psi_2(\mathbf{r}) = (n_s/2)^{1/2} e^{i\theta_2(\mathbf{r})}$, where $n_s$ is the density of superconducting electrons, and $\theta_1$ and $\theta_2$ are the phases. The order parameters decay in the insulator, as shown schematically in Fig. 2.15. If the insulator is not very thick, the two order parameters will overlap, resulting in the onset of phase coherence across the junction. Since the superconductors are identical, their Fermi levels are identical too. Set the potential difference between the two superconductors, $V = (E_1 - E_2)/2e$ (the pair charge is $2e$). Then, the order parameters will evolve according to the following equations:

$$i\hbar \frac{\partial \Psi_1}{\partial t} = E_1 \Psi_1 + C\Psi_2 \text{ and}$$
$$i\hbar \frac{\partial \Psi_2}{\partial t} = -E_2 \Psi_2 + C\Psi_1,$$
(2.36)

where $C$ is a coupling constant that measures the interaction of the two order parameters (depends mainly on the thickness of the insulator). Substituting the order parameters into the above equations, and separating real and imaginary parts, we have

$$\frac{\partial n_s}{\partial t} = \frac{2}{\hbar} Cn_s \sin\theta,$$
$$\frac{\partial \theta_1}{\partial t} = \frac{C}{\hbar}\cos\theta - \frac{eV}{\hbar},$$
$$\frac{\partial \theta_2}{\partial t} = \frac{C}{\hbar}\cos\theta + \frac{eV}{\hbar},$$
(2.37)

where $\theta = \theta_1 - \theta_2$. The first equation means that the current $I = \frac{4eC}{\hbar}n_s \sin\theta = I_c \sin\theta$ circulates between the two superconductors at zero bias. The critical current density $I_c$ in

$$I = I_c \sin\theta$$
(2.38)

is the maximum dissipation-free current through the junction. Since the coupling constant $C$ is unknown, $I_c$ cannot be obtained explicitly. Subtracting the second equation of Eqs. (2.37) from the last one, we get

$$\frac{\partial \theta}{\partial t} = \frac{2e}{\hbar} V.$$ (2.39)

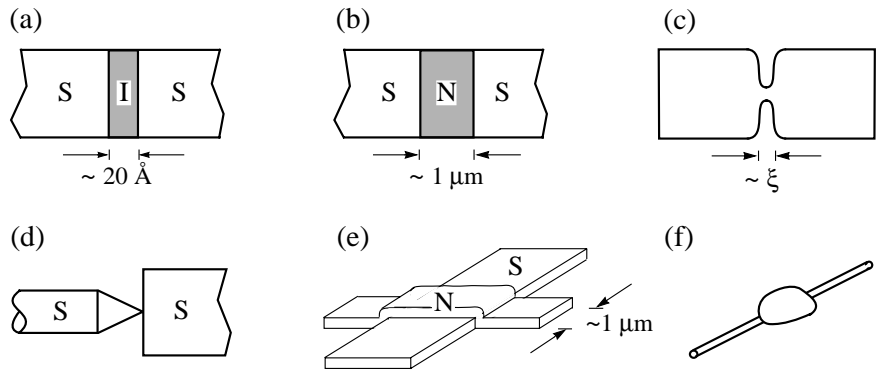

*Figure 2.14.* Different types of weak links: (a) SIS tunneling junction; (b) SNS sandwich; (c) microbridge formed by a narrow constriction; (d) point-contact junction; (e) and (f) weak links due to the proximity effect: (e) a normal film N causes local suppression of the order parameter of a superconducting film S, and (f) small drop of solder on a superconducting wire.

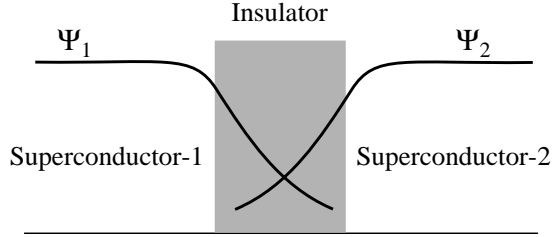

*Figure 2.15.* Superconductor-insulator-superconductor junction: sketch of the decay of the order parameters in the insulator.

In the case when the superconductors in the junction shown in Fig. 2.15 are not identical, Equations (2.37) will be slightly different, and the reader can easily derive these equations independently.

Equations (2.38) and (2.39) respectively represent the *dc* and *ac* Josephson effects (also known as stationary and nonstationary, respectively). The first equation implies that a direct superconducting current can flow through a junction of weakly coupled superconductors with no applied potential difference. The magnitude of this zero-bias current depends on the phase difference across the junction, $\theta = \theta_1 - \theta_2$. The amplitude of the *dc* Josephson current depends on temperature. For a tunneling junction with identical *conventional* superconductors, the temperature dependence of the critical Josephson current was derived in the framework of the BCS theory by Ambegaokar and Baratoff,

$$I_c(T) = \frac{\pi \Delta(T)}{2eR_n} \tanh \frac{\Delta(T)}{2k_B T},$$ (2.40)

where $R_n$ is the junction resistance in the normal state, and $\Delta(T)$ is the energy gap. The Josephson current is maximal at $T = 0$:

$$I_c(0) = \frac{\pi \Delta(0)}{2eR_n}.$$ (2.41)

Figure 2.16 shows the $I(V)$ characteristic of a tunneling junction at $T = 0$. Let us estimate the value of $I_c(0)$ for a conventional superconductor. In conventional superconductors, $\Delta(0) \approx 1$ meV. For an oxide junction of 1 mm$^2$ area, with $R_n \simeq 1\,\Omega$, $I_c(0)$ is of the order of 1 mA. Then, the current density through the junction is about $10^3$ A m$^{-2}$. At high temperatures, as $T \to T_c$, the amplitude of $I_c(T)$ decreases, so that $I_c \propto \Delta^2 \sim (T - T_c)$.

The second Josephson equation, Eq. (2.39), implies that if a *constant* voltage is applied across the barrier, then an *alternating* supercurrent of Cooper pairs with a characteristic frequency $\omega = 2eV/\hbar$ will flow across the junction. An applied *dc* current of 1 mV will produce a frequency $\nu = \omega/2\pi = 2eV/h = 483.6$ GHz, which lies in the far infrared region. Every time a Cooper pair crosses the barrier (obviously, resistanceless), it emits (or absorbs) a photon of energy $\hbar\omega = 2eV$. This radiation is observed experimentally. The latter expression involves twice the electron charge due to electron pairing. This very simple relation between the radiation frequency and the applied voltage is now used to verify the fact of the electron pairing, every time a new superconductor is discovered. This is done in the following way. A tunneling junction is placed in a microwave cavity and, in measured $I(V)$ characteristics, one can then observe the appearance of equidistant steps, called the *Shapiro steps*. The steps appear at voltages $V_{0,n} = n\,\hbar\omega_{\mu w}/2e$, where $\omega_{\mu w}$ is the microwave frequency (the Josephson current at zero bias is reduced when microwaves are applied, enabling one to detect the Shapiro steps). The relation $\hbar\omega = 2eV$ was also used to derive a value for the ratio $e/h$, which is the most accurate determination of this ratio so far.

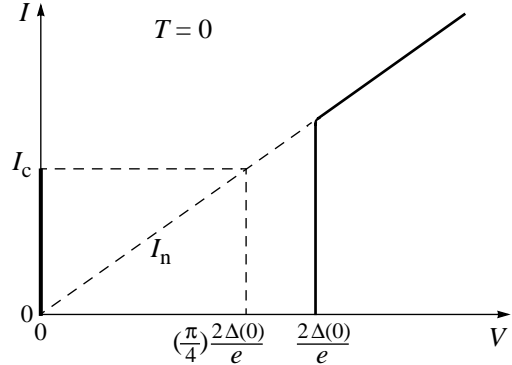

*Figure 2.16.* $I(V)$ characteristic for a Josephson junction at $T = 0$. $I_c$ at zero voltage is the maximum Josephson current, and $I_n$ is the normal-state current.

It is important to emphasize that the effects of weak superconductivity have their origin in the *quantum nature* of the superconducting state. The superconducting condensate is a *Bose condensate*, and is similar to a Bose-Einstein condensate (see Chapter 4). Therefore, the Josephson effects will manifest themselves in every Bose-Einstein condensate, even if the bosons have no charge. In the later case, it is not easy to detect the current of chargeless particles. (In a chargeless Bose-Einstein condensate, the energy $2eV$ in the ac Josephson effect in Eq. (2.39) is represented by another energy scale).

The so-called *superconducting quantum interference devices* (SQUIDs), consisting of two parallel tunneling junctions connected in parallel (dc SQUIDs), as shown in Fig. 2.17, are the most sensitive device for measuring the value of a magnetic field. The actual resolution of such a device can be much better than a single flux quantum, $\sim 10^{-5}\Phi_0$. The most celebrated examples are SQUID magnetometers, which are able to resolve flux increments of $\sim 10^{-10}$ G, and precision voltmeters with the sensitivity of $\sim 10^{-15}$ V. SQUIDs based on a single point-contact junction incorporated in a loop (rf SQUIDs), can measure only the change of flux, variations of magnetic field or of its gradient.

Finally, let us consider briefly how $I_c$ is affected by an applied magnetic field and the size of junction. It turns out that by applying a magnetic field to a tunneling junction, the magnitude of $I_c$ is found to be a nonmonotonic function of the field strength, as shown in Fig. 2.18. This is due to a quantum interfer-ence effect caused by the phases of the order parameters. The dependence of the amplitude of the *dc* Josephson current on the magnetic field is

$$I_c(\Phi) = I_c(0) \left| \frac{\sin(\pi\Phi/\Phi_0)}{\pi\Phi/\Phi_0} \right|, \qquad (2.42)$$

where $\Phi$ is the magnetic flux threading through the insulating layer in the junction, and $\Phi_0$ is the flux quantum. When the total magnetic flux is a multiple of the flux quantum, $\Phi = n\,\Phi_0$, $n = 1, 2, 3, \ldots$, the Josephson current vanishes, as shown in Fig. 2.18. This result is a commonly used criterion for the uniformity of tunneling current in a Josephson junction: one measures the maximum zero-bias current as a function of magnetic field, and the extent to which the dependence fits Eq. (2.42) is a measure of the uniformity.

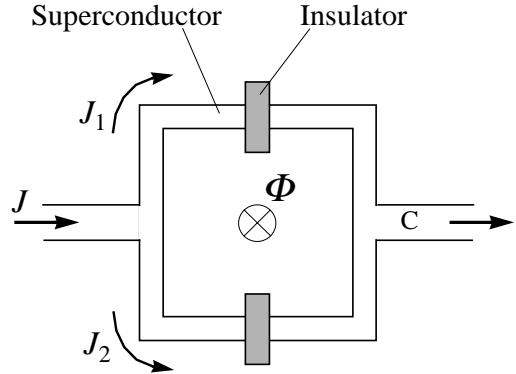

*Figure 2.17.* SQUID magnetometer consisting of two Josephson junctions connected in parallel. A magnetic flux Φ which is threaded through the interior of the SQUID loop changes the combined current that emerges at C.

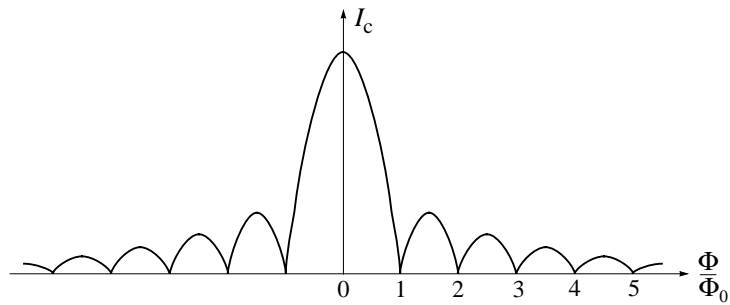

*Figure 2.18.* Maximum supercurrent through a *dc* Josephson junction versus the external magnetic field applied parallel to the plane of the junction.

In the above discussion, we neglected entirely the fact that the tunneling current can also produce a magnetic field that affects the behavior of the junction. If the size of the junction is small, the effect of this self-field can be neglected; however in *long Josephson junctions*, the self-field effect leads to the appearance of a nonlinear term in the equation describing the behavior of $I_c$. For a long Josephson junction, the nonlinear equation exactly coincides with the so-called *sine-Gordon equation* describing the behavior of a chain of pendulums coupled by torsional springs. For the chain of pendulums, the solution of the sine-Gordon equation represents the propagation of soliton along the chain of pendulums. In a long Josephson junction, the sine-Gordon solitons describe quanta of magnetic flux expelled from the superconductors, that travel back and forth along the junction. Their presence, and the validity of the soliton description, can be easily checked by the microwave emission which is associated with their reflection at the ends of the junction. For the long Josephson junction, the quantity $\lambda_J$, called the *Josephson penetration length*, gives a measure of the typical distance over which the phase (or magnetic flux) changes:

$$\lambda_J = \left(\frac{\Phi_0}{2\pi\mu_0 I_c(d + 2\lambda_L)}\right)^{1/2}, \qquad (2.43)$$

where $\Phi_0 = 2.07 \times 10^{-15}$ Wb, $\mu_0 = 4\pi \times 10^{-7}$ H m$^{-1}$, $I_c$ is the density of the critical current through the junction in A m$^{-2}$, $\lambda_L$ is the London penetration depth in m, and $d$ is the thickness of the insulator (oxide layer) in m. The quantity $(d + 2\lambda_L)$, or $(d + \lambda_{L,1} + \lambda_{L,2})$ if two superconductors in the junction are not identical, is the width of the region penetrated by the magnetic field. The Josephson penetration length allows one to define precisely a *small* and a *long* junction. A junction is said to be long if its geometric dimensions are large compared with $\lambda_J$. Otherwise, the junction is small. Let us estimate $\lambda_J$. Taking typical parameters for a Josephson junction, $(d + 2\lambda_L) \sim 10^{-5}$ cm, $I_c \sim 10^2$ A cm$^{-2}$, we get $\lambda_J \sim 0.1$ mm, i.e. it can be a macroscopic length.

A very useful feature of the Josephson solitons is that they are not difficult to operate by applying bias and current to the junction. Then, long Josephson junctions can be used in computers. One of the most useful properties of such devices would be very high performance speed. Indeed, the characteristic time may be as small as $10^{-10}$ sec, while the size of the soliton may be less than 0.1 mm. The main problem for using the Josephson junctions in electronic devices is the cost of cooling refrigerators. For commercial use in electronics, the long Josephson junctions await for the availability of room-temperature superconductors.

## 4.5    Energy gap in the excitation spectrum

At $T = 0$, the elementary excitation spectrum of a superconductor has an energy gap. In conventional superconductors, however, at some special con-ditions, there may exist *gapless* superconductivity, since conventional superconductors have only one energy gap—the pairing one. We shall consider this case at the end of this subsection.

As already discussed above, the energy gap in a superconductor is carried by the Fermi surface, and occurs on either side of the Fermi level $E_F$, as shown in Fig. 2.11. The excited states of a superconductor are altered from the normal state. If in the normal state, it costs energy $|E_k - E_F|$ to put electron into an excited one-electron state, where $E_k$ is the single particle energy spectrum; in the superconducting state, the energy cost is $\sqrt{(E_k - E_F)^2 + \Delta^2}$. Thus, the minimum energy cost in the normal state is zero, whereas in the superconducting state, it is instead the smallest value of $\Delta$. As a result, the electronic system in the superconducting state is unable to absorb arbitrary small amounts of energy. At $T = 0$ all electrons are accommodated in states below the energy gap, and a minimum energy $2\Delta(0)$ must be supplied to produce an excitation across the gap, as shown in Fig. 2.11. The BCS temperature dependence of the energy gap for a conventional superconductor is depicted in Fig. 2.12. In conventional superconductors, the value of the energy gap $\Delta(0)$ is of the order of 1 meV ($\simeq 12$ K).

It is worth to recall that the superconducting state requires the electron pairing *and* the onset of long-range phase coherence. Superconductors in which the long-range phase coherence occurs due to a mechanism different from the overlap of wavefunctions, have two distinct energy gaps—the pairing gap $\Delta_p$ and phase-coherence gap $\Delta_c$. As a consequence, in the superconducting state the magnitude of total energy gap in the elementary excitation spectrum of such unconventional superconductors is equal to $\sqrt{\Delta_p^2 + \Delta_c^2}$.

Experimental evidence of the existence of the energy gap in the elementary excitation spectrum of superconductors comes from many different types of measurements, such as tunneling, infrared, microwave, acoustic, specific-heat measurements etc. The most direct way of examining the energy gap is by tunneling measurements. The experiment consists in examining the current-voltage characteristics obtained in a tunneling junction, $I(V)$. Let us briefly discuss the basics of tunneling measurements.

The phenomenon of tunneling has been known for more than sixty five years—ever since the formulation of quantum mechanics. As one of the main consequences of quantum mechanics, a particle such as an electron, which can be described by a wave function, has a finite probability of entering a classically forbidden region. Consequently, the particle may tunnel through a potential barrier which separates two classically allowed regions. The tunneling probability was found to be exponentially dependent on the potential barrier width. Therefore the experimental observation of tunneling events is measurable only for barriers that are small enough. Electron tunneling was for the first time observed experimentally in junctions between two semiconductors by Esaki in 1957. In 1960, tunneling measurements in planar metal-oxide-metal junctions were performed by Giaever. The first tunneling measurements between a normal metal and a superconductor were also carried out in 1960. The direct observation of the energy gap in the superconductor in these and the following tunneling tests provided strong conformation of the BCS theory.

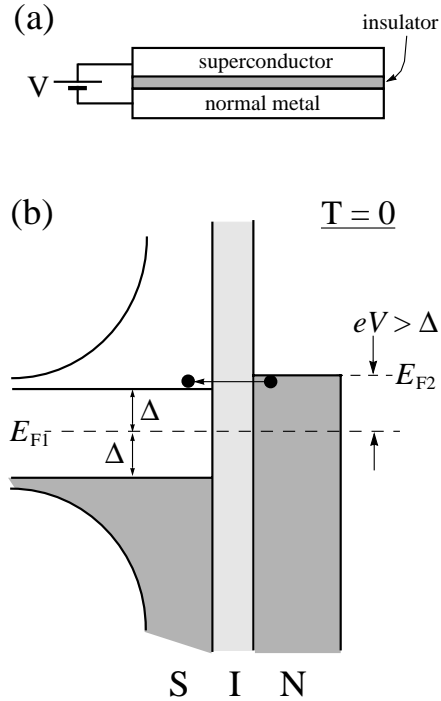

*Figure 2.19.* (a) Superconductor-insulator-normal metal tunneling junction, and (b) corresponding energy diagram at $T = 0$ in the presence of an applied voltage: quasiparticles can tunnel when $|V| \geq \Delta/e$.

Consider the flow of electrons across a thin insulating layer having the thickness of a few nanometers, which separates a normal metal from a *conventional* superconductor. Figure 2.19a shows a superconductor-insulator-normal metal (SIN) tunneling junction. At $T = 0$, no tunneling current can appear if the absolute value of the applied voltage (bias) in the junction is less than $\Delta(0)/e$. Tunneling will become possible when the applied bias reaches the value of $\pm\Delta(0)/e$, as shown in Fig. 2.19b. Figure 2.20 shows schematically three current-voltage $I(V)$ characteristics for an SIN junction at $T = 0$, $0 < T < T_c$ and $T_c < T$. At $T = 0$, the absence of a tunneling current at small voltages constitutes an experimental proof of the existence of a gap in the elementary excitation spectrum of a superconductor. At $0 < T < T_c$, there are always ex-cited electrons due to thermal excitations, as shown in Fig. 2.21, and one can measure some current for any voltage. In other words, at finite temperatures, quasiparticles tend "to fill the gap." As shown in Fig. 2.20, the $I(V)$ curves, measured below $T_c$, approach at high bias the $I(V)$ characteristic measured above $T_c$ (thus corresponding to tunneling between two normal metals). In conventional superconductors, the gap vanishes completely at $T_c$, as shown in Fig. 2.12. This, however, is not the case for high-$T_c$ superconductors—there is a pseudo-gap in the elementary excitation spectrum of the cuprates above $T_c$. We shall analyze tunneling measurements in the cuprates in Chapter 6. Tunneling $I(V)$ characteristics for a superconductor-insulator-superconductor (SIS) junction are similar to those for a SIN junction. However, in SIS junctions, the tunneling current is absent at $T = 0$ between $-2\Delta(0)/e$ and $2\Delta(0)/e$, with exception of the Josephson current at zero bias, as shown in Fig. 2.16.

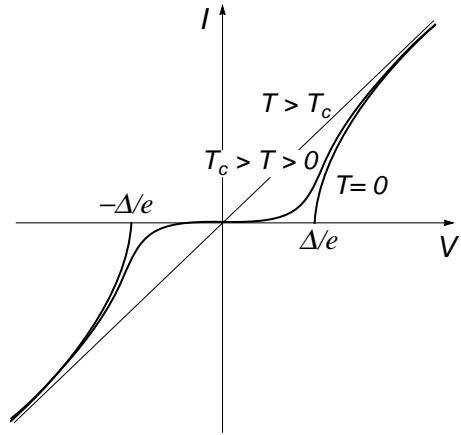

*Figure 2.20.* Tunneling $I(V)$ characteristics for an SIN junction at different temperatures: $T = 0$; $0 < T < T_c$, and $T > T_c$ (the latter case corresponds to a NIN junction). At $0 < T < T_c$, quasiparticle excitations exist at any applied voltage.

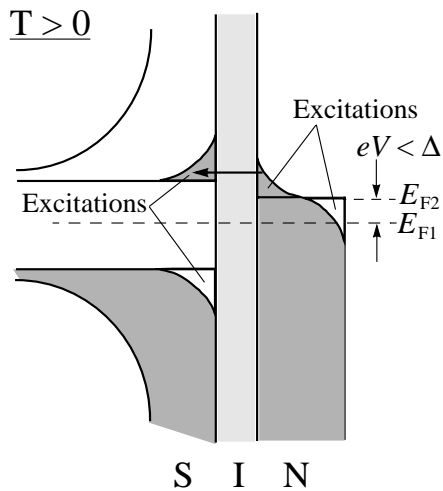

*Figure 2.21.* The density of states near the Fermi level $E_F$ in a superconductor and a normal metal in an SIN junction at $0 < T < T_c$. Due to thermal excitations, there are states above the gap in the superconductor and above the Fermi level in the metal. Quasiparticles can tunnel at any applied voltage, as shown in Fig. 2.20.

The energy gap in the elementary excitation spectrum of a superconductor can also be measured directly in acoustic (ultrasound) measurements. Consider absorption of high-frequency sound waves in a conventional superconductor. Ultrasound waves are scattered in a superconductor by normal electrons, not by Cooper pairs, so that their attenuation is a measure of the fraction of normal electrons. As a consequence, superconductors absorb sound waves more weakly than normal metals. Ultrasound attenuation in a superconductor is described by the following expression

$$\frac{\alpha_s}{\alpha_n} = \frac{2}{e^{\Delta/k_B T} + 1}, \tag{2.44}$$

where $\alpha_s$ and $\alpha_n$ are the absorption coefficients in the normal and superconducting states, respectively. This formula is valid if $\hbar\omega < 2\Delta$, where $\omega$ is the sound frequency. In practice, the sound frequency is less than 1 GHz ($= 10^9$ Hz), so that $\hbar\omega < 10^{-2}\Delta$. The ratio of the attenuation measured in conventional superconductors as a function of temperature indeed follows the prediction of the BCS theory shown in Fig. 2.12, confirming the validity of the principal ideas of the theory.

In conventional superconductors, the energy gap in the elementary excitation spectrum at some conditions can be absent. Consider this particular case. In Chapter 5, we shall discuss how magnetic and non-magnetic impurities affect the superconducting state in conventional superconductors: doping a conventional superconductor with non-magnetic atoms does not affect strongly the critical temperature and the energy gap. Only a very pure superconductor will suffer a small decrease in $T_c$ (about 1%). This decrease comes to an end when the mean free path $\ell$ becomes equal to the size of a Cooper pair, $\xi$ (at this moment, the energy gap becomes isotropic). At the same time, doping a *conventional* superconductor with magnetic impurities drastically affect the superconducting properties: a marked change in the critical temperature and the energy gap is always observed when magnetic impurities are introduced. Even a small impurity concentration (a few percent) can lead to a complete destruction of the superconducting state. Experimentally, it turns out that as one keeps adding the magnetic impurities, the energy gap in the elementary excitation spectrum of a conventional superconductor decreases faster than $T_c$, and when the impurity concentration reaches $n_0 = 0.91 \times n_{cr}$, the gap vanishes while the sample remains superconducting, i.e. there is still no electrical resistivity. $n_{cr}$ is the impurity concentration at which superconductivity completely disappears.

How does gapless superconductivity occur? Having a magnetic moment, magnetic impurities destroy the electron pairs. At the impurity concentration $n_0$, some fraction of the Cooper pairs are broken up. Then even at $T = 0$, the situation is similar to that in the two-fluid model: the Cooper pairs and the free electrons, created by the partial breakup of the Cooper pairs, coexist. The Cooper pairs can still sustain resistanceless current flow, while the free electrons can absorb radiation of arbitrary low frequency, so that the energy gap in the elementary excitation spectrum disappears, as seen for example, in tunneling measurements. Gapless superconductivity can occur in conventional superconductors not only in the presence of magnetic impurities, but in the presence of any external "force," for example, a sufficiently strong magnetic field or an applied current, which is able to destroy the superconducting order. In unconventional superconductors, the occurrence of gapless superconductivity is only possible *locally*, and we shall discuss this case in Chapter 6. On a macroscopic scale, the occurrence of gapless superconductivity in unconventional superconductors is impossible because unconventional superconductors have two energy gaps—pairing and phase-coherence.

## 4.6    Thermodynamic properties

The transition from the normal state to the superconducting state is the *second-order phase transition*. At a second-order phase transition, the *first* derivatives of the *Gibbs free energy* are always continuous, while the *second* derivatives have finite-step discontinuities. The Gibbs free energy $G$ for a system in thermal equilibrium is defined (in CGS units) as

$$G \equiv U - TS - \mathbf{B} \cdot \mathbf{H}/4\pi + pV \equiv F - \mathbf{B} \cdot \mathbf{H}/4\pi + pV, \qquad (2.45)$$

where $U$ is the total internal energy of the system; $T$ is the temperature of the system; $S$ is the entropy per unit volume; $p$ is the pressure in the system; $V$ is the volume of the system; $\mathbf{H}$ and $\mathbf{B}$ are the applied magnetic field and flux, respectively. The function $F \equiv U - TS$ is the Helmholtz free energy, already discussed above. The Gibbs free energy is also called the *Gibbs potential*.

As obtained above, the Helmholtz free energy of the superconducting state $F_s$ is lower than that of the normal state $F_n$ by the value

$$F_n - F_s = \frac{H_c^2}{8\pi} \qquad (2.46)$$

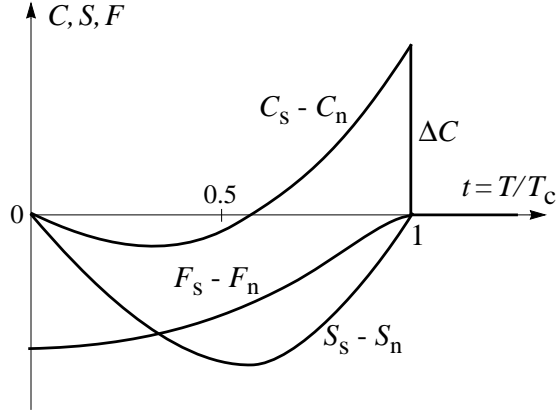

*Figure 2.22.* Temperature dependences of the specific heat $C_s$, the entropy $S_s$ and the free energy $F_s$ of a superconductor in $H = 0$ with respect to their values in the normal state, $C_n$, $S_n$ and $F_n$.

called the condensation energy. The magnetic field $H_c$ is the thermodynamic critical field. The condensation energy $-(F_n - F_s)$ is shown in Fig. 2.22 as a function of reduced temperature $t = T/T_c$.

Let us now derive the difference in entropy between the normal and superconducting states. By the first law of thermodynamics, we have

$$\delta Q = \delta R + \delta U, \qquad (2.47)$$

where $\delta Q$ is the element of the thermal energy density for the body under consideration, and $\delta R$ is the work done by the body on external bodies, per unit volume. For the Helmholtz free energy, one obtains $\delta F = \delta U - T\delta S - S\delta T$. Since for a reversible process $\delta Q = T\delta S$, we get

$$\delta U = T\delta S - \delta R \quad \text{and} \quad \delta F = -\delta R - S\delta T. \qquad (2.48)$$

From the last equation, it follows that

$$S = -\left(\frac{\partial F}{\partial T}\right)_R. \qquad (2.49)$$

We can use Eq. (2.46) to calculate the difference in entropy between the normal and superconducting states. Substituting Eq. (2.46) into Eq. (2.49), we obtain

$$S_s - S_n = \frac{H_c}{4\pi}\left(\frac{\partial H_c}{\partial T}\right)_R. \qquad (2.50)$$

The third law of thermodynamics states that as $T \to 0$, the entropy of a system approaches a limit $S_0$ that is independent of all its parameters. Therefore, $S_s - S_n \to 0$ as $T \to 0$. Since the superconducting transition is the second-order phase transition, then $S_s - S_n = 0$ at $T_c$. From Eq. (2.20), we have $(\partial H_c / \partial T) < 0$. Therefore, $S_s < S_n$ at $0 < T < T_c$, meaning that the superconducting state is a more ordered state than the normal one. Therefore, the superconducting transition can be considered as the order-disorder transition. A sketch of the dependence $S_s - S_n$ is shown in Fig. 2.22 as a function of temperature.

The *latent heat of transformation* $L_{1,2}$ is defined as the heat absorbed by the system from the reservoir as a transformation takes place from phase 1 to phase 2. Since for a reversible process $\delta Q = T \delta S$, then the heat added to the system at constant temperature is

$$L_{1,2} = Q = T(S_2 - S_1) \tag{2.51}$$

Because $S_s - S_n$ is zero at $T = 0$ and $T_c$, therefore the latent heat of transformation is zero at $T = 0$ and $T_c$. So, the transition is the second order not only at $T_c$ but also at $T = 0$. However at $0 < T < T_c$, the transition is the first order since the latent heat is not zero.

Consider now the difference in *specific heat* per unit volume between the superconducting and normal states. The specific heat of matter can be defined as $C = T(\partial S / \partial T)$. By taking the derivative of Eq. (2.50), we can write the difference in specific heat per unit volume between the superconducting and normal states as

$$C_s - C_n = \frac{T}{4\pi} \left[ \left( \frac{\partial H_c}{\partial T} \right)^2 + H_c \frac{\partial^2 H_c}{\partial T^2} \right]. \tag{2.52}$$

At $T_c$, one obtains that there is a specific-heat discontinuity

$$C_s - C_n = \frac{T_c}{4\pi} \left( \frac{\partial H_c}{\partial T} \right)^2_{T_c}. \tag{2.53}$$

The last expression is known as the Rutgers formula, and defines the height of the specific-heat jump at $T_c$. In the framework of the BCS theory for conventional superconductors (the weak electron-phonon coupling approximation), this jump is given by

$$\beta \equiv \frac{\Delta C}{C_n} \bigg|_{T_c} \equiv \frac{C_s - C_n}{C_n} \bigg|_{T_c} = 1.43. \tag{2.54}$$

The temperature dependence of the difference $C_s - C_n$ is plotted in Fig. 2.22. In the normal state, thus above $T_c$, the specific heat $C_n$ *linearly* decreases as the temperature decreases, $C_n = \gamma T$, as shown in Fig. 2.23. Such a linear dependence of specific heat is typical for normal metals, and represents the *electronic* specific heat. In the superconducting state, thus below $T_c$, the specific heat falls exponentially, as the temperature decreases:

$$C_s \propto \exp\left(-\frac{\Delta(T)}{k_B T}\right), \tag{2.55}$$

as schematically shown in Fig. 2.23. In the language of the two-fluid model, the exponential temperature dependence means that below $T_c$, only the normal component transports heat. The Cooper-pair condensate does not contribute to the energy transfer. In the strong electron-phonon coupling regime, i.e. when $2\Delta > 3.52 \, k_B \, T_c$, the value of specific-heat jump increases, i.e. $\beta > 1.43$ (for example, in Pb $\beta = 2.7$).

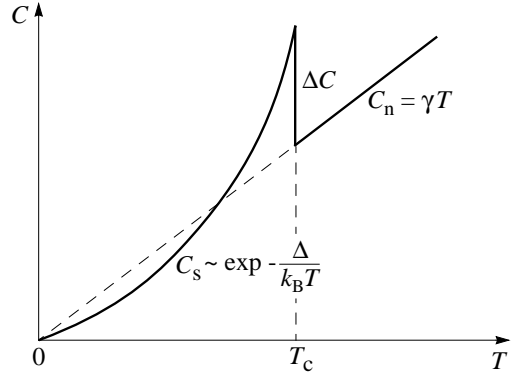

*Figure 2.23.* Temperature dependence of the specific heat of a superconductor. The characteristic jump $\Delta C$ occurs at $T_c$.

Thus, it is worth to emphasize that *the specific-heat jump at $T_c$ is a universal property of all superconductors*; however, its magnitude varies.

Finally, it is necessary to note that in deriving the above formulas for specific heat, we did not take into account the contribution from the lattice. In general, the specific heat of a metal is made up of the electronic and lattice contributions: $C = C_{el} + C_{lat}$. From the theory of normal metals, the temperature dependence of the lattice specific heat at low temperature is given by $C_{lat} \propto T^3$. In metallic superconductors, the superconducting transition has practically no effect on the lattice, therefore, the lattice specific heat is not changed below $T_c$ (in contrast to the electronic specific heat which changes drastically below $T_c$). As a consequence, the difference in specific heat between the superconducting and normal states in metals, $C_s - C_n$, is mainly determined by the electronic component. However, it may be not the case for non-metallic superconductors, for example, for oxides. Therefore, generally speaking, it is possible that at the superconducting transition of some exotic superconductors, one can observe the apparent absence of specific-heat jump, or even, a negative specific-heat jump due to a negative contribution from the lattice. In a sense, this effect is similar to the occurrence of negative isotope effect in conventional superconductors.

Heat transfer in superconductors is also characterized by a peculiar behavior. The thermal conductivity of a normal metallic alloy decreases as the temperature decreases. In a superconductor, however, this is not the case. Following the superconducting transition, the thermal conductivity below $T_c$ rises sharply, then passes through a maximum, and after that, begins to drop. This is due to the fact that in addition to the electronic heat transfer, the lattice can contribute to the flow of thermal energy. This makes the phenomenon of heat conduction more complicated than electrical conduction.

## 4.7    Proximity effect

*Every superconductor exhibits the proximity effect.* The proximity effect occurs when a superconductor S is in contact with a normal metal N. If the contact between the superconductor and normal metal is of a sufficiently good quality, the order parameter of the superconductor close to the interface, $\Psi$, will be altered. The superconductor, however, does not "react passively" to this "intrusion." Instead, it induces superconductivity into the metal which was in the normal state before the contact. Of course, this induced superconductivity exists only in a thin surface layer of the normal metal near the NS interface. The distances measured from the NS interface, along which the properties of the superconductor and the normal metal are modified, are of the order of the coherence length, i.e. $\sim 10^4$ Å.

Thus, when a normal metal and a superconductor are in good contact, the Cooper pairs from the superconductor penetrates into the normal metal, and "live" there for some time. This results in the reduction of the Cooper-pair density in the superconductor. This also means that in a material which by itself is not a superconductor, one can, under certain conditions, induce the superconducting state. So, the proximity effect gives rise to induced superconductivity. The proximity effect is strongest at temperatures $T \ll T_c$, i.e. close to zero.

Let us consider an interface between a normal metal and a superconductor. Assume that the interface between the two materials is flat and coincides with the plane $x = 0$, as shown in Fig. 2.24. The superconductor occupies the semispace $x > 0$, and the normal metal the semispace $x < 0$. The order parameter penetrates the normal metal to a certain depth $\xi_N$, called the *effective coherence length*. In a first approximation, the decay of the order parameter in the normal metal is exponential, $\Psi_n \propto \exp(-|x|/\xi_N)$. Rigorous calculations based on the microscopic theory give the following expressions for $\xi_N$. In a pure N metal, that is, when the electron mean free path is much larger than the effective coherence length, $\ell_n \gg \xi_N$ (the clean limit), the effective coherence

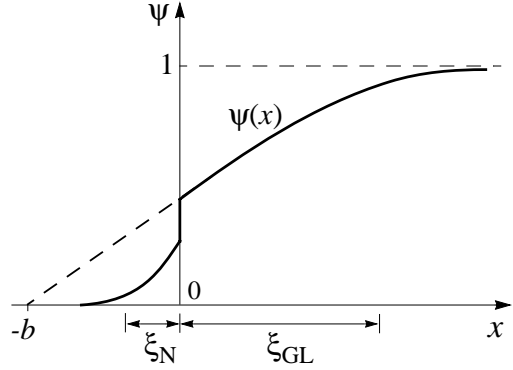

*Figure 2.24.* Order parameter $\Psi(x)$ near the interface between a superconductor ($x > 0$) and a normal metal ($x < 0$) at $T \ll T_c$.

length is

$$\xi_n = \frac{\hbar v_{F,n}}{2\pi k_B T}, \qquad (2.56)$$

where $v_{F,n}$ is the Fermi velocity in the normal metal. From this expression, one can see that when $T \rightarrow 0$, $\xi_N \rightarrow \infty$. In such a limit, the decay of the order parameter in the N region is much slower than the exponential one. In the so-called dirty limit, when $\ell_n \ll \xi_N$, the effective coherence length in the N metal is

$$\xi_N = \left(\frac{\hbar v_{F,n} \ell_n}{6\pi k_B T}\right)^{1/2}. \qquad (2.57)$$

Evaluations by Eqs. (2.56) and (2.57) give values for $\xi_N$ in the range of $10^3$–$10^4$ Å.

The behavior of the order parameter in the general case is sketched in Fig. 2.24. The length $b$ in Fig. 2.24 is called the *extrapolation length*. This length is a measure of extrapolation to the point outside of the boundary at which $\Psi$ would go to zero if it maintained the slope it had at the surface. In the dirty limit, the value of $b$ is

$$b \simeq \frac{\sigma_s}{\sigma_n} \xi_N, \qquad (2.58)$$

where $\sigma_s$ and $\sigma_n$ are the conductivities in the S and N regions, respectively, and $\xi_N$ is defined by Eq. (2.57). For a superconductor-*insulator* interface, the microscopic theory gives

$$b \sim \frac{\xi_0^2}{a_0}, \qquad (2.59)$$

where $\xi_0$ is the intrinsic coherence length of the superconductor, and $a_0$ is the interatomic distance in the insulator.

If the thickness of a superconductor which is in contact with a normal metal, is sufficiently large, i.e. when $d \gg \xi_{GL}$, the critical temperature of the superconductor is practically unaffected. However, when a superconducting thin film is deposited onto the surface of a normal metal, and the thickness of this film is small $d \ll \xi_{GL}$, the critical temperature of the whole system decreases, depending on $d$, and in principle, can fall to zero.

The proximity effect is utilized in SNS Josephson junctions in which the phase coherence between the two superconducting electrodes is established via a normal layer that can be quite thick ($\sim 10^4$ Å). It is worth to mention that any Bose-Einstein condensate will also exhibit the proximity effect.

As stated above, the proximity effect involves Cooper pairs entering into the normal metal. The transition of a Cooper pair from the superconductor into the normal metal can be considered as a reflection off the NS interface, with two electrons incident on the interface and two holes reflected back. Indeed, the disappearance of an electron is equivalent to the creation of a hole. Contrary to this, we are interested in what happens at the NS interface to an electron in the normal metal moving towards the superconductor, when it encounters the NS interface. If the electron energy is less than the energy gap of the superconductor, the electron is reflected back from the interface. The propagation of a negative charge in the normal metal from the interface is equivalent to propagation of a positive charge in the superconductor in the opposite direction. Therefore, the process of the electron reflection gives rise to a charge transfer from the normal metal to the superconductor, i.e. to an electrical current. This process was first proposed theoretically by Andreev and is now called the *Andreev reflection*.

In a sense, the electron tunneling and the Andreev reflection are two "inverse" processes. In an superconductor-insulator-normal metal (SIN) junction at $T = 0$, in the tunneling regime (high values of the normal resistance of the junction, $R_n \gg 0$), the current is absent at bias $|V| < \Delta/e$ (see Fig. 2.20), whereas in the Andreev-reflection regime (small $R_n$), the current at $|V| < \Delta/e$ can be twice as large as the current at high bias (in *unconventional* superconductors, usually lower than 2). In addition, it is worth noting that tunneling spectroscopy is a *phase-insensitive* probe (*at least*, for s-wave superconductors), whereas the Andreev reflection is sensitive exclusively to coherence properties of the superconducting condensate.

## 4.8 Isotope effect

It was experimentally found that different isotopes of the same superconducting metal have different critical temperatures, and

$$T_c M^\alpha = \text{constant}, \qquad (2.60)$$

where $M$ is the isotope mass. For the majority of superconducting elements, $\alpha$ is close to the classical value 0.5.

The vibrational frequency of a mass $M$ on a spring is proportional to $M^{-1/2}$, and the same relation holds for the characteristic vibrational frequencies of the atoms in a crystal lattice. Thus, the existence of the isotope effect indicated that, although superconductivity is an electronic phenomenon, it is nevertheless related in an important way to the vibrations of the crystal lattice in which the electrons move. The isotope effect provided a crucial key to the development of the BCS microscopic theory of superconductivity for *conventional* superconductors. Luckily, not until after the development of the BCS theory was it discovered that the situation is more complicated than it had appeared to be. For some conventional superconductors, the exponent of $M$ is not -1/2, but near zero, as listed in Table 2.3.

*Table 2.3.*   Isotope effect ($T_c \propto M^{-\alpha}$)

| Element | $\alpha$ |
|---------|----------|
| Mg | 0.5 |
| Sn | 0.46 |
| Re | 0.4 |
| Mo | 0.33 |
| Os | 0.21 |
| Ru | 0 ($\pm 0.05$) |
| Zr | 0 ($\pm 0.05$) |

So, the isotope effect is not a universal phenomenon, and can be absent even in *conventional* superconductors. In unconventional superconductors, the situation is very peculiar. For example in copper oxides, varying the doping level, the isotope effect is almost absent in the optimally-doped region ($\alpha \approx 0.03$). At the same time in the underdoped region, the exponent $\alpha$ is about 1 (see Fig. 6.28), thus its value is two times larger than the classical value 0.5! This fact was initially taken as evidence against the BCS mechanism of high-$T_c$ superconductivity (*true*), and against the phonon pairing mechanism (*false*). We shall consider the mechanism of electron pairing in unconventional superconductors in Chapter 6.

## 4.9    Type-II superconductors: Properties of the mixed state

The absolute majority of all superconductors is of type-II. In the mixed state, their behavior has the same pattern. So, the basic properties of the mixed state are worth to be considered in detail. The term "type-II superconductors" was first introduced by Abrikosov in his phenomenological theory of these materials. As defined by Abrikosov, a superconductor is of type-II if $k =$ $\lambda/\xi_{GL} > 1/\sqrt{2} \simeq 0.71$. Then it follows that the magnetic penetration depth in type-II superconductors is much larger than the coherence length, $\xi_{GL} < \lambda$. This case is schematically shown in Fig. 2.7b. The electrodynamics in type-II superconductors is local, i.e. of the London type.

The mixed state occurs at magnetic fields having a magnitude between $H_{c1}(T)$ (the lower critical field) and $H_{c2}(T)$ (the upper critical field), as shown in Fig. 2.9. This state of a type-II superconductor is referred to as the mixed state because it is characterized by a partially penetration of the magnetic field in the interior of the superconducting sample. The field penetrates the superconductor in microscopic filaments called *vortices* which form a regular *triangular* lattice, as schematically shown in Fig. 2.25. Each vortex consists of a normal core in which the magnetic field is large, surrounded by a superconducting region, and can be approximated by a long cylinder with its axis parallel to the external magnetic field. Inside the cylinder, the superconducting order parameter $\Psi$ is zero. The radius of the cylinder is of the order of the coherence length $\xi_{GL}$. The supercurrent circulates around the vortex within an area of radius $\sim \lambda$. The spatial variations of the magnetic field and the order parameter inside and outside an isolated vortex are sketched in Fig. 2.10. Each vortex carries one magnetic flux quantum.

As an example, Figure 2.26 shows the magnetization curve of a type-II superconductor in the form of a long cylinder placed in a parallel magnetic field. At $H < H_{c1}$, the average field in the interior of the cylinder is $B = 0$. If $H_{c1} < H < H_{c2}$, a steadily increasing field $B$ penetrates the superconductor. The magnitude of this field always remains below the external field. As long as the external field $H < H_{c2}$, the cylinder superconducts. At $H = H_{c2}$, the average field in the interior becomes equal to the external field, thus to $H_{c2}$, and the bulk superconductivity disappears. The transition into the normal state at $H_{c2}(T)$ is a second-order phase transition.

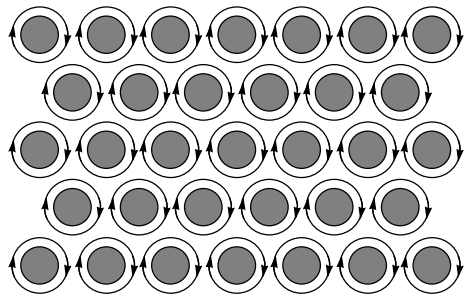

*Figure 2.25.* Normal-state vortices (grey areas) in the mixed state of a type-II superconductor form a regular triangular lattice. Arrows shows the supercurrent circulating around the vortices at $\sim \lambda$ from the centers of vortices. The radius of the vortices is about $\xi_{GL}$.

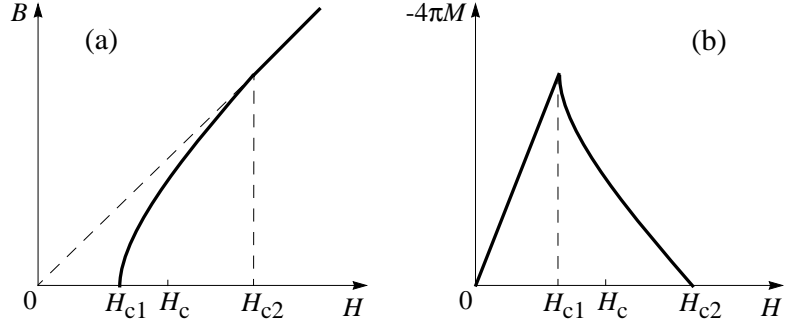

*Figure 2.26.*  Magnetization curves for a type-II superconductor: (a) flux $B$ as a function of an external magnetic field $H$, and (b) magnetic moment $M$ as a function of $H$.

Why does an external magnetic field penetrate a type-II superconductor in the form of vortices, and not uniformly? In the next section, while considering the Ginzburg-Landau theory, we shall see that for type-II superconductors, the surface energy of the interface between a normal metal and a superconductor is negative, $\sigma_{ns} < 0$. This implies that, under certain circumstances, it is energetically favorable for type-II superconductors, when placed in an external magnetic field, to become subdivided into alternating normal and superconducting domains in order to minimize the total free energy of the system. The energy in the $|\nabla\Psi|^2$ term limits the variations of the order parameter. By definition, significant variations of the order parameter cannot exist on a scale smaller than the coherence length $\xi_{GL}$. Thus at $H > H_{c1}$, penetration of vortices with a size $\sim \xi_{GL}$ into the interior of a type-II superconductor becomes thermodynamically favorable. Once inside the superconductor, the vortices arrange themselves at distance $\sim \lambda$ from each other. Energetically, the most favorable regular vortex lattice is triangular. As the external field increases, the vortex lattice period steadily decreases, and the density of the vortices rises. At $H = H_{c2}$, the vortex lattice grows so dense that the distance between the neighboring vortices, i.e. the vortex lattice period, becomes of the order $\xi_{GL}(T)$. When it happens, the normal cores of the vortices come into contact with each other, and the order parameter $\Psi$ becomes zero over the entire volume of the superconductor. At this moment, the superconducting state is suppressed, and the superconductor becomes normal.

### 4.9.1    The lower and upper critical fields

How do the lower and upper critical fields, $H_{c1}$ and $H_{c2}$, relate to $H_c$, the thermodynamic critical field of bulk material?  In the framework of the Ginzburg-Landau theory, the thermodynamic critical field in CGS units is given

by

$$H_c = \frac{\Phi_0}{2\sqrt{2}\pi\lambda\xi_{GL}} = \frac{\Phi_0}{2\sqrt{2}\pi\lambda^2}\,k. \tag{2.61}$$

[see also Eq. (2.21)]. The lowest magnetic field, at which formation of vortices in a type-II superconductor becomes thermodynamically favorable, is

$$H_{c1} = \frac{\Phi_0}{4\pi\lambda^2}(\ln k - 0.18). \tag{2.62}$$

This expression is obtained in the limit $k \gg 1$. From Eqs. (2.61) and (2.62), one obtains that at $k \gg 1$

$$H_{c1} \simeq H_c \frac{\ln k}{\sqrt{2}k}. \tag{2.63}$$

It follows from this expression that at $k \gg 1$, the lower critical field is always $H_{c1} \ll H_c$. The exact calculations of the upper critical field give

$$H_{c2} = \sqrt{2}kH_c. \tag{2.64}$$

Thus, at $k \gg 1$, $H_{c2}$ is always $H_c \ll H_{c2}$. Let us estimate the values of $H_{c1}$ and $H_{c2}$. Taking $k \sim 100$ and $H_c \sim 10^3$ Oe, we get $H_{c1} \sim 30$ Oe and $H_{c2} \sim 10^5$ Oe.

Combining Eqs. (2.63) and (2.64), an interesting relation can be obtained for the product $H_{c1}H_{c2}$:

$$H_{c1}H_{c2} = H_c^2 \ln k, \tag{2.65}$$

which indicates that if $H_{c1}$ is very small, then $H_{c2}$ must be very large. Thus, this means that in type-II superconductors with very high $H_{c2}$, the lower critical field is always very small. And vice versa, in superconductors with low $H_{c2}$, the lower critical field is sufficiently large.

What does the thermodynamic critical field $H_c$ mean for a type-II superconductor? In a type-I superconductor, it is the field at which the superconductor goes to the normal state. What happens to a type-II superconductor at $H_c$? The answer is: nothing special. For a type-II superconductor, the quantity $H_c$ should be considered as a measure of the extent to which the superconducting state of a particular material is favored over its normal state in the absence of magnetic field: $F_n - F_s = H_c^2/8\pi$.

### 4.9.2 Surface superconductivity

All the above expressions, obtained in the framework of the Ginzburg-Landau theory, are rigorously valid only for an infinite sample. As we have discussed above, superconductivity in type-II superconductors appears within the volume of the sample *below* $H_{c2}$. For a finite sample, it turns out that, at the surface of a superconductor, the superconducting state can exist in much higher fields than $H_{c2}$, provided the surface is parallel to the external field. Resolving the first linearized Ginzburg-Landau equation together with the boundary condition (see the following section), one can obtain that superconductivity in a thin surface layer can survive in the external magnetic field up to

$$H_{c3} = 1.69 H_{c2}, \qquad (2.66)$$

if the field is parallel to the surface of the superconductor. Thus, even if the bulk of a superconductor remains normal, superconductivity can exist in a thin surface layer with a thickness of the order $\sim \xi_{GL}(T)$. What is interesting is that the phenomenon of surface superconductivity can also be observed in some type-I superconductors with $k > 0.42$.

### 4.9.3    Anisotropy in layered superconductors

All layered superconductors are of type-II. In these superconductors, the critical magnetic fields $H_{c1}$ and $H_{c2}$, as well as $\lambda$ and $\xi_{GL}$, are different in different directions—parallel and perpendicular to the layers. For example, the upper critical field applied perpendicular to the layers, $H_{c2,\perp}$, is determined by vortices whose screening currents flow parallel to the planes. Then, from Eq. (2.23), we have

$$H_{c2,\perp} = \frac{\Phi_0}{2\pi \xi_{GL,ab}^2}, \qquad (2.67)$$

where the letters "ab" indicate that the direction of the screening currents is in the ab-plane. All the formulas above must be adjusted for layered superconductors in the same manner. Some of them, however, assume a slightly unusual form. For instance, the same expression for the upper critical field applied parallel to the layers, $H_{c2,\parallel}$, becomes

$$H_{c2,\parallel} = \frac{\Phi_0}{2\pi \xi_{GL,ab} \xi_{GL,c}}, \qquad (2.68)$$

where $\xi_{GL,c}$ is the coherence length perpendicular to the planes. Then from Eqs. (2.67) and (2.68), one can obtain the anisotropy ratio for a layered type-II superconductor:

$$\frac{H_{c2,\parallel}}{H_{c2,\perp}} = \frac{\xi_{GL,ab}}{\xi_{GL,c}}. \qquad (2.69)$$

In some layered unconventional superconductors, this anisotropy ratio is extremely large. For example, in highly underdoped cuprates, $H_{c2,\parallel}/H_{c2,\perp} \sim 50$. The order parameter in layered superconductors, $\Psi$, is also anisotropic.

### 4.9.4    Vortices and their interactions

What is the magnetic field at the center of an isolated vortex? In other words, what is the field at $r = 0$ in Fig. 2.10? The value of this field is determined by

the material:

$$H(r = 0) = \frac{\Phi_0}{2\pi\lambda^2}(\ln k - 0.18).\tag{2.70}$$

Comparing Eqs. (2.62) and (2.70), one can see that at $k \gg 1$, the field at the center of an isolated vortex is twice larger than the lower critical field $H_{c1}$. Far from the center of the vortex, the field goes exponentially to zero. At large $k \gg 1$, the numerical term 0.18 in Eqs. (2.62) and (2.70) can be dropped. The penetration of the magnetic field in the interior of a thin superconducting film in the form of vortices is schematically shown in Fig. 2.27.

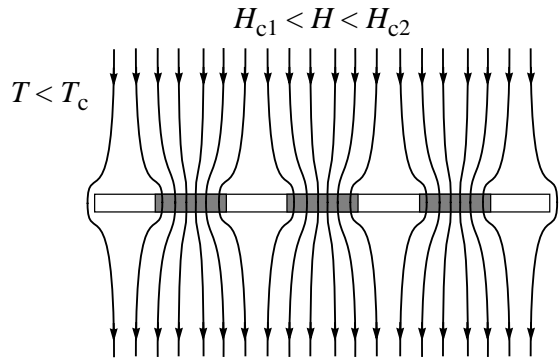

*Figure 2.27.* The mixed state of a thin superconducting film in a perpendicular magnetic field $H$. Vortices shown in grey are normal.

Do the vortices interact with each other? Yes, two neighboring parallel vortices of the same orientation strongly repel each other, and the repulsion force acts only on the vortex core. The same orientation of vortices means that the directions of magnetic field in the vortex cores are the same. How do they interact with one another? As we already know, the radius of a vortex is of the order of the coherence length, $\xi_{GL}$. The supercurrent circulates around the vortex within an area of radius $\sim \lambda$. As long as the distance between the vortices exceeds $\lambda$, they do not "feel" each other. However, when the distance between them becomes less than $\lambda$, the core of one vortex moves into the area where the supercurrent of the other vortex circulates, and vice versa. Since the vortices have the same orientation, the directions of the supercurrents circulating around the vortices also coincide. Then in the area between the vortices, the supercurrents cancel each other, resulting in a difference of Bernoulli pressures exerted on the cores of vortices. Therefore, they repel each other.

In equilibrium, the mutual repulsion of vortices gives rise to a regular vortex lattice with the minimum free energy. As mentioned above, the most favorable regular vortex lattice is triangular, as shown in Fig. 2.25. However, the ideal triangular vortex lattice can only occur in absolutely homogeneous superconductor. As a matter of fact, the free-energy difference between various regular lattice configurations is relatively small; in practice, the material structure will have a greater influence on the vortex pattern. For example, vortices can easily be trapped or "pinned" by defects in the material or by impurities. The impurities and defects, such as grain boundary, dislocation walls, dislocation tangles, voids, second-phase precipitates etc., are often referred to as pinning centers. However, not every defect can interact with vortices effectively. For example, in *conventional* type-II superconductors, vacancies, individual second-phase atoms, or other similar tiny defects are not effective as pinning centers because the characteristic size of a vortex ($\sim \xi_{GL}$) exceeds by far the atomic size and, therefore, the characteristic size of such a defect. A vortex simply does not notice them; they are too small. In contrast, structural defects with dimensions $\sim \xi_{GL}$ and larger are very effective, and can be the cause of very large critical current densities which we shall discuss further below.

How do the vortices in a finite-size superconductor interact with the surface of the superconductor? In the absence of an external magnetic field *at* the surface of a superconductor, vortices are *attracted* to the surface. The interaction of a vortex with the surface can be interpreted as its interaction with its image "existing" at the other side of the surface, thus in vacuum. Since a vortex and its image have opposite orientations, the vortex is attracted to its image and, thus, to the surface. In the presence of an external magnetic field parallel to the surface of the superconductor, the Meissner supercurrent generated by the field will push the vortex away from the surface. Thus, in the presence of a magnetic field, the vortex is repelled from the surface by the Meissner current. At the same time, the vortex is still attracted by its image to the surface. The sign of the net force depends on the value of the external magnetic field $H$. The exact calculations by de Gennes showed that if $H < H_c$, the vortex will be attracted to the surface. Otherwise, the vortex will be repelled from the surface if $H > H_c$.

How do the pinning centers affect a transport current? Consider a type-II superconductor in the mixed state to which a transport current, i.e. a current from an external source, is applied in the direction perpendicular to the vortices. The transport current gives rise to a Lorentz force $[\mathbf{j} \times \mathbf{h}]$ which acts to move the vortices. Their movements lead to the development of a longitudinal potential gradient in the superconductor or, equivalently, to the onset of resistance, thus to dissipative loses in the superconductor. There are several dissipation mechanisms. The main one is connected with the normal phase (the vortex cores) moving through the superconductor. The normal-phase electrons are scattered by the thermal lattice vibrations, resulting in Joule losses. There is also the so-called thermal mechanism of dissipation caused by the fact that vortex motion is always accompanied by energy absorption in the region of the forward boundary of the vortices (the superconducting phase changes into the normal phase). This leads to the appearance of microscopic thermal gradients accompanied by heat flow and energy dissipation. Thus, the transport current through a superconductor in the mixed state is accompanied by generation of heat, which is equivalent to saying that the critical current is vanishingly small.

However, the picture changes drastically in the presence of pinning centers which trap the vortices. Then, the vortices will start moving only if the Lorentz force becomes strong enough to overcome pinning and tear the vortices off the centers. In other words, the Lorentz force must exceed a certain value. The current density corresponding to the initiation of vortex break-off from the pinning centers is the so-called critical current density $j_c$.

The critical current density is a structure-sensitive property and can vary by as much as several orders of magnitude as a result of thermal or mechanical treatment of the material. At the same time, the critical temperature $T_c$ and the upper critical field $H_{c2}$ can remain virtually unaffected. These specially prepared type-II superconductors are called the *hard* superconductors in which the pinning force is sufficiently large to prevent flux motion, resulting in the resistanceless current flow. The hard superconductors are also called the type-III superconductors.

To summarize, owing to the presence of vortices, hard superconductors can withstand large magnetic fields, while structural inhomogeneities make it possible to pass large currents through them. If the transport current exceeds $j_c$, the Lorentz force becomes stronger than the pinning force, and the vortices are depinned. Even though this state is no longer dissipationless as in a usual superconductor, the resistivity of a superconductor at such conditions is still lower than that of the same sample in the normal state. Interestingly, at $H \to H_{c2}$, instead of monotonically vanishing, the critical current has a local peak near $H_{c2}$. This is the so-called *peak effect*. One of possible explanations of this strange effect is that at $H \to H_{c2}$, the elastic moduli of the vortex lattice decreases, that is, the lattice becomes softer and pinning becomes stronger [27].

## 4.10    Suppression of the superconducting state

The superconducting state requires the electron pairing *and* the onset of long-range phase coherence. They are two different and *independent* phenomena. As we already know, superconductivity can be suppressed, for example, by a sufficiently strong magnetic field. Then, a question rises: what does a sufficiently strong magnetic field destroy first, the electron pairing or the long-range phase coherence? This question has never been considered in earlier textbooks on the physics of superconductivity because, for *conventional* superconductors, this question has no sense. In conventional superconductors, it is always the electron pairing which is suppressed. The long-range phase coherence in conventional superconductors is mediated by the overlap of Cooper-pair wavefunctions, and the density of Cooper pairs is relatively high. There-fore, as long as the Cooper pairs exist, and their density is high, the onset of long-range phase coherence arises *automatically* (and at the same temperature as the electron pairing occurs). So, in conventional superconductors, it is impossible to discontinue the spread of phase coherence without breaking Cooper pairs. The suppression of the superconducting state in conventional superconductors can *only* be achieved by breaking up the Cooper pairs.

However, this is not the case for unconventional superconductors, specially in layered ones. In these superconductors, the onset of long-range phase coherence occurs due to a mechanism which different from the overlap of Cooper-pair wavefunctions. Therefore, in unconventional superconductors, the onset of long-range phase coherence is *always* the weakest link, and will be suppressed first, for example, by a sufficiently strong magnetic field. Of course, the electron pairing in unconventional superconductors can be suppressed too, but for this, one must increase the magnitude of magnetic field in comparison with that needed to destroy the phase coherence.

In the context of the above discussion, the difference between conventional and unconventional superconductors can be illustrated by the following example. In conventional superconductors, by applying a sufficiently strong magnetic field, the part of resistivity curve corresponding to the transition into the superconducting state (see, for example, Fig. 2.1) remains steplike but is only shifted to lower temperatures. The same takes place in half-conventional superconductors (see Chapter 7). Contrary to this, the transition width in unconventional superconductors becomes broader with increasing magnetic fields, meaning that at temperatures just below $T_c(H = 0)$, there are large phase fluctuations. In *layered* unconventional superconductors, the transition widths in resistivity become broader in both directions, along and perpendicular to the layers. The onset of long-range phase coherence perpendicular to the layers is usually affected by an applied magnetic field to a higher degree than that along the layers because the *in-plane* phase coherence is usually established by two independent processes, one of which is the same as that perpendicular to the layers, and the second is the direct hopping, overlap of Cooper-pair wavefunctions. We shall discuss the mechanism of phase coherence in layered unconventional superconductors in detail in Chapter 6.

In unconventional superconductors, knowing the magnitude of a magnetic field at which the onset of long-range phase coherence is discontinued, one can estimate the value of the *coherence* length by using Eq. (2.23). By increasing the magnitude of magnetic field (if the laboratory conditions allow to do so), one can then estimate the value of the coherence length of electron pairing, thus the *size* of Cooper pairs. One can now understand why, while defining the coherence length and the size of Cooper pairs (see above), it was underlined that generally speaking, the two notions—the coherence length and Cooper-pair size—are not the same. From Eq. (2.23), one can also grasp that in unconventional superconductors, at any temperature $T < T_c$,

$$\xi_0 < \xi_{GL}, \tag{2.71}$$

where $\xi_{GL}$ is the coherence length, and $\xi_0$ is the Cooper-pair size. It is worth to recall that $\xi_{GL}$ depends on temperature, while $\xi_0$ is temperature-independent.

To finish this subsection, it is worth noting that superconductivity is not always destroyed by a sufficiently strong magnetic field. In fact, superconductivity can be induced by a very strong magnetic field! As an example, in the quasi-two-dimensional organic conductor $\lambda$-$(BETS)_2FeCl_4$, where BETS stands for bis-(ethylenedithio)tetraselenafulvalene, the superconducting phase is induced by a magnetic field exceeding 18 Tesla [26, 27]. Crystalline $\lambda$-$(BETS)_2FeCl_4$ consists of layers of highly conducting BETS sandwiched between insulating layers of iron chloride $FeCl_4$. The field is applied parallel to the conducting layers. This is particularly remarkable since this compound at zero field is an antiferromagnetic insulator below 8.5 K. Field-induced superconductivity was earlier reported for $Eu_xSn_{1-x}Mo_6S_8$ but this compound is paramagnetic above $T_c = 3.8$ K. In $\lambda$-$(BETS)_2FeCl_4$, the $Fe^{3+}$ ions within the $FeCl_4$ molecules are responsible for a long-range antiferromagnetic order.

Field-induced superconductivity in magnetic materials is usually discussed in terms of the *Jaccarino-Peter compensation*, in which the applied field compensates the internal magnetic field provided by the magnetic ions. By increasing the magnitude of applied magnetic field above $B = 41$ Tesla, the compound $\lambda$-$(BETS)_2FeCl_4$ becomes metallic above 0.8 K [27]. The dependence $T_c(H)$ has a bell-like shape with a maximum $T_c \simeq 4.2$ K near 33 Tesla. Interestingly, a magnetic field of only 0.1 Tesla, applied perpendicular to the conducting BETS planes, destroys the superconducting state. This indicates that, in this organic conductor, superconductivity is robust along the $c$ axis and, at the same time, week in the planes (in the cuprates, it is the other way round).

## 5.    Universal theory of the superconducting state

We discuss here the Ginzburg-Landau theory which is in fact universal in the sense that it is applicable to any superconductor independently of the material *and* the mechanism of superconductivity. A major early triumph of the Ginzburg-Landau theory was in handling the mixed state of superconductors, in which superconducting and normal domains coexist in the presence of $H \approx H_c$. Thus, the Ginzburg-Landau theory is able to describe the behavior of spatially inhomogeneous superconductors in which the spatial variations of the order parameter $\Psi(\mathbf{r})$ and the vector potential $\mathbf{A}$ are not too rapid. However, the main disadvantage of the Ginzburg-Landau theory is that it can only be applied at temperatures sufficiently near the critical temperature: $T_c - T \ll T_c$. The range of validity of the Ginzburg-Landau theory will be discussed in more detail at the end of this section.

The Ginzburg-Landau theory is phenomenological and based on the general theory of second-order phase transitions developed by Landau. According to this theory, a phase transition of the second order occurs when the state of a body changes gradually while its symmetry changes discontinuously at the transition temperature. Furthermore, the low-temperature phase with the reduced symmetry is a more ordered one. One of the examples of a second-order phase transition is the ferromagnetic transition at the Curie temperature $T_{Cur}$, that is, the transition from paramagnetic to ferromagnetic state. The spontaneous magnetization $M$ of the sample appears at $T_{Cur}$, and its magnitude increases on cooling. Close to $T_{Cur}$, the thermodynamics of such a system can be described by expanding the Helmholtz free energy $F(T, M)$ in powers of magnetization $M$ which is small near $T_{Cur}$:

$$F(T, M) = F(T, 0) + a(T - T_{Cur})M^2 + bM^4 + c|\nabla M|^2, \qquad (2.72)$$

where $a$, $b$ and $c$ are the expansion coefficients. By minimizing the free energy with respect to $M$, one can obtain that

$$M = 0 \text{ for } T > T_{Cur}, \qquad (2.73)$$

$$M \neq 0 \text{ for } T < T_{Cur}. \qquad (2.74)$$

Using this analysis as the starting point, one can describe the magnetic transition. Assuming that any second-order phase transition can be described in the same manner, Landau suggested that the magnetization $M$ in Eq. (2.72) can be replaced by another quantity, and in the case of a superconducting transition, by the order parameter $\Psi(\mathbf{r})$. This assumption inspired Ginzburg and Landau to develop a simple and exact description of superconducting properties near the critical temperature.

### 5.0.1   Equations of the Ginzburg-Landau theory

In the framework of the two-fluid model, in a superconductor below $T_c$, there are superconducting and normal electrons. Assume that in an inhomogeneous superconductor, the superconducting electrons are described by an order parameter $\Psi(\mathbf{r}) = |\Psi(\mathbf{r})|\, e^{i\theta}$, so that $|\Psi(\mathbf{r})|^2$ gives the *local* density of the Cooper pairs, $n_s(\mathbf{r})/2$, where $n_s(\mathbf{r})$ is the local density of superconducting electrons. The basic postulate of the Ginzburg-Landau theory is that if $\Psi$ of an inhomogeneous superconductor in a uniform external magnetic field is small and varies slowly in space, the Helmholtz free-energy density $F_s(\mathbf{r}, T)$ can be expanded in a series of the form

$$F_s = F_n + \alpha|\Psi|^2 + \frac{\beta}{2}|\Psi|^4 + \frac{1}{2m^*}\Big|\left(-i\hbar\nabla - \frac{e^*}{c}\mathbf{A}\right)\Psi\Big|^2 + \frac{\mathbf{h}^2}{8\pi}, \quad (2.75)$$

where $F_n$ is the free energy of the superconductor in the normal state; $\mathbf{h}$ is the local magnetic field; $\mathbf{A}$ is the local vector potential and $\mathbf{h} = \text{curl}\mathbf{A}$; $e^* =$

$2e$ and $m^* = 2m$ are respectively the effective charge and mass of Cooper pairs, and $\alpha$ and $\beta$ are the phenomenological expansion coefficients which are characteristics of the material. The order parameter is normalized that $|\Psi(\mathbf{r})|^2$ gives the density of Cooper pairs, or a half of the density of superconducting electrons, $n_s/2$.

The total free energy of the superconductor is

$$F_s(T) = \int\limits_V F_s(\mathbf{r}, T)\, \mathrm{d}^3 r, \tag{2.76}$$

where $V$ is the volume of the sample.

Evidently, if $\Psi = 0$, Equation (2.75) reduces properly to the free energy of the normal state $F_n + \mathbf{h}^2/8\pi$. In the absence of fields and gradients, we have

$$F_s = F_n + \alpha|\Psi|^2 + \frac{\beta}{2}|\Psi|^4. \tag{2.77}$$

Let us find the value of $|\Psi|^2$ for which the free energy in a homogeneous superconductor is minimum. This value is the solution of the equation

$$\frac{\mathrm{d}F_s}{\mathrm{d}|\Psi|^2} = 0. \tag{2.78}$$

Carrying out elementary calculations we obtain

$$|\Psi_{min}|^2 = -\frac{\alpha}{\beta}. \tag{2.79}$$

Substituting this expression into Eq. (2.77), we find the difference in energy

$$F_n - F_s = \frac{\alpha^2}{2\beta}. \tag{2.80}$$

Recalling that, from Eq. (2.27), this difference equals $H_c^2/8\pi$, we have

$$H_c^2 = \frac{4\pi\alpha^2}{\beta}. \tag{2.81}$$

Let us discuss the temperature dependence of the coefficients $\alpha$ and $\beta$. Since the order parameter must be zero at $T = T_c$, and finite at $T < T_c$, it follows from Eq. (2.79) that $\alpha = 0$ at $T = T_c$ and $\alpha < 0$ at $T < T_c$. Therefore, in a first approximation, we can write

$$\alpha \propto (T - T_c). \tag{2.82}$$

This temperature dependence of $\alpha$ correlates Eq. (2.81) with the empirical formula for $H_c$ near $T_c$, given by Eq. (2.20).

It follows from Eqs. (2.79) and (2.81) that the coefficient $\beta$ must be positive. From Eqs. (2.20) and (2.82), we obtain that, in a *first approximation*, $\beta$ is independent of temperature. So, the coefficient $\beta$ is a positive constant. Then, from Eqs. (2.79) and (2.82), we get

$$|\Psi|^2 \propto (T - T_c) \tag{2.83}$$

for temperatures near, but below, $T_c$.

Let us go back to Eq. (2.75). Taking into account that $\Psi = |\Psi|\,e^{i\theta}$, we rewrite Eq. (2.75) as

$$F_s = F_n + \alpha|\Psi|^2 + \frac{\beta}{2}|\Psi|^4 + \frac{\hbar}{2m}(\nabla|\Psi|)^2 + \frac{1}{2}|\Psi|^2 m \mathbf{v}_s^2 + \frac{\mathbf{h}^2}{8\pi}, \tag{2.84}$$

where we introduced

$$\mathbf{v}_s = \frac{1}{m}(\hbar\nabla\theta - \frac{2e}{c}\mathbf{A}). \tag{2.85}$$

In Eq. (2.84), one can see that we have obtained the Landau expansion given by Eq. (2.72), plus the free energy of the magnetic field and the current. If the order parameter does not vary in space, one gets back exactly to the London free energy and the London equations by carrying out the minimization. Thus, the Ginzburg-Landau free energy is the way to introduce the London idea in the usual second-order phase transition.

In order to determine the order parameter $\Psi(\mathbf{r})$ and the vector potential $\mathbf{A}(\mathbf{r})$, we minimize the Helmholtz free energy with respect to $\Psi$ and $\mathbf{A}$. By this double minimization, one gets two equations named after their authors, the *Ginzburg-Landau equations*

$$\frac{1}{4m}\left(i\hbar\nabla + \frac{2e}{c}\mathbf{A}\right)^2 \Psi + \beta|\Psi|^2\Psi = -\alpha(T)\Psi, \tag{2.86}$$

$$\mathbf{j}_s = -\frac{ie\hbar}{2m}(\Psi^{ast}\nabla\Psi - \Psi\nabla\Psi^*) - \frac{2e^2}{mc}|\Psi|^2\mathbf{A}. \tag{2.87}$$

These two equations are coupled and should therefore be solved simultaneously. The first equation gives the order parameter while the second enables one to describe the supercurrent that flows in the superconductor ($\mathbf{j}_s = c/4\pi \times$ curl $\mathbf{h}$). It is worth noting that the first equation is analogous to the Schrödinger equation for a free particle, but with an additional nonlinear term $\beta|\Psi|^2\Psi$.

In carrying through the variational procedure, it is necessary to provide boundary conditions. One possible choice, which assures that no supercurrent passes through the surface, is

$$\left(i\hbar\nabla\Psi + \frac{2e}{c}\mathbf{A}\,\Psi\right) \cdot \mathbf{n} = 0, \tag{2.88}$$

where $\mathbf{n}$ is the unit vector normal to the surface of the superconductor. This boundary condition used by Ginzburg and Landau is also appropriate at an insulating surface. Using the microscopic theory, de Gennes showed that for a normal metal-superconductor interface with no current, Equation (2.88) must be generalized to

$$\left(i\hbar\nabla\Psi + \frac{2e}{c}\mathbf{A}\Psi\right)\cdot\mathbf{n} = \frac{\hbar}{i\,b}\Psi, \tag{2.89}$$

where $b$ is a real constant. If $A_n = 0$, $b$ is the extrapolation length shown in Fig. 2.24. The value of $b$ depends on the nature of the material to which contact is made, approaching zero for a magnetic material and infinity for an insulator, with normal metals lying in between.

### 5.0.2 Two characteristic lengths

If we introduce two characteristic lengths

$$\xi_{GL}^2 = \frac{\hbar^2}{4m|\alpha|} \quad \text{and} \tag{2.90}$$

$$\lambda^2 = \frac{mc^2}{4\pi n_s e^2} = \frac{mc^2\beta}{8\pi e^2|\alpha|}, \tag{2.91}$$

the Ginzburg-Landau equations can be written in a more concise and convenient form

$$\xi_{GL}^2\left(i\nabla + \frac{2\pi}{\Phi_0}\mathbf{A}\right)^2\psi - \psi + \psi|\psi|^2 = 0, \tag{2.92}$$

$$\operatorname{curl}\operatorname{curl}\mathbf{A} = -i\frac{\Phi_0}{4\pi\lambda^2}(\psi^*\nabla\psi - \psi\nabla\psi^*) - \frac{|\psi|^2}{\lambda^2}\mathbf{A}, \tag{2.93}$$

where $\psi(\mathbf{r}) = \Psi(\mathbf{r})/\Psi(\infty)$ is a dimensionless wavefunction, and $\Phi_0 \equiv \pi\hbar c/e$ is the flux quantum. Furthermore, taking into account that the order parameter has the form $\psi = |\psi|\,\mathrm{e}^{i\theta}$, the second Ginzburg-Landau equation for a simply connected (not a multiple connected) superconductor becomes

$$\operatorname{curl}\operatorname{curl}\mathbf{A} = \frac{|\psi|^2}{\lambda^2}\left(\frac{\Phi_0}{2\pi}\nabla\theta - \mathbf{A}\right). \tag{2.94}$$

Let us now find the physical significance of the two characteristic lengths $\xi_{GL}$ and $\lambda$. We start with $\xi_{GL}$. Consider a normal metal in contact with a clean flat surface of a superconductor, as shown in Fig. 2.24. Let us take the $x$ axis perpendicular to the surface of the superconductor, with the origin ($x = 0$) at the surface. Then it is obvious that $\psi$ can vary only along the $x$ axis, i.e. $\psi = \psi(x)$. In the absence of external magnetic field, $\mathbf{A} = 0$, the differential equation Eq. (2.92) has only real coefficients. Then, this equation can be reduced to a simple form

$$\xi_{GL}^2 \frac{d^2\psi}{dx^2} + \psi - \psi^3 = 0. \tag{2.95}$$

Suppose that the normal layer at the surface is so thin that the magnitude of $\psi$ at the surface is not very different from 1, that is,

$$\psi(x) = 1 - \varepsilon(x) \quad \text{and} \quad \varepsilon(x) \ll 1. \tag{2.96}$$

Substituting this expression into Eq. (2.95) and keeping only linear terms in $\varepsilon(x)$, we get

$$\xi_{GL}^2 \frac{d^2\varepsilon(x)}{dx^2} - 2\varepsilon(x) = 0. \tag{2.97}$$

Taking into account that $\psi \to 1$ as $x \to \infty$, we have $\varepsilon(\infty) = 0$. Then the solution of Eq. (2.97) is

$$\varepsilon(x) = \varepsilon_0 e^{-\sqrt{2}x/\xi_{GL}} \tag{2.98}$$

which shows that a small disturbance of $\psi$ from 1 will decay in a characteristic length of order $\xi_{GL}$. Then, we can call this length the coherence length.

The other quantity, $\lambda$, is already known to us [see Eq. (2.9)]. This is the penetration depth for a weak magnetic field. Both $\xi_{GL}$ and $\lambda$ are temperature-dependent. Taking into account Eq. (2.82), we find that in the vicinity of $T_c$,

$$\xi_{GL}^2 = \frac{\hbar^2}{4m|\alpha|} \propto (T_c - T)^{-1}, \tag{2.99}$$

$$\lambda^2 = \frac{mc^2}{4\pi n_s e^2} = \frac{mc^2\beta}{8\pi e^2|\alpha|} \propto (T_c - T)^{-1}. \tag{2.100}$$

As a consequence, in the vicinity of $T_c$, the Ginzburg-Landau parameter $k = \lambda/\xi_{GL}$ is temperature-independent. One can demonstrate that the Ginzburg-Landau free energy depends only on $k$. This means that the Ginzburg-Landau parameter $k$ phenomenologically characterizes completely a given superconductor.

Combining Eqs. (2.19), (2.99) and (2.100), one can obtain an important relationship among characteristic quantities of a superconductor,

$$H_c(T)\lambda(T)\xi_{GL}(T) = \frac{\Phi_0}{2\sqrt{2}\pi}, \tag{2.101}$$

which was already discussed above in SI units [see Eq. (2.21)].

The relation between the Ginzburg-Landau theory and the BCS microscopic theory derived for conventional superconductors will be considered in Chapter 5.

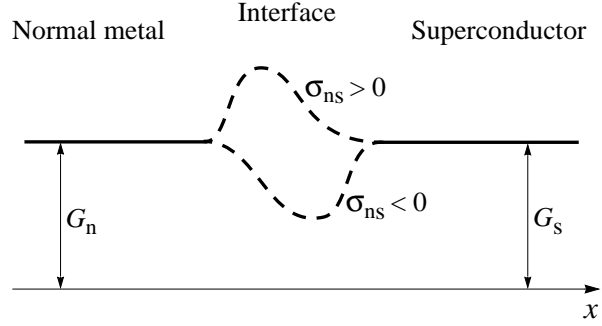

*Figure 2.28.* Density of the Gibbs free energy $G_{ns}$ in the vicinity of a normal metal-superconductor interface in an external magnetic field. In type-I superconductors, the surface energy of the interface, shown by broken lines, is positive, $G_{ns} - G_n = \sigma_{ns} > 0$, while in type-II superconductors, $\sigma_{ns}$ is negative.

### 5.0.3  Surface energy at the NS interface

Consider now the normal metal-superconductor (NS) interface, shown in Fig. 2.24, in an external magnetic field, $\mathbf{A} \neq 0$. As we already know, type-I and type-II superconductors can show different responses to an applied magnetic field. The reason is that the surface energy of the interface between a normal metal and a superconductor, $\sigma_{ns}$, is positive for type-I superconductors and negative for type-II superconductors. To show this, one must calculate the Gibbs free energy deep inside the superconductor, $G_s$, and that deep inside the normal metal, $G_n$. In equilibrium, one easily finds that $G_s = G_n$, thus the density of the Gibbs free energy far to the right of the NS interface equals the energy density far to the left, as schematically shown in Fig. 2.28. At the same time, the calculation of the Gibbs free energy at the NS interface, $G_{ns}$, shows that in general, $\sigma_{ns} = G_{ns} - G_n \neq 0$. Thus, the density of the Gibbs free energy at the NS interface differs from those inside the superconductor and normal metal. In order to evaluate $\sigma_{ns}$, one must also resolve the Ginzburg-Landau equations given by Eqs. (2.92) and Eq. (2.94).

In the case $k \ll 1$ (i.e. $\lambda \ll \xi_{GL}$), one obtains that

$$\sigma_{ns} = 1.89 \frac{H_c^2}{8\pi} \xi_{GL} > 0, \qquad (2.102)$$

thus, the surface energy of the NS interface in type-I superconductors is always positive.

In the case $k \gg 1$ (i.e. $\lambda \gg \xi_{GL}$), the exact calculations yield that the surface energy of the NS interface in type-II superconductors,

$$\sigma_{ns} = -\frac{H_c^2}{8\pi} \lambda < 0, \qquad (2.103)$$

is always negative.

Let us now interpret the results physically.

(1) In the case of type-I superconductors, $\lambda \ll \xi_{GL}$ (or $k \ll 1$), the variations of the order parameter $\psi$ and the magnetic field $H$ in the vicinity of the interface are sketched in Fig. 2.7a. The former falls off over a distance $\xi_{GL}$ and the latter over a distance $\lambda$. In the vicinity of the interface, there is a region of thickness $\sim \xi_{GL}$ where the order parameter is sufficiently small, and the magnetic field is kept out. Since $\psi$ is small, in order to keep the magnetic field out of this region, work must be done on the magnetic field to expel it from this region. This means that one must overcome a magnetic pressure $H_c^2/8\pi$ and shift its boundary by the distance $\xi_{GL}$. Then, this work is about $(H_c^2/8\pi)\xi_{GL}$, in accord with Eq. (2.102).

(2) In the case of type-II superconductors, $\lambda \gg \xi_{GL}$ (or $k \gg 1$), the variations of the order parameter $\psi$ and the magnetic field $H$ in the vicinity of the interface are shown in Fig. 2.7b. In the vicinity of the interface, there is a region of thickness $\sim \lambda$ with $\psi \sim 1$ which is penetrated by the magnetic field. It implies that the free energy of this region must be smaller in comparison with that far from the NS interface (on the left or on the right) to shift the magnetic field $H_c$ by the distance $\lambda$. Then, this difference in free energy is about $(H_c^2/8\pi)\lambda$, in agreement with Eq. (2.103).

Obviously, at some value $k \sim 1$, the energy $\sigma_{ns}$ must be zero. The exact calculations made by Abrikosov show that this occurs at $k = 1/\sqrt{2}$. This value "separates" type-I and type-II superconductors.

### 5.0.4   The range of validity

Let us establish the range of validity of the Ginzburg-Landau theory. In the series expansion of the Helmholtz free energy density in powers of $|-i\hbar\nabla\Psi - (2e/c)\mathbf{A}\Psi|^2$, given by Eq. (2.75), only the first term has been kept. This means that only slow changes of $\Psi$ and $\mathbf{A}$ are assumed over distances comparable with the characteristic size of an inhomogeneity in the superconductor, that is, over the size of the Cooper pair. Consider the cases of type-I and type-II superconductors separately.

In the clean limit when the electron mean free path is much larger than the size of the Cooper pair, $\ell \gg \xi_0$, the Ginzburg-Landau theory is valid if $\xi_{GL}(T) \gg \xi_0$ and $\lambda(T) \gg \xi_0$. Since $\xi_{GL}(T) \sim \xi_0(1 - T/T_c)^{-1/2}$ [see Eq. (2.16)], then the coherence length $\xi_{GL}(T)$ always exceeds the Cooper-pair size $\xi_0$. So, at $T \sim T_c$, the first condition, $\xi_{GL}(T) \gg \xi_0$, is automatically satisfied. The second condition, $\lambda(T) \gg \xi_0$, in fact, represents the requirement that the local electrodynamics is applicable, or in other words, that the superconductor is of the London type. Since $\lambda(T) \propto (1 - T/T_c)^{-1/2}$ and $\lambda(T) \to \lambda_L$ at $T \to 0$, and $k \sim \lambda_L/\xi_0$, then the condition $\lambda(T) \gg \xi_0$ reduces to $k^2 \gg (1 - T/T_c)$, which is a rather strict condition because $k$ in type-I superconductors can be very small. For example, in Al $k = 0.01$.

In the dirty limit ($\ell \ll \xi_0$), the validity interval for the Ginzburg-Landau theory is much wider than that for clean superconductors. For "dirty" superconductors, the characteristic scale of inhomogeneity is the mean free path $\ell$. This means that the Ginzburg-Landau theory can be applied if $\xi_{GL}(T) \gg \ell$ and $\lambda(T) \gg \ell$. Since $\xi_{GL}(T) \sim (\xi_0 \ell)^{1/2}(1 - T/T_c)^{-1/2}$ [see Eq. (2.17)], the condition $\xi_{GL}(T) \gg \ell$ reduces to $\xi_0/\ell \gg (1 - T/T_c)$. In addition, since $\xi_0 \gg \ell$, this condition is much less strict than the general condition for the Ginzburg-Landau theory, namely, $T_c - T \ll T_c$. For the second condition, $\lambda(T) \gg \ell$, recalling that $\lambda(T) \propto (1 - T/T_c)^{-1/2}$ and $\lambda \to \lambda_L(\xi_0/\ell)^{1/2}$ at $T \to 0$ [see Eq. (2.13)], and $k \sim \lambda_L/\ell$, then it can be rewritten as $k^2(\xi_0/\ell) \gg (1 - T/T_c)$. Even if $k \sim 1$, one can find again that it is less strict than the general condition $T_c - T \ll T_c$. Thus, in the case of type-II superconductors, the Ginzburg-Landau theory is valid within a rather wide temperature interval, and automatically provided the general condition $T_c - T \ll T_c$ to be satisfied.

## APPENDIX: Books recommended for further reading

# Chapter 3

# SUPERCONDUCTING MATERIALS

This book deals with a problem which is directly connected with the materials physics. Therefore, it is worthwhile to review materials that superconduct. In this chapter, we shall only give a brief introduction to such materials. At present, there are about 7000 known superconducting compounds, and it is impossible in one chapter to give a detailed description of all these superconducting materials or even to cover only their principle classes. Indeed, all superconducting materials can be classified into several groups according to their crystal structure and their properties.

In addition to such a classification of superconducting materials, they can also be sorted according to the mechanism of superconductivity in each compound. Recently, it was shown that the mechanisms of superconductivity in various compounds are different and, basically, there are three types of superconducting mechanisms [19]. In Chapters 5, 6 and 7, we shall discuss these three types of superconducting mechanisms in detail. However, already in this chapter, superconducting materials are divided into these three groups. In a first approximation, these three groups consist of the following superconducting materials:

   **1) metals and some of their alloys,**
   **2) low-dimensional, non-magnetic compounds,**
   **3) low-dimensional, magnetic compounds.**

The phenomenon of superconductivity occurs in solids, and they exist in the form of single crystals, thin films and polycrystalline ceramics (consisting of a large number of micron-size single crystals). Interestingly, some materials superconduct only in one of these forms but not in the other. Some compounds become superconducting exclusively under high pressure or when irradiated. In a few unconventional superconductors, the superconducting state is induced by an applied magnetic field.

## 1.    First group of superconducting materials

The first group of superconductors incorporates non-magnetic elemental superconductors and some of their alloys. The superconducting state in these materials is well described by the BCS theory of superconductivity presented in Chapter 5. Thus, this group of superconductors includes all classical, conventional superconductors. The critical temperature of these superconductors does not exceed 10 K. Most of them are type-I superconductors. As a consequence, superconductors from this group are not suitable for applications because of their low transitional temperature and low critical field. The phenomenon of superconductivity was discovered by Kamerlingh Onnes and his assistant Gilles Holst in 1911 in mercury—a representative of this group.

Ironically, many superconductors, discovered mainly before 1986, were assigned to this group by mistake. In fact, they belong to either the second or third group of superconductors. For example, the so-called A-15 superconductors, during a long period of time, were considered as conventional; in reality, they belong to the second group. The so-called Chevrel phases were first assigned also to the first group; however, superconductivity in Chevrel phases is of unconventional type, and they are representatives of the third group of superconductors.

In the periodic table of chemical elements, over half of the elements can exhibit the superconducting state. However, sixteen of them (at the moment of writing) superconduct when made into thin films, under high pressure, or irradiated. Most metallic elements are superconductors, and some of them are listed in Table 2.1. However, noble metals—copper, silver and gold—which are excellent conductors of electricity at room temperature never become superconducting. So, superconductivity occurs rather in "bad" metals than in the best conductors. This fact will be explained in Chapter 5. One of the magnetic metals, iron, exhibits superconductivity under extremely high pressure (superconductivity in Fe is of unconventional type). The semiconductors Si and Ge become superconducting under a pressure of $\sim 2$ kbar with $T_c = 7$ and 5.3 K, respectively. At a pressure of 15.2 kbar, the critical temperature of Si increases to $T_c = 8.2$ K. Other elements that superconduct under pressure include As, Ba, Bi, Ce, Cs, Li, P, Sb, Se, Te, U and Y.

The critical temperature of some elements is raised dramatically by preparing them in thin films. For example, $T_c$ of tungsten (W) was increased from its bulk value of 0.015 K to 5.5 K in a thin film; molybdenum (Mo) exhibits an increase from 0.92 K to 7.2 K and titanium (Ti) from 0.42 K to 2.52 K. At ambient pressure, chromium (Cr) superconducts only in the thin-film state; other non-superconductors, such as bismuth (Bi), cesium (Cs), germanium (Ge), lithium (Li) and silicon (Si) can be converted into superconductors by either applying pressure or preparing them as thin films.

A very limited number of metallic alloys belong to this group. For example, the alloy NbTi listed in Table 2.2 belongs most likely to the second group of superconductors. So, the number of superconductors in this group is very small, and they are not suitable for applications. The classical type of superconductivity (the BCS type) occurs only in superconductors of this group.

## 2. Second group of superconducting materials

The second group of superconductors incorporates low-dimensional, non-magnetic compounds. The superconducting state in these materials is characterized by the presence of two interacting superconducting subsystems. One of them is low-dimensional and exhibits genuine superconductivity of unconventional type, while superconductivity in the second subsystem which is three-dimensional is induced by the first one and of the BCS type. So, superconductivity in this group of materials can be called half-conventional (or alternatively, half-unconventional). We shall discuss the mechanism of superconductivity in these half-conventional superconductors in Chapter 7. The critical temperature of these superconductors is limited by $\sim 40$ K and, in some of them, $T_c$ can be tuned. All of them are type-II superconductors with an upper critical magnetic field usually exceeding 10 T. Therefore, many superconductors from this group are suitable for different types of practical applications.

### 2.1 A-15 superconductors

Intermetallic compounds of transition metals of niobium (Nb) and vanadium (V) such as $Nb_3B$ and $V_3B$, where B is one of the nontransitional metals, have the structure of beta-tungsten ($\beta$-W) designated in crystallography by the symbol A-15. As a consequence, superconductors having the structure $A_3B$  (A = Nb, V, Ta, Zr and B = Sn, Ge, Al, Ga, Si) are called *the A-15 superconductors*.

Figure 3.1 shows the crystal structure of the binary $A_3B$ compounds. The atoms B form a body-centered cubic sublattice, while the atoms A are situated on the faces of the cube forming three sets of non-interacting orthogonal *one-dimensional* chains. The distance between atoms A on the chains is about 22% shorter than the distance between chains.

The first A-15 superconductor $V_3Si$ was discovered by Hardy and Hulm in 1954. Nearly 70 different A-15 superconductors were already known in 1985. Before the discovery of superconductivity in cuprates, the A-15 superconductors had the highest $T_c$. Table 3.1 lists some characteristics of six A-15 superconductors with the highest values of $T_c$. The critical temperature of A-15 superconductors is very sensitive to changes in the 3:1 stoichiometry. These materials are also very sensitive to the effects of radiation damage.

In addition to unusually high $T_c$ values, the A-15 superconductors display several superconducting properties which cannot be explained in the frame-work of the BCS theory. The coherence length of the A-15 superconductors is very short, $\xi_0 \simeq 35$–200 Å. They have extraordinary soft acoustic and optical phonon modes. A lattice instability preceding a structural phase transition and then followed by superconductivity is typical for the A-15 compounds. This structural phase transition is called *martensite*. The presence of densely packed one-dimensional chains is believed to be responsible for this crystalline instability. The phase-transition temperature $T_m$ for $V_3Si$ and $Nb_3Sn$ is 20.5 K and 43 K, respectively. In $V_3Ga$ and $Nb_3Al$, this transition occurs at about 50 K and 80 K, respectively. The symmetry transition, from cubic to tetragonal, is accompanied only by a rearrangement of the crystal lattice, with the volume of the crystal remaining unchanged. Immediately after the transition, there is a softening of the elastic coefficients of the lattice, as observed in acoustic measurements. The Debye temperature of the A-15 superconductors is moderate, 300–500 K. It has been established that the more metastable lattice exhibits the higher $T_c$ value. One of the main factors in increasing the $T_c$ value is the softening of the phonon spectrum, i.e. the phonon spectrum shifts to lower phonon frequencies.

*Table 3.1.*   Characteristics of A-15 superconductors: the critical temperature $T_c$; the upper critical magnetic field $H_{c2}$; and the gap ratio $\frac{2\Delta}{k_B T_c}$ inferred from infrared measurements.

| Compound | $T_c$ (K) | $H_{c2}$ (T) | $\frac{2\Delta}{k_B T_c}$ |
|----------|-----------|--------------|----------------------------|
| $Nb_3Ge$ | 23.2 | 38 | 4.2 |
| $Nb_3Ga$ | 20.3 | 34 | - |
| $Nb_3Al$ | 18.9 | 33 | 4.4 |
| $Nb_3Sn$ | 18.3 | 24 | 4.2-4.4 |
| $V_3Si$ | 17.1 | 23 | 3.8 |
| $V_3Ga$ | 15.4 | 23 | - |

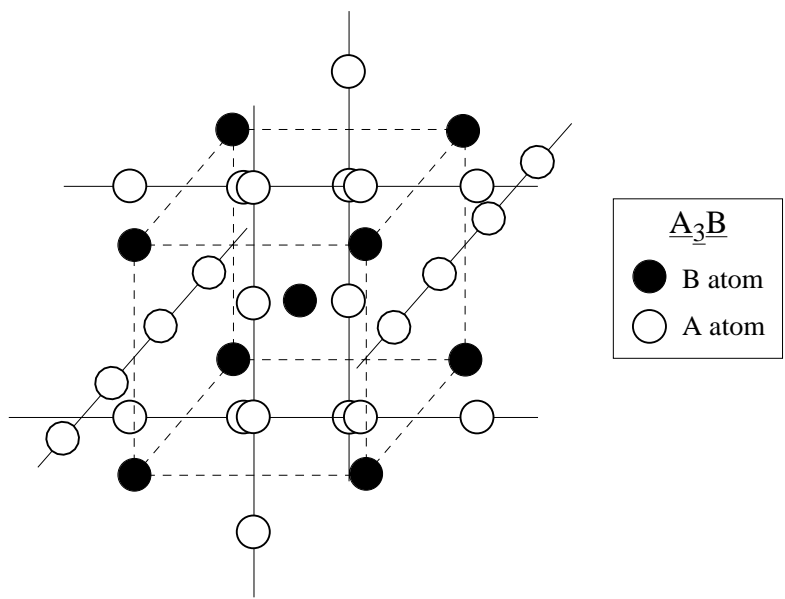

*Figure 3.1.*   Crystal structure of $A_3B$ compounds (A-15 superconductors). The atoms A form one-dimensional chains on each face of the cube. Chains on the opposite faces are parallel, while on the neighboring faces orthogonal to each other.

It is worth noting that the search of high-temperature superconductors in materials exhibiting structural instabilities was already proposed in 1971 [28]. Even at that time, it was already known that metals with strong electron-lattice coupling tend towards structural instabilities.

In spite of the fact that the A-15 compounds exhibit high critical temperatures and upper magnetic fields (see Table 3.1), they are not widely used in applications because they are too brittle and therefore not flexible enough to be drawn into wires. In contrast, the alloy niobium-titanium (NbTi) with lower $T_c = 9.5$ K is easily drawn into wires; hence it is much more useful for applications. This problem was however solved for $Nb_3Sn$ and $V_3Ga$ by the use of a technique called the bronze process. Magnets made from these wires can produce magnetic fields of about 20 T at 4.2 K.

## 2.2    Metal oxide $Ba_{1-x}K_xBiO_3$

In 1975, Sleight and co-workers discovered superconductivity in the metal oxide $BaPb_{1-x}Bi_xO_3$ with a maximum $T_c \simeq 13.7$ K at $x = 0.25$. Other members of this family, $BaPb_{0.75}Sb_{0.25}O_3$ ($T_c = 0.3$ K) and $Ba_{1-x}K_xBiO_3$ (BKBO), were discovered in 1988. The metal oxide BKBO is an exceptionally interesting material and the first oxide superconductor without copper with a critical temperature above that of all the A-15 compounds. Its critical temperature is $T_c \simeq 32$ K at $x = 0.4$. At the moment of writing, BKBO still exhibits the highest $T_c$ known for an oxide other than the cuprates. For the potassium concentration $x \geq 0.35$, this compound has the regular cubic *perovskite* structure sketched in Fig. 3.2. However, in one recent study the crystal structure of BKBO has been found to be non-cubic and of the layered nature, having the lattice parameters $a \approx a_0$ and $c \approx 2a_0$, where $a_0$ is a simple cubic perovskite cell parameter shown in Fig. 3.2.

Materials called perovskites are minerals (hard ceramics) whose chemical formula is $ABX_3$ or $AB_2X_3$. Thus, perovskites contain three elements A, B, X in the proportion 1:1:3 or 1:2:3. The atoms A are metal cations, and the atoms B and X are nonmetal anions. The element X is often represented by oxygen. The compound BKBO is a perovskite of type 1:1:3 with a part of the barium atoms replaced by potassium atoms.

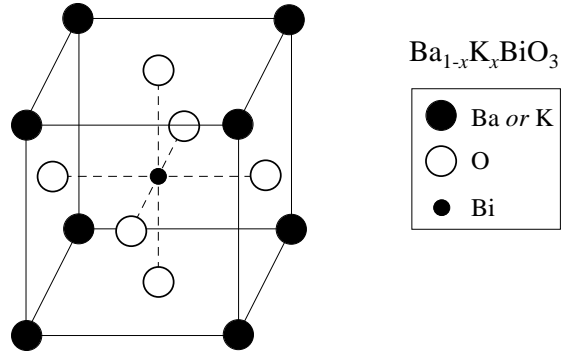

$Ba_{1-x}K_xBiO_3$

● Ba *or* K
○ O
• Bi

*Figure 3.2.* Cubic perovskite unit cell of $Ba_{1-x}K_xBiO_3$.

Superconductivity in BKBO is observed near a metal-insulator transition. The undoped parent compound for BKBO is $BaBiO_3$ containing an insulating charge-density-wave phase formed of an ordered arrangement of non-equivalent bismuth ions referred to as $Bi^{3+}$ and $Bi^{5+}$. Superconducting BKBO with low potassium content also exhibits a charge-density-wave ordering. The density of charge carriers in BKBO is very low (see Fig. 3.6). Depending on $x$, the effective mass of charge carriers in BKBO is small, and can be even smaller than the electron mass, $m^* < m_e$. Various evidence suggests that the electron-phonon coupling is responsible for superconductivity in BKBO. For example, the isotope effect in BKBO is sufficiently large: upon partial replacement of $^{16}O$ with $^{18}O$ in BKBO, the critical temperature is shifted down, and the isotope exponent is about $\alpha \approx 0.4$. Above and below $T_c$, BKBO exhibits a normal-state pseudogap caused most likely by the electron-phonon coupling. A two-band model applied to BKBO accounts very well for all the available data on BKBO.

Acoustic measurements performed in BKBO show that many physical properties of BKBO are quite similar to those of the A-15 superconductors. Examples are some structural instabilities just above the superconducting state, soft phonon modes (acoustic in A-15 compounds and optical in BKBO), an anharmonicity of some phonons and large softening of the elastic constants. In the A-15 compounds and BKBO, there is a structural phase transition slightly above $T_c$ (martensite transition). For comparison, Table 3.2 lists some characteristics of $Nb_3Ge$ (A-15 compound) and BKBO. In Table 3.2, one can see that the characteristics of these superconductors have similar values.

## 2.3    Magnesium diboride $MgB_2$

In January 2001, magnesium diboride $MgB_2$ was found to superconduct at $T_c = 39$ K. The discovery was made by the group of Akimitsu in Tokyo. At the moment of writing, the intermetallic $MgB_2$ has the highest critical temper-ature at *ambient* pressure among all superconductors with the exception of the cuprates. As shown in Fig. 3.3, the crystal structure of MgB$_2$ is very simple: it is composed of layers of boron and magnesium, alternating along the $c$ axis. Each boron layer has a hexagonal lattice similar to that of graphite. The magnesium atoms are arranged between the boron layers in the centers of the hexagons.

*Table 3.2.* Characteristics of Nb$_3$Ge (A-15 compound), BKBO ($x \simeq 0.4$) and MgB$_2$: the critical temperature $T_c$; the energy of the highest phonon peak $\omega_{ph}$ in the Eliashberg function $\alpha^2 F(\omega)$; the Fermi velocity v$_F$; the coherence length $\xi_0$; the gap ratio $\frac{2\Delta}{k_B T_c}$ and the upper critical magnetic field $H_{c2}$

| Compound | $T_c$ (K) | $\omega_{ph}$ (meV) | v$_F$ ($10^7$ cm/s) | $\xi_0$ (Å) | $\frac{2\Delta}{k_B T_c}$ | $H_{c2}$ (T) |
|---|---|---|---|---|---|---|
| Nb$_3$Ge | 23 | 25 | 2.2 | 35–50 | 4.2 | 38 |
| BKBO | 32 | 70 | 3 | 35–50 | 4.5 | 32 |
| MgB$_2$ | 39 | 90 | 4.8 | 35–50 | 4.5 | 39 |

The physical properties of MgB$_2$ are also quite unique. The density of states in MgB$_2$ is small. MgB$_2$ has a very low normal-state resistance: at 42 K the resistivity of MgB$_2$ is more than 20 times smaller than that of Nb$_3$Ge (A-15 compound) in its normal state. In the superconducting state, MgB$_2$ has a highly anisotropic critical magnetic field ($\sim$ 7 times), and exhibits two energy gaps. The gap ratio $2\Delta/(k_B T_c)$ for the larger gap $\Delta_L$ is given in Table 3.2. For the smaller gap $\Delta_s$, this ratio is around 1.7, so that $\Delta_L/\Delta_s \simeq 2.7$. Seemingly, both the energy gaps have s-wave symmetries: the larger gap is highly anisotropic, while the smaller one is either isotropic or slightly anisotropic. Band-structure calculations of MgB$_2$ show that there are at least two types of bands at the Fermi surface. The first one is a heavy hole band, built up of boron $\sigma$ orbitals.

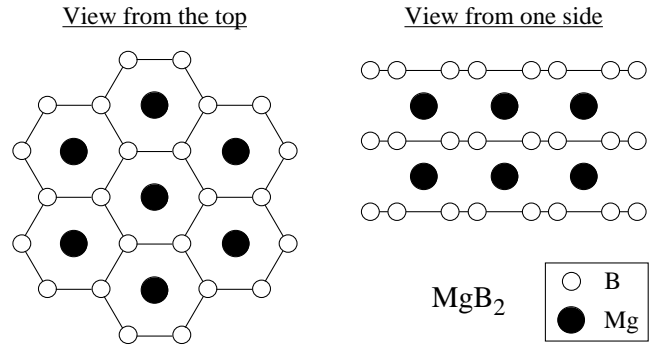

Underlined: View from the top    View from one side

MgB$_2$

○ B
● Mg

*Figure 3.3.* Crystal structure of MgB$_2$. Boron atoms form honeycomb planes, and magnesium atoms occupy the centers of the hexagons in between boron planes.

The second one is a broader band with a smaller effective mass, built up mainly of $\pi$ boron orbitals. The larger energy gap $\Delta_L$ occurs in the $\sigma$-orbital band, while $\Delta_s$ in the $\pi$-orbital band.

In $MgB_2$, superconductivity occurs in the boron layers. The electron-phonon interaction seems to be responsible for the occurrence of superconductivity in $MgB_2$. For example, the boron isotope effect is sufficiently large, $\alpha \simeq 0.3$ (the Mg isotope effect is very small). The muon relaxation rate in $MgB_2$ is about 8–10 $\mu s^{-1}$. So, in the Uemura plot (see Fig. 3.6), $MgB_2$ is literally situated between the large group of unconventional superconductors and the conventional superconductor Nb.

In Table 3.2 which is presented for comparison of some characteristics of the A-15 compounds, BKBO and $MgB_2$, one can see that the characteristics of these superconductors have similar values.

From the standpoint of practical application, magnesium diboride is very attractive because $MgB_2$ has a very high critical temperature, and it is inexpensive to produce in large quantities since it is made from elements that are abundant in nature. Because magnesium and diboride atoms are light, $MgB_2$ is light-weight.

## 2.4    Binary compounds

There are a large number of binary superconductors. Non-magnetic binary compounds exhibiting high values of $T_c$ and $H_{c2}$ most likely belong to the second group of superconductors.

### 2.4.1    Nitrides and carbides

There is a number of superconducting binary compounds AB with the sodium chloride structure shown in Fig. 3.4. The NaCl structure is a cubic face-centered structure with alternating A and B elements in all directions. In crystallography, such a structure is denoted as B1. In AB superconductors with the NaCl structure, the A atom is one of the transition elements of the III, IV, V and VI subgroups of the periodic table, and the B atom is a nontransitional element. The highest critical temperature is observed in the binary compounds with transition metals of the IV, V and VI subgroups: Zr, Nb, Mo, Ta and W, which have incomplete $4d$- and $5d$-shells when they join nitrogen (nitrides) or carbon (carbides). Like the A-15 compounds, these nitrides and carbides have extraordinary properties in the normal and superconducting states. Table 3.3 gives the values of $T_c$ for some nitrides and carbides.

In fact, some nitrides and carbides do not have precisely the 1:1 stoichiometry. For example, the NbN nitride listed in Table 3.3 cannot be prepared with 1:1 stoichiometry. Its exact formula is $NbN_{0.92}$, so the structure has many vacancies. Another example from Table 3.3 is vanadium nitride which is in reality $VN_{0.75}$. Vanadium carbide also has the non-exact 1:1 stoichiometry, $VC_{0.84}$. Interestingly, if the vacancies in $NbN_{0.92}$ are filled by carbon to form $NbC_{0.1}N_{0.9}$, the critical temperature increases to 17.8 K. The latter B1 compound was synthesized by Matthias in 1953. During the fourteen years that followed, it was one of the available superconductors with the highest $T_c$. It is worth to mention that the first nitride discovered to superconduct was NbN, in 1941.

*Table 3.3.* Critical temperature $T_c$ for some nitrides and carbides

| Nitride | $T_c$ (K) ‖ | Carbide | $T_c$ (K) |
|---------|-------------|---------|-----------|
| NbN     | 17.3        | MoC     | 14.3      |
| ZrN     | 10.7        | NbC     | 12.0      |
| HfN     | 8.8         | TaC     | 10.4      |
| VN      | 8.5         | WC      | 10.0      |
| TaN     | 6.5         | TlC     | 3.4       |

As all compounds of the second group of superconductors, the B1 compounds have two conduction bands formed by the $d$-electrons and $sp$-electrons. The $d$-electron band is very narrow, while the $sp$-electron band is sufficiently wide. Tunneling measurements carried out in some B1 compounds showed that the electron-phonon interaction is mainly responsible for the occurrence of superconductivity in these materials, and the dominant contribution to the electron-phonon interaction parameter comes from acoustic phonons. The neutron scattering studies of the phonon spectrum performed in the stoichiometric carbides HfC ($T_c = 0.25$ K) and TaC ($T_c = 10.3$ K) confirmed the dominant role of the electron-phonon interaction for the occurrence of superconductivity in these materials.

As opposed to the A-15 superconductors, the binary compounds with the NaCl lattice are very stable and less sensitive to mechanical defects: the B1 materials are far more resistant to radiation and disorder than the A-15 compounds. Nonetheless, the nitrides and carbides are also quite brittle and difficult to fabricate into wires, but they have been successfully exploited as thin films for superconducting electronics.

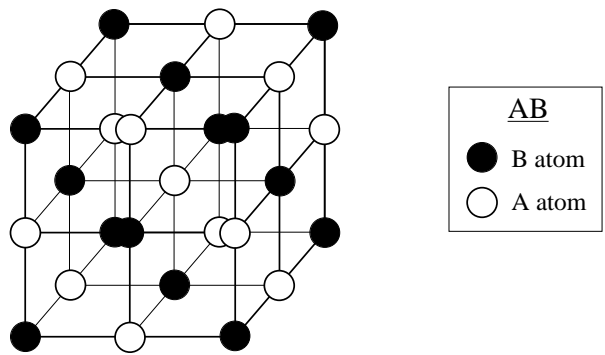

*Figure 3.4.* Crystal structure of AB compounds (B1 superconductors). NaCl has the same crystal structure.

### 2.4.2    Laves phases

There are several dozen metallic $AB_2$ compounds called the *laves phases* that superconduct. Some of them have a critical temperature above 10 K and a high critical field $H_{c2}$. For example, $Zr_{0.5}Hf_{0.5}V_2$ has $T_c = 10.1$ K and $H_{c2} = 24$ T, and the same compound with a different $Zr : Hf$ ratio has similar $T_c$ and $H_{c2}$ values and the critical current density $J_c \approx 4 \times 10^5$ A/cm$^2$. The critical temperatures of the laves phases $CaIr_2$ and $ZrV_2$ are 6.2 K and 9.6 K, respectively. The $AB_2$ materials also have the advantage of not being so hard and brittle as some other compounds and alloys with comparable critical temperatures.

## 2.5    Semiconductors

There are a few semiconductors that become superconducting at very low temperatures. The carrier concentration in semiconductors is much lower than that in metals. Since semiconductors are not magnetic, they most likely belong to the second group of superconductors. GeTe was the first superconducting semiconductor discovered in 1964 , having a very low critical temperature $T_c \simeq 0.1$ K. This was followed by the discovery of superconductivity in the perovskite $SrTiO_3$ with $T_c \simeq 0.3$ K. The crystal structure of $SrTiO_3$ is cubic-perovskite and similar to that of BKBO (see Fig. 3.2). However, by lowering the temperature, it undergoes some phase transitions which destroy its cubic symmetry. Doping $SrTiO_3$ by carriers, one can tune its critical temperature, and the dependence $T_c(p)$ has a shape similar to that of the hyperbolic sine, $\sinh p$, with a sharp maximum of 0.3 K located at $p \approx 10^{20}$ cm$^{-3}$. The critical temperature of some superconducting semiconductors can dramatically be increased by applying an external pressure. In $SrTiO_3$, the Ginzburg-Landau parameter $k$ is of the order of $k \sim 10$, thus, $SrTiO_3$ is a type-II superconductor and has $H_{c2}(0) \sim 400$ Oe.

The tunneling studies carried out in Nb-doped $SrTiO_3$ with different Nb concentrations clearly showed the presence of two-band superconductivity in this compound.

The semiconductor SnTe also superconducts at low temperatures.

## 3.    Third group of superconducting materials

The third group of superconductors is the largest and incorporates so-called unconventional superconductors. A distinctive characteristic of unconventional superconductors is that they are low-dimensional and magnetic, or at least, these compounds have strong magnetic correlations. Furthermore, the density of charge carriers in these superconductors is very low. In unconventional superconductors, spin fluctuations mediate the onset of long-range phase coherence. In the majority of unconventional superconductors, the magnetic correlations favor an antiferromagnetic ordering. In contrast to antiferromagnetic superconductors, ferromagnetic ones usually have a low critical temperature. Independently of the type of magnetic ordering, the pairing mechanism in unconventional superconductors is due to the electron-phonon interaction which is moderately strong and non-linear. We shall discuss the mechanism of unconventional superconductivity in detail in Chapter 6. In all superconductors belonging to this group, the coherence length is very short, while the penetration depth is very large, so that all unconventional superconductors are of type-II. They have a very large upper critical magnetic field. As a consequence, many superconductors from this group are used for practical applications. We start with the so-called Chevrel phases.

## 3.1 Chevrel phases

In 1971, Chevrel and co-workers discovered a new class of ternary molybdenum sulfides, having the general chemical formula $M_x Mo_6 S_8$, where M stands for a large number of metals and rare earths (nearly 40), and $x = 1$ or 2. They were called the *Chevrel phases*. The Chevrel phases with S substituted by Se or Te also display superconductivity. Before the discovery of high-$T_c$ superconductivity in cuprates in 1986, the A-15 superconductors had the highest values of $T_c$, but the Chevrel phases were the record holders in exhibiting the highest values of upper critical magnetic field $H_{c2}$, listed in Table 3.4. The Chevrel phases are of great interest, largely because of their striking superconducting properties.

*Table 3.4.* Critical temperature and the upper critical magnetic field of Chevrel phases

| Compound | $T_c$ (K) | $H_{c2}$ (T) |
|---|---|---|
| $PbMo_6S_8$ | 15 | 60 |
| $LaMo_6S_8$ | 7 | 44.5 |
| $SnMo_6S_8$ | 12 | 36 |
| $LaMo_6Se_8$ | 11 | 5 |
| $PbMo_6Se_8$ | 3.6 | 3.8 |

The crystal structure of Chevrel phases, shown in Fig. 3.5, is quite interesting. These compounds crystallize in a hexagonal-rhombohedral structure. The building blocks of the Chevrel-phase crystal structure are the M elements and $Mo_6X_8$ molecular clusters. Each $Mo_6X_8$ is a slightly deformed cube with X atoms at the corners, and Mo atoms at the face centers (these distortions of the cubes are not shown in Fig. 3.5). Such a crystal structure leads to materials particularly brittle, which give problems in the fabrication of wires. The electronic and superconducting properties of these compounds depend mainly on the $Mo_6X_8$ group, with the M ion having very little effect.

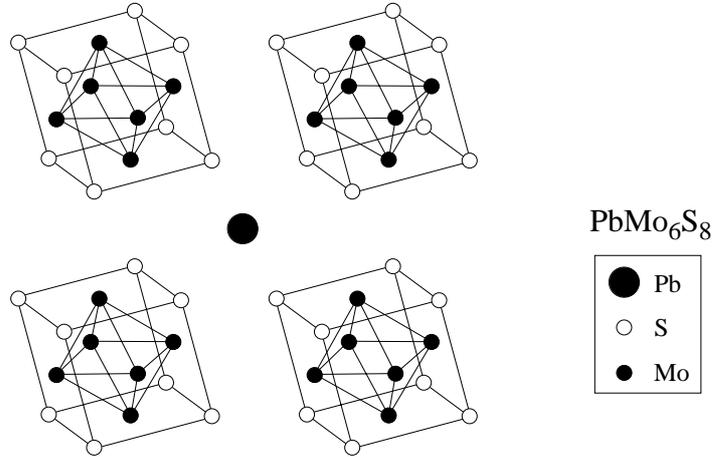

*Figure 3.5.* Crystal structure of the Chevrel phase $PbMo_6S_8$. Each lead atom is surrounded by eight $Mo_6S_8$ units (only four $Mo_6S_8$ units are shown).

Superconductivity in the Chevrel phases coexists with antiferromagnetism of the rare earth elements (in fact, superconductivity in the Chevrel phases is mediated by magnetic fluctuations). For example, a long-range antiferromagnetic order of the rare earth elements RE = Gd, Tb, Dy and Er in $(RE)Mo_6X_8$, setting in respectively at $T_N = 0.84$, 0.9, 0.4 and 0.15 K, coexists with superconductivity occurring at $T_c = 1.4$, 1.65, 2.1 and 1.85 K, respectively. $T_N$ is the Néel temperature of an antiferromagnetic ordering. In $HoMo_6S_8$, for example, the magnetic correlations of rare earth elements result in a long-range ferromagnetic ordering. First, a non-uniform ferromagnetic phase appears in the superconducting state of $HoMo_6S_8$. Then, on further cooling, a long-range ferromagnetic order develops, destroying superconductivity. $HoMo_6S_8$ is superconducting only between two critical temperatures 2 K and 0.65 K. This is called reentrant superconductivity. Below 0.65 K, the material is ferromagnetic.

The superconductivity in the Chevrel phases is primarily associated with the mobile $4d$-shell electrons of Mo, while the magnetic order involves the localized $4f$-shell electrons of the rare earth atoms which occupy regular positions throughout the lattice. In the normal state, the Chevrel phases exhibit a strong softening in the elastic constants as a function of temperature for the longitudinal and transverse modes.

The Chevrel phases and all superconductors of this group have one distinctive characteristic: they all have a very low superfluid density. Furthermore, the critical temperature of unconventional superconductors depends linearly on superfluid density, and this dependence is universal for all superconductors of this group, including the Chevrel phases. Let us consider this dependence.

From Eq. (2.9), the superfluid density $n_s$ directly relates to the penetration depth, $n_s \propto 1/\lambda^2$. So, penetration-depth measurements are able to provide the value of superfluid density. Carrying out muon-Spin-Relaxation ($\mu$SR) measurements, Uemura and co-workers showed that the critical temperature in unconventional superconductors first depends linearly on the superfluid density, as shown in Fig. 3.6. However, on further increasing the superfluid density, $T_c$ follows a "boomerang" path shown in the lower inset of Fig. 3.6. The Uemura plot shows also that the concentration of charge carriers in unconventional superconductors is more than one order of magnitude lower than that in the metallic Nb superconductor. We shall continue to discuss the Uemura plot to the end of this section.

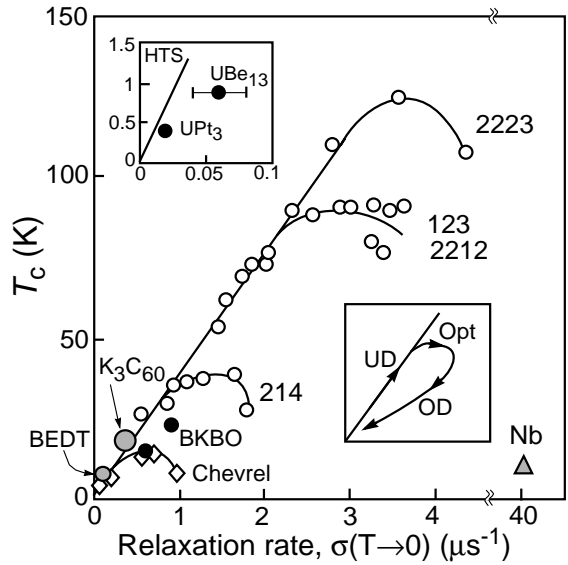

*Figure 3.6.* Critical temperature versus muon-spin-relaxation rate $\sigma(T \rightarrow 0)$ for various superconductors ($\sigma \propto 1/\lambda^2 \propto n_s/m^*$). The cuprates are marked by 214, 123, 2212 and 2223 (see Table 3.5). BEDT is a layered organic superconductor. Heavy fermions are shown in the upper inset (HTS = high-$T_c$ superconductors). For the cuprates, the lower inset shows the "boomerang" path with increasing doping: the underdoped (UD), optimally doped (Opt) and overdoped (OD) regions [29].

## 3.2    Copper oxides

A compound is said to belong to the family of copper oxides (cuprates) if it has the $CuO_2$ planes. Cuprates that superconduct are also called high-$T_c$ superconductors. The first high-$T_c$ superconductor was discovered in 1986 by Bednorz and Müller at IBM Zurich Research Laboratory [5]. Without doubt, this discovery was revolutionary because it showed that, contrary to a general belief of that time, superconductivity can exist above 30 K, and it can occur in very bad conductors.

The parent compounds of superconducting cuprates are antiferromagnetic Mott insulators. A Mott insulator is a material in which the conductivity vanishes as temperature tends to zero, even though the band theory would predict it to be metallic. A Mott insulator is fundamentally different from a conventional (band) insulator. If, in a band insulator, conductivity is blocked by the Pauli exclusion principle, in a Mott insulator charge conduction is blocked by the electron-electron repulsion. Quantum charge fluctuations in a Mott insulator generate the so-called *superexchange interaction* which favors antiparallel alignment of neighboring spins. Thus, a Mott insulator has a charge gap of $\sim 2$ eV, whereas the spin wave spectrum extends to zero energy. When cuprates are slightly doped by holes or electrons (the hole/electron concentration is changed from one per cell), on cooling they become superconducting. At the moment of writing, the cuprates are the only Mott insulators known to superconduct.

The cuprates are materials with the strong electron correlation. What does this mean? Electrons in a metal can be treated in a mean-field approximation. In the framework of this approach, it is assumed that an electron in the crystal moves in an average field created by other electrons. Thus, it is not necessary to know the exact positions of all the other electrons. In cuprates and other strongly-correlated electron materials, the mean-field approach breaks down. The position and motion of each electron in these materials are correlated with those of all the others. Furthermore, in this class of materials, the electron-phonon interaction is much stronger than that in metals. As a consequence, the combination of the electron-electron correlation and electron-phonon interaction results in a strong coupling of electronic, magnetic and crystal structures, so that, depending on temperature, they interact strongly with each other. This gives rise to many fascinating phenomena, such as superconductivity, colossal magnetoresistance, spin- and charge-density waves.

There are many cuprates which become superconducting at low temperature. They can be classified in several groups according to their chemical formulas which are sufficiently complicated; therefore, it is useful to use abbreviations. The abbreviations which will be used further are summarized in Table 3.5. In addition, Table 3.5 indicates the number of the $CuO_2$ planes per unit cell and the critical temperature of these cuprates. From these data, one can see that for members of the same family of cuprates, the critical temperature of a cuprate with double $CuO_2$ layer per unit cell is always higher than that of a single-layer cuprate, and a cuprate with the triple $CuO_2$ layer has a higher $T_c$ than that with the double layer. With the exception of NCCO which is electron-doped, all the cuprates in Table 3.5 are hole-doped.

*Table 3.5.* Abbreviations for some cuprates

| Cuprate | $CuO_2$ planes | $T_c$ (K) | abbreviation |
|---|---|---|---|
| $La_{2-x}Sr_xCuO_4$ | 1 | 38 | LSCO |
| $Nd_{2-x}Ce_xCuO_4$ | 1 | 24 | NCCO |
| $YBa_2Cu_3O_{6+x}$ | 2 | 93 | YBCO |
| $Bi_2Sr_2CuO_6$ | 1 | $\sim$12 | Bi2201 |
| $Bi_2Sr_2CaCu_2O_8$ | 2 | 95 | Bi2212 |
| $Bi_2Sr_2Ca_2Cu_3O_{10}$ | 3 | 110 | Bi2223 |
| $Tl_2Ba_2CuO_6$ | 1 | 95 | Tl2201 |
| $Tl_2Ba_2CaCu_2O_8$ | 2 | 105 | Tl2212 |
| $Tl_2Ba_2Ca_2Cu_3O_{10}$ | 3 | 125 | Tl2223 |
| $TlBa_2Ca_2Cu_4O_{11}$ | 3 | 128 | Tl1224 |
| $HgBa_2CuO_4$ | 1 | 98 | Hg1201 |
| $HgBa_2CaCu_2O_8$ | 2 | 128 | Hg1212 |
| $HgBa_2Ca_2Cu_3O_{10}$ | 3 | 135 | Hg1223 |

The crystal structure of cuprates is of a perovskite type, and it is highly anisotropic. Such a structure defines most physical properties of the cuprates. In conventional superconductors, there are no important structural effects since the coherence length is much longer than the penetration depth. This, however, is not the case for the cuprates.

The simplest copper-oxide perovskites are insulators. To become superconducting they have to be doped by charge carriers. The effect of doping has the most profound influence on the superconducting properties of the cuprates. Introducing charge carriers into the $CuO_2$ planes of cuprates, their lattice becomes very unstable, especially at low temperatures. On lowering the temperature, all cuprates undergo a number of structural phase transitions. Generally speaking, it is the *unstable* lattice that is responsible for the occurrence of superconductivity in the cuprates.

Superconductivity in the cuprates occurs in the copper-oxide planes. Therefore, the structural parameters of the $CuO_2$ planes affect the critical temperature the most. In the $CuO_2$ planes, the structure of the $CuO_2$ layers of cuprates is basically tetragonal. In the $CuO_2$ planes, each copper ion is strongly bonded to four oxygen ions separated by a distance of approximately 1.9 Å. At a fixed doping level, the highest $T_c$ is observed in cuprates having *flat* and *square* $CuO_2$ planes. The $CuO_2$ layers in the cuprates are always separated by layers of other atoms such as Bi, O, Y, Ba, La etc., which provide the charge carriers into the $CuO_2$ planes. These layers are often called *charge reservoirs*.

In conventional superconductors, the critical temperature rises monotonically with the rise of charge-carrier concentration, $T_c(p) \propto p$. In the cuprates, the $T_c(p)$ dependence is nonmonotonic. In most hole-doped cuprates, the $T_c(p)$ dependence has a bell-like shape and can be approximated by the empirical expression

$$T_c(p) \simeq T_{c,max}[1 - 82.6(p - 0.16)^2], \tag{3.1}$$

where $T_{c,max}$ is the maximum critical temperature for a certain compound. Superconductivity occurs within the limits $0.05 \leq p \leq 0.27$ which vary slightly in different cuprates. Different doping regions of the superconducting phase are mainly known as *underdoped*, *optimally doped* and *overdoped*. The insulating phase at $p < 0.05$ is usually called the *undoped* region. These designations are used in the remainder of the book. Above $p = 0.27$, cuprates are practically metallic.

Let us now discuss several cuprates in more detail.

### 3.2.1   LSCO

This compound was the first high-$T_c$ superconductor discovered. The maximum value of $T_c$ is 38 K. The tetragonal unit cell of LSCO is shown in Fig. 3.7a. The lattice constants are $a \approx 5.35$ Å, $b \approx 5.40$ Å and $c \approx 13.15$ Å. This compound is often termed the *214 structure* because it has two La (Sr), one Cu and four O atoms. Upon examining the unit cell shown in Fig. 3.7a, one can clearly see that the basic 214 structure is doubled to form a unit cell. Therefore a more proper label might be 428. The reason for this doubling is that every other $CuO_2$ plane is offset by one-half a lattice constant, so that the unit cell would not be truly repetitive if we stopped counting after one cycle of the atoms.

In LSCO, the conducting $CuO_2$ planes are $\sim 6.6$ Å apart, separated by two LaO planes which form the charge reservoir that captures electrons from the conducting planes upon doping. In the crystal, oxygen is in an $O^{2-}$ valence state that completes the $p$ shell. Lanthanum loses three electrons and becomes $La^{3+}$ which is in a stable closed-shell configuration. To conserve charge neutrality, the copper atoms must be in a $Cu^{2+}$ state which is obtained by losing the $(4s)$ electron and also one $d$ electron. This creates a hole in the $d$ shell, and thus $Cu^{2+}$ has a net spin of 1/2 in the crystal. Along the $c$ direction, each copper atom in the conducting planes has an oxygen above and below. These oxygen atoms are called *apical*. Thus, in LSCO, the copper ions are surrounded by octahedra of oxygens, as shown in Fig. 3.7a. However, the distance between a Cu atom and an apical O is $\sim 2.4$ Å, which is considerably larger than the distance Cu–O in the planes (1.9 Å). Consequently, the dominant bonds are those on the plane, and the bonds with apical oxygens are much less important. Many

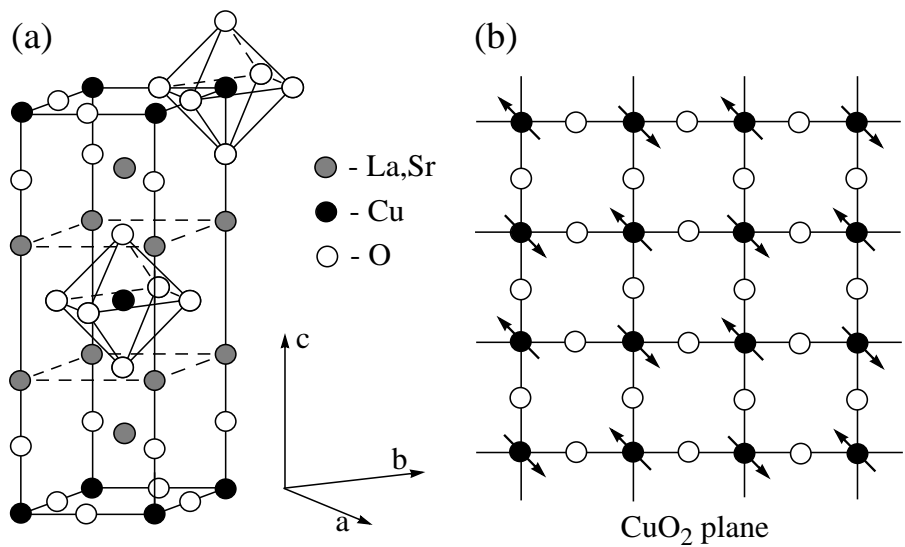

*Figure 3.7.* (a) Crystal tetragonal structure of LSCO. (b) Schematic of a CuO$_2$ plane, the crucial subunit for high-$T_c$ superconductivity. The arrows indicate a *possible* alignment of spins in the antiferromagnetic ground state.

high-$T_c$ compounds also have apical oxygens which are always separated from the conducting planes by a distance of about 2.4 Å.

Figure 3.7b schematically shows a CuO$_2$ plane, the crucial subunit for high-$T_c$ superconductivity. In Fig. 3.7b, the arrows indicate a possible alignment of spins in the antiferromagnetic ground state of La$_2$CuO$_4$. In reality, however, the copper spin are not fully in the planes—they are oriented slightly out of the planes, i.e. along the $c$ axis.

The phase diagram of this material is shown in Fig. 3.8. Near half-filling, the antiferromagnetic order is clearly observed. For higher Sr doping, $0.02 \leq x \leq 0.08$, there is no long-range antiferromagnetic order, but at very low temperatures there is a spin-glass-like phase. This phase is not a conventional spin glass. The insulator-metal transition occurs at about $x \sim 0.04$–0.05. For Sr doping between $x \sim 0.05$ and $\sim 0.27$, a superconducting phase is found at low temperatures. The maximum $T_c$ is observed at the "optimal" doping $x \sim 0.16$. As one can see in Fig. 3.8, the Sr substitution for La in LSCO induces a structural phase transition from the high-temperature tetragonal (HTT) to low-temperature orthorhombic (LTO) and, at low temperatures, from the LTO phase to the low-temperature tetragonal (LTT) phase (not shown).

In Fig. 3.8, the $T_c(x)$ dependence has a nearly bell-like shape. However, at doping $x = \frac{1}{8}$, the curve has a dip. This dip is the so-called $\frac{1}{8}$ *anomaly* and inherent exclusively to LSCO.

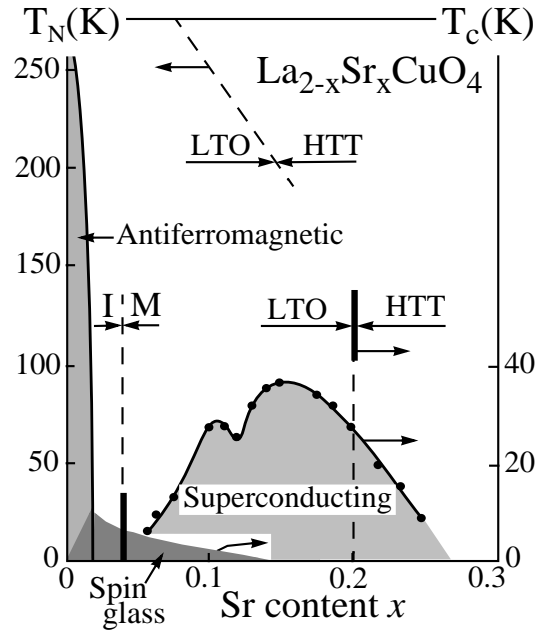

*Figure 3.8.* Phase diagram of LSCO.

In the Uemura plot shown in Fig. 3.6, one can see that the superfluid density in LSCO (marked by 214) and other cuprates is very low. The main superconducting characteristics of LSCO and some other cuprates are listed in Table 3.6. From these data, one can conclude that the superconducting properties of cuprates are very anisotropic.

*Table 3.6.* Characteristics of optimally doped cuprates: the critical temperature $T_c$; the coherence length $\xi_i$; the penetration depth $\lambda_i$, and the upper critical magnetic fields $H_{c2}^i$ ($i = ab$ or $c$)

| Compound | $T_c$ (K) | $\xi_{ab}$ (Å) | $\xi_c$ (Å) | $\lambda_{ab}$ (Å) | $\lambda_c$ (Å) | $H_{c2}^{ab}$ (T) | $H_{c2}^c$ (T) |
|---|---|---|---|---|---|---|---|
| NCCO | 24 | 58 | 3.5 | 1200 | 260 000 | 10 | - |
| LSCO | 38 | 33 | 2.5 | 2000 | 20 000 | 62 | 15 |
| YBCO | 93 | 15 | 2 | 1450 | 6000 | 120 | 40 |
| Bi2212 | 95 | 20 | 1 | 1800 | 7000 | 100 | 30 |
| Bi2223 | 110 | 15 | 1 | 2000 | 10 000 | 250 | 30 |
| Tl1224 | 128 | 14 | 1 | 1500 | – | 160 | - |
| Hg1223 | 135 | 13 | 2 | 1770 | 30 000 | 190 | - |

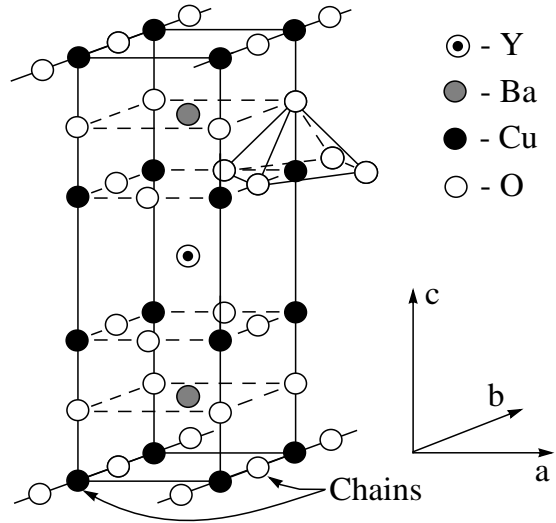

*Figure 3.9.* Crystal orthorhombic structure of YBCO. Note that the orthorhombic lattice vectors are at 45° relatively to the tetragonal lattice vectors (see Fig. 3.7a).

### 3.2.2   YBCO

YBCO is the first superconductor found to have $T_c > 77$ K, and is commonly termed "123". The orthorhombic unit cell of YBCO is shown in Fig. 3.9. The lattice constants are $a \approx 3.82$ Å, $b \approx 3.89$ Å and $c \approx 11.7$ Å. The two $CuO_2$ layers are separated by a single yttrium atom. The role of yttrium is very minor, it just holds the two $CuO_2$ layers apart. In the crystal, Y has a valence of +3. The replacement of Y by many of the lanthanide series of rare-earth elements causes no appreciable change in the superconducting properties. Outside the $CuO_2$–Y–$CuO_2$ sandwich, the BaO planes and CuO chains are located, as shown in Fig. 3.9. In the crystal, Ba has a valence of +2. The distance Cu–O in the chains is $\sim 1.9$ Å, as that in the planes. Each copper ion in the $CuO_2$ planes is surrounded by a pyramid of five oxygen ions.

YBCO is the only high-$T_c$ compound having the one-dimensional CuO chains. In $YBCO_6$, there are no CuO chains, and the compound is an antiferromagnetic insulator, as shown in Fig. 3.10. It has to be doped to gradually become a metallic conductor and a superconductor at low temperature. The doping is achieved by adding additional oxygen atoms which form the CuO chains. So, the oxygen content can be changed reversibly from 6.0 to 7.0 simply by pumping oxygen in and out of the parallel CuO chains running along the $b$ axis. Thus, the CuO chains play the role of charge reservoirs. At an oxygen content of 6.0, the lattice parameters $a \neq b$, and the unit cell is orthorhombic. The increase of oxygen content causes the unit cell to have square symmetry, i.e. $a = b$. So, the crystal becomes tetragonal at low temperatures, as shown in Fig. 3.10. At an oxygen content of 6.4, the antiferromagnetic long-range order disappears and the superconducting phase starts to develop. The maximum value of $T_c$ is achieved at a doping level of about 6.95 (the optimal doping). Unfortunately, it is not possible to explore the phase diagram above the oxygen content of 7.0, since the CuO chains are completed. YBCO$_7$ is a stoichiometric compound with the highest $T_c$.

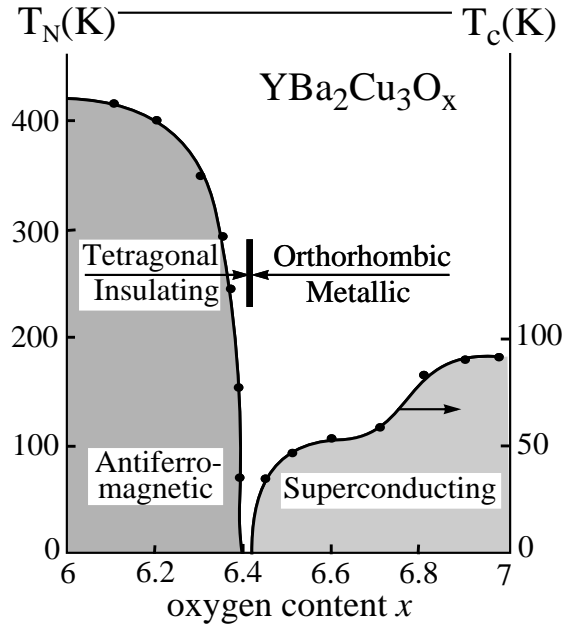

In Fig. 3.10, one can see that, at $x \simeq 6.7$, there is a plateau at $T_c \simeq 60$ K. This is the so-called *60 K plateau*. There are two different explanations for the origin of the 60 K plateau: the hole concentration in the CuO$_2$ planes at $x \simeq 6.7$ remains unchanged, and the change of hole concentration occurs exclusively in the CuO chain layers. The second explanation is that this plateau is directly related to the $\frac{1}{8}$ anomaly observed in LSCO.

### 3.2.3    Bi2212

Figure 3.11 shows the unit cell of Bi2212 which has a maximum $T_c$ of 95 K. The dimensions of the tetragonal lattice constants of Bi2212 are $a \simeq b \approx 5.4$ Å and $c \approx 30.89$ Å. In addition to the CuO$_2$ double layer intercalated by Ca, the unit cell also contains two semiconducting BiO and two insulating SrO layers, as shown in Fig. 3.11. In the crystal, Bi and Sr have a valence of +3 and +2, respectively. In Bi2212 crystals, there is a lattice modulation along the $b$ axis, with the period of 4.76 $b$. The family of the bismuth cuprates consists of three members: Bi2201, Bi2212 and Bi2223 with the unit having 1, 2 and 3 $CuO_2$ planes, respectively. The maximum $T_c$ increases with increasing number of $CuO_2$ planes.

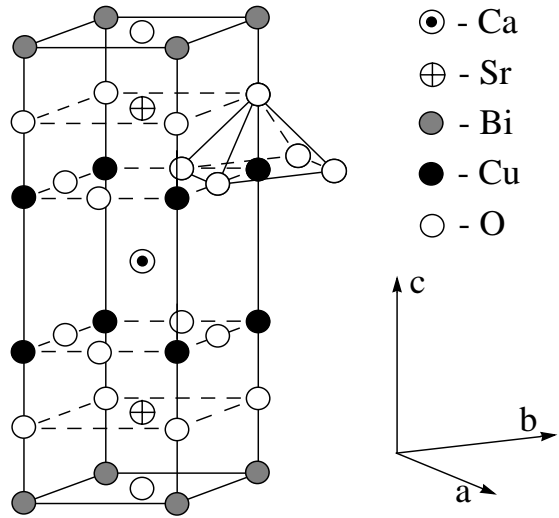

*Figure 3.11.* Crystal structure of Bi2212.

The structure of the bismuth cuprates is very similar to the structure of thallium cuprates such as Tl2201, Tl2212 and Tl2223, with bismuth replaced by thallium, and strontium replaced by barium. In spite of similar structural features of bismuth and thallium compounds, there are differences in the superconducting and normal-state properties.

The bismuth cuprates are very suitable for tunneling and other measurements: the oxygen content in Bi2212, Bi2201 and Bi2223 at room temperature is stable, unlike that in YBCO. Secondly, the bonds between BiO layers in the crystal are weak, so it is easy to cleave a Bi2212 crystal. After the cleavage, the Bi2212 crystal has a BiO layer on the surface. However, it is generally agreed that Bi2212 samples have not reached the degree of purity and structural perfection attained in YBCO.

Since the bismuth, thallium and mercury cuprates have the lattice constants $a = b$, there is no twinning within a crystal.

### 3.2.4 NCCO

The structure of the electron-doped NCCO cuprate, shown in Fig. 3.12, is body-centered tetragonal like that of LSCO. The difference between the two lies in the position of the oxygen atoms of the charge reservoirs. The tetragonal lattice constants of NCCO are $a \simeq b \approx 5.5$ Å and $c \approx 12.1$ Å. In the crystal, Nd and Ce have a valence of +3 and +4, respectively. When a $Nd^{3+}$ ion is replaced by $Ce^{4+}$, the $CuO_2$ planes get an excess of electrons. It is believed that an added electron occupies a hole in the $d$ shell of copper, producing an $S = 0$ closed-shell configuration.

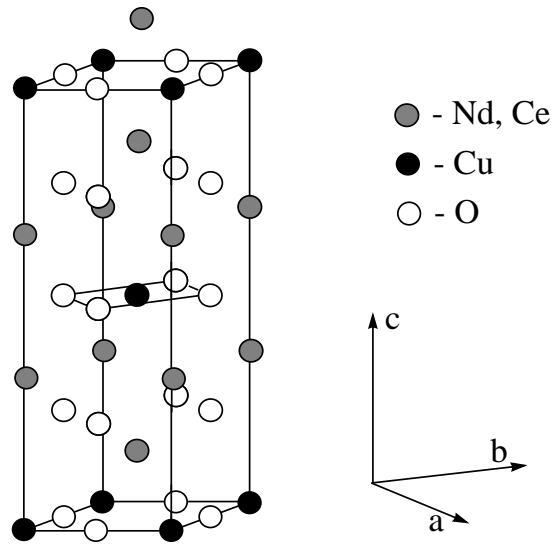

*Figure 3.12.* Crystal structure of NCCO.

Superconductivity in NCCO is observed when the Ce content varies between $x \simeq 0.14$ and 0.18. The phase diagram of NCCO is compared in Fig. 3.13 with that of the hole-doped LSCO cuprate. As one can see in Fig. 3.13, the two diagrams are very similar. Both present an antiferromagnetic phase with similar Néel temperature. When $x$ is increased further, a superconducting phase appears close to antiferromagnetism, although the width of the superconducting phase in the two cases differs by a factor of 3. The maximum value of $T_c$ in NCCO is 24 K, so it is almost twice as small as that of LCSO.

### 3.2.5    Applications of high-$T_c$ superconductors

Soon after the discovery of high-$T_c$ superconductors, it was realized that, at liquid nitrogen temperatures, enormous savings are possible. For example, if cryogenic liquids are used for cooling, a litre of liquid helium costs approximately \$25, as opposed to \$0.60 for a litre of liquid nitrogen. The difference in the cost of electricity, if electrical cryocoolers are used at 4.2 K and 77 K, is similar.

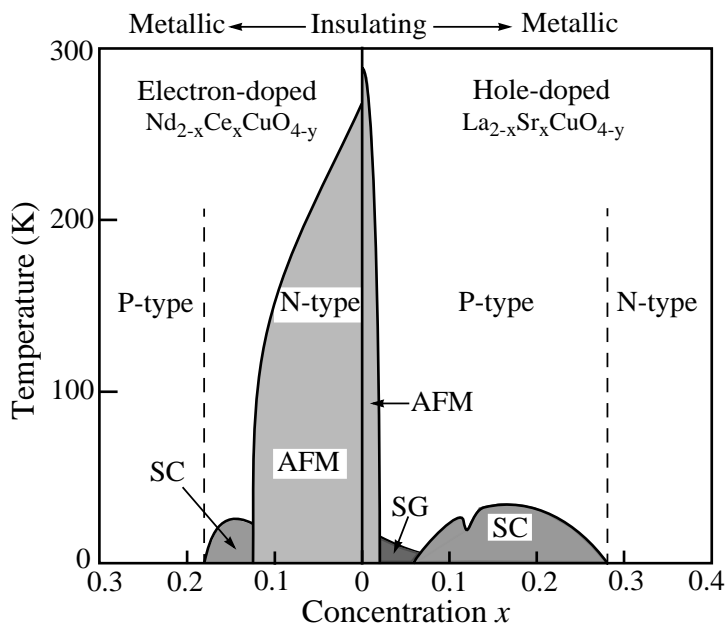

Figure 3.13. Schematic phase diagram of NCCO and LSCO shown together for better comparison.

The main problem in using cuprates for applications is that, like all ceramics, high-$T_c$ superconductors are very brittle and very difficult to shape and handle, while long, flexible, superconducting wires are necessary for many large-scale applications. Large supercurrents can only flow along $CuO_2$ planes, and only a small fraction of the material in a completed device is likely to be correctly oriented. The grain boundaries attract impurities, leading to weak links, which reduce the inter-grain current density and provide an easy path for flux vortices to enter the material. Flux creep or vortex penetration into high-$T_c$ superconductors is unusually rapid. The coherence length or diameter of a vortex core tends to be very small. This is a problem because pinning is most effective if the defect or impurity is of the same size as the coherence length.

*Small-scale applications.* Small-size devices based on thin films of cuprates are now commercially available. Progress is now largely limited by refrigeration packages, not by materials, films or junctions. Thin films of YBCO are widely used for small-size high-$T_c$ superconducting devices because YBCO has a high critical temperature and can accommodate a high current density. Most success has been achieved with SQUIDs which are based on the Josephson effect. The SQUID sensitivity is so high that it can detect magnetic fields 100 billion times smaller than the Earth magnetic field. Other devices that have reached commercial availability are high-$T_c$ superconducting passive RF (radio frequency) and microwave filters for wide-band communications and radars. These are based on conventional microstrip and cavity designs. They have the advantages of very low noise and much higher selectivity and efficiency than conventional filters. They are now used at mobile-phone base stations. Finally, high-$T_c$ superconducting thin films are also widely used for bolometric detection of radiation.

*Large-scale applications.* Large-scale applications for high-$T_c$ superconductors present a major challenge to the materials scientists. Compared with the small-scale applications, a large-scale application generally requires much larger currents and lengths of superconductor in a working environment where the magnetic field may be several Teslas. The most important applications under consideration are in magnets, power transmission cables, current leads, fault current limiters, transformers, generators, motors, and energy storage. Applications related to magnet technology are probably among the most significant that are under development at the present time. These include magnetic energy storage, Maglev trains (relaying on repulsion between magnets mounted on the train and the guideway) and magnets for MRI (Magnetic Resonance Imaging) and other medical applications. In all these cases the superconductor must not only carry a large current with zero resistance under a high magnetic field, but it also must be possible to fabricate it in long lengths with high flexibility and a high packing density. Research on large-scale applications of high-$T_c$ superconductors has focused on the Bi-based family because it is difficult to grow YBCO in bulk. Bi2212/Bi2223 powder is packed into a silver tube, which is drawn fine and goes through a sintering, rolling and annealing process. The major remaining barrier to wider use is cost. However, in some cases the extra cost is justified: high-$T_c$ superconducting underground power transmission cables, which can carry 3 to 5 times the current of a copper cable of the same diameter, are already coming into commercial use in cities such as Detroit. The use of high-$T_c$ superconductors is now a multi-billion-dollar growing business: more than 50 companies around the world have set out to commercialize high-$T_c$ superconductors over the past 14 years.

## 3.3    Charge transfer organics

Organic compounds and polymers are usually insulators, but it is now known that some of them form good conductors. These conducting organics were widely studied during the 1970s. It turns out that some of them superconduct at low temperatures. All organic superconductors are layered. So, all these materials are basically two-dimensional. However, the electron transport in some of them is not quasi-two-dimensional but quasi-one-dimensional.

The first organic superconductor was discovered in 1979 by Bechgaard and Jerome: the compound $(TMTSF)_2PF_6$ was found to superconduct below $T_c = 0.9$ K under a pressure of 12 kbar. TMTSF denotes tetramethyltetraselenaful-valene, and $PF_6$ is the hexafluorophosphate. However, the ten years following this discovery saw a remarkable increase in $T_c$. In 1990, an organic superconductor with $T_c \approx 12$ K was synthesized. In only 10 years, $T_c$ increased over a factor 10!

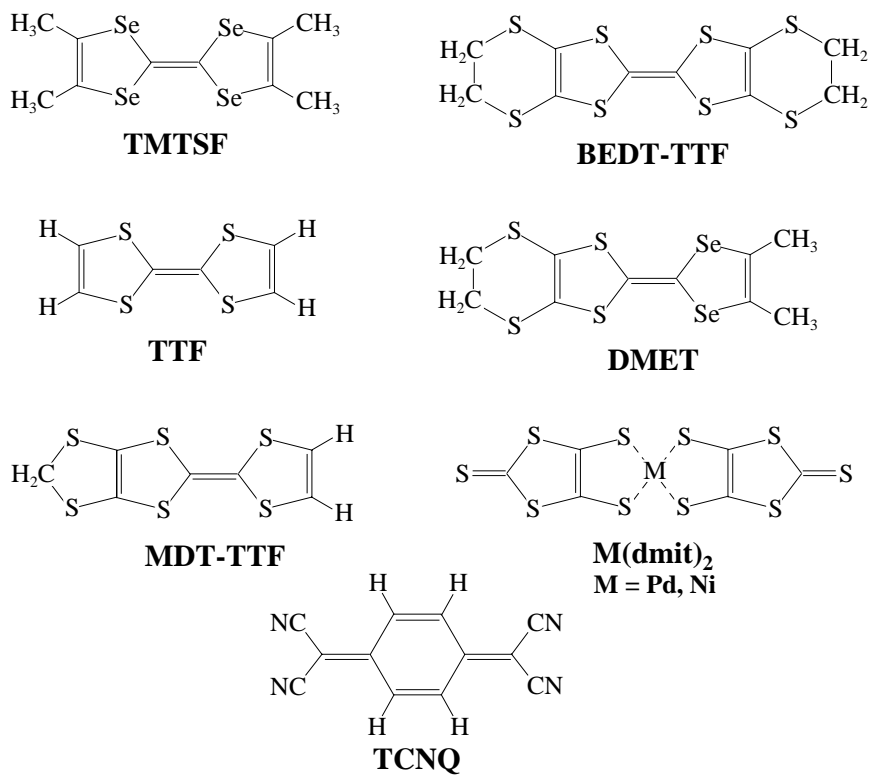

*Figure 3.14.* Structure of organic molecules that form superconductors. Abbreviations of their names are shown below each molecule.

Figure 3.14 shows several organic molecules that form superconductors. In general, they are flat, planar molecules. Among other elements, these molecules contain sulfur or selenium atoms. In a crystal, these organic molecules are arranged in stacks. The chains of other atoms (Cs or I) or molecules ($PF_6$, $ClO_4$ etc.) are aligned in these crystals parallel to the stacks. As an example, the crystal structure of the first organic superconductor $(TMTSF)_2PF_6$, a representative of the *Bechgaard salts*, is schematically shown in Fig. 3.15. The planar TMTSF molecules form stacks along which the electrons are most conducting (the $a$ axis). The chains of $PF_6$ lie between the stacks, aligned parallel to them. Two molecules TMTSF donate one electron to an anion $PF_6$:

$$(TMTSF)_2 + PF_6 \longrightarrow (TMTSF)_2^+ + PF_6^- .$$

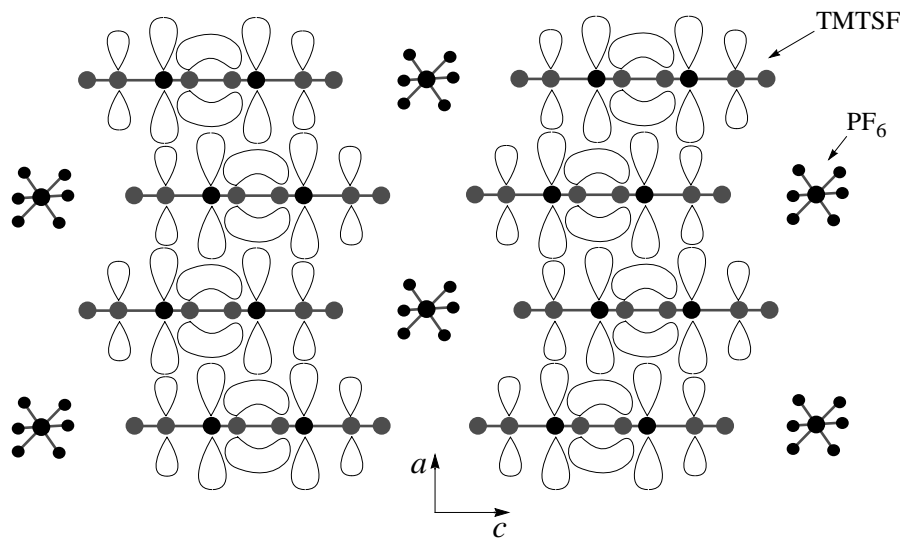

*Figure 3.15.* A side view of the crystal structure of the Bechgaard salt $(TMTSF)_2PF_6$. Each TMTSF molecule is shown with the electron orbitals (the hydrogen atoms are not shown). The chemical structure of the TMTSF molecules is depicted in Fig. 3.14. The organic salt $(TMTSF)_2PF_6$ is the most conductive along the TMTSF stacks (along the $a$ axis).

The separation of charge creates electrons and holes that can become delocalized to render the compound conducting and, at low temperatures, superconducting (under pressure).

After 1979, several more organic superconductors of similar structure were discovered. In all cases, some anion $X^-$ is needed to affect charge balance in order to obtain metallic properties and, at low temperature, superconductivity. So, the anions are mainly charge-compensating spacers; the conductivity is in the organic molecules. There are six different classes of organic superconductors. Two of them are the most studied—the Bechgaard salts $(TMTSF)_2X$ and the organic salts $(BEDT-TTF)_2X$ based on the compound BEDT-TTF shown in Fig. 3.14. BEDT-TTF denotes bis-ethylenedithio-tetrathiafulvalene. The members of the $(BEDT-TTF)_2X$ family exhibit the highest values of $T_c$, and have a rich variety of crystalline structures. In contrast to the flatness of the TMTSF molecules shown in Fig. 3.15, the $CH_2$ groups in the BEDT-TTF molecule lie outside the plane of the remaining part of this molecule. Furthermore, the arrays of BEDT-TTF stacks form conducting layers separated by insulating anion sheets. So, in contrast to the Bechgaard salts which exhibit quasi-one-dimensional electron transport, the electronic structure of the BEDT-TTF family is of two-dimensional nature which appears in the anisotropy of the conductivity and superconducting properties. Also in contrast to the Bechgaard salts, one molecule BEDT-TTF, not two, donates one electron to an anion $X^-$. The highest values of $T_c$ are observed in the (BEDT-TTF)$_2$X salts with the anions X = Cu(NCS)$_2$; Cu[N(CN)$_2$]Br and Cu[N(CN)$_2$]Cl. Their critical temperatures are respectively $T_c = 10.4$, 11.6 and 12.8 K. The first two compounds superconduct at ambient pressure, while the last one with Cu[N(CN)$_2$]Cl becomes superconducting under a pressure of 0.3 kbar.

Interestingly, the hydrogen isotope effect in BEDT-TTF is negative. In conventional superconductors, the critical temperature of a metal is always higher than that of its isotope with a heavier mass. It is just the opposite for the (BEDT-TTF)$_2$Cu(NCS)$_2$ compound: in 1989, Japanese researchers replaced some hydrogen atoms in BEDT-TTF molecules by deuterium, and its critical temperature rose to 11.0 K. Such an isotope effect is called *negative* or *inverse*.

Organic superconductors with the same chemical formula can exist in a variety of crystal phases. This is because the electronic properties of organic conductors depend on the preparation method. For example, there are at least five known phases of the (BEDT-TTF)$_2$I$_3$ compound that differ considerably in their critical temperatures. It is necessary to emphasize that the conditions in which the single crystals of organic conductors are synthesized differ drastically from those at which the crystals of the cuprates are grown. While the single crystals of cuprates are prepared at temperatures near 950 C, the single crystals of organic superconductors are grown at ambient temperatures. Above 100 C, the crystals of organic conductors decompose, melt or change composition. To make an organic charge-transfer salt, including the (BEDT-TTF)$_2$X series, the *electrocrystallization* synthesis process is generally used. Solutions of the cation and the anion are placed in a container, separated by a porous glass plug (a "frit") that allows ions to pass only when electrical current flows. Applying a small current (0.1–0.5 $\mu$A/cm$^2$) causes small crystals of (BEDT-TTF)$_2$X to form on the anode. Typical crystal masses are 140-280 $\mu$g. The crystals are very thin, about 1 to 2 mm long, and black in color. So, at this stage, no one regards the organic superconductors as practical materials.

However, organic superconductors attract a lot of attention because they are in many respects similar to the cuprates. They have reduced dimensionality, low superfluid density, low values of the Fermi energy, magnetic correlations, unstable lattice and numerous phase transitions above $T_c$. Indeed, as discussed above, the Bechgaard salts and salts based on the TCNQ molecules shown in Fig. 3.14 are quasi-one-dimensional conductors, while the BEDT-TTF family is quasi-two-dimensional. Obviously, their superconducting properties are also highly anisotropic. For example, the values of in-plane and out-of-plane coherence lengths in (BEDT-TTF)$_2$Cu[N(CN)$_2$]Br are $\xi_{0,\parallel} \simeq 37$ Å and $\xi_{0,\perp} \simeq 4$ Å, respectively (compare with those for LSCO in Table 3.6). In the Uemura plot shown in Fig. 3.6, one can see that the superfluid density obtained in $k$-(BEDT-TTF)$_2$Cu(NCS)$_2$ (marked in Fig. 3.6 by BEDT) is very low, and comparable with that in the cuprates (the prefix $k$ indicates one of the five crystal phases of the (BEDT-TTF)$_2$X family). Depending on pressure, organic superconductors exhibit a long-range antiferromagnetic ordering. If, in the phase diagram of the Bechgaard salts, the superconducting phase evolves out of the antiferromagnetic phase, in $k$-(BEDT-TTF)$_2$Cu[N(CN)$_2$]Br, these two phases overlap. The latter fact suggests that antiferromagnetic fluctuations—short-lived excitations of the hole-spin arrangements—are important in the mechanism of unconventional superconductivity in organic salts. The unconventional character of superconductivity in organics manifests itself in the gap ratio, $2\Delta/(k_B T_c) \sim$ 6.7, obtained in tunneling measurements in $k$-(BEDT-TTF)$_2$Cu(NCS)$_2$. Such a value of the gap ratio is too large for the conventional type of superconductivity.

As discussed in Chapter 2, in the quasi-two-dimensional organic conductor $\lambda$-(BETS)$_2$FeCl$_4$, superconductivity is induced by a very strong magnetic field, $18 \leq H \leq 41$ T. The dependence $T_c(H)$ has a bell-like shape with a maximum $T_c \simeq 4.2$ K near 33 T. At zero field, this organic compound is an antiferromagnetic insulator below 8.5 K. The other two-dimensional compound, $\alpha$-(BEDT-TTF)$_2$KHg(NCS)$_4$, at low magnetic fields is a charge-density-wave insulator. Thus, in these organic salts, the magnetic and electronic degrees of freedom are coupled. Furthermore, the fact that the electronic and magnetic properties of organic superconductors strongly depend on pressure indicates that their electronic, magnetic and crystal structures are strongly coupled, as those in the cuprates.

Finally, let us consider the longitudinal $\rho_a$ and transverse $\rho_c$ resistivities measured in (TMTSF)$_2$PF$_6$ and the in-plane $\rho_{ab}$ and out-of-plane $\rho_c$ resistivities obtained in an *undoped* TmB$_2$Cu$_3$O$_{6.37}$ (TmBCO) single crystal. Figure 3.16 depicts these two sets of resistivities as functions of temperature. A visual inspection of Figs. 3.16a and 3.16b shows a striking similarity between the two plots. In Fig. 3.16a, the steep rises in $\rho_a$ and $\rho_c$ at low temperatures are due to a metal-insulator transition (and occur at different temperatures, $T_{\rho_a} < T_{\rho_c}$). In Fig. 3.16b, only the out-of-plane resistivity $\rho_c$ exhibits this rise; the in-plane resistivity $\rho_{ab}$ in Fig. 3.16b does not show the rise at low temperature because the minimum temperature available in these measurements was not sufficiently low to observe it [30]. As shown elsewhere [30], this insulating phase at low temperatures in $(TMTSF)_2PF_6$ and TmBCO occurs mainly due to a charge-density-wave ordering. This fact is important and will be discussed in Chapter 6. From the data in Fig. 3.16, one can also conclude that, below 327 K, the electron transport in TmBCO is in fact quasi-one-dimensional.

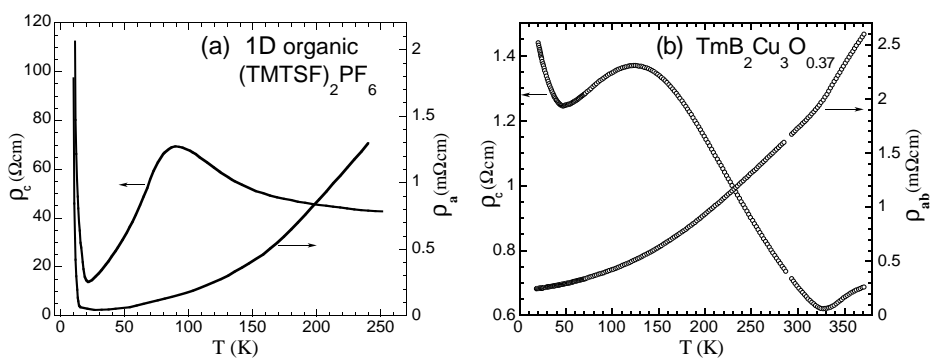

*Figure 3.16.* (a) Temperature dependences of longitudinal $\rho_a$ (see the axes in Fig. 3.15) and transverse $\rho_c$ resistivities measured in one-dimensional (TMTSF)$_2$PF$_6$ organic conductor. (b) The temperature dependences of in-plane $\rho_{ab}$ and out-of-plane $\rho_c$ resistivities obtained in an undoped TmBCO single crystal (after [30]).

## 3.4 Fullerides

Historically, any allotrope based on the element carbon has been classed as organic, but a new carbon allotrope stretches that definition. The pure element carbon forms not only graphite and diamond but a soccer-ball shaped molecule containing 60 atoms, sketched in Fig. 3.17. Because the structure of $C_{60}$ is a mixture of five-sided and six-sided polygons, reminiscent of the geodesic dome designed by architect R. Buckminister Fuller, the molecule $C_{60}$ has been affectionately named "buckminster-fullerene" (without one i), or just "fullerene" for short. Due to its resemblance to a soccer ball, the molecule $C_{60}$ is also called "buckyball." There are also lower and higher molecular weight variations such as $C_{20}$, $C_{28}$, $C_{70}$, $C_{72}$, $C_{100}$ and so forth, which share many of the same properties. The word "fullerenes" is now used to denote all these molecules and other closed-cage molecules consisting of only carbon atoms. The alkali-doped fullerenes are called "fullerides."

The $C_{60}$ molecules were officially discovered in 1985; however, their presence was first seen by astrophysicists a few years earlier in the interstellar dust. The light transmitted through interstellar dust had an increased extinction in the ultraviolet region at a wavelength of 2200 Å (5.6 eV) caused by the $C_{60}$ molecules. Only since 1990 has $C_{60}$ been available to many laboratories in large enough quantities to make solids of a size that allowed traditional solid-state experiments. Very soon, in 1991 it was found that intercalation of alkali-metal atoms in solid $C_{60}$ leads to metallic behavior. Shortly afterwards, also in 1991, it was discovered that some of these alkali-doped $C_{60}$ compounds are superconducting with a transition temperature that is only surpassed by that in the cuprates. In fullerides, the maximum critical temperature of 33 K is observed at ambient pressure in $RbCs_2C_{60}$, and $T_c = 40$ K in $Cs_3C_{60}$ under a pressure of 12 kbar.

The $C_{60}$ molecule has a great stability because it has an incredibly large number of resonant structures. Large organic molecules, for example those in

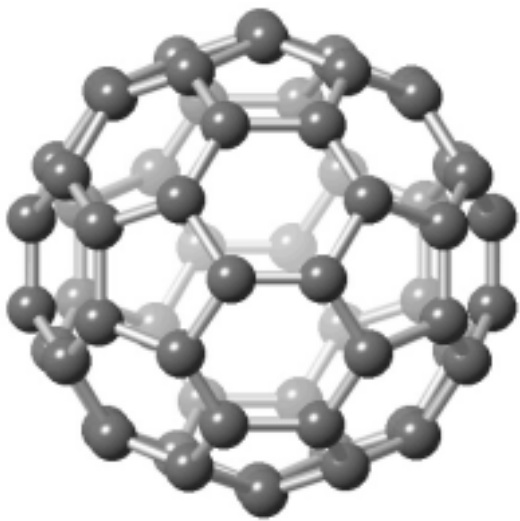

*Figure 3.17.* One $C_{60}$ molecule.

Fig. 1.3, have alternating single and double bonds between the carbon atoms—the conjugate bonds. Stability of such molecules arises from the possibility of having different arrangements of single and double bonds. Each such arrangement is called a *resonant structure*. The more an organic compound has these resonance structures, the more it is stable. For example, extensive sheets of graphite have a virtually infinite number of resonance structure. So, from this standpoint, the $C_{60}$ molecule is very stable. The $C_{60}$ molecule, as well as a soccer ball, has 12 pentagonal (5-sided) and 20 hexagonal (6-sided) faces. The mean diameter of a $C_{60}$ ball is 7.1 Å. The average C-C distance in a $C_{60}$ molecule is 1.43 Å. There are 90 C-C bonds in a $C_{60}$ molecule.

The $C_{60}$ molecules bind with each other in the solid state to form a crystal lattice with a face-centered cubic structure (see Fig. 3.18). The lattice constant $a$ of the $C_{60}$ crystal is 14.161 Å. In such a lattice, the distance between centers of two neighboring $C_{60}$ molecules is 10 Å. These $C_{60}$ molecules are held together by weak *van der Waals* forces. Because $C_{60}$ is soluble in benzene, single crystals of it can be grown by slow evaporation from benzene solutions. In the face-centered cubic fullerene structure, about 26% of the volume of the unit cell is empty. So, when doped by alkali atoms, these easily fit into empty space between molecular balls of the materials, as schematically shown in Fig. 3.18. Unfortunately, the fullerides are extremely unstable in air, burning spontaneously, so they must be prepared and kept in an inert atmosphere. When $C_{60}$ crystals and, for example, potassium metal are placed in evacuated tubes, then heated to 400 C, an atmosphere of potassium vapor diffuses into the empty space between the $C_{60}$ molecules, forming the compound $K_3C_{60}$. This compound is no longer an insulator but becomes superconducting at 19.5 K. In this fulleride, the potassium atoms become ionized to form the positive ion $K^+$, while each $C_{60}$ molecule accepts three electrons:

$$3K + C_{60} \longrightarrow 3K^+ + C_{60}^{3-}.$$

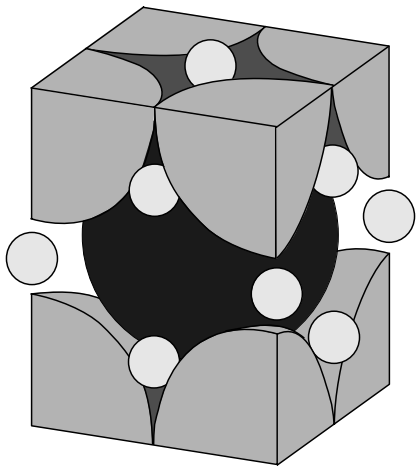

*Figure 3.18.* Unit cell of $A_3C_{60}$. The large spheres represent the $C_{60}$ molecules, and the small spheres are alkali ions. In a given unit cell, there are two ions with tetrahedral coordination and one ion with octahedral coordination.

Thus, each fullerene molecule has three extra delocalized electrons. These extra electrons not only wander around their respective $C_{60}$ molecules, but they can also jump from one to another $C_{60}$ molecule and thereby carry electrical current. The fullerides are magnetic due to spins of alkali atoms, which are ordered antiferromagnetically at low temperatures.

There are a few dozens of fullerides $M_3C_{60}$ known to become superconducting at low temperature. The critical temperatures for several superconducting fullerides with the highest $T_c$ are listed in Table 3.7. As already noted, $Cs_3C_{60}$ also superconducts below $T_c = 40$ K but exclusively under pressure ($\sim 12$ kbar). The crystal structure of the superconducting phase for $Cs_3C_{60}$ is believed to be not face-centered cubic but a mixed A-15 and body-centered tetragonal. Table 3.7 also gives the values of lattice constant for these superconducting fullerides. When a $C_{60}$ single crystal is doped by alkali metals, its lattice constant slightly increases in comparison with that of the pristine $C_{60}$ crystal. The degree of this lattice expansion depends on the radius of the dopant alkali atom. In Table 3.7, one can see that the critical temperature increases as the cubic C$_{60}$ lattice expands. Thus, there is a correlation between the critical temperature $T_c$ and the lattice constant $a$. Nevertheless, it is generally believed that a relation between $T_c$ and the electronic density of states at the Fermi level $N(E_F)$ is more fundamental than between $T_c$ and $a$. On the other hand, there is considerable uncertainty regarding the magnitude of the experimental electronic density of states for specific fullerides, while the lattice constants can be more reliably measured. It is for this reason that plots of $T_c$ versus $a$ are more commonly used in the literature.

*Table 3.7.*   Critical temperature $T_c$ and the lattice constant $a$ for some $M_3C_{60}$ fullerides [31]

| $M_3$ in $M_3C_{60}$ | $a$ ($\overset{\circ}{A}$) | $T_c$ ($K$) |
|---|---|---|
| RbCs$_2$ | 14.555 | 33 |
| Rb$_2$Cs | 14.431 | 31.3 |
| Na$_2$Cs(NH$_3$)$_4$ | 14.473 | 29.6 |
| Rb$_3$ | 14.384 | 29 |
| Rb$_2$K | 14.323 | 27 |
| K$_2$Cs | 14.292 | 24 |
| KRb$_2$ | 14.243 | 23 |
| K$_2$Rb | 14.243 | 23 |
| K$_3$ | 14.240 | 19.5 |

Let us now discuss superconducting properties of the fullerides. First of all, it is necessary to emphasize that the fullerides are electron-doped superconductors, not hole-doped as the cuprates and organic salts. Experimentally, the critical temperature of hole-doped superconductors is usually a few times higher than that of electron-doped superconductors. So, it is possible that the temperature $T_c \sim 40$ K can be a maximum for *electron*-doped fullerides.

It is by now generally agreed that the electron-phonon interaction is the dominant pairing mechanism in the fullerides [31, 32]. At the same time, antiferromagnetic spin fluctuations participate also in mediating superconductivity in the fullerides [19]. For example, the $T_c(p)$ dependence in the fullerides, where $p$ is the carrier concentration, has a bell-like shape [32], typical for the cuprates and organic salts. Furthermore, the Néel temperature in antiferromagnetic non-superconducting fullerides as a function of crystal volume also has a bell-like shape [19]. Generally speaking, a bell-like shape of the $T_c(p)$ dependence is the "fingerprint" left by spin fluctuations participating in superconductivity. Therefore, such a bell-like $T_c(p)$ dependence is in fact typical for all compounds of the third group of superconductors. For the fullerides, this means that superconductivity in alkali-doped C$_{60}$ is unconventional.

The unconventional type of superconductivity in the fullerides manifests itself through most superconducting characteristics. For example, the carbon isotope effect in some $C_{60}$ compounds is not of the BCS type. In spite of the fact that the isotope-mass exponent $\alpha$ in most of the fullerides is around 0.3, in some of them $\alpha$ is much larger than 0.5 (the BCS value). For instance, the carbon-isotope-mass exponent in $Rb_3C_{60}$ is larger than 2. Such an exponent value is similar to that of oxygen isotope effect in underdoped cuprates. Secondly, the superfluid density $n_s$ in the fullerides is very low: in the Uemura plot shown in Fig. 3.6, $K_3C_{60}$ is situated among other unconventional superconductors. As a consequence of low values of $n_s$, the Fermi energy $E_F$ in the fullerides is also low ($\sim 0.25$ eV) and comparable with that of the cuprates. The values of the coherence length in alkali-doped $C_{60}$ are small, $\sim 30$ Å, while the penetration depth is very large, $\sim 4000$ Å. So, the fullerides are type-II superconductors. Table 2.2 presents some superconducting characteristics for $K_3C_{60}$ and $Rb_3C_{60}$. The values of $H_{c1}$ in the fullerides are very small, $\sim 100$–$200$ Oe, whilst those of $H_{c2}$ are sufficiently large for electron-doped superconductors, $\sim 30$–$50$ T. The gap ratio obtained in $Rb_3C_{60}$ in tunneling measurements is also sufficiently large for electron-doped compounds, $2\Delta/(k_B T_c) \simeq 5.4$ (thus, $\Delta \simeq 7$ meV).

## 3.5 Graphite intercalation compounds

The first observation of superconductivity in a doped graphite goes back to 1965, when superconductivity was observed in the potassium graphite intercalation compound $C_8K$ having a critical temperature of 0.55 K. Later, superconductivity was observed in other graphite intercalation compounds (GICs) [33]. A single layer of three-dimensional graphite is defined as a *graphene* layer. In GICs, the graphene layers are separated by the layers of intercalant atoms. The crystal structure of graphite is shown in Fig. 3.19. The interlayer spacing in graphite is about 3.354 Å, and the length of C–C bonds in the graphene is 1.421 Å. The bonds between adjacent layers in graphite are weak.

According to the preparation method, the superconducting GICs can be divided into two subgroups: the stage 1 and stage 2 GICs. The stage 2 GICs are synthesized in two stages, and so they are referred to as the stage 2 compounds. The structures of the stage 1 and 2 GICs are different along the $c$ axis. In the stage 1 GICs, the adjacent intercalant layers are separated from one another by *one* graphene layer, while in the stage 2 GICs, the neighboring intercalant layers are separated by *two* graphene layers. The stage 1 GICs consist of the binary $C_8M$, ternary $C_4MHg$ and $C_4MTl_{1.5}$ compounds, and the stage 2 GICs are represented by the ternary $C_8MHg$ and $C_8MTl_{1.5}$, where M = K, Rb and Cs, i.e. the same alkali atoms which are used to dope the fullerene $C_{60}$ (see the previous subsection). This means that the superconducting GICs are magnetic due to spins of the alkali atoms. In the superconducting GICs, as well as in the fullerides, the charge carriers are electrons, not holes.

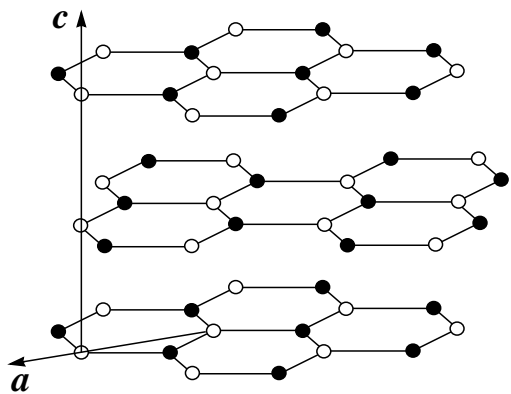

*Figure 3.19.* Crystal structure of hexagonal graphite. Only three planes of carbon (graphene layers) are shown. The nearest-neighbor carbon distance in graphene layer is 1.421 Å, and the distance between the layers is 3.354 Å.

At ambient pressure, the critical temperatures of the superconducting GICs discovered before 1986 are low, $T_c < 3K$. After 1986 when high-$T_c$ superconductivity was discovered in cuprates, most groups suspended the search for superconductivity in GICs. For binary $C_8M$ compounds, the highest critical temperatures reported for M = K, Rb and Cs are 0.55, 0.15 and 0.135 K, respectively. In the alkali metal amalgam GICs $C_8KHg$ and $C_8RbHg$, the critical temperatures are 1.93 and 1.44 K, respectively. In the potassium thallium GICs $C_4KTl_{1.5}$ and $C_8KTl_{1.5}$, respectively $T_c = 2.7$ and 1.3 K. With the potassium thallium GICs excluded, the critical temperature of the stage 2 GICs is in general higher than that of the stage 1 GICs. Under pressure, the sodium graphite intercalation compound $C_2Na$ superconducts below $T_c \sim 5$ K.

The physical properties of superconducting GICs, in many respects, are similar to those of the fullerides and $MgB_2$. The latter material is a representative of the second group of superconductors, similar to graphite both electronically and crystallographically (compare Figs. 3.3 and 3.19). So, it is possible that nonmagnetic GICs that superconduct at low temperature will be discovered in the near future. This family of superconducting GICs will already belong to the second group of superconductors.

All the GICs are two-dimensional. As a consequence, their superconducting properties are anisotropic as those in the cuprates, organic salts and $MgB_2$. Thus, the GICs are type-II superconductors. The anisotropy in most GICs, $\xi_\parallel/\xi_\perp$, is between 10 and 50. In low-$T_c$ GICs, the values of $\xi_\parallel$ and $\xi_\perp$ are of the order of 3000 and 100 Å, respectively, and $H_{c2,\perp} \sim 0.2$ T. In $C_4KTl_{1.5}$ which has the highest $T_c$ at ambient pressure (= 2.7 K), $\xi_\parallel \simeq 280$ Å, $\xi_\perp \simeq 40$ Å, and $H_{c2,\perp} \simeq 3$ T [33]. So, the anisotropy in $C_4KTl_{1.5}$, $\xi_\parallel/\xi_\perp \simeq 7$, is not very large, and $\xi_\perp \simeq 40$ Å is much larger than the interlayer distance in the superconducting GICs, $\sim 10$ Å.

Recently, considerable scientific interest in graphite and graphite-based superconducting materials has been renewed after the discovery of superconductivity in $MgB_2$. In 2001, superconductivity at $T_c = 35$ K was observed in graphite-sulfur composites [35]. In this work, however, the structure of the sulfur intercalant layers was not identified. As a result, it is not clear to what group this C-S composite belongs, and whether it is an electron-doped or hole-doped superconductor.

Finally, let us briefly discuss how the stage 1 and 2 GICs are synthesized. The stage 1 GICs are prepared similarly to the superconducting fullerides: a single crystal of highly oriented pyrolytic graphite is placed in an evacuated tube, then heated to $\sim 300°C$ in an atmosphere of intercalant vapor for couple of days. The intercalant pristine is however heated in another tube to a much lower temperature, $\sim 150$–$200°C$, so that the intercalant vapor can reach the graphite single crystal to diffuse between the graphene layers. This technique is called the two-temperature method [34]. The stage 2 GICs are synthesized in two stages. As an example, let us consider the preparation procedure of the stage 2 compound $C_8MHg$. In the first step, the binary compound $C_8M$ is prepared by the same two-temperature method, and then transferred to a new tube and exposed to mercury vapor at about $100°C$. As the reaction proceeds, the stage 1 binary $C_8M$ changes into the stage 2 ternary $C_8MHg$. As all the fullerides, the alkali metal GICs are extremely unstable in air and, therefore, must be kept in an inert atmosphere.

## 3.6 Polymers

At present, no organic polymer yet discovered exhibits superconductivity. In contrast to solid crystals, conducting organic polymers like polyacetylene are very flexible. So, a superconducting organic polymer with a high critical temperature will have an enormous potential for practical applications, first of all, for making superconducting wires. However, one *inorganic* polymer, $(SN)_x$, is already known to superconduct below $T_c = 0.3$ K.

Superconductivity in $(SN)_x$ was discovered in 1975. It is the first superconductor found among quasi-one-dimensional conductors and, moreover, the first that contained no metallic elements. $(SN)_x$ is a chain-like polymer in which sulphur and nitrogen atoms alternate along the chain. Single crystals have a *dc* electrical conductivity of about $1.7 \times 10^5 \ \Omega^{-1} \mathrm{m}^{-1}$ along the chains, and the anisotropy is of the order of $10^3$. A remarkable property of $(SN)_x$ is that it does not undergo a metal-insulator (Peierls) transition at low temperatures but turns instead into a superconductor below 0.3 K.

In a sense, carbon nanotubes, which we shall discuss in a moment, can be considered as *organic* polymers since they can be viewed as giant conjugated molecules with a conjugated length corresponding to the whole length of the tube. They, in fact, can already be used in superconducting chips.

## 3.7  Carbon nanotubes and DNA

In addition to ball-like fullerenes, it is possible to synthesize tubular fullerenes. By rolling a graphene sheet (see Fig. 3.19) into a cylinder and capping each end of the cylinder with a half of a fullerene molecule, a *fullerene-derived tubule*, one atomic layer, is formed, which we shall call a *carbon nanotube*, or just a *nanotube* for short. According to their structure, one can have three types of the nanotubes. If one rolls up a graphene sheet along the *a* axis, shown in Fig. 3.19, one will obtain a nanotube called *zigzag*. By rolling a graphene sheet in the direction $\theta = 30°$ relative to the *a* axis, one obtains an *armchair* nanotube. In the case $0° < \theta < 30°$, a nanotube called *chiral* will be formed. Figure 3.20 shows a piece of armchair nanotube. The armchair nanotubes are usually metallic, while the zigzag ones are semiconducting. The carbon nanotubes and fullerenes have a number of common features and also many differences.

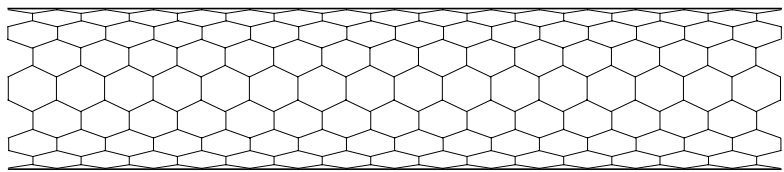

*Figure 3.20.*  A piece of armchair nanotube.

Carbon nanotubes were first observed in 1991 by Iijima in Japan. In fact, they were multi-walled carbon nanotubes consisting of several concentric single-walled nanotubes nested inside each other, like a Russian doll. Two years later, single-walled nanotubes were observed for the first time. They had just 10–20 Å in diameter. But the field really took off a few years later when various groups found ways to mass-produce high-quality nanotubes. At the present, carbon-nanotube research is probably the most active research field in carbon science.

The nanotubes have an impressive list of attributes. They can behave like metals or semiconductors, can conduct electricity better than copper, can transmit heat better than diamond. They rank among the strongest materials known, and they can superconduct at low temperatures—not bad for structures that are just a few nanometers across. These remarkable properties of carbon nanotubes suggest enormous opportunities for practical applications which, undoubtedly, will follow in the near future.

In 1999, proximity-induce superconductivity below 1 K was observed in single-walled carbon nanotubes, followed by the observation of genuine superconductivity with $T_c = 0.55$ K. In the latter case, the diameter of single-walled nanotubes was of the order of 14 Å. Soon afterwards, superconductivity below $T_c \simeq 15$ K was seen in single-walled carbon nanotubes with a diameter of $4.2 \pm 0.2$ Å [36]. So, the nanotubes with a smaller diameter (4.2 Å < 14 Å) exhibit a higher $T_c$ (15 K > 0.55 K). The nanotube diameter of 4.2 Å is very small, and can be at, or very close to, the theoretical limit. In the case of $T_c = 15$ K, the coherence length estimated along the tube direction is about $\xi_0 \approx 42$ Å. The effective mass of charge carriers, obtained in calculations, is $m^* = 0.36\,m$, where $m$ is the free electron mass.

One of the main problems to study carbon nanotubes, as well as DNA (deoxyribonucleic acid), is not only their structure and possible defects along them, but also the quality of electrical contacts between a nanotube and leads. For example, it is impossible to solder metal leads onto carbon nanotubes in the conventional sense of this expression because metals do not wet the tubes. Therefore, a new laser-based technique was developed for solving the problem [37].

There is a report suggesting the observation of superconductivity at 645 K in single-walled carbon nanotubes which contain a small amount of the magnetic impurities Ni and Co [38]. It is assumed that the nanotubes are only partially in the superconducting state, and the normal charge carriers are also present at such high temperatures. We shall discuss these data in Chapters 8 and 10.

In 2001, proximity-induced superconductivity was observed below 1 K in DNA [37]. The observation of a proximity effect in DNA molecules signifies that they are in a state near a metal-insulator transition point. The double helix of DNA has a diameter of 20 Å. It is assumed that if one can find a technique to dope DNA, it is most likely that it will exhibit genuine superconductivity.

## 3.8 Heavy-fermion systems

This family of superconductors includes superconducting compounds which consist of one magnetic ion with $4f$ or $5f$ electrons (usually Ce or U) and other constituent or constituents being $s$, $p$, or $d$ electron metals. The principal feature of these materials is reflected in their name: below a certain coherence temperature ($\sim$ 20–100 K), the effective mass of charge carriers in these compounds become gigantic, up to several hundred times greater than that of a free electron. A large number of heavy fermions superconduct exclusively under pressure. The $T_c$ values of superconducting heavy fermions are in general very low; however, the family of these intermetallic compounds is one of the best examples of highly correlated condensed matter systems.

The first such superconductor, $CeCu_2Si_2$, was discovered in 1979 by Steglich and co-workers, and some time passed before the heavy-fermion phenomenon was confirmed by the discovery of $UBe_{13}$ and then $UPt_3$, with critical temperatures of $T_c = 0.65$, 0.9 and 0.5 K, respectively. Since then many new heavy-fermion systems that superconduct at low temperatures have been found. A few characteristics for five heavy fermions are given in Table 2.2. The crystal structure of these compounds does not have a common pattern, but varies from case to case. For example, the crystal structure of the first discovered superconducting heavy fermions—$CeCu_2Si_2$, $UBe_{13}$ and $UPt_3$—is tetragonal, cubic and hexagonal, respectively.

These systems display a rich variety of phenomena both in the normal and superconducting states. Let us start with their anomalous normal-state properties. At room temperatures, the $f$-electrons of the magnetic ions behave as localized spins; the conduction electrons are the $s$, $p$ or $d$ electrons and have quite ordinary effective masses. As the temperature is lowered, the $f$-electrons begin to couple to the conduction electrons, resulting in very large effective masses for the hybridized carriers. Due to the strong electron correlation, these materials have several characteristics that distinguish them from ordinary metals. The electronic heat capacities are $10^2$–$10^3$ times larger than that observed in ordinary metals, and the magnetic (Pauli) susceptibility at low temperatures is $\sim 100$ times larger. Both these abnormalities are consequences of a very large effective mass of the charge carriers. For example, the values of effective mass in $UBe_{13}$, $UPt_3$ and $URu_2Si_2$ are respectively $m^*/m \simeq 300$, 180 and 25, where $m$ is the electron mass. The Fermi velocity of such heavy quasiparticles is very small. If in ordinary metals $v_F \sim 10^8$ cm/s, the values of the Fermi velocity in $CeCu_2Si_2$, $UPt_3$, $UBe_{13}$ and $CeAl_3$ are $1.0 \times 10^5$, $6.6 \times 10^6$, $3.4 \times 10^6$ and $1.2 \times 10^5$ cm/s, respectively.

The temperature dependence of resistivity in heavy fermions is similar to those measured along the $c$ axis in the Bechgaard salt and underdoped cuprates, shown in Fig. 3.16. Unlike the ordinary metals where the resistance falls with decreasing temperature, in heavy fermions it first rises, attains a maximum, and then falls, vanishing at $T_c$. Ultrasound measurements carried out in the normal state show that, between room temperature and $T_c$, heavy fermions undergo structural phase transitions. In $UPt_3$, the attenuation ultrasound measurements in the normal state reveal a $T^2$ dependence of the attenuation down to the lowest temperature, consistent with electron-electron scattering.

The superconducting state in heavy fermions also displays some anomalous properties. The enormous value of the Sommerfeld constant $\gamma = C/T$, where $C$ is the specific heat capacity, and the jump in $C$ at $T_c$ (see Chapter 2) reveal that the heavy electrons participate in superconducting pairing. Furthermore, the temperature dependence of the heat capacity below $T_c$ is not exponential. Instead, it follows a power law, indicating that the energy gap at the Fermi surface has nodes in certain directions. Thus, the energy gap is highly anisotropic. Ultrasound measurements performed in a large number of heavy fermions be-low $T_c$ confirm also the existence of nodes in the gap. In $UBe_{13}$, the gap ratio obtained in Andreev-reflection measurements, $2\Delta/(k_B T_c) \simeq 6.7$, uncovers the unconventional type of superconductivity. A zero-bias conductance peak observed in these measurements is theoretically an indicator of a d-wave energy gap. At the same time, tunneling measurements show that $UBe_{13}$ is an s-wave superconductor. Tunneling measurements performed in $UPt_3$ indicate that the s-wave order parameter is anisotropic (see references in [19]). Such a conflicting situation concerning the gap symmetry is similar to that for superconducting cuprates.

The superfluid density in heavy fermions is very low. In the inset of Fig. 3.6, only $UPt_3$ and $UBe_{13}$ are shown. Recent $\mu$SR measurements performed in $UPd_2Al_3$, $URu_2Si_2$ and $U_6Fe$ show that, in the Uemura plot, these compounds are also situated in the group of all unconventional superconductors. From Fig. 3.6, the heavy fermions, in fact, have relatively high $T_c$ as scaled with their low superfluid density, which is a consequence of their large effective mass $m^*$. The phase diagram of many superconducting heavy fermions is very complex. For example, specific-heat capacity measurements show that, in zero magnetic field, $UPt_3$ has two superconducting phase transitions at $\sim 0.475$ K and $\sim 0.520$ K (a similar phenomenon is observed in superfluid $^3$He). Furthermore, $UPt_3$ has three distinct superconducting phases in the magnetic field-temperature plane. As in all unconventional superconductors, the superconducting properties of heavy fermions are very anisotropic. For example, the values of the upper critical field of tetragonal $URu_2Si_2$ are $\sim 2$ T for $H\|c$ and 8 T for $H\perp c$. The electrical resistivity in heavy fermions also varies with direction in the crystal.

Probably, the most interesting characteristic of superconducting heavy fermion materials is the interplay between superconductivity and magnetism. The magnetic ions are responsible for the magnetic properties of heavy fermions. For example, in the heavy fermions $UPt_3$, $URu_2Si_2$, $UCu_5$ and $CeRhIn_5$, magnetic correlations lead to an itinerant spin-density-wave order, while, in $UPd_2Al_3$ and $CeCu_2Si_2$, to a localized antiferromagnetic order. In the latter two heavy fermions, the antiferromagnetic order appears first, followed by the onset of superconductivity. In these compounds, as well as in other superconducting heavy fermions with long-range antiferromagnetic order, the Néel temperature is about $T_N \sim 10T_c$. For instance, in $CeRh_{0.5}Ir_{0.5}In_5$ and $CeRhIn_5$, the bulk superconductivity coexists *microscopically* with small-moment magnetism ($\leq 0.1\mu_B$). In the heavy fermion $CeIrIn_5$, the onset of a small magnetic field ($\sim 0.4$ Gauss) sets in exactly at $T_c$. In the heavy fermions, superconductivity and antiferromagnetic order do not compete, since superconductivity is mediated by spin fluctuations.

Recently, superconductivity was discovered in $PuCoGa_5$ [39], the first superconducting heavy fermion based on plutonium. What is even more interest-ing is that the superconductivity survives up to an astonishingly high temperature of 18 K. Such a high critical temperature indicates that in $PuCoGa_5$ the effective mass of quasiparticles is much lower than that in other heavy fermion compounds. The crystal structure of $PuCoGa_5$ is layered and tetragonal. The estimated value of $H_{c2}$ is around 35 T.

All the superconducting heavy fermion systems considered up to now have *antiferromagnetic* correlations. All experimental facts known before 2000 supported a point of view that superconductivity and *ferromagnetism* are mutually hostile and cannot coexist. For example, in the boride $ErRh_4B_4$ and Chevrel-phase $HoMo_6S_8$, superconductivity is destroyed by the onset of a first-order ferromagnetic phase transition. So, it was a surprise when in 2000 the coexistence of superconductivity and ferromagnetism was discovered in an alloy of uranium and germanium, $UGe_2$. At ambient pressure, $UGe_2$ is known as a metallic ferromagnet with a Curie temperature of $T_C = 53$ K. However, as increasing pressure is applied to the ferromagnet, $T_C$ falls monotonically, and appears to vanish at a critical pressure of $P_c \simeq 16$–17 kbars. In a narrow range of pressure below $P_c$ and thus *within* the ferromagnetic state, the superconducting phase appears in the millikelvin temperature range below the critical temperature. Above $P_c$, $UGe_2$ is paramagnetic.

As a matter of fact, magnetic fluctuations are strongest when magnetic order is about to form or disappear, a point known as *the quantum critical point*. Quantum critical points have attracted a great deal of attention because the large slow spin fluctuations that occur near the critical pressure (critical density) play a key role in the making and breaking of Cooper pairs.

Soon after the discovery of superconductivity in itinerant ferromagnet $UGe_2$, two new itinerant ferromagnetic superconductors were discovered—zirconium zinc $ZrZn_2$ and uranium rhodium germanium $URhGe$. $ZrZn_2$ superconducts only when it is ferromagnetic, i.e. below the critical pressure $P_c \simeq 21$ kbars. Above $P_c$, it is a paramagnet showing no trace of superconductivity. In $ZrZn_2$, the maximum critical temperature is slightly less than 3 K at ambient pressure, and decreases with increasing pressure. $URhGe$ is also a superconductor at ambient pressure, and has many similar properties of high-pressure $UGe_2$—it loses its resistance below 9.5 K, exhibits the Meissner effect and has a large specific-heat anomaly at the superconducting critical temperature.

The archetypal ferromagnet, iron, is found to superconduct at high pressure between 15 and 30 kbars. Albeit, at such pressures, iron ceases to be ferromagnetic, and there is evidence that, at low temperature, it is weakly antiferromagnetic.

The mechanism of superconductivity in the ferromagnetic heavy fermions $UGe_2$, $ZrZn_2$ and $URhGe$ is most likely the same as that in the antiferromagnetic heavy fermions and cuprates, with the exception of the symmetry of the order parameter. In the ferromagnetic heavy fermions, it has a p-wave symmetry, not a d-wave.

## 3.9 Nickel borocarbides

The nickel borocarbide class of superconductors has the general formula $R\text{Ni}_2\text{B}_2\text{C}$, where $R$ is a rare earth which is either magnetic (Tm, Er, Ho, or Dy) or nonmagnetic (Lu and Y). In the case when $R$ = Pr, Nd, Sm, Gd or Tb in $R\text{Ni}_2\text{B}_2\text{C}$, the Ni borocarbides are not superconducting at low temperatures but antiferromagnetic. In the Ni borocarbides with a magnetic rare earth, superconductivity coexists at low temperatures with a long-range antiferromagnetic order. Interestingly, while in the superconducting heavy fermions with a long-range antiferromagnetic order $T_N \sim 10 T_c$, in some Ni borocarbides it is just the opposite, $T_c \sim 10 T_N$. Thus, antiferromagnetism appears deeply in the superconducting state. Furthermore, if in the superconducting antiferromagnetic Ni borocarbides $T_c \sim 15$ K, in the non-superconducting antiferromagnetic Ni borocarbides with $R$ = Pr, Nd, Sm, Gd or Tb, the Néel temperature is also $T_N \sim 15$ K. This fact indicates that there exists a direct connection between magnetism and superconductivity in the Ni borocarbides. Indeed, in the Ni borocarbides the study of an interplay between superconductivity and antiferromagnetism shows that they do **not** compete [40].

Superconductivity in the Ni borocarbides was discovered in 1994 by Eisaki and co-workers. Transition temperatures in these quaternary intermetallic compounds can be as high as 17 K. Some characteristics for the antiferromagnetic $\text{TmNi}_2\text{B}_2\text{C}$ and nonmagnetic (i.e. without a long-range magnetic order) $\text{LuNi}_2\text{B}_2\text{C}$ borocarbides can be found in Table 2.2. The Ni borocarbides have a layered-tetragonal structure alternating $R\text{C}$ sheets and $\text{Ni}_2\text{B}_2$ layers. As a consequence, the superconducting properties of the Ni borocarbides are also anisotropic, $\xi_c < \xi_{ab}$. It is agreed that the phonon-electron interaction plays an important role in mediating superconductivity in these compounds. At the same time, in the normal state, electrical resistivity shows a $T^2$ dependence implying the presence of a strong electron-electron correlation in the Ni borocarbides.

Many different types of measurements carried out in the Ni borocarbides show that the gap ratio $2\Delta/(k_B T_c)$ is between 3.3 and 5.3. So, the coupling strength in $R\text{Ni}_2\text{B}_2\text{C}$ seems to be not very strong. At the same time, there is complete disagreement in the literature about the shape of the energy gap. In photoemission and microwave measurements, the energy gap in some Ni borocarbides was found to be an s-wave but highly anisotropic. On the other hand, in specific-heat, thermal-conductivity and Raman-scattering measurements carried out in the Ni borocarbides with $R$ = Y and Lu, the energy gap was found to be a highly anisotropic gap, most likely with nodes. Furthermore, in other thermal-conductivity measurements, the gap appears to have *point* nodes along the [100] and [010] directions, thus along the $a$ and $b$ axes. Recent tunneling measurements performed in the antiferromagnetic $TmNi_2B_2C$ show unambiguously that this Ni borocarbide is a fully gapped s-wave superconductor with a gap being slightly anisotropic. To reconcile all these data, one should assume that different measurements probe different energy gaps, either $\Delta_p$ or $\Delta_c$.

For the Ni borocarbides, there are still many open questions. For example, in the Ni borocarbide $ErNi_2B_2C$, besides the presence of incommensurate spin-density-wave order at low temperatures, the *microscopic* coexistence of spontaneous weak ferromagnetism with superconductivity was found by neutron diffraction. The other borocarbide $YbNi_2B_2C$ is unique in its behavior as a heavy fermion system. Some of its normal-state characteristics are similar to those of the heavy fermions. This borocarbide is not superconducting nor antiferromagnetic.

The layered borocarbides $DyB_2C$ and $HoB_2C$ without Ni also superconduct, with $T_c = 8.5$ and 7.1 K, respectively. At the same time, $ErB_2C$ is an antiferromagnet below $T_N = 16.3$ K.

Other related compounds, such as the Ni boronitride $La_3Ni_2B_2N_3$, are also found to superconduct.

## 3.10    Strontium ruthenate

Nearly 40 years ago it was found that $SrRuO_3$ is a ferromagnetic metal with a Curie temperature of 160 K. In its cousin, $Sr_2RuO_4$, the superconducting state with $T_c \approx 1.5$ K was discovered in 1994 by Maeno and his collaborators. The crystal structure of $Sr_2RuO_4$ is layered perovskite, and almost isostructural to the high-$T_c$ parent compound $La_2CuO_4$ (see Fig. 3.7), in which the $CuO_2$ layers are substituted by the $RuO_2$ ones. Below 50 K, electrical resistivity—both in the $RuO_2$ planes and perpendicular to the planes—shows a $T^2$ dependence implying that the electron-electron correlations in the Sr ruthenate are important. Therefore, the Fermi-liquid approach is appropriate for this compound. While searching for optimal crystal growth conditions, a eutectic solidification system, Ru metal embedded in the primary phase of $Sr_2RuO_4$, was found. An intriguing observation was that the critical temperature of this eutectic system was enhanced up to 3 K.

The superconducting properties of $Sr_2RuO_4$ are highly anisotropic: $\xi_{ab} \simeq 660$ Å and $\xi_c \simeq 33$ Å. In the mixed state, the vortex lattice has a square structure, not triangular. Different types of measurements show that the energy gap in $Sr_2RuO_4$ has line nodes. Furthermore, there is a consensus that spin fluctuations mediate superconductivity in $Sr_2RuO_4$; however, there is no agreement on the type of these fluctuations—antiferromagnetic or ferromagnetic. This issue is still widely debated because it is directly related to another important question: does the energy gap in $Sr_2RuO_4$ have a p-wave or d-wave symmetry? Let us briefly discuss this issue.

The problem is that, in analogy with $^3$He, it is assumed that the p-wave pairing is mediated via ferromagnetic spin fluctuations. Since the compounds related to $Sr_2RuO_4$ are dominated by ferromagnetic interactions—$SrRuO_3$ becomes ferromagnetic below 160 K and $Sr_3Ru_2O_7$ orders ferromagnetically at 100 K under pressure—it was initially suggested that superconductivity in $Sr_2RuO_4$ is mediated by ferromagnetic spin fluctuations. Therefore, it was immediately assumed that the energy gap in $Sr_2RuO_4$ has a p-wave symmetry. However, quite astonishingly there is not much experimental evidence for ferromagnetic spin fluctuations in $Sr_2RuO_4$. On the contrary, in inelastic neutron scattering and NMR measurements, it was found that spin fluctuations have significant antiferromagnetic character (superconducting $Sr_2RuO_4$ is extremely close to an incommensurate spin-density-wave instability). Furthermore, its cousin $Ca_2RuO_4$ was found to be an antiferromagnetic insulator with $T_N \approx 113$ K. On the other hand, the other cousin $Sr_2IrO_4$ turned out to be a weakly ferromagnetic insulator. In recent Andreev-reflection measurements performed in $Sr_2RuO_4$, a zero-bias conductance peak was observed in the superconducting state. In analogy with the cuprates, the presence of this peak in conductances indicates that the gap has a d-wave symmetry.

Recently, bilayer and trilayer strontium ruthenates have been synthesized: $Sr_3Ru_2O_7$ is an enhanced paramagnetic metal, and $Sr_4Ru_3O_{10}$ is ferromagnetic with a Curie temperature of 105 K.

## 3.11 Ruthenocuprates

Ruthenocuprates are in a sense a hybrid of the superconducting cuprates and strontium ruthenate. As a consequence, they have a number of common features with the cuprates, but also many differences. Basically, there are two ruthenocuprates that superconduct at low temperatures. The general formulas of these ruthenocuprates are $RuSr_2RCu_2O_8$ and $RuSr_2R_2Cu_2O_{10}$ with $R =$ Gd, Eu and Y. The second ruthenocuprate was discovered first in 1997. The crystal structure of $RuSr_2RCu_2O_8$ is similar to that of YBCO except for the replacement of one-dimensional CuO chains by two-dimensional $RuO_2$ layers (see Fig. 3.9). It is assumed that the $RuO_2$ layers act as charge reservoirs for the $CuO_2$ layers. The principal feature of the ruthenocuprates is that they are magnetically ordered below $T_m \sim 130$ K, and become superconducting at $T_c \sim 40$ K. For $RuSr_2RCu_2O_8$, $T_m = 130$–150 K and $T_c = 30$–45 K, while for $RuSr_2R_2Cu_2O_{10}$, $T_m = 90$–180 K and $T_c = 30$–40 K. It is believed that the magnetic order arises from ordering of Ru ions in the $RuO_2$ layers, while the transport occurs in the $CuO_2$ layers.

As in the cuprates, the superconducting properties of the ruthenocuprates are highly anisotropic: $\xi_{ab} \sim 60$–75 Å and $\xi_c \sim 10$ Å. Superconductivity and the magnetic order are found to be *homogeneous*. Chu and his collaborates suggested that a bulk Meissner effect does not exist in the ruthenocuprates. Indeed, the value of the penetration depth is very large, $\lambda \sim 30$–$50$ $\mu$m. In principle, this can happen if metallic $RuO_2$ layers remain normal below $T_c$.

The situation with the ruthenocuprates is even worse than that with $Sr_2RuO_4$. If in the Sr ruthenate, there is disagreement only on the type of spin fluctuations that mediate superconductivity, in the ruthenocuprates there are two major problems. First, the type of ordering at $T_m$ still remains controversial. Earlier experimental studies suggested a homogeneous ferromagnetic ordering of the Ru moments, while the latest ones report that the magnetic order of the Ru spins is predominantly antiferromagnetic. Second, from the beginning it was assumed that superconductivity occurs exclusively in the $CuO_2$ planes. However, recent NMR studies reveal that the superconducting gap develops also at the magnetically ordered $RuO_2$ planes with a ferromagnetic component.

There is a consensus that in the ruthenocuprate, there is a small ferromagnetic component; however, there is no agreement on its origin. It may originate not only from the Ru moments but also, for example, from the Gd spins. There are many reports on this issue, which often contradict one another. In an attempt to reconcile these discrepancies, it was suggested that the $RuO_2$ layers ordered ferromagnetically couple antiferromagnetically. In $RuSr_2RCu_2O_8$ with $R = Gd$ and Y, neutron scattering studies found that these two compounds have an antiferromagnetic ground state with a very small canting ferromagnetic component, and that an external magnetic field can tune the field-induced ferromagnetic component that coexists with superconductivity in a high field.

## 3.12    MgCNi$_3$

$MgCNi_3$ is the second most recent superconductor described in this chapter, after $Cd_2Re_2O_7$ (see the following subsection). Superconductivity in $MgCNi_3$ was discovered in 2001 by Cava and co-workers, a few months later than that in $MgB_2$. The crystal structure of $MgCNi_3$ is cubic-perovskite, and similar to that of BKBO (see Fig. 3.2). The perovskite $MgCNi_3$ is special in that it is neither an oxide nor does it contain any copper. Since Ni is ferromagnetic, the discovery of superconductivity in $MgCNi_3$ was surprising. The critical temperature is near 8 K. $MgCNi_3$ is metallic, and the charge carriers are electrons which are derived predominantly from Ni.

The estimated values of the coherence length and upper critical field in $MgCNi_3$ are $\xi \approx 46$ Å and $H_{c2} \simeq 15$ T, respectively. Penetration-depth measurements at microwave frequencies show unambiguously that superconductivity in $MgCNi_3$ is not of the BCS type, and $\lambda(0) = 2480$ Å. In Andreev-reflection measurements performed on polycrystalline samples (single crystals of $MgCNi_3$ are not yet available), a zero-bias conductance peak was observed. In analogy with the cuprates, the presence of this peak in a conductance indi-cates a d-wave symmetry of the energy gap. The gap ratio obtained in these Andreev-reflection measurements, $2\Delta/k_B T_c \sim 10$, is very large. However, this value is only an estimate because it is obtained in polycrystalline samples. The Debye temperature obtained from specific-heat measurements is $\Theta \simeq 284$ K, and the value of specific-heat jump at $T_c$ is $\beta \simeq 2.1$. At low temperature, the Ginzburg-Landau parameter is $k = 54$.

Structural studies of $MgCNi_3$ reveal structural inhomogeneity. Apparently, the perovskite cubic structure of $MgCNi_3$ is modulated locally by the variable stoichiometry on the C sites.

## 3.13  $Cd_2Re_2O_7$

$Cd_2Re_2O_7$ is the most recent superconductor described in this chapter. Although $Cd_2Re_2O_7$ was synthesized in 1965, its physical properties remained almost unstudied. Unexpectedly, superconductivity in $Cd_2Re_2O_7$ was discovered in the second half of 2001 by Sakai and co-workers. The critical temperature of $Cd_2Re_2O_7$ is low, $T_c = 1$–1.5 K. This compound is the first superconductor found among the large family of pyrochlore oxides with the formula $A_2B_2O_7$, where A is either a rare earth or a late transition metal, and B is a transition metal. In this structure, the A and B cations are 4- and 6-coordinated by oxygen anions. The A-$O_4$ tetrahedra are connected as a pyrochlore lattice with straight A-O-A bonds, while B-$O_6$ octahedra form a pyrochlore lattice with the bent B-O-B bonds with an angle of 110–140°. Assuming that electronic structure in $Cd_2Re_2O_7$ as formally $Cd^{2+}$ $4d^{10}$ and $Re^{5+}$ $4f^{14}5d^2$, the electronic and magnetic properties are primarily dominated by the Re $5d$ electrons. $Cd_2Re_2O_7$ shows an anomaly at 200 K in electrical resistivity, magnetic susceptibility, specific heat and Hall coefficient: there is a structural phase transition near 200 K. Another structural phase transition occurs around 1.5 K, just above the superconducting transition.

Oxide superconductors with non-perovskite structure are rare. Previous studies indicate that the pyrochlores, like the spinels, are geometrically frustrated. The effect of geometric frustration on the physical properties of spinel materials is drastic, resulting in, for example, heavy-fermion behavior in $LiV_2O_4$. Another spinel compound $LiTi_2O_4$ is a superconductor below $T_c = 13.7$ K. Indeed, x-ray diffraction studies performed under high pressure showed that superconductivity in $Cd_2Re_2O_7$ is detected only for the phases with a structural distortion. It was suggested that the charge fluctuations of Re ions play a crucial role in determining the electronic properties of $Cd_2Re_2O_7$.

Between 2 and 60 K, the resistivity in $Cd_2Re_2O_7$ exhibits a $T^2$ dependence indicative of the Fermi-liquid behavior. The value of the specific-heat jump at $T_c$, 1.29, is close to the weak coupling BCS value. In comparison with other pyrochlores, the value of the Sommerfeld constant $\gamma$ for $Cd_2Re_2O_7$ is large, suggesting that the electrons in $Cd_2Re_2O_7$ are strongly correlated with the en-hanced effective mass, resulting possibly from geometric frustration. The Re nuclear quadrupole resonance (NQR) measurements performed in zero magnetic field below 100 K rule out any magnetic or charge order. Specific heat and Re NQR measurements suggest that the superconducting gap in $Cd_2Re_2O_7$ is almost isotropic.

The value of the coherence length in $Cd_2Re_2O_7$ is $\xi_0 \sim 260$ Å. The value of the penetration depth is very large, $\lambda(0) \sim 7500$ Å. The lower and upper magnetic fields are $H_{c1} \leq 0.002$ T and $H_{c2} \approx 0.85$ T, respectively.

## 3.14    Hydrides and deuterides

In addition to the nitrides and carbides from the second group of superconductors, another class of superconducting compounds that also has the NaCl structure are hydrides and deuterides (i.e. compounds containing hydrogen or deuterium). However, in contrast to the nitrides and carbides, superconducting hydrides and deuterides are magnetic. In the seventies it was discovered that some metals and alloys, not being superconducting in pure form, become relatively good superconductors when they form alloys or compounds with hydrogen or deuterium. These metals include the transition elements palladium (Pd) and thorium (Th) that have unoccupied $4d$- and $5f$-electron shells, respectively.

In 1972, Skoskewitz discovered that the transition element Pd which has a small magnetic moment normally preventing the pairing of electrons, joins hydrogen and forms the PdH compound that superconducts at $T_c = 9$ K. This compounds has the NaCl cubic structure, thus it is a B1 compound. Later on, it was found that by doping such a system with noble metals the critical temperature increases up to 17 K. Interestingly, the palladium-deuterium compound also superconducts, and its critical temperature equal to 11 K is higher than that of PdH. So the hydrogen isotope effect in PdH is reverse (negative), which is similar to that observed in some organic superconductors. In contrast, the critical temperatures of the ThH and ThD compounds do not differ drastically from each other like those of PdH and PdD. Probably, the higher atomic mass of thorium is the cause of this discrepancy.

The experimental studies of the $Pd_{1-x}M_xH_y$ hydrides, where M = Al, Pb, In and Cu, showed that, as one increases the M concentration from zero, their critical temperatures as a function of $x$ first increase and then drop quite sharply. This study was performed under the most favorable hydrogen concentrations that correspond to the maximum value of $T_c$ with fixed $x$. The critical temperature of the $Pd_{1-x}Ag_xD_y$ deuteride as a function of $x$ also exhibits the same tendency as those of the doped palladium hydrides. However, in the doped hydrides and deuterides, their critical temperatures reach maximums at different dopant concentrations. This is due to the fact that different metals have different abilities to donate electrons. However, such a bell-like shape of the $T_c(p)$ dependence, where $p$ is the doping level, is typical for all the compounds of the third group of superconductors.

The hydrides and deuterides have two conduction bands as the heavy fermions do: the wide valence band of $s$- and $p$-electrons and the very narrow band generated by electrons in the inner $4d$- and $5f$-subshells incompletely occupied by electrons. The existence of two conduction bands is also typical for the second-group superconductors.

## 3.15 Oxides

Superconducting oxides are a special family of superconductors. The class of superconducting compounds containing the element oxygen O is probably the largest family among all superconducting materials. Furthermore, they exhibit the highest critical temperatures (cuprates). Representatives of this family of superconductors are members of either the second (BKBO, $SrTiO_3$) or third group of superconductors (cuprates, ruthenates, etc.). NbO was the first superconducting oxide discovered in 1965 by Miller and his collaborators. The oxide series, beginning from NbO, is shown in Fig. 1.2.

Usually, oxides are associated with insulators, while superconductors with the best conductors such as Cu, Ag and Au. However, the opposite is true. In materials with a weak phonon-electron interaction, like Cu, Ag and Au, superconductivity is absent, while materials with a moderately strong phonon-electron interaction, like oxides, exhibit sometimes superconductivity. This fact shows the importance of the electron-lattice interaction for superconductivity.

# Chapter 4

# PRINCIPLES OF SUPERCONDUCTIVITY

The issue of room-temperature superconductivity is the main topic of this book. Even if this subject was raised for the first time before the development of the BCS theory and later by Little in 1964 [2], from the standpoint of practical realization, this issue is still a new, "untouched territory." To go there, we need to know Nature's basic rules for arrangement of matter over there. Otherwise, this journey will face a *fiasco*. To have the microscopic BCS theory in a bag is very useful, but not enough. It is clear to everyone by now that a room-temperature superconductor can not be of the BCS type. Therefore, we need to know more general rules, principles of superconductivity that incorporate also the BCS-type superconductivity as a particular case.

The purpose of this chapter is to discuss the main principles of superconductivity as a phenomenon, valid for every superconductor independently of its characteristic properties and material. The underlying mechanisms of superconductivity can be different for various materials, but certain principles must be satisfied. One should however realize that the principles of superconductivity are not limited to those discussed in this chapter: it is possible that there are others which we do not know yet about.

The first three principles of superconductivity were introduced in [19].

## 1.     First principle of superconductivity

The microscopic theory of superconductivity for *conventional* superconductors, the BCS theory, is based on Leon Cooper's work published in 1956. This paper was the first major breakthrough for understanding the phenomenon of superconductivity on a *microscopic* scale. Cooper showed that electrons in a solid would always form pairs if an attractive potential was present. It did not matter if this potential was very weak. It is interesting that, during his cal-culations, Cooper was not looking for pairs—they just "dropped out" of the mathematics. Later it became clear that the interaction of electrons with the lattice allowed them to attract each other despite their mutual Coulomb repulsion. These electron pairs are now known as Cooper pairs.

An important note: in this chapter and further throughout the book, we shall use the term "a Cooper pair" more generally than its initial meaning. In the framework of the BCS theory, the Cooper pairs are formed in momentum space, not in real space. Further, we shall consider the case of electron pairing in real space. For simplicity, we shall sometimes call electron pairs formed in real space also as Cooper pairs.

In solids, superconductivity as a quantum state cannot occur without the presence of bosons. Fermions are not suitable for forming a quantum state since they have spin and, therefore, they obey the Pauli exclusion principle according to which two identical fermions cannot occupy the same quantum state. Electrons are fermions with a spin of 1/2, while Cooper pairs are already composite bosons since the value of their total spin is either 0 or 1. Therefore, the electron pairing is an inseparable part of the phenomenon of superconductivity and, in any material, superconductivity cannot occur without electron pairing.

In some unconventional superconductors, the charge carriers are not electrons but holes with a charge of $+|e|$ and spin of 1/2. The reasoning used above for electrons is valid for holes as well. Thus, in the general case, it is better to use the term "quasiparticles" which also reflects the fact that the electrons and holes are in a medium.

The first principle of superconductivity:

Principle 1:          **Superconductivity requires quasiparticle pairing**

In paying tribute to Cooper, the first principle of superconductivity can be called the *Cooper principle*.

In the framework of the BCS theory, the quasiparticle (electron) pairing occurs in momentum space, not in real space. Indeed in the next section, we shall see that the electron pairing in conventional superconductors cannot occur in real space because the onset of long-range phase coherence in classical superconductors occurs due to the overlap of Cooper-pair wavefunctions, as shown in Fig. 4.1. As a consequence, the order parameter and the Cooper-pair wavefunctions in conventional superconductors are the same: the order parameter is a "magnified" version of the Cooper-pair wavefunctions. However, in unconventional superconductors, the electron pairing is not restricted by the momentum space because the order parameter in unconventional superconductors has nothing to do with the Cooper-pair wavefunctions. *Generally speaking*, the electron pairing in unconventional superconductors may take place not only in momentum space but also in real space. We shall discuss such a possibility in the following section.

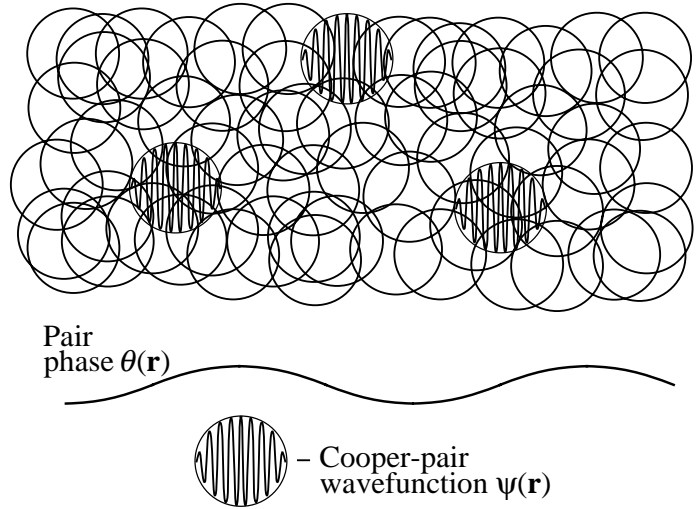

*Figure 4.1.* In conventional superconductors, the superconducting ground state is composed by a very large number of overlapping Cooper-pair wavefunctions, $\psi(\mathbf{r})$. To avoid confusion, only three Cooper-pair wavefunctions are shown in the sketch; the other are depicted by open circles. The phases of the wavefunctions are locked together since this minimizes the free energy. The Cooper-pair phase $\Theta(\mathbf{r})$, illustrated in the sketch, is also the phase of the order parameter $\Psi(\mathbf{r})$.

The electron pairing in momentum space can be considered as a *collective* phenomenon, while that in real space as *individual*. We already know that the density of free (conduction) electrons in conventional superconductors is relatively high ($\sim 5 \times 10^{22}$ cm$^{-3}$); however, only a small fraction of them participate in electron pairing ($\sim 0.01\%$). In unconventional superconductors it is just the other way round: the electron density is low ($\sim 5 \times 10^{21}$ cm$^{-3}$) but a relatively large part of them participate in the electron pairing ($\sim 10\%$). Independently of the space where they are paired—momentum or real—two electrons can form a bound state **only if** *the net force acting between them is attractive*.

## 2. Second principle of superconductivity

After the development of the BCS theory in 1957, the issue of long-range phase coherence in superconductors was not discussed widely in the literature because, in conventional superconductors, the pairing and the onset of phase coherence take place simultaneously at $T_c$. The onset of phase coherence in conventional superconductors occurs due to the overlap of Cooper-pair wave-functions, as shown in Fig. 4.1. Only after 1986 when high-$T_c$ superconductors were discovered, the question of electron pairing above $T_c$ appeared. So, it was then realized that it is necessary to consider the two processes—the electron pairing and the onset of phase coherence—separately and independently of one another [12].

In many unconventional superconductors, quasiparticles become paired above $T_c$ and start forming the superconducting condensate only at $T_c$. Superconductivity requires both the electron pairing and the Cooper-pair condensation. Thus, the second principle of superconductivity deals with the Cooper-pair condensation taking place at $T_c$. This process is also known as the onset of long-range phase coherence.

Principle 2:        **The transition into the superconducting state is the Bose-Einstein-like condensation and occurs in momentum space**

Let us first start with one main difference between fermions and bosons. Figure 4.2 schematically shows an ensemble of fermions and an ensemble of bosons at $T \gg 0$ and $T = 0$. In Fig. 4.2 one can see that, at high temperatures, both types of particles behave in a similar manner by distributing themselves in their energy levels somewhat haphazardly but with more of them toward lower energies. At absolute zero, the two types of particles rearrange themselves in their lowest energy configuration. Fermions obey the Pauli exclusion principle. Therefore, at absolute zero, each level from the bottom up to the Fermi energy $E_F$ is occupied by two electrons, one with spin up and the other with spin down, as shown in Fig. 4.2. At absolute zero, all energy levels above the Fermi level are empty. In contrast to this, bosons do not conform to the exclusion principle, therefore, at absolute zero, they all consolidate in their lowest energy state, as shown in Fig. 4.2. Since all the bosons are in the same quantum state, they form a quantum condensate (which is similar to a superconducting condensate). In practice, however, absolute zero is not accessible.

We are now ready to discuss the so-called *Bose-Einstein condensation*. In the 1920s, Einstein predicted that if an ideal gas of identical atoms, i.e. bosons, at thermal equilibrium is trapped in a box, at sufficiently low temperatures the particles can in principle accumulate in the lowest energy level (see Fig. 4.2). This may take place only if the quantum wave packets of the particles overlap. In other words, the wavelengths of the matter waves associated with the particles—the *Broglie waves*—become similar in size to the mean particle distances in the box. If this happens, the particles condense, almost motionless, into the lowest quantum state, forming a Bose–Einstein condensate. So, the Bose–Einstein condensation is a macroscopic quantum phenomenon and, thus, similar to the superconducting condensation.

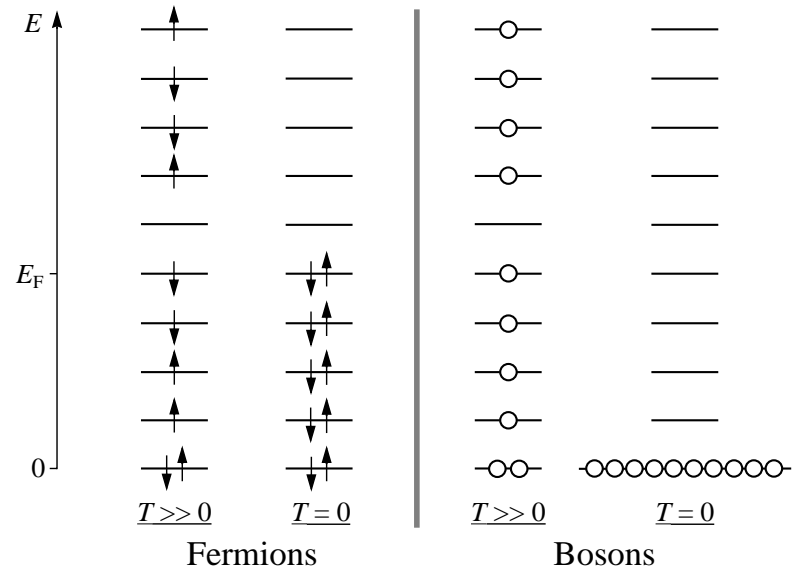

*Figure 4.2.* Sketch of the occupation of energy levels for fermions and bosons at high temperatures and absolute zero. Arrows indicate the spin direction of the fermions. For simplicity, the spin of the bosons is chosen to be zero. $E_F$ is the Fermi level for the fermions.

For many decades physicists dreamt of cooling a sufficiently large number of ordinary atoms to low enough temperatures to undergo the Bose–Einstein condensation spontaneously. During 1995 this was accomplished by three groups acting independently. The first Bose–Einstein condensate was formed by using rubidium atoms cooled to $2 \times 10^{-9}$ K.

The superconducting and Bose–Einstein condensates have much in common but also a number of differences. Let us start with their similarities. Firstly, the superconducting and Bose–Einstein condensations are both quantum phenomena occurring on a macroscopic scale. Thus, every Bose–Einstein condensate exhibits most of the superconducting-state properties described in Chapter 2. Secondly, the superconducting and Bose–Einstein condensations both occur in momentum space, not in real space. What is the difference between a condensation in momentum space and one in real space? For example, the vapor-liquid transition is a condensation in ordinary space. After the transition, the average distance between particles (atoms or molecules) is changed—becomes smaller when the vapor condenses and larger when the liquid evaporates. So, if a condensation takes place in real space, there may be some noticeable changes in the system (second-order phase transitions occurring in real space, if such exist, are not accompanied by changes in real space). On the other hand, if a condensation occurs in momentum space there are no changes in ordinary space. In the aforementioned example of the Bose-Einstein condensation occurring in the box, after the condensation, the mean distance between particles remains the same.

The superconducting and Bose-Einstein condensates have two major differences. In spite of the fact that the superconducting and Bose-Einstein condensates are both quantum states, they, however, have "different goals to achieve." Through the Bose-Einstein condensation bosons assume to reach the lowest energy level existing in the system (see Fig. 4.2). At the same time, the Cooper pairs try to descend below the Fermi level as deeply as possible, generating an energy gap (see Fig. 2.11). The second difference is that a Bose-Einstein condensate consists of real bosons, while a superconducting condensate comprises composite bosons. To summarize, the two condensates—superconducting and Bose-Einstein—have common quantum properties, but also, they have a few differences.

In conventional superconductors, the onset of phase coherence occurs due to the overlap of Cooper-pair wavefunctions. In a sense, it is a passive process because the overlap of wavefunctions does not generate an order parameter— it only makes the Cooper-pair wavefunctions be in phase. This means that in order to form a superconducting condensate, the Cooper pairs in conventional superconductors must be paired in momentum space, not in ordinary space. However, this may not be the case for unconventional superconductors where the onset of long-range phase coherence occurs due to not the overlap of Cooper-pair wavefunctions but due to another "active" process. As a consequence, if the onset of phase coherence in unconventional superconductors takes place in momentum space, it relieves the Cooper pairs of the duty to be paired in momentum space. This means that, in unconventional superconductors, the Cooper pairs may be formed in real space. Of course, they are not required to, but they may.

If the Cooper pairs in some unconventional superconductors are indeed formed in real space, this signifies that the BCS theory and the future theory for unconventional superconductors can hardly be unified.

Let us go back to the second principle of superconductivity. After all these explanations, the meaning of this principle should be clear. The transition into the superconducting state always occurs in momentum space, and this condensation is similar to that predicted by Einstein.

## 3.    Third principle of superconductivity

If the first two principles of superconductivity, in fact, are just the ascertaining of facts and can hardly be used for future predictions, the third and fourth principles are better suited for this purpose, and we shall use them further in Chapters 8 and 9.

The third principle of superconductivity is:

Principle 3: **The mechanism of electron pairing and the mechanism of Cooper-pair condensation must be different**

The validity of the third principle of superconductivity will be evident after the presentation of the fourth principle. Historically, this principle was introduced first [19].

It is worth to recall that, in conventional superconductors, phonons mediate the electron pairing, while the overlap of wavefunctions ensures the Cooper-pair condensation. In the unconventional superconductors from the third group of superconductors (see Chapter 3), such as the cuprates, organic salts, heavy fermions, doped $C_{60}$ etc., phonons also mediate the electron pairing, while spin fluctuations are responsible for the Cooper-pair condensation. So, in all superconductors, the mechanism of electron pairing differs from the mechanism of Cooper-pair condensation (onset of long-range phase coherence). Generally speaking, if in a superconductor, the same "mediator" (for example, phonons) is responsible for the electron pairing and for the onset of long-range phase coherence (Cooper-pair condensation), this will simply lead to the collapse of superconductivity (see the following section).

Since in solids, phonons and spin fluctuations have two channels—acoustic and optical (see Chapter 5)—*theoretically*, it is possible that one channel can be responsible for the electron pairing and the other for the Cooper-pair condensation. The main problem, however, is that these two channels—acoustic and optical—usually compete with one another. So, it is very unlikely that such a "cooperation" will lead to superconductivity.

## 4. Fourth principle of superconductivity

If the first three principles of superconductivity do not deal with numbers, the forth principle can be used for making various estimations.

Generally speaking, a superconductor is characterized by a pairing energy gap $\Delta_p$ and a phase-coherence gap $\Delta_c$ (see Chapter 2). For genuine (not proximity-induced) superconductivity, the phase-coherence gap is proportional to $T_c$:

$$2\Delta_c = \Lambda\, k_B T_c, \tag{4.1}$$

where $\Lambda$ is the coefficient proportionality [not to be confused with the phenomenological parameter $\Lambda$ in the London equations, given by Eq. (2.7)]. At the same time, the pairing energy gap is proportional to the pairing temperature

$T_{pair}$:

$$2\Delta_p = \Lambda' \, k_B T_{pair}. \tag{4.2}$$

Since the formation of Cooper pairs must precede the onset of long-range phase coherence, then in the general case, $T_{pair} \geq T_c$.

In conventional superconductors, however, there is only one energy gap $\Delta$ which is in fact a pairing gap but proportional to $T_c$:

$$2\Delta = \Lambda \, k_B T_c, \tag{4.3}$$

This is because, in conventional superconductors, the electron pairing and the onset of long-range phase coherence take place at the same temperature—at $T_c$. In all known cases, the coefficients $\Lambda$ and $\Lambda'$ lie in the interval between 3.2 and 6 (in one heavy fermion, $\sim 9$). Thus, we are now in position to discuss the fourth principle of superconductivity:

**Principle 4:**
> **For genuine, homogeneous superconductivity,**
> $\Delta_p > \Delta_c > \frac{3}{4} k_B T_c$ **always**
> **(in conventional superconductors, $\Delta > \frac{3}{4} k_B T_c$)**

Let us start with the case of conventional superconductors. The reason why superconductivity occurs exclusively at low temperatures is the presence of substantial thermal fluctuations at high temperatures. The thermal energy is $\frac{3}{2}k_B T$. In conventional superconductors, the energy of electron binding, $2\Delta$, must be larger than the thermal energy; otherwise, the pairs will be broken up by thermal fluctuations. So, the energy $2\Delta$ must exceed the energy $\frac{3}{2}k_B T_c$. In the framework of the BCS theory, the ratio between these two energies, $2\Delta/(k_B T_c) \simeq 3.52$, is well above 1.5.

In the case of unconventional superconductors, the same reasoning is also applicable for the phase-coherence energy gap: $2\Delta_c > \frac{3}{2}k_B T_c$.

We now discuss the last inequality, namely, $\Delta_p > \Delta_c$. In unconventional superconductors, the Cooper pairs condense at $T_c$ due to their interaction with some bosonic excitations present in the system, for example, spin fluctuations. These bosonic excitations are directly coupled to the Cooper pairs, and the strength of this coupling with each Cooper pair is measured by the energy $2\Delta_c$. If the strength of this coupling will exceed the pairing energy $2\Delta_p$, the Cooper pairs will immediately be broken up. Therefore, the inequality $\Delta_p > \Delta_c$ must be valid.

What will happen with a superconductor if, at some temperature, $\Delta_p = \Delta_c$? Such a situation can take place either at $T_c$, defined *formally* by Eq. (4.1), or below $T_c$, i.e. inside the superconducting state. In both cases, the temperature at which such a situation occurs is a critical point, $T_{cp}$. If the temperature remains constant, locally there will be superconducting fluctuations due to thermal fluctuations, thus, a kind of inhomogeneous superconductivity. If the temperature falls, two outcomes are possible (as it usually takes place at a critical point). In the first scenario, superconductivity will never appear if $T_{cp} = T_c$, or will disappear at $T_{cp}$ if $T_{cp} < T_c$. In the second possible outcome, homogeneous superconductivity may appear. The final result depends completely on bosonic excitations that mediate the electron pairing and that responsible for the onset of phase coherence. The interactions of these excitations with electrons and Cooper pairs, respectively, vary with temperature. If, somewhat below $T_{cp}$, the strength of the pairing binding increases *or/and* the strength of the phase-coherence adherence decreases, homogeneous superconductivity will appear. In the opposite case, superconductivity will never appear, or disappear at $T_{cp}$. It is worth noting that, in principle, superconductivity may reappear at $T < T_{cp}$.

The cases of disappearance of superconductivity below $T_c$ are well known. However, it is assumed that the cause of such a disappearance is the emergence of a ferromagnetic order. As discussed in Chapter 3, the Chevrel phase $HoMo_6S_8$ is superconducting only between 2 and 0.65 K. The erbium rhodium boride $ErRh_4B_4$ superconducts only between 8.7 and 0.8 K. The cuprate Bi2212 doped by Fe atoms was seen superconducting only between 32 and 31.5 K [30]. The so-called $\frac{1}{8}$ anomaly in the cuprate LSCO, discussed in Chapter 3, is caused apparently by *static* magnetic order [19] which may result in the appearance of a critical point where $\Delta_p \simeq \Delta_c$.

It is necessary to mention that the case $\Delta_p = \Delta_c$ must not be confused with the case $T_{pair} = T_c$. There are unconventional superconductors in which the electron pairing and the onset of phase coherence occur at the same temperature, i.e. $T_{pair} \simeq T_c$. This, however, does not mean that $\Delta_p = \Delta_c$ because $\Lambda \neq \Lambda'$ in Eqs. (4.1) and (4.2). Usually, $\Lambda' > \Lambda$. For example, in hole-doped cuprates, $2\Delta_p/k_B T_{pair} \simeq 6$ and, depending on the cuprate, $2\Delta_c/k_B T_c = 5.2$–5.9.

Finally, let us go back to the third principle of superconductivity to show its validity. The case in which the same bosonic excitations mediate the electron pairing **and** the phase coherence is equivalent to the case $\Delta_p = \Delta_c$ discussed above. Since, in this particular case, the equality $\Delta_p = \Delta_c$ is independent of temperature, the occurrence of homogeneous superconductivity is impossible.

## 5. Proximity-induced superconductivity

The principles considered above are derived for genuine superconductivity. By using the same reasoning as that in the previous section for proximity-induced superconductivity, one can obtain a useful result, namely, that $2\Delta \sim \frac{3}{2} k_B T_c$, meaning that the energy gap of proximity-induced superconductivity should be somewhat larger than the thermal energy. Of course, to observe this gap for example in tunneling measurements may be not possible if the density of induced pairs is low. This case is reminiscent of gapless superconductivity discussed in Chapter 2. Hence, we may argue that

**For proximity-induced superconductivity,
at low temperature, $2\Delta_p \geq \frac{3}{2}k_B T_c$**

One should however realize that this is a general statement; the final result depends also upon the material and, in the case of thin films, on the thickness of the normal layer.

What is the maximum critical temperature of BCS-type superconductivity? In conventional superconductors, $\Lambda = 3.2$–4.2 in Eq. (4.3). Among conventional superconductors, Nb has the maximum energy gap, $\Delta \simeq 1.5$ meV. Then, taking $\Delta_{max}^{BCS} \approx 2$ meV and using $\Lambda = 3.2$, we have $T_{c,max}^{BCS} = 2\Delta_{max}^{BCS}/3.2 k_B \approx 15$ K for conventional superconductors. Let us now estimate the maximum critical temperature for induced superconductivity of the BCS type in a material with a strong electron-phonon interaction. In such materials, genuine superconductivity (if exists) is in the strong coupling regime and characterized by $\Lambda \simeq 4.2$ in Eq. (4.3). Assuming that the same strong coupling regime is also applied to the induced superconductivity with $2\Delta \sim 1.5 k_B T_c^{ind}$ and that, in the superconductor which induces the Cooper pairs, $\Delta_p \gg 2$ meV, one can then obtain that $T_{c,max}^{ind} \sim 15$ K $\times \frac{4.2}{1.5} \simeq 42$ K.

If the superconductor which induces the Cooper pairs is of the BCS type, the value $\Delta_{max}^{ind} = 2$ meV can be used to estimate $T_{c,max}^{ind}$ independently. Substituting the value of 2 meV into $2\Delta \sim 1.5 k_B T_c^{ind}$, we have $T_{c,max}^{ind} \simeq 31$ K which is lower than 42 K.

In second-group superconductors which are characterized by the presence of two superconducting subsystems, the critical temperature never exceeds 42 K. For example, in MgB$_2$, $T_c = 39$ K and, for the smaller energy gap, $2\Delta_s \simeq 1.7 k_B T_c$ (see Chapter 3). At the same time, for the larger energy gap in MgB$_2$, $2\Delta_L \simeq 4.5 k_B T_c$ or $\Delta_L \simeq 7.5$ meV. Then, on the basis of the estimation for $T_{c,max}^{ind}$, it is more or less obvious that, in MgB$_2$, one subsystem with genuine superconductivity (which is low-dimensional), having $\Delta_L \simeq 7.5$ meV, induces superconductivity into another subsystem and the latter one controls the bulk $T_c$.

The charge carriers in compounds of the first and second groups of superconductors are electrons. Is there hole-induced superconductivity? Yes. At least one case of hole-induced superconductivity is known: in the cuprate YBCO, the CuO chains (see Fig. 3.9) become superconducting due to the proximity effect. The value of the superconducting energy gap on the chains in YBCO is well documented; in optimally doped YBCO, it is about 6 meV [19].

Using $T_{c,max} = 93$ K for YBCO and $\Delta \sim 6$ meV, one obtains $2\Delta/k_B T_c \simeq 1.5$. This result may indicate that the bulk $T_c$ in YBCO is controlled by induced superconductivity on the CuO chains.

Chapter 5

# FIRST GROUP OF SUPERCONDUCTORS:
# MECHANISM OF SUPERCONDUCTIVITY

The first group comprises classical, conventional superconductors. This group incorporates non-magnetic elemental metals and some of their alloys. The phenomenon of superconductivity was discovered by Kamerlingh Onnes and his assistant Gilles Holst in 1911 in mercury, a representative of this group.

This Chapter does not set out to cover all aspects of the BCS theory of superconductivity in metals; here we present only the main results of this theory. There are many excellent books devoted exclusively to the BCS mechanism of superconductivity, and the reader who is interested in following all calculations leading to the main formulas of the BCS theory is referred to the books (see Appendix in Chapter 2).

## 1.    Introduction

In 1957, Bardeen, Cooper and Schrieffer showed how to construct a wavefunction in which the electrons are paired. The wavefunction which is adjusted to minimize the free energy is further used as the basis for a complete microscopic theory of superconductivity in metals. Thus, they showed that the superconducting state is a peculiar correlated state of matter—a quantum state on a macroscopic scale, in which all the electron pairs move in a single coherent motion. The success of the BCS theory and its subsequent elaborations are manifold. One of its key features is the prediction of an energy gap.

In Landau's concept of the Fermi liquid, excitations called *quasiparticles* are bare electrons dressed by the medium in which they move. Quasiparticles can be created out of the superconducting ground state by breaking up the pairs, but only at the expense of a minimum energy of $\Delta$ per excitation. This minimum energy $\Delta$, as we already know, is called the energy gap. The BCS theory predicts that, for any superconductor at $T = 0$, $\Delta$ is related to the critical temperature by $2\Delta = 3.52k_BT_c$, where $k_B$ is the Boltzmann constant. This turns out to be nearly true, and where deviations occur they can be understood in terms of modifications of the BCS theory. The manifestation of the energy gap in tunneling provided strong conformation of the theory.

The key to the basic interaction between electrons which gives rise to superconductivity was provided by the isotope effect which was discussed in Chapter 2. The interaction of electrons with the crystal lattice is one of the basic mechanisms of electrical resistance in an ordinary metal. It turns out that it is precisely the electron-lattice interaction that, under certain conditions, leads to an absence of resistance, i.e. to superconductivity. This is why, in excellent conductors such as copper, silver and gold, a rather weak electron-lattice interaction does not lead to superconductivity; however, it is completely responsible for their nonvanishing resistance near absolute zero.

We start with a qualitative description of superconductivity in metals.

## 2.    Interaction of electrons through the lattice

Superconductivity is not universal phenomenon. It shows up in materials in which the electron attraction overcomes the repulsion. This attractive force occurs due to the interaction of electrons with the crystal lattice. Thus, the *electron-phonon interaction* in solids is responsible for the electron attraction, leading to the electron pairing. Phonons are quantized excitations of the crystal lattice.

The effective interaction of two electrons via a phonon can be visualized as the emission of a "virtual" phonon by one electron, and its absorption by the other, as shown in Fig. 5.1. An electron in a state $\mathbf{k}_1$ (in momentum space) emits a phonon, and is scattered into a state $\mathbf{k}'_1 = \mathbf{k}_1 - \mathbf{q}$. The electron in a state $\mathbf{k}_2$ absorbs this phonon, and is scattered into a state $\mathbf{k}'_2 = \mathbf{k}_2 + \mathbf{q}$. The diagram shown in Fig. 5.1 is the simplest way of calculating the force acting on the two electrons. We shall consider this diagram in a moment; let us first discuss the spectrum of lattice vibrations in a solid.

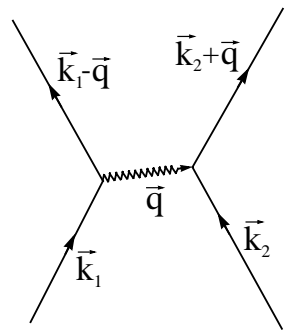

*Figure 5.1* Diagram illustrating electron-electron interaction via exchange of a virtual phonon of momentum $\hbar\mathbf{q}$.

Phonons are quantized and, in different solids, propagate with different frequencies, $\omega = E/\hbar$. Because the lattice in solids are periodic, one unit cell is interchangeable with another, and the lattice vibrations can propagate from one cell to the next without change. Thus, it is unnecessary to consider the crystal lattice of a whole sample; it is enough to study just one unit cell. This unit cell can be described not only in ordinary space but also in momentum space. The simplest model for studying the spectrum of lattice vibrations is the one-dimensional model: it gives a useful picture of the main features of the mechanical behavior of a periodic array of atoms. The simplest model among one-dimensional ones is the model corresponding to a monatomic crystal, which can be visualized as a linear chain of masses $m$ with the same spacing $a$, and connected to each other by massless springs. However, it is more practical to consider the one-dimensional model for a diatomic crystal in which a unit cell contains two different atoms. In this model, the linear chain consists of two different masses, $M$ and $m$, which alternate along the chain, as shown in Fig. 5.2.

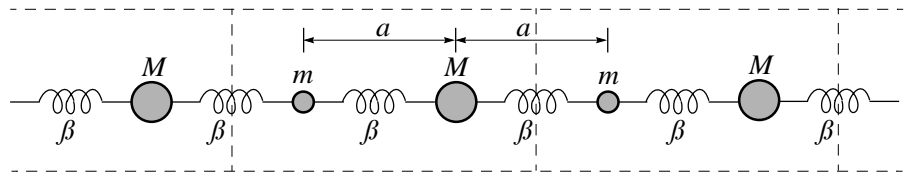

*Figure 5.2.* One-dimensional mass-spring model for lattice vibrations in a diatomic crystal.

Figure 5.3 shows schematically the energy-momentum relation $E(k)$, obtained in the framework of the one-dimensional model for a diatomic crystal depicted in Fig. 5.2. The $E(k)$ relation is generally known as *the dispersion relation*. The momentum space in the range $\pm\pi/2a$, where $2a$ is the periodicity of the lattice, is known as the *Brillouin zone*. In Fig. 5.3, the higher-energy oscillations are conventionally called *optical modes* (or branches), and the lower-energy oscillations *acoustic modes*. In Fig. 5.3, there are two optical and two acoustic branches, corresponding to longitudinal and transverse vibrations of atoms. The situation in three dimensions becomes more complicated and, in general, there are different dispersion relations for waves propagating in different directions in a crystal as a result of anisotropy of the force constants.

In the BCS theory, the Debye spectrum of phonon frequencies is used to determine a critical temperature $T_c$. The Debye model assumes that the energies available are insufficient to excite the optical modes, so the BCS theory considers only low-energy (acoustic) phonons. In the Debye model, the Brillouin zone, which bounds the allowed values of **k**, is replaced by a sphere of

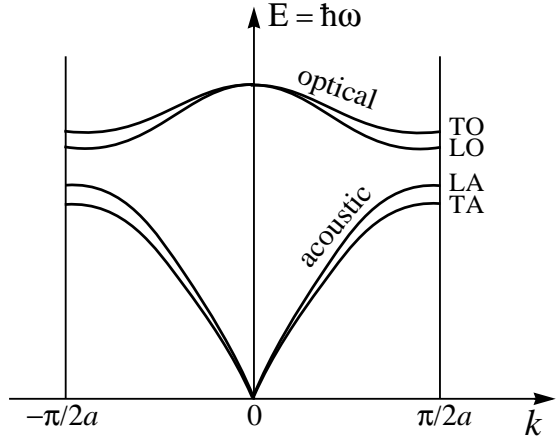

*Figure 5.3.* Vibration frequencies of diatomic chain shown in Fig. 5.2 (L = longitudinal and T = transverse).

the same volume in **k**-space. The Debye temperature $\Theta$ is defined by

$$k_B\Theta = \hbar\omega_D, \qquad (5.1)$$

where $\omega_D$ is the phonon frequency at the edge of the Debye sphere. Thus, $k_B\Theta$ (or $\hbar\omega_D$) is the energy of the highest-energy phonon in the Debye sphere.

Let us go back to our electrons shown in Fig. 5.1. To enable an electron to scatter from the state $\mathbf{k}_1$ into the state $\mathbf{k}'_1$, the latter must be free (in accordance with the Pauli exclusion principle). This is possible only in the vicinity of the Fermi surface which is represented in momentum space by a sphere of radius $\mathbf{k}_F$, as shown in Fig. 5.4. Now we are ready to formulate the law of phonon-mediated interaction between electrons which forms the foundation of the BCS theory: *Electrons with energies that differ from the Fermi energy by no more than $\hbar\omega_D$ are attracted to each other.* Thus, in the BCS model, only those electrons that occupy the states within a narrow spherical layer near the Fermi surface experience mutual attraction. The thickness of the layer $2\Delta k$ is determined by the Debye energy:

$$\frac{\Delta k}{k_F} \sim \frac{\hbar\omega_D}{E_F}, \quad \text{where} \quad E_F = \frac{\hbar^2 k_F^2}{2m}, \qquad (5.2)$$

and $m$ is the electron mass. As we shall see, the attraction is greatest for electrons with opposite spins ($\mathbf{s}_1 = -\mathbf{s}_2$) and equal and opposite wave vectors ($\mathbf{k}_1 = -\mathbf{k}_2$).

The electron-electron attraction mediated by the background crystal lattice can crudely be pictured as follows. An electron tends to create a slight distortion of the elastic lattice as it moves because of the Coulomb attraction between the negatively charged electron and the positively charged lattice, as illustrated in Fig. 5.5. If the distortion persists for a brief time (retardation), a second passing electron will feel the distortion and will be affected by it. Under certain circumstances, this can give rise to a weak indirect attractive interaction between the two electrons which may more than compensate their Coulomb repulsion. Thus, as shown in Fig. 5.5, the process of electron pairing in conventional superconductors is *local* in space, but *non-local* in time.

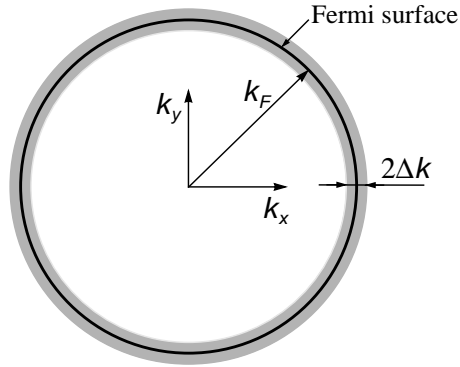

*Figure 5.4.* In the BCS picture, only the electrons within the $2\Delta k$ layer near the Fermi surface interact via phonons.

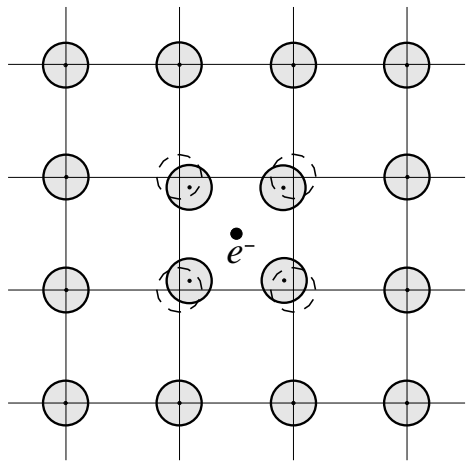

*Figure 5.5.* Polarization of a lattice near a moving electron, which in turn attracts another electron.

## 3.    Main results of the BCS theory

The main idea of the BCS theory is based on the Cooper work. In 1956, Cooper showed that two electrons with an attractive interaction can bind together in the momentum space to form a bound pair, if they are in the presence of a high-density fluid of other electrons, no matter how weak the interaction is. This bound state of two electrons is today known as the Cooper pair.

### 3.1    Instability of the Fermi surface in the presence of attractive interaction between electrons

Let us consider two electrons in a metal, added at the Fermi surface. In the absence of interactions, their wavefunction can be written as

$$\psi(\mathbf{r_1}, \mathbf{r_2}) = e^{i(\mathbf{k_1}\mathbf{r_1} + \mathbf{k_2}\mathbf{r_2})} = e^{i[\mathbf{q}(\mathbf{r_1} + \mathbf{r_2}) + \mathbf{k}(\mathbf{r_1} - \mathbf{r_2})]} \tag{5.3}$$

where

$$\mathbf{q} = \frac{1}{2}(\mathbf{k_1} + \mathbf{k_2}), \tag{5.4}$$

$$\mathbf{k} = \frac{1}{2}(\mathbf{k_1} - \mathbf{k_2}).$$

In the center of mass, $\mathbf{q} = 0$. Then, in the presence of an interaction between electrons (assumed attractive), the wavefunction can be presented as

$$\psi(\mathbf{r_1}, \mathbf{r_2}) = \sum_k g(\mathbf{k}) e^{i\mathbf{k}(\mathbf{r_1} - \mathbf{r_2})}, \tag{5.5}$$

where $|g(\mathbf{k})|^2$ is the probability of finding one electron with momentum $\mathbf{k}$ and the other one with momentum $-\mathbf{k}$. Of course,

$$g(\mathbf{k}) \equiv 0 \ \ \text{for} \ \ |\mathbf{k}| < |\mathbf{k}_F|, \tag{5.6}$$

because all the electronic states $|\mathbf{k}| < |\mathbf{k}_F|$ are completely filled with electrons and, in accordance with the Pauli exclusion principle, the two electrons cannot occupy these states.

The Schrödinger equation for these two electrons is

$$-\frac{\hbar^2}{2m}(\nabla_1^2 + \nabla_2^2)\psi + V(\mathbf{r_1}, \mathbf{r_2})\psi = (E + 2E_F)\psi, \tag{5.7}$$

where $E_F$ is the Fermi energy, and $V(\mathbf{r_1}, \mathbf{r_2})$ is the potential energy of electron-electron interaction. Substituting the wavefunction obtained in the center of mass into the Schrödinger equation, we have the following equation for $g(\mathbf{k})$:

$$\frac{\hbar^2}{m}g(\mathbf{k}) + \sum_{k'} g(\mathbf{k}')V_{kk'} = (E + 2E_F)g(\mathbf{k}), \tag{5.8}$$

where

$$V_{kk'} = \frac{1}{L^3} \int V(\mathbf{r}) e^{i(\mathbf{k}-\mathbf{k'})\mathbf{r}} \, d^3\mathbf{r}, \tag{5.9}$$

is a matrix element of the electron-electron interaction, and $L^3$ is the volume. To solve this equation, it is necessary to know $V_{kk'}$ explicitly. Taking into account that the energies of electrons participating in pairing is $|\varepsilon_k|, |\varepsilon_{k'}| \leq \hbar\omega_D$, where $\omega_D$ is the Debye frequency, we choose a simple form of the electron-electron interaction

$$V_{kk'} = \begin{cases} -V & \text{for } |\varepsilon_k|, |\varepsilon_{k'}| \leq \hbar\omega_D \\ 0 & \text{otherwise,} \end{cases} \tag{5.10}$$

where

$$\varepsilon_k = \frac{\hbar^2 k^2}{2m} k^2 - E_F = \frac{\hbar^2 k^2}{2m} - \frac{\hbar^2 k_F^2}{2m}, \tag{5.11}$$

and $k_F$ is the wave vector on the Fermi surface. In this case, Equation (5.8) transforms into

$$g(\mathbf{k}) \left[ E + 2E_F - \frac{\hbar^2}{m} k^2 \right] = -V \sum_{k'} g(\mathbf{k'}) = C, \tag{5.12}$$

where $C$ is a constant independent of $\mathbf{k}$. From Eq. (5.12), one can easily obtain the following self-consistent equation

$$1 = V \sum_k \frac{1}{\hbar^2 k^2/m - E - 2E_F}. \tag{5.13}$$

If we introduce the density of states per spin direction

$$N(\varepsilon) = \frac{4\pi}{(2\pi)^3} k^2 \frac{dk}{d\varepsilon}, \tag{5.14}$$

we obtain

$$1 = V \int\limits_0^{\hbar\omega_D} N(\varepsilon) \frac{1}{2\varepsilon - E} \, d\varepsilon. \tag{5.15}$$

Since, in metals $\hbar\omega_D \ll E_F$ ($E_F \sim 5$ eV and $\hbar\omega_D \sim 25$ meV), then $N(\varepsilon) \simeq N(0)$, and we can write

$$1 = \frac{N(0) V}{2} \ln \frac{E - 2\hbar\omega_D}{E}. \tag{5.16}$$

The last equation can be re-written as

$$E = -\frac{2\hbar\omega_D}{\left[ \exp\frac{2}{N(0)V} - 1 \right]} \simeq -2\hbar\omega_D \exp -\frac{2}{N(0) V} \text{ if } N(0) V \ll 1. \tag{5.17}$$

The energy gain $E$ (because $E < 0$) indicates that the two electrons form a bound state and, as a consequence, other electrons can also condense into this state.

This result is obtained under the assumption that $\mathbf{q} = 0$ [see Eq. (5.5)]. In the case if $\mathbf{q} \neq 0$, the energy gain is $E(q) = E(0) + \hbar q v_F$. This means that the energy gain is a maximum when $\mathbf{q} = 0$. The energy gain depends also on the orientation of spin of each electron. The absolute value of $E$ is a maximum when the spins of the two electrons are oppositely directed ($\mathbf{s}_1 = -\mathbf{s}_2$). Since the distance between two electrons in a pair is sufficiently large,

$$\frac{\hbar v_F}{E} \simeq 10^3 - 10^4 \quad \text{Å}, \tag{5.18}$$

and since the density of conduction electrons in metals is relatively high, the wavefunctions of different pairs are largely overlap. To estimate the distance between electrons in a pair, we used $\hbar \omega_D \simeq 300$ K and $N(0)V \simeq 0.3$. The sketch of electron-pair wavefunction is shown in Fig. 2.3, and their overlap in Fig. 4.1.

## 3.2     Electron-electron attraction via phonons

How can electrons in a solid attract each other? The Coulomb force acting between electrons is always repulsive, so that, the matrix element of the Coulomb interaction, $V^C$, is always positive, $V^C > 0$. In order to obtain an attraction between electrons, it is necessary that they interact with a "third party," leading to the formation of electron pairs. In a solid, electrons constantly interact with lattice vibrations.

Considering the interaction of two electrons via a virtual phonon, as shown in Fig. 5.1, the element of the total matrix can in general be presented as

$$V^P(\mathbf{k}, \mathbf{k}', \mathbf{q}) = \langle \mathbf{k}, \mathbf{k}' | V | \mathbf{k} - \mathbf{q}, \mathbf{k}' + \mathbf{q} \rangle = \frac{M_{k,k-q} M_{k',k'+q}}{\varepsilon_{k-q} - \varepsilon_k - \hbar \omega_q}, \tag{5.19}$$

where $M_{k,k-q}$ is the matrix element of the electron-phonon interaction, and $\hbar \omega_q$ is the energy of a phonon with the wave vector $\mathbf{q}$ shown in Fig. 5.1 [not be confused with the vector $\mathbf{q}$ in Eq. (5.4)]. If we add to this process another one in which the electron $\mathbf{k}'$ emits a phonon with the wave vector $-\mathbf{q}$, we obtain that the element of the total matrix becomes

$$V_{kk'}^P = \frac{2 \hbar \omega_q |M_{k,k-q}|^2}{(\varepsilon_{k-q} - \varepsilon_k)^2 - (\hbar \omega_q)^2} \tag{5.20}$$

in which the element of the electron-phonon matrix $M$ depends weakly on $\mathbf{k}$.

The matrix element $V_{kk'}^P$ is negative if $|\varepsilon_{k-q} - \varepsilon_k| < \hbar \omega_q$. In this case, the electron-electron interaction is attractive. This interaction becomes repul-sive when the characteristic frequencies of electrons exceed the frequencies of lattice vibrations. This will happen when heavy ions cannot follow the movements of electrons. The interaction is a maximum when the two frequencies—electronic and lattice vibrations—coincide. Taking into account the Coulomb repulsion between the electrons, the total interaction $(V^P + V^C)$ will be attractive (i.e. negative) if $|V^P| > V^C$.

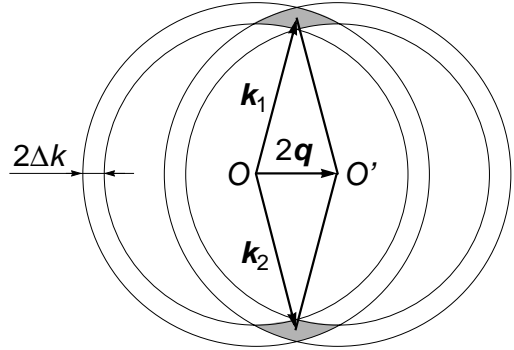

*Figure 5.6.* If electron pairs have a total momentum 2**q**, the interaction involves only the electrons occupying **k** states within the dashed areas.

The attraction is greatest for electrons with equal and opposite wave vectors $(\mathbf{k}_1 = -\mathbf{k}_2)$. Why? A transition of an electron pair from the state $(\mathbf{k}_1, \mathbf{k}_2)$ to the state $(\mathbf{k}'_1, \mathbf{k}'_2)$, as shown in Fig. 5.1, must obey the low of momentum conservation:

$$\mathbf{k}_1 + \mathbf{k}_2 = \mathbf{k}'_1 + \mathbf{k}'_2. \tag{5.21}$$

For example, if $\mathbf{k}_1 + \mathbf{k}_2 = 2\mathbf{q}$ as illustrated in Fig. 5.6, only the electrons occupying the **k** states in the dashed areas of momentum space are allowed to participate in the transitions. The dashed area is maximum when $\mathbf{q} = 0$, and all states within a band of width $\sim 2\hbar\omega_D$ near the Fermi surface are available.

### 3.3 Electron distribution in the ground state of a superconductor

In this subsection, our objective is to investigate the state of a superconductor at $T = 0$, i.e. when its energy is a minimum.

As shown by Cooper, in the presence of an attractive interaction, the electrons in a solid will condense at low temperature in pairs. In its turn, the BCS theory supplies a formalism capable to treat the correlations of the pairs when the number of interacting electrons is macroscopic.

In the framework of the BCS theory, it is first assumed that the element $V_{kk'q}$ of the matrix $V = |V^P| - V^C$ (thus, $V > 0$) has a simple form, namely,

$$V_{kk'q} = \begin{cases} -V & \text{for } |\varepsilon_k|, \, |\varepsilon_{k'}| \leq \hbar\omega_D \\ 0 & \text{otherwise.} \end{cases} \qquad (5.22)$$

As was already discussed above, this means that *only electrons with energies that differ from the Fermi energy by no more than the Debye energy $\hbar\omega_D$ are attracted to each other*. This layer is schematically shown in Fig. 5.4.

The second postulate of the BCS theory is that the difference in energy between the normal and the superconducting states originates exclusively from the energy gain of electron pairing, and other forms of energy are not affected by the superconducting transition. Each transition $(\mathbf{k}, -\mathbf{k}) \rightarrow (\mathbf{k'}, -\mathbf{k'})$ is accompanied by a contribution of $-V$ to the condensation energy. Below $T_c$, the normal electrons are also present, but all become paired at $T = 0$. So, the condensation energy is a maximum at $T = 0$. If the pair state $(\mathbf{k}, -\mathbf{k})$ is occupied only by one electron, all the transitions $(\mathbf{k'}, -\mathbf{k'}) \rightarrow (\mathbf{k}, -\mathbf{k})$ are forbidden.

Let us introduce two new functions of $\mathbf{k}$, namely, $v_k^2$ and $f_k$. Suppose that $v_k^2$ gives the probability that the pair state $(\mathbf{k}, -\mathbf{k})$ is occupied. Then, the probability for the pair state $(\mathbf{k}, -\mathbf{k})$ being empty is $u_k^2 = 1 - v_k^2$. Suppose that $f_k$ is the probability for the pair state $(\mathbf{k}, -\mathbf{k})$ is occupied by one (normal) electron. Then, the probability for the electronic states $\mathbf{k}$ and $-\mathbf{k}$ both being empty, i.e. being not occupied by single electrons simultaneously, is $(1 - 2f_k)$. Finally, the probability for the pair state $(\mathbf{k}, -\mathbf{k})$ being occupied and the pair state $(\mathbf{k'}, -\mathbf{k'})$ being empty is $[v_k^2(1 - v_{k'}^2)]^{1/2} = v_k u_{k'}$. It is worth noting that, in the pair wavefunction in Eq. (5.5), $g(\mathbf{k}) \equiv v_k$.

Let us now express the first three terms in the Hemholtz free energy

$$F = E - TS = E_c + E_p - TS + NE_F \qquad (5.23)$$

through the probability functions $v_k^2$, $u_k^2$ and $f_k$, where $N$ is total number of electrons and $E_F$ is the Fermi energy. The kinetic energy is

$$E_c = 2\sum_k [\varepsilon_k f_k + (1 - 2f_k)v_k^2 \varepsilon_k], \qquad (5.24)$$

where $\varepsilon_k$ is the energy of an electron in the state $\mathbf{k}$ measured from the Fermi level, given by Eq. (5.11).

The potential energy is

$$E_p = V\sum_{kk'} v_k u_{k'} u_k v_{k'} (1 - 2f_k)(1 - 2f_{k'}). \qquad (5.25)$$

In

$$TS = -2k_B T \sum_k [f_k \log f_k + (1 - f_k) \log(1 - f_k)], \qquad (5.26)$$

we assume that the entropy comes exclusively from normal electrons which are fermions.

In equilibrium, the free energy $F$ has a minimum. Then, the equilibrium values of $v_k^2$ and $f_k$ can be obtained by minimizing the free energy with respect to $v_k^2$ and $f_k$:

$$\frac{\partial F}{\partial v_k^2} = 0 \ \text{ and } \ \frac{\partial F}{\partial f_k} = 0. \tag{5.27}$$

Taking these derivations and introducing the quantity

$$\Delta_0 = V \sum_{k'} u_{k'} v_{k'} (1 - 2f_{k'}) \tag{5.28}$$

which has the dimensions of energy, one obtains the following equations

$$E_k u_k = \varepsilon_k u_k + \Delta_0 v_k \tag{5.29}$$
$$E_k v_k = -\varepsilon_k v_k + \Delta_0 u_k,$$

where

$$E_k = \sqrt{\varepsilon_k^2 + \Delta_0^2}. \tag{5.30}$$

The solutions of these two equations are

$$v_k^2 = \frac{1}{2}\left(1 - \frac{\varepsilon_k}{E_k}\right) \tag{5.31}$$

$$u_k^2 = \frac{1}{2}\left(1 + \frac{\varepsilon_k}{E_k}\right). \tag{5.32}$$

The minus sign in Eq. (5.31) stems from a general argument that, as $\mathbf{k} \to 0$, we ought to have $v_k^2 \to 1$ while $\varepsilon_k \to -E_F$. The dependence of $v_k^2$ on $k$ is illustrated in Fig. 5.7. As one can see, for a normal metal at $T = 0$, $v_k^2$ has a discontinuity at $k_F$, while the total energy of a superconductor reaches its minimum when the electron distribution in the vicinity of the Fermi level is "smeared out" over the energy interval $\sim 2\Delta_0$. It is important to emphasize that this occurs at absolute zero! Such a ground state of the superconductor is a consequence of the interaction between electrons.

From Fig. 5.7, one can make a very important conclusion. At $T = 0$, the kinetic energy of electrons near the Fermi level, forming the superconducting condensate, is larger than that of a normal metal. Thus, the superconducting condensation leading to a reduction in potential energy is accompanied by an increase in kinetic energy. Nevertheless, this reduction in potential energy is more than enough to compensate the Coulomb repulsion and the increase in kinetic energy. The total energy of the superconducting condensate is thereby reduced relative to that of the conventional one-electron description of states in a metal. Thus, in the framework of the BCS theory, *superconductivity in conventional superconductors is driven by potential energy.*

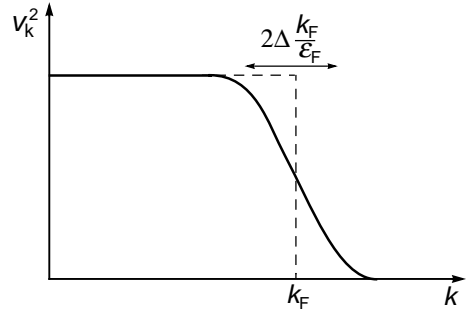

*Figure 5.7.* Dependence of $v_k^2$ on $k$ at $T = 0$, or the probability of pair occupancy in the superconducting ground state.

The second condition for equilibrium in Eq. (5.27) gives

$$f_k = \frac{1}{\exp \frac{E_k}{k_B T} + 1}. \tag{5.33}$$

At $T = 0$, $f_k = 0$, meaning that all the electrons occupy the lowest energy levels, and there are no excitations in the system.

### 3.4    Energy gap

In Eq. (5.30), $E_k$ presents a new spectrum of elementary excitations in a superconductor, which is separated from the ground-state energy level by an energy gap $\Delta_0$. The new spectrum of a superconductor is sketched in Fig. 5.8. At $T = 0$, the Cooper pairs occupy their ground-state level separated by $\Delta_0$ from the next energy level of elementary excitations. However, the lowest amount of energy which can be absorbed by the superconductor is $2\Delta_0$, and not $\Delta_0$. Why? Because if one electron becomes excited and jumps at the first energy level above the ground-state level, its ex-partner is still at the ground-state level. Such a situation is forbidden, so the two electrons must be excited simultaneously. As a consequence, the minimum energy needed for this process is $2\Delta_0$. In other words, the energy $2\Delta_0$ is necessary to break up a Cooper pair.

It is worth to emphasize that, in a conventional superconductor, *the Cooper pairs cannot be excited*. They are either at the ground-state energy level, **or** they are already broken up. There is nothing in between. This, however, is not the case for unconventional superconductors where the Cooper pairs can be excited, being still paired.

Since the energy gap depends on temperature, $\Delta(T)$, the minimum energy needed for breaking up a Cooper pair varies with temperature as well. Let us determine the $\Delta(T)$ dependence. Substituting Eqs. (5.31), (5.32) and (5.33)

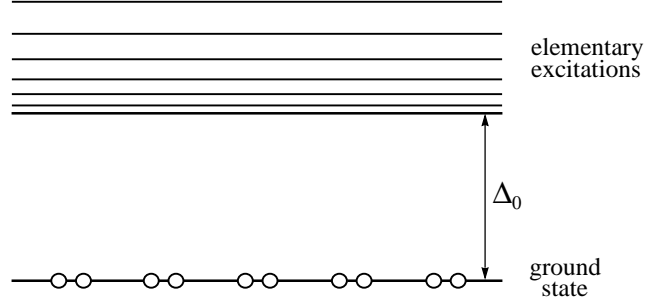

*Figure 5.8.* Elementary excitation spectrum of a superconductor at $T = 0$. The energy gap $\Delta_0$ separates the first excited level from the ground-state level.

into Eq. (5.28), the latter one reduces to

$$\Delta(T) = V \sum_k \frac{\Delta(T)}{2E_k} \left( 1 - \frac{2}{\exp(E_k/k_B T) + 1} \right). \tag{5.34}$$

The replacement of summation with integration yields, after simple algebra,

$$\frac{1}{N(0)V} = \int\limits_0^{\hbar \omega_D} \frac{\mathrm{d}\varepsilon}{\sqrt{\varepsilon^2 + \Delta^2(T)}} \tanh \frac{\sqrt{\varepsilon^2 + \Delta^2(T)}}{2k_B T}, \tag{5.35}$$

where $N(0)$ is the density of states of a superconductor at the Fermi level at $T = 0$ (see the following subsection). This temperature dependence of the energy gap, obtained in the framework of the BCS theory, is illustrated in Fig. 2.12. As we shall see, near $T_c$ the gap varies with temperature as $\Delta(T) \propto (T_c - T)^{1/2}$.

At $T = 0$, Equation (5.35) becomes

$$\frac{1}{N(0)V} = \int\limits_0^{\hbar \omega_D} \frac{\mathrm{d}\varepsilon}{\sqrt{\varepsilon^2 + \Delta_0^2}}. \tag{5.36}$$

Carrying out the integration, we have

$$\Delta_0 = \frac{\hbar \omega_D}{\sinh \frac{1}{N(0)V}} \simeq 2\hbar \omega_D \exp\left( -\frac{1}{N(0)V} \right) \text{ if } N(0)V \ll 1. \tag{5.37}$$

In reality, $N(0)V \leq 0.3$. Let us estimate $\Delta_0$. Taking the Debye temperature $\Theta = \hbar \omega_D/k_B \sim 280$ K and $N(0)V = 0.3$, one obtains $\Delta_0 \sim 10$ K $= 0.86$ meV.

## 3.5    Density of states of elementary excitations

The lower part of the elementary excitation spectrum of a superconductor is depicted in Fig. 5.8. By combining the two equations, Eqs. (5.11) and (5.30), we have

$$E_k = \sqrt{\varepsilon_k^2 + \Delta_0^2} = \sqrt{\left(\frac{\hbar^2 k^2}{2m} - \frac{\hbar^2 k_F^2}{2m}\right)^2 + \Delta_0^2}. \qquad (5.38)$$

This dependence of $E_k$ on $k$ is illustrated in Fig. 5.9a. As one can see from this plot and Fig. 5.8, the energy levels of elementary excitations become denser at $E_k \to \Delta_0$. In other words, the density of states near the Fermi level is the highest.

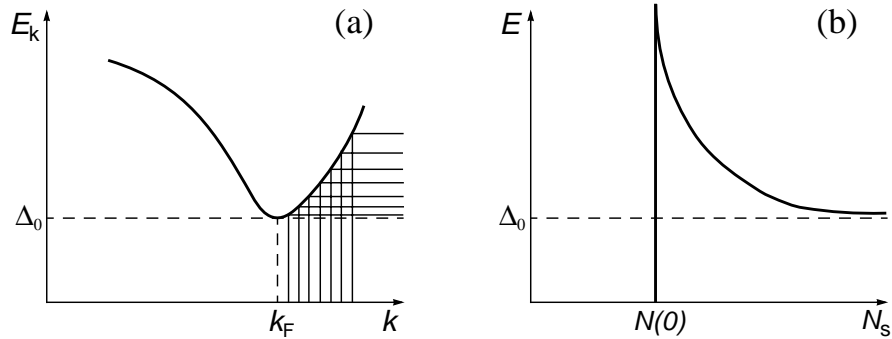

*Figure 5.9.*    (a) Spectrum of elementary excitations of a superconductor, $E_k$ (see also Fig. 5.8), and (b) its density of states $N_s$ at $T = 0$.

In a superconductor, the density of states is

$$N(E) = N(\varepsilon)\frac{\mathrm{d}\varepsilon}{\mathrm{d}E} = N(\varepsilon)\frac{\Delta(T)}{\sqrt{E^2 - \Delta^2(T)}}, \qquad (5.39)$$

where $N(\varepsilon)$ is the density of states of a superconductor in the normal state. The density of states $N(E)$ diverges at $E \to \Delta_0$, as shown in Fig. 5.9b.

Figure 5.9b presents the density of states of a superconductor at $T = 0$. At finite temperatures $0 < T < T_c$, however, the density of states does not diverge at $E \to \Delta_0$ because, at $0 < T$, there are always excitations inside the gap, as illustrated in Fig. 5.10. This effect can directly be observed in tunneling measurements. It is worth noting that, the tunneling conductances $dI(V)/dV$ obtained in a superconductor-insulator-normal metal (SIN) junction corresponds directly to the density of states of a superconductor if normalized to a normal-state conductance $G_n$ [thus, to $N(\varepsilon)$]. The current-voltage tunneling characteristics for a conventional superconductor are shown in Fig. 2.20. The conduc-

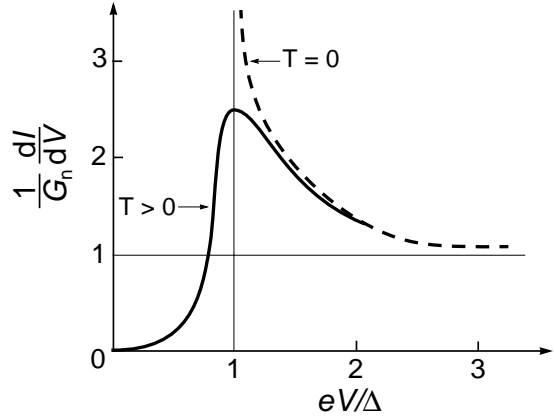

*Figure 5.10.* Tunneling $dI(V)/dV$ characteristic for a SIN junction at $T > 0$. The dashed line is the density of states of a superconductor at zero temperature (see Fig. 5.9b). $G_n$ is the normal-state conductance.

tances measured in a superconductor-insulator-superconductor (SIS) tunneling junction are proportional to the convolution of the density-of-states function of a superconductor with itself.

## 3.6 Critical temperature

In the framework of the BCS theory, one can derive an expression for the critical temperature. At $T = T_c$, the gap is $\Delta(T_c) = 0$. Then, replacing $T$ in Eq. (5.35) with $T_c$ and setting $\Delta(T) = 0$ yields an equation with respect to $T_c$:

$$\frac{1}{N(0)V} = \int\limits_0^{\hbar\omega_D} \frac{d\varepsilon}{\varepsilon} \tanh \frac{\varepsilon}{2k_B T_c}. \tag{5.40}$$

Carrying out this integration, one gets

$$k_B T_c = 1.14\hbar\omega_D \exp\left(-\frac{1}{N(0)V}\right). \tag{5.41}$$

Taking Eq. (5.37) into account, we have

$$2\Delta_0 = 3.52 k_B T_c. \tag{5.42}$$

The last two relations are in good quantitative agreement with numerous experiments. Moreover, Equation (5.41) provides an explanation for the isotope effect (see Chapter 2): since the Debye frequency varies as $\omega_D \propto M^{-1/2}$, where $M$ is the isotope mass, one immediately obtains from Eq. (5.41) that the product $T_c M^{1/2}$ is constant for a given superconductor.

### 3.7    Condensation energy

Here we calculate the condensation energy of the superconducting state at $T = 0$, i.e. the difference in energy between the superconducting and normal states:

$$W \equiv E_s - E_n = (E_s^{kin} - E_n^{kin}) + (E_s^{pot} - E_n^{pot}), \qquad (5.43)$$

where $E_i^{kin}$ and $E_i^{pot}$ are respectively the kinetic and potential energies. In the normal state at $T = 0$, all states below the Fermi level are completely occupied and, therefore, $E_n^{pot} = 0$ and

$$E_n^{kin} = \sum_{k < k_F} 2\varepsilon_k. \qquad (5.44)$$

Here the coefficient 2 appears because the sum is taken over pairs of states $(\mathbf{k}, -\mathbf{k})$. Then, taking into account that, at $T = 0$, $f_k = 0$, and using Eq. (5.24), the difference of kinetic energies is

$$E_s^{kin} - E_n^{kin} = 2 \sum_{k < k_F} \varepsilon_k (v_k^2 - 1) + 2 \sum_{k > k_F} \varepsilon_k v_k^2. \qquad (5.45)$$

Using Eqs. (5.30), (5.31) and (5.37), one can easily transform Eq. (5.45) into

$$E_s^{kin} - E_n^{kin} = N(0)\Delta_0^2 \left[ \frac{1}{N(0)V} - \frac{1}{2}(1 - \mathrm{e}^{-2/N(0)V}) \right]. \qquad (5.46)$$

Similarly, the potential energy of the superconducting state from Eq. (5.25) at $T = 0$ is

$$E_s^{pot} = E_p = -\frac{\Delta_0^2}{V}. \qquad (5.47)$$

Then, the condensation energy is the sum of the last two equations:

$$W = -\frac{1}{2}N(0)\Delta_0^2 \left[ 1 - \mathrm{e}^{-2/N(0)V} \right] \simeq -\frac{1}{2}N(0)\Delta_0^2 \ \text{if} \ N(0)V \ll 1. \quad (5.48)$$

Since $N(0) \simeq N/E_F$, where $N$ is the total number of free conduction electrons, then the condensation energy per electron is about $\Delta_0^2/(2E_F)$. A similar result was obtained in Chapter 2 by using common sense. Also in Chapter 2, we established that the difference in free energy between the superconducting and normal state equals $H_c^2/8\pi$, where $H_c$ is the thermodynamic critical field. It then follows that at $T = 0$

$$\frac{H_c^2}{8\pi} = \frac{1}{2}N(0)\Delta_0^2. \qquad (5.49)$$

From Eq. (2.20), $H_c(T)$ varies with temperature as $[1 - (T/T_c)^2]$. As $T \to T_c$, this dependence becomes linear with $T$: $H_c \propto (T_c - T)$. As a consequence, from Eq. (5.49), $\Delta(T)$ varies with temperature at $T \to T_c$ as $(T_c - T)^{1/2}$.

## 3.8 Coherence length

As defined in Chapter 2, the coherence length $\xi_{GL}$ is determined by variations of the order parameter $\Psi(\mathbf{r})$, whilst the Cooper-pair size $\xi$ is related to the wavefunction of a Cooper pair, $\psi(\mathbf{r})$. Furthermore, the coherence length depends on temperature, $\xi_{GL}(T)$, while the Cooper-pair size is temperature-independent. Since the order parameter in conventional superconductors is a "magnified" version of Cooper-pair wavefunctions, the values of coherence length and Cooper-pair size in conventional superconductors coincide at $T = 0$: $\xi_{GL}(0) = \xi(0) = \xi_0$.

The superconducting ground state can be represented by the distribution of electron pairs in momentum space given by the function $v_k^2$. The sketch of the dependence of $v_k^2$ on $k$ is shown in Fig 5.7. In this plot, one can see that large variations of $v_k^2$ at $T = 0$ can occur only within the region

$$\Delta k \sim k_F \frac{2\Delta_0}{E_F}, \qquad (5.50)$$

where $E_F = \hbar^2 k_F^2/2m$ is the Fermi energy. Then in real space, large variations of the order parameter of the ground state can be expected within the interval $\Delta x$ defined by the uncertainty relation

$$\Delta x \Delta k \sim 1. \qquad (5.51)$$

Then, it follows that

$$\Delta x \sim \frac{E_F}{2\Delta_0 k_F} = \frac{1}{2\Delta_0 k_F} \frac{\hbar^2 k_F^2}{2m} = \frac{\hbar p_F}{4m\Delta_0} = \frac{\hbar v_F}{4\Delta_0}, \qquad (5.52)$$

where $p_F$ and $v_F$ are respectively the electron momentum and velocity on the Fermi surface. By definition, $\Delta x$ is the coherence length at $T = 0$, thus, the intrinsic coherence length, $\xi_0 \equiv \Delta x$. A rigorous calculation yields

$$\xi_0 = \frac{\hbar v_F}{\pi \Delta_0}. \qquad (5.53)$$

The difference between the two expressions is only in the numerical coefficient of $\pi/4 \simeq 0.785$. As mentioned above, $\xi_0$ is also the size of the Cooper pairs at $T = 0$.

The estimation of $\xi_0$ for a metallic superconductor, made in Chapter 2, yielded $\xi_0 \simeq 3 \times 10^3$ Å. The values of the intrinsic coherence length in conventional superconductors can be found in Table 2.1. In spite of the fact that two electrons in a Cooper pair in metallic superconductors are far apart from each other, the other Cooper pairs are only a few tens of Å away, as shown schematically in Fig. 4.1. Such a high concentration of the Cooper pairs in conventional superconductors automatically leads to the onset of phase coherence.

### 3.9    Specific-heat jump

As discussed in Chapter 2, the appearance of the superconducting state is accompanied by quite drastic changes in both the thermodynamic equilibrium and thermal properties of a superconductor. The normal–superconducting transition is a second-order phase transition accompanied by a jump in heat capacity. On cooling, the heat capacity of a superconductor has a discontinuous jump at $T_c$ and then falls exponentially to zero, as illustrated in Fig. 2.23. In the framework of the BCS theory, the value of this jump in heat capacity equals $\beta = 1.43$, as specified by Eq. (2.54). Experimentally, the value of the jump in specific heat in conventional superconductors with a strong electron-phonon coupling can be much larger than 1.43.

### 3.10    Relation between the BCS and Ginzburg-Landau theory

The Ginzburg-Landau theory of the superconducting state, which was discussed in Chapter 2, is phenomenological. Soon after the development of the BCS theory, Gor'kov showed that the two theories—microscopic BCS and phenomenological Ginzburg-Landau—are basically the same at $T \to T_c$. However, the relation between the BCS and Ginzburg-Landau theory differs in the so-called clean and dirty limits. To recall, the case when the mean electron free path is larger than the Cooper-pair size, $\ell \gg \xi$, is known as the clean limit. In the dirty limit, $\ell \ll \xi$. In what follows, all quantities corresponding to "clean" superconductors will be labeled with an index "cl" and those to "dirty" superconductors with an index "d".

First of all, the energy-gap function in the BCS theory, $\Delta(\mathbf{r})$, is proportional to the order parameter $\Psi(\mathbf{r})$ in the Ginzburg-Landau theory at $T \to T_c$ as

$$\Psi_{cl}(\mathbf{r}) = \left[ \frac{7mv_F^2 N(0)}{2\pi^2 k_B^2 T_c^2} \zeta(3) \right]^{1/2} \Delta(\mathbf{r}), \qquad (5.54)$$

$$\Psi_d(\mathbf{r}) = \left[ \frac{\pi mv_F N(0) l}{12\hbar k_B T_c} \right]^{1/2} \Delta(\mathbf{r}), \qquad (5.55)$$

where $\zeta(3) \simeq 1.202$ is the Riemann function. The temperature dependence of the energy gap at $T \to T_c$ is

$$\Delta(T) \simeq 3.06 \, k_B T_c \left( 1 - \frac{T}{T_c} \right)^{1/2}. \qquad (5.56)$$

The temperature dependences of the coherence length and magnetic penetration depth in the two limits are

$$\xi_{GL,cl} = 0.74 \, \xi_0 \left( 1 - \frac{T}{T_c} \right)^{1/2}, \qquad (5.57)$$

$$\xi_{GL,d} = 0.85 \, (\xi_0 l)^{1/2} \left(1 - \frac{T}{T_c}\right)^{1/2}, \tag{5.58}$$

$$\lambda_{cl} = \frac{\lambda(0)}{\sqrt{2}} \left(1 - \frac{T}{T_c}\right)^{1/2}, \tag{5.59}$$

$$\lambda_d = 0.615 \, \lambda(0) \left(\frac{\xi_0}{l}\right)^{1/2} \left(1 - \frac{T}{T_c}\right)^{1/2}, \tag{5.60}$$

where $\xi_0$ is the intrinsic coherence length defined in Eqs. (5.53) and (2.15), and

$$\lambda^2(0) \equiv \frac{3c^2}{8\pi e^2 v_F^2 N(0)}. \tag{5.61}$$

The coefficients $\alpha$ and $\beta$ in the Ginzburg-Landau theory can be expressed in terms of $\xi_0$ and $\lambda(0)$ as

$$\alpha_{cl} = -1.83 \, \frac{\hbar^2}{2m\xi_0^2} \left(1 - \frac{T}{T_c}\right)^{1/2}, \tag{5.62}$$

$$\alpha_d = -1.36 \, \frac{\hbar^2}{2m\xi_0 l} \left(1 - \frac{T}{T_c}\right)^{1/2}, \tag{5.63}$$

$$\beta_{cl} = \frac{0.35}{N(0)} \left(\frac{\hbar^2}{2m\xi_0^2 k_B T_c}\right)^2, \tag{5.64}$$

$$\beta_d = \frac{0.2}{N(0)} \left(\frac{\hbar^2}{2m\xi_0 l k_B T_c}\right)^2. \tag{5.65}$$

Finally, the Ginzburg-Landau parameter $k$ in these two limits is

$$k_{cl} = 0.96 \, \frac{\lambda(0)}{\xi_0} \quad \text{and} \quad k_d = 0.725 \, \frac{\lambda(0)}{l}. \tag{5.66}$$

## 4. Extensions of the BCS theory

As is natural with a development of such importance, the appearance of the original paper by Bardeen, Cooper and Schrieffer was followed rapidly by a number of papers giving reformulation of the calculations and some corrections to the original results. Several of these reformulations and corrections proved important in the later development.

### 4.1 Critical temperature

Here we discuss an important correction to the critical temperature obtained for the case of strong electron-phonon coupling. For convenience, we introduce a new parameter $\lambda = N(0)V$ called the electron-phonon coupling con-stant (not to be confused with the penetration depth). In the weak-coupling limit, i.e. when $\lambda \ll 1$, the influence of the Coulomb interaction on $T_c$ in Eq. (5.41) can be taking into account by introducing the Coulomb pseudo-potential:

$$k_B T_c = 1.14 \hbar \omega_D \exp \left( -\frac{1}{\lambda - \mu^*} \right), \qquad (5.67)$$

where $\mu^*$ is $N(0)$ times the Coulomb pseudo-potential. Later, McMillan extended this result for the case of strong-coupling superconductors, and obtained

$$T_c = \frac{\Theta}{1.45} \exp \left[ -\frac{1.04(1 + \lambda)}{\lambda - \mu^*(1 + 0.62\lambda)} \right], \qquad (5.68)$$

where $\Theta$ is the Debye temperature. So, when the Coulomb interaction is taken into account, the isotope-mass dependence of the Debye frequency appearing in $\mu^*$ modifies the isotope effect, and thus explains the deviation of the isotope effect in some superconductors from the ideal value of 0.5 (see Table 2.3).

In BCS-McMillan's expression for $T_c$, the Debye temperature $\Theta$ occurs not only in the pre-exponential factor in Eq. (5.68), but also in the electron-phonon coupling constant $\lambda$ which can be presented as $\lambda \approx C/M \langle \omega^2 \rangle$, where $C$ is a constant for a given class of materials, $M$ is the isotope mass, and $\langle \omega^2 \rangle$ is the mean-square average phonon frequency, and $\langle \omega^2 \rangle \propto \Theta$. As a consequence, in BCS-type superconductors, $T_c$ increases as $\Theta$ decreases. In other words, $T_c$ increases with lattice softening. This, however, is not the case for high-$T_c$ superconductors (copper oxides) in which $T_c$ increases with lattice stiffening (see Fig. 6.29), thus, contrary to the case of conventional superconductors. This fact clearly indicates that the mechanism of superconductivity in high-$T_c$ superconductors, which will be discussed in the next chapter, is not of the BCS type.

## 4.2    Strength of the electron-phonon interaction

The electron-phonon coupling constant $\lambda$ in Eq. (5.68) can be determined experimentally. In a given material, the strength of the electron-phonon interaction depends on the function $\alpha^2(\omega)F(\omega)$, where $F(\omega)$ is the density of states of lattice vibrations (the phonon spectrum); $\alpha^2(\omega)$ describes the interaction between the electrons and the lattice, and $\omega$ is the phonon frequency. The spectral function $\alpha^2(\omega)F(\omega)$ is the parameter of the electron-phonon interaction in the Eliashberg equations, and if this function is known explicitly, one can calculate the coupling constant $\lambda$ with the help of the following relation

$$\lambda = 2 \int \alpha^2(\omega)F(\omega) \frac{d\omega}{\omega}. \qquad (5.69)$$

The phonon spectrum $F(\omega)$ can be determined by inelastic neutron scattering. The product $\alpha^2(\omega)F(\omega)$ can be obtained in tunneling measurements.

A comparison of experimental data obtained by these two different techniques reveals in many superconducting materials a remarkable agreement of the spectral features. The function $\alpha(\omega)$ can be determined explicitly from these data, which is usually smooth relative to $F(\omega)$. As an example, Figure 5.11 shows the phonon spectrum $F(\omega)$ and the function $\alpha^2(\omega)F(\omega)$ for Nb, obtained by neutron and tunneling spectroscopies, respectively. In Fig. 5.11, one can see that there is good agreement between the two spectra. It is worth to mention that the phonon spectrum in metals often has two peaks, as that in Fig. 5.11. These peaks originate from longitudinal and transverse phonons.

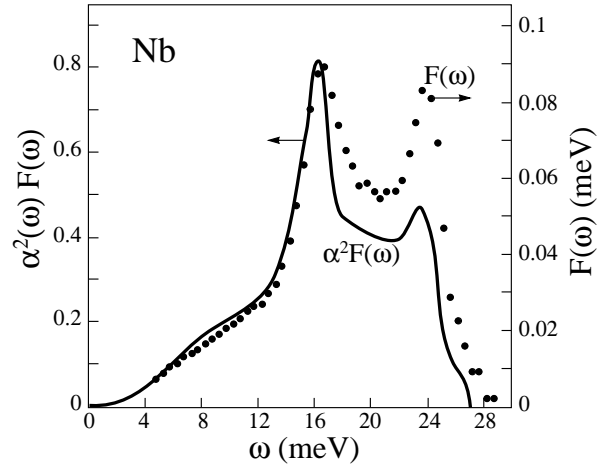

*Figure 5.11.* Tunneling and neutron spectroscopic data for Nb (taken from [19]).

## 4.3 Tunneling

As we already know, tunneling $dI(V)/dV$ characteristics obtained in an SIN junction correspond to the density of states of quasiparticle excitations in a superconductor. Let us derive this result. Assuming that the normal metal has a constant density of states near the Fermi level and the transmission of the barrier (insulator) is independent of energy, the tunneling conductance $dI(V)/dV$ is proportional to the density of states of the superconductor, broadened by the Fermi function $f(E, T) = [\exp(E/k_B T) + 1]^{-1}$. Thus, at low temperature,

$$\frac{dI(V)}{dV} \propto \int\limits_{-\infty}^{+\infty} N_s(E) \left[ -\frac{\partial}{\partial(eV)} f(E + eV, T) \right] dE \cong N_s(eV), \quad (5.70)$$

where $V$ is the bias; $N_s(E)$ is the quasiparticle density of states in a superconductor, and the energy $E$ is measured from the Fermi level of a superconductor.

So, the differential conductance at negative (positive) voltage reflects the density of states below (above) the Fermi level $E_F$.

In order to smooth the gap-related structures in the density of states $N_s$ shown in Fig. 5.10 at $T \to 0$, a phenomenological smearing parameter $\Gamma$ was introduced, which accounts for a lifetime broadening of quasiparticles ($\Gamma = \hbar/\tau$, where $\tau$ is the lifetime of quasiparticle excitations). The energy $E$ in the density-of-state function is replaced by $E - \mathrm{i}\Gamma$ as

$$N_s(E, \Gamma) \propto Re \left\{ \int \frac{E - \mathrm{i}\Gamma}{\sqrt{(E - \mathrm{i}\Gamma)^2 - \Delta^2(\mathbf{k})}} \, \mathrm{d}\mathbf{k} \right\}, \qquad (5.71)$$

where, in the general case, the energy gap $\Delta(\mathbf{k})$ is $\mathbf{k}$-dependent. (In a two-dimensional case, the integration is reduced to integrating over the in-plane angle $0 \le \theta < 2\pi$.)

In SIN tunneling junctions of conventional superconductors, there is good agreement between the theory and experiment. However, for SIS-junction characteristics, the correspondence between the smoothed BCS density of states and experimental data is poor. This issue was raised for the first time elsewhere [19].

## 4.4    Effect of impurities on $T_c$

How do magnetic and non-magnetic impurities affect the critical temperature in conventional superconductors? It turns out that, in superconductors described by the BCS theory, non-magnetic impurities do not alter $T_c$ much, whereas magnetic impurities drastically suppress the superconducting transition temperature.

The effect of non-magnetic impurities on $T_c$ was first explained by Anderson in his theorem. In the normal state, the electrons may be described by wave functions $\phi_{n\uparrow}(\mathbf{r})$ and $\phi_{n\downarrow}(\mathbf{r})$, where $\phi_n$ is supposed to include the effects of the impurity scattering. The quantum number $n$ replaces the wave number $\mathbf{k}$ we used for the *pure* metal. The Cooper pairs in the pure metal is composed of the states $(\mathbf{k}, \uparrow)$ and $(-\mathbf{k}, \downarrow)$. The latter state is in fact the former one but with momentum and current reversed in time. Anderson argued that, in the impure metal, one should equally pair time-reversed states, namely, $\phi_{n\uparrow}(\mathbf{r})$ and $\phi_{n\downarrow}^*(\mathbf{r})$: the complex conjugate $\phi_{n\downarrow}^*$ is the time reverse of $\phi_{n\uparrow}$ just as $\mathrm{e}^{-\mathrm{i}\mathbf{k}\cdot\mathbf{r}}$ is the time reverse of $\mathrm{e}^{\mathrm{i}\mathbf{k}\cdot\mathbf{r}}$. With non-magnetic impurities, $\phi_n^*$ and $\phi_n$ have the same energy, and all the BCS calculations go through unmodified. In fact, the impurity scattering washes out the effects of Fermi-surface anisotropy, so that the BCS results apply rather better to alloys than to pure metals. Experimentally, when non-magnetic impurities are added to a pure superconductor, the critical temperature first drops sharply and then varies only slowly as a function of the impurity concentration. The first drop is related to the destruction of anisotropy effects, and the subsequent slow change is caused by a change in $N(0)$.

In the case of magnetic impurities, $\phi_n^*$ and $\phi_n$ have different energies. As a consequence, this leads to a difference in energy between the two electrons in a Cooper pair. When the mean value of this energy difference exceeds the binding energy of the pairs, the electron pairing and, thus, the superconducting state, can no longer occur.

The effect of suppression of the superconducting state by magnetic and non-magnetic impurities is always used to determine the nature of the electron-electron attraction in superconducting materials. For example, the effect of impurities on $T_c$ in superconductors in which superconductivity is mediated by magnetic fluctuations should be opposite to that in conventional superconductors.

## 4.5 High-frequency residual losses

The vanishing of the *dc* resistance is the most striking feature of the superconducting state. However, the losses in a superconductor are non-zero for an *ac* current which flows in a thin surface layer having a thickness of $\lambda$ (penetration depth). The surface resistance depends on the frequency $\omega$ of an *ac* current (or electromagnetic field), as well as on the temperature and the energy gap of a superconductor. At microwave frequencies and at temperatures less than half the transition temperature, the surface resistance has the following approximate BCS form:

$$R_s(T, \omega) = \frac{C\omega^2}{T} e^{-\Delta/k_B T} + R_0(\omega), \qquad (5.72)$$

where $C$ is a constant that depends on the penetration depth in the material, and $R_0$ is the residual resistance. Although it is not entirely clear where the residual losses originate, it has been determined experimentally that trapped flux and impurities are principal causes. An additional source of the residual losses can be imperfection of the surface of a superconductor, which may not be perfectly flat or contains non-superconducting regions (for example, oxides etc.). Nevertheless, in conventional superconductors, the residual losses which are determined by $R_0$ are small in comparison, for example, with those in high-$T_c$ superconductors.

# Chapter 6

# THIRD GROUP OF SUPERCONDUCTORS: MECHANISM OF SUPERCONDUCTIVITY

*True laws of Nature cannot be linear.*

—Albert Einstein

Here we consider the third group of superconductors; the second group will be discussed in the following chapter. Such a sequence simplifies the presentation of the second-group superconductors.

The third group of superconductors incorporates unconventional superconductors which are *low-dimensional* and *magnetic* (at least, these compounds have local magnetic correlations). These superconductors are also known as materials with strongly-correlated electrons (holes). This means that the position and motion of each electron in these compounds are correlated with those of all the others.

At the moment of writing, there is no exact theory of unconventional superconductivity. However, the combination of two theories can describe, in a first approximation, some pairing and phase-coherence characteristics, for example, in superconducting cuprates. With respect to the BCS mechanism of superconductivity which can be considered as *linear*, the mechanism of unconventional superconductivity is *nonlinear*, meaning that the electron-phonon interaction in these compounds is moderately strong and nonlinear. In the absence of an exact theory of unconventional superconductivity, we shall discuss a general description of the mechanism of unconventional superconductivity fully based on experimental data obtained mainly in the cuprates. These data are presented in [19]. The two theoretical models mentioned above will also be discussed.

One could ask why it is that at present there exists no complete theory for unconventional superconductivity given that its mechanism is *in general* un-derstood. The answer is very simple. There is a lack of understanding of the normal-state properties of materials with strongly-correlated electrons. This knowledge is crucial for understanding the mechanism of unconventional superconductivity *on the nanoscale*, and thus, for a theoretical description of this phenomenon.

The development of the BCS theory was possible only because, by that time, the normal-state properties of ordinary metals were very well understood. The development of the exact theory of unconventional superconductivity will not be possible until we understand the normal-state properties of materials with strongly-correlated electrons, including the cuprates, as well as understanding those of ordinary metals. It is **most likely** that a room-temperature superconductor will be available earlier than the exact theory of unconventional superconductivity. It may sound pessimistic with respect to the theory; at the same time, it reflects difficulties that are confronted by physicists working in this field. On the other hand, it may sound too optimistic with respect to room-temperature superconductivity. The truth is probably somewhere in between.

Since the mechanism of unconventional superconductivity was described in detail in [19], the references related to this chapter can be found there. Only a few articles will be mentioned here. We start this chapter with some important remarks concerning materials with strongly-correlated electrons, some of which superconduct.

## 1.    Systems with strongly-correlated electrons

Electrons in an ordinary metal can be treated in a mean-field approximation. Such an approach is not applicable to materials with strongly correlated electrons, in which the position and motion of each electron are correlated with those of all the others. In this class of materials, the electronic, magnetic and crystal structures are strongly coupled, and they actively interact with each other. This gives rise to many fascinating phenomena, such as superconductivity, colossal magnetoresistance, interacting spin- and charge-density waves, as schematically illustrated in Fig. 6.1. Depending on temperature, all these states in Fig. 6.1 may exist simultaneously (except coexistence of colossal magnetoresistance and superconductivity—they are mutually exclusive). The Fermi-liquid approach is not applicable to most of these materials.

At present it may be too early to say that in *all* compounds with strongly correlated electrons, *quasiparticles are solitons*; nevertheless, for most of these materials, this statement is correct. This would imply that electrical current in some materials with strongly-correlated electrons is carried by solitons, not by electrons/holes. As discussed in Chapter 1, the essence of the soliton concept is that a soliton is a "compromise" achieved between two "destructive" forces— nonlinearity and dispersion. It is then no wonder why solitons appear in such a "harsh" environment that exists in materials with strongly-correlated electrons. We shall consider solitons in more detail in the following subsections.

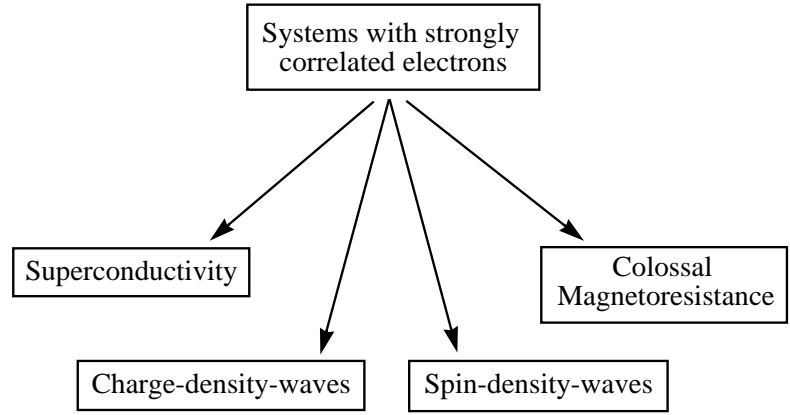

*Figure 6.1.*  Diagram for materials with strongly-correlated electrons.

## 2.    General description of the mechanism

Among superconductors of the third group, the cuprates are the most studied. For this reason the cuprates are chosen for presentation of the mechanism of unconventional superconductivity. We shall also discuss superconductivity in $C_{60}$.

The undoped cuprates are antiferromagnetic Mott insulators. To become superconducting they must be doped by either electrons or holes. The doping can be achieved chemically, by pressure, or in a field-effect transistor configuration. The crystal structure of the cuprates is layered and highly anisotropic. The doped carriers are accumulated in the two-dimensional copper-oxide planes (see Fig. 3.7). The $CuO_2$ layers are always separated by layers of other atoms, which are usually insulating or semiconducting. In chemical doping these layers provide charge carriers in the $CuO_2$ planes; therefore, they are often called the charge reservoirs. Superconductivity occurs at low temperature when the doping level is not too high nor too low, approximately one doped carrier per three $Cu^{2+}$ ions ($\sim 16\%$). For simplicity, we shall further discuss the hole-doped cuprates; the electron-doped cuprates will be considered separately at the end of this chapter.

What is the main cause of the onset of superconductivity in the cuprates? After the charge-carrier doping which is absolutely necessary, the main cause of the occurrence of superconductivity in the cuprates is their **unstable lattice**. Experimentally, the lattice in the cuprates is very unstable especially at low temperatures. Upon lowering the temperature, all superconducting cuprates undergo a number of structural phase transitions. In the cuprates the unstable lattice provokes the phase separation taking place in the $CuO_2$ planes on the nanoscale ($\sim$ a few nanometers). Because of a lattice mismatch between different layers in the *doped* cuprates, below a certain temperature in the $CuO_2$ planes there appear, *at least*, two different phases which fluctuate. The doped charges prefer to join one of these phases, avoiding the other(s). By doing so, the charge presence on one phase enlarges the difference between the two phases. Thus, the phase separation is self-sustaining. Clusters containing the hole-poor phase in the $CuO_2$ planes remain antiferromagnetically ordered. This phase separation taking place in the normal state of the cuprates on the nanoscale is the main key point for the understanding of the mechanism of unconventional superconductivity.

What is probably even more important is that the doped holes in the $CuO_2$ planes, gathered in clusters having a pancake-like shape, are not distributed homogeneously in these clusters but they form quasi-one-dimensional charge stripes. Such a type of doping is called *topological*. Thus, the phase separation into the $CuO_2$ planes takes place not only on the nanoscale but also on a sub-nanoscale ($\sim$ several Å). The charge stripes are a manifestation of self-trapped states. The combination of the electron-phonon, Coulomb and magnetic interactions result in the appearance of the charge stripes. The electron-phonon interaction in the cuprates is moderately strong and nonlinear. In *superconducting* cuprates, the stripes are quoter-filled (i.e. one hole per two Cu sites) and run along –O–Cu–O–Cu– bonds (see Fig. 3.7b). The charge stripes are separated by two-dimensional insulating antiferromagnetic stripes, as shown in Fig. 6.2. The charge stripes are *dynamic*: they can meander and move in the transverse direction. Intrinsically the charge stripes are insulating, i.e. there is a charge gap on the stripes. However, the presence of soliton-like excitations on the stripes makes them conducting. These soliton-like excitations are *one-dimensional polarons* called also *polaronic solitons*, *Davydov solitons* or *electrosolitons*. They propagate in the middle of the charge gap. On cooling, the polaronic solitons give rise to pairs coupled in a singlet state due to local deformation of the lattice. Thus, the moderately strong, nonlinear electron-phonon interaction is responsible for electron pairing in the cuprates. The pairs of polaronic solitons—*bisolitons*—are formed above the critical temperature, at $T_{pair} > T_c$, and reside on the charge stripes. They represent the Cooper pairs in the cuprates.

In the cuprates, the long-range phase coherence is established at $T_c$ due to local spin fluctuations in the antiferromagnetic stripes shown in Fig. 6.2. The fluctuating charge stripes locally induce spin excitations which mediate the long-range phase coherence. In other words, in the cuprates the phase coherence is locked at $T_c$ via a spin wave oscillating in space and time. Such a spin wave is locally commensurate with the lattice, as shown in Fig. 6.2; however, it manifests itself in inelastic neutron scattering (INS) spectra by four incommensurate peaks because of phase jumps by $\pi$ at a periodic array of domain walls (charge stripes).

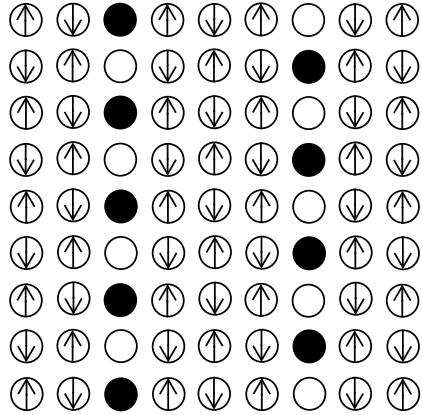

*Figure 6.2.* Idealized diagram of the spin and charge stripe pattern within a CuO$_2$ plane with hole density of 1/8. Only Cu atoms are shown; the oxygen atoms which are located between the copper atoms have been omitted. Arrows indicate the spin orientations. Holes (filled circles) are situated at the anti-phase domain boundaries [13].

Below $T_c$ and somewhat above $T_c$, the charge, spin and lattice structures in the CuO$_2$ planes are coupled. At any doping level, the pairing $\Delta_p$ and phase-coherence $\Delta_c$ energy gaps relate to one another as $\Delta_p > \Delta_c$.

In a first approximation, unconventional superconductivity in the cuprates can be described by a combination of two theoretical models: the bisoliton [9, 10] and spin-fluctuation [4] theories. Some pairing characteristics of superconducting cuprates can be estimated by using the bisoliton model, while the spin-fluctuation model describes phase-coherence characteristics.

## 3. Detailed description of the mechanism

In this section, we discuss in detail important elements of superconductivity in the cuprates and the *physics* of high-$T_c$ superconductors. We start with the normal-state properties of the cuprates which are abnormal in comparison with those of ordinary metals. In a sense, it is ironic to use the word "normal" in the phrase "the normal-state properties of the cuprates."

### 3.1 Structural phase transitions

In the cuprates, *the lattice is very unstable*. As a consequence, the cuprates exhibit several structural phase transitions which finally result in the occurrence of superconductivity.

Structural phase transitions are probably the best documented in LSCO and La$_{2-x}$Ba$_x$CuO$_4$ (LBCO) because of their relatively simple crystal structure and the availability of large high-quality single crystals. As shown in Fig. 6.3 (see also Fig. 3.8), the Sr (Ba) substitution for La in LSCO (LBCO) induces a structural phase transition from the high-temperature tetragonal (HTT) to low-temperature orthorhombic (LTO). The HTT → LTO transition which had been studied a decade before the discovery of superconductivity in LBCO [5], is driven by soft phonons. At high temperatures, LSCO has the same body-centered tetragonal structure as K$_2$NiF$_4$. With cooling, the HTT phase transforms into the LTO phase, because of a staggered tilt of the CuO$_6$ octahedra (see Fig. 3.7a). The tilt angle of the octahedra is about 3.6° and uniform, as shown in Fig. 6.4. In LBCO, on further cooling, the LTO phase transforms into new phases with doubled unit cells, one of which is called the low-temperature tetragonal (LTT) phase. In the LTT phase, the CuO$_6$ octahedra rotate about alternate orthogonal axes in successive layers with no change in the magnitude of the tilt, as shown in Fig. 6.4.

The underlying interaction that gives rise to the low-temperature structural phases, including the LTT phase, is the mismatch in preferred lattice constants of the CuO$_2$ layer and the intervening rare-earth oxide layers. At high Ba doping ($0.2 < x$), there is no well defined LTO → LTT phase transition below

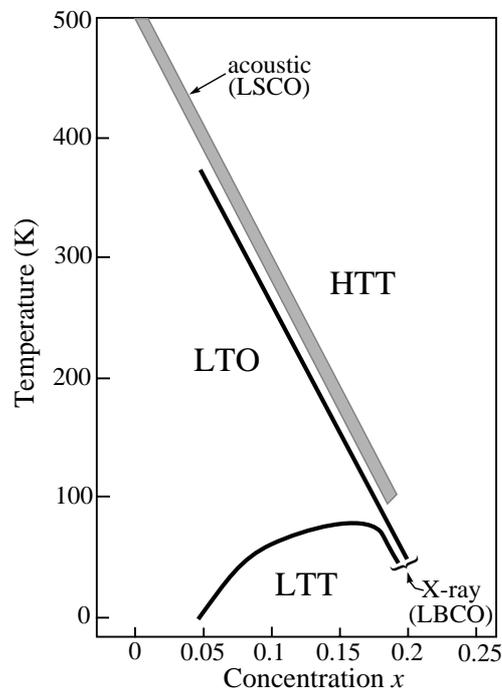

*Figure 6.3* Structural phase diagram of LSCO and LBCO obtained by X-ray and acoustic measurements (see references in [19]).

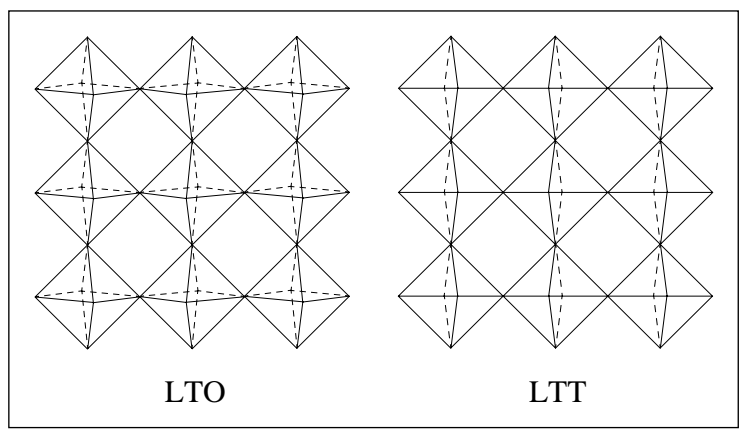

*Figure 6.4.* Tilt pattern of $CuO_6$ octahedra in the LTO and LTT phases.

300 K, thus a crystal is a mixture of the LTO and LTT phases. Even at low Ba (Sr) doping level, neutron and X-ray scattering measurements performed in LBCO and Nd-doped LSCO show that the LTO and LTT phases coexist well below the onset of the LTO → LTT structural transition. As shown in Fig. 6.3, the HTT → LTO transition in LSCO is also observed in acoustic measurements; however, acoustic measurements alone cannot identify the type of a structural transition. The small structural differences between the LTO and LTT phases have a drastic influence on electronic properties of LSCO and LBCO. Results of some transport measurements suggest that the LTO → LTT transition in LSCO induces intra-gap electronic states in the middle of normal-state gap (pseudogap).

In YBCO, there are at least three structural transitions which occur at $T_c$, 140–150 K and 220–250 K. The transition at 220–250 K is close to that shown in Fig. 3.10; therefore it can be associated with the HTT → LTO transition. Unfortunately, the other two structural transitions in YBCO are unknown: the measurements performed in YBCO with different dopings ($0.55 \leq x \leq 1$) by ion channeling spectrometry cannot identify the type of these structural transitions. Heat-capacity measurements in YBCO ($x = 0.85$–$0.95$) show the presence of three anomalies in the temperature dependence of heat capacity, occurring in each of the regular intervals 100–200 K, 205–230 K and 260–290 K. In near optimally doped YBCO, a structural phase transition was even observed deep in the superconducting state: a lattice distortion taking place near 60 K induces a redistribution of holes in the $CuO_2$ planes.

The CuO-chain oxygen ions in $YBCO_8$ (124) undergo a correlated displacement in the $a$ direction (perpendicular to the chains) of about 0.1 Å, with the onset of correlations occurring near 150 K. A similar effect is observed in YBCO.

In slightly overdoped Bi2212, three structural phase transitions are observed in acoustic measurements at 95 K, 150 K and 250 K. Again, acoustic measurements cannot determine the type of the structural transitions, even one is able to observe them in acoustic measurements.

In underdoped and optimally doped regions, some of these transitions are almost doping level $p$ independent. Acoustic measurements performed in *undoped* and underdoped LSCO, YBCO, NCCO, Bi- and Tl-based compounds show that the elastic coefficients display some kind of structural transition at maximum $T_c$ for each compound, although some of these cuprates either are not superconducting or have low $T_c$, i.e. $T_c \ll T_{c,max}$. This fact suggests that this structural transition at $T_{c,max}$ does not require the presence of superconductivity. One may then conclude that $T_{c,max}$ is determined in each compound by the underlying (unstable) lattice.

In the cuprates, the charge-stripe phase shown in Fig. 6.2, which will be discussed in detail in the following subsection, is also induced by a structural phase transition. As an example, Figure 6.5 shows the neutron-diffraction data obtained in Nd-doped LSCO. In the plot, one can see that, upon cooling, the charge order appears immediately after the lattice transformation, $T_{CO} \leq T_d$, where $T_{CO}$ is the onset temperature of charge ordering, and $T_d$ is the structural-phase transition temperature. Spins located between the charge stripes become antiferromagnetically aligned at much lower temperature $T_{MO} < T_{CO}$, as depicted in Fig. 6.5. In the striped phase of cuprates, the charge ordering always precedes the magnetic ordering. In other strongly-correlated electron systems—in nickelates and manganites, the magnetic order always arrears after the charge ordering. For example, in the nickelate $La_{2-x}Sr_xNiO_4$ with $x = 0.29$, 0.33 and 0.39, these two temperatures are $T_{MO} \simeq 115$, 180 and 150 K and $T_{CO} \simeq 135$, 230 and 210 K, respectively. In $La_{2-x}Sr_xNiO_4$, a clear charge gap is formed below $T_{CO}$. In the manganite $La_{0.35}Ca_{0.65}MnO_3$, $T_{MO} \simeq 140$ K and $T_{CO} \simeq 260$ K. In the latter case, the magnitude of the charge gap $2\Delta(0)/k_BT_{CO} \sim 13$ is too large for a conventional charge-density-wave (CDW) order. In the nickelates and manganates, a structural phase transition, observed in acoustic measurements, also precedes the charge ordering. Thus, one can conclude that the striped phase in all these compounds is induced by a structural phase transition. Secondly, it is charge driven and the spin order between charge stripes is subsequently enslaved.

## 3.2 Phase separation and the charge distribution into the CuO$_2$ planes

In the cuprates and some other compounds with strongly-correlated electrons, the distribution of charge carriers is not homogeneous in comparison with that in ordinary metals: the doped charge carriers in the $CuO_2$ planes of the cuprates are distributed inhomogeneously. Depending on the doping level, this inhomogeneous charge distribution takes place on the nanoscale *and* a sub-nanoscale, resulting in the appearance of charge clusters and charge stripes, respectively.

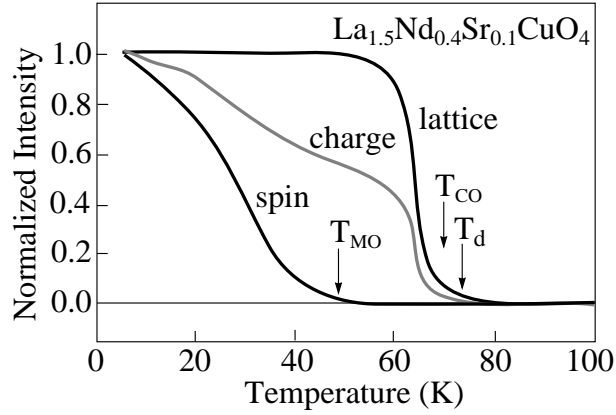

*Figure 6.5.* Temperature evolutions of lattice, charge and spin superlattice peaks in neutron diffraction measurements in Nd-doped LSCO. The magnetic, charge and lattice orderings are marked by $T_{MO}$, $T_{CO}$ and $T_d$, respectively. The backgrounds are subtracted [13].

In the *undoped* region ($p < 0.05$) of hole-doped cuprates, the doped holes form nanoscale clusters in the $CuO_2$ planes containing *diagonal* charge stripes (i.e. they run along the diagonal –Cu–Cu–Cu– direction [see Fig. 3.7b]). The sketch in Fig. 6.6 depicts the charge distribution in a $CuO_2$ plane as a function of doping level. At $0 < p < 0.05$, the nanoscale clusters with diagonal charge stripes are embedded in an antiferromagnetic matrix. On lowering the temperature, unconventional spin glass occurs in these clusters. In undoped cuprates, the distance between diagonal charge stripes is sufficiently large, $\sim 8a$, where $a \simeq 3.85$ Å is the distance between adjacent copper sites in the $CuO_2$ planes.

At $p \simeq 0.05$, the charge stripes in most of the nanoscale clusters change their orientation by $45°$ relative to that in the undoped region. So, they now run along the –Cu–O–Cu– bonds (see Fig. 3.7b). Therefore, they are called vertical (or horizontal). However, in a very small fraction of the clusters, the stripes remain diagonally oriented.

In the underdoped region ($0.05 < p < 0.13$), the nanoscale clusters with vertical charge stripes are still embedded into the antiferromagnetic matrix. As schematically shown in Fig. 6.6, the average distance between vertical charge stripes $d_s$ decreases in comparison with that between diagonal charge stripes. At $0.05 < p < 0.125$, $1/d_s \propto p$, and $d_s$ saturates above $p = \frac{1}{8}$, as illustrated in Fig. 6.7. Due to a balanced interplay among the charge, spin and lattice structures, the average distance between vertical charge stripes in the CuO$_2$ planes above $p \simeq 0.125$ remains constant, $d_s \simeq 4a$, in correspondence with that shown in Fig. 6.2. The quasi-two-dimensional spin stripes that separate the dynamic charge stripes can be considered as a local memory effect of the antiferromagnetic insulating phase at low doping.

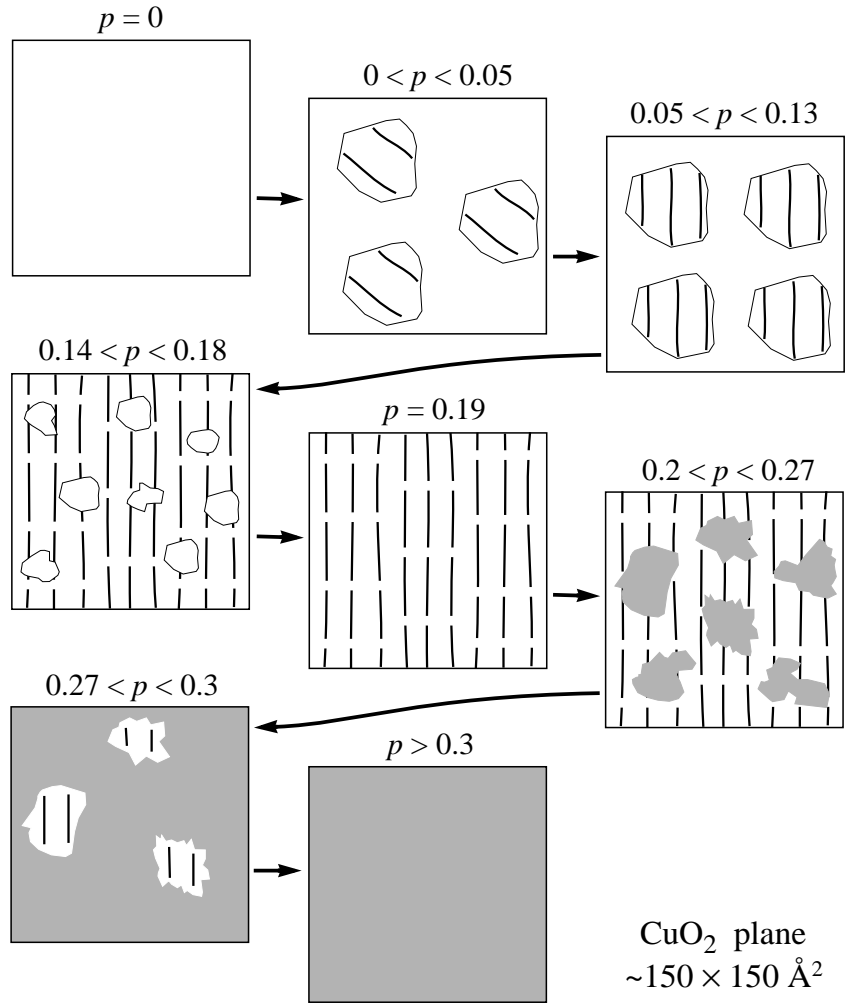

*Figure 6.6.* Sketch of charge distribution into a CuO$_2$ plane as a function of doping. Anti-ferromagnetic and metallic phases are shown in white and grey, respectively. The lines show charge stripes (see text for more details).

The vertical charge stripes are dynamic in the clusters: they meander and can move in the transverse direction. Therefore, they are not strictly one-dimensional but *quasi*-one-dimensional. The mean length of a separate vertical charge stripe is about 30–40 Å, i.e. $\sim 10a$. The striped phase will be discussed in detail in the following subsection.

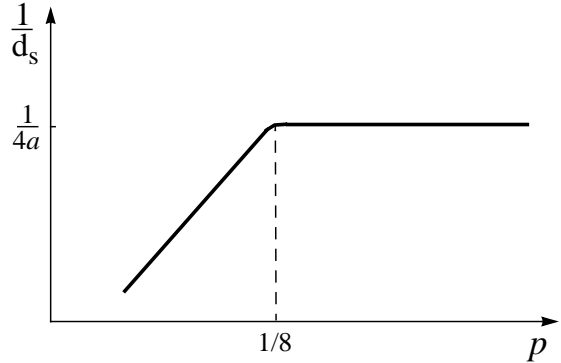

*Figure 6.7.* Doping dependence of incommensurability $2\delta$ ($\propto 1/d_s$) of magnetic ordering, where $d_s$ is the distance between charge stripes (see references in [19]).

In the optimally doped ($p \sim 0.16$) and overdoped ($0.2 \leq p \leq 0.27$) regions, the average distance between charge stripes remains unchanged, as shown in Fig. 6.7. Therefore, as the doping level increases, new doped holes take over the virginal antiferromagnetic matrix. In the optimally doped region, the picture is diametrically opposed to that in the underdoped region: now clusters with intact antiferromagnetic order are embedded in a matrix with vertical charge stripes, as illustrated in Fig. 6.6. The clusters with intact antiferromagnetic order disappears completely at $p = 0.19$, as shown in Fig. 6.8a. The point $p = 0.19$ is a quantum critical point in the cuprates. At $p = 0.19$, the $CuO_2$ planes, in a first approximation, consist only of the striped phase shown in Fig. 6.2. The fraction of this phase as a function of doping level is sketched in Fig. 6.8b. Since the Cooper pairs in the cuprates originate from the striped phase, it is the most robust at $p = 0.19$, and not at $p = 0.16$ where $T_c$ is a maximum for most of the cuprates.

Finally, let us consider the overdoped region ($0.2 \leq p \leq 0.27$). Above $p = 0.19$, new doped holes gather between stripes, forming clusters with the Fermi sea in which the hole distribution is more or less homogeneous, as schematically depicted in Fig. 6.6. The Fermi-sea clusters are embedded in a matrix with vertical charge stripes. As $p \to 0.27$, the Fermi-sea clusters grow in size and start coalescing. The fraction of the Fermi-see clusters (matrix) in the $CuO_2$ planes is depicted in Fig. 6.8a. At low temperature, superconductivity in the cuprates collapses when the doping level reaches $p \simeq 0.27$. However, the hole distribution in the $CuO_2$ planes, apparently, is not yet homogeneous: it is most likely that in the interval $0.27 < p < 0.3$, there are clusters with vertical charge stripes imbedded in a Fermi-see matrix, as shown in Fig. 6.6.

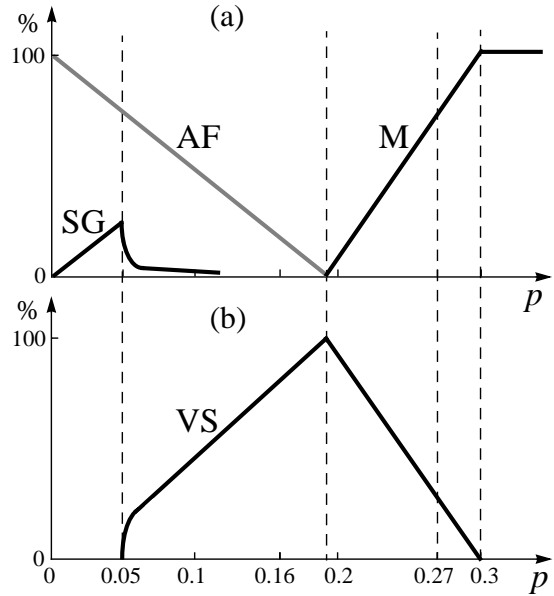

*Figure 6.8.*   Fraction of a certain matrix or clusters in the CuO$_2$ planes of cuprates at $T \ll T_c$ as a function of doping: (a) antiferromagnetic matrix or clusters (AF), Fermi-sea clusters or matrix (M), and spin-glass clusters (SG); and (b) clusters or matrix with vertical charge stripes (VS). Both plots are shown schematically.

The hole distribution in the CuO$_2$ planes becomes homogeneous only above $p = 0.3$. The question of hole distribution in the interval $0.27 < p < 0.3$ is not very important for the understanding of the mechanism of unconventional superconductivity in the cuprates.

As was noted above, in undoped cuprates, unconventional spin glass occurs at low temperatures in the clusters with diagonal charge stripes which are embedded into an antiferromagnetic matrix. The fraction of this spin glass as a function of doping level is sketched in Fig. 6.8a. At $p \simeq 0.05$, the charge-stripe orientation in most of the nanoscale clusters becomes vertical (or horizontal). However, in a very small fraction of the clusters, the charge stripes remain diagonally oriented. Therefore, this unconventional spin glass can still be observed in the cuprates at very low temperatures and $p > 0.05$, for example, by muon spin relaxation/rotation ($\mu$SR) measurements. The clusters with diagonal charge stripes in the CuO$_2$ planes disappear completely somewhere between 0.16 and 0.19.

In Fig. 6.8a, the disappearance of the matrix (clusters) with local antiferromagnetic order at $p = 0.19$ does not imply the disappearance of magnetic correlations above this doping level. The insulating quasi-two-dimensional stripes that are located between vertical charge stripes remain magnetically ordered at low temperature. Experimentally, even in the highly overdoped region, magnetic relaxation is still dominant.

### 3.2.1 Nickelates and manganites

A similar stripe order is observed in nickelates with the $NiO_2$ planes [13]. In the nickelates, however, the charge orientation is always diagonal as that in the undoped cuprates. Doped holes in the nickelates are less destructive to the antiferromagnetic background than those in the isostructural cuprates. The phase diagram of $La_{2-x}Sr_xNiO_4$ (LSNO) is very similar to the phase diagram of the isostructural LSCO, having a series of phases which are closely related to the HTT, LTO and LTT phases of LSCO. Thus, there are many similarities between the striped phases of the cuprates and nickelates.

The stripe order is also observed in manganites with the $MnO_2$ planes. Depending on the doping level, the manganites can exhibit either antiferromagnetic or ferromagnetic ordering. A mesoscopic phase separation into the $MnO_2$ planes of the manganites is an experimental fact [41].

Charge stripes in the nickelates and manganites fluctuate slowly in time and space, while in the cuprates very quickly. Therefore, it is much easier to observe a charge inhomogeneity in the nickelates and manganites than that in the cuprates. As a consequence, there is ample evidence in the literature for charge stripes in the nickelates and manganites (see references in [19]), and much less for the cuprates. Nevertheless, there is direct evidence for dynamic charge stripes in superconducting cuprates as well, for example, in LSCO, YBCO (see references in [19]) and Bi2212 [42].

## 3.3    The striped phase

In the cuprates, the length of a separate charge stripe can vary. The mean length is about 30–40 Å, or $\sim 10a$. The charge stripes in the cuprates are dynamic: they can meander and move in the transverse direction. In the cuprates, the charge stripes are quoter-filled: that is one positive electron charge per two $Cu^{2+}$ sites along the stripes (in the nickelates, the charge density along the stripes is one hole per each $Ni^{2+}$ site). The pattern of charge stripes in real space shown in Fig. 6.2, in fact, is not correct for the $CuO_2$ planes. In Fig. 6.2, such an alternating order along the stripes has a periodicity of $4k_F$ in momentum space, where $k_F$ is the wave number at the Fermi surface. Analysis of INS data obtained in some cuprates show that, in the cuprates, the periodicity of charge stripes in momentum space is not $4k_F$ but $2k_F$. A real-space sketch of a $2k_F$ quoter-filled charge stripe is shown in Fig. 6.9. One can see that this pattern is different from that in Fig. 6.2. The chains in YBCO have also the $2k_F$ charge modulation observed by tunneling spectroscopy. It is worth noting that traditional charge-ordered states such as sinusoidal CDWs, for example, in Peierls compounds have also a $2k_F$ charge modulation; however, they are

## $2k_F$ charge ordering

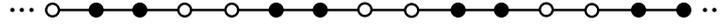

*Figure 6.9.* Schematic ordering pattern on a $2k_F$ quoter-filled charge stripe. In plot, ● (○) denotes the presence (absence) of a hole.

quasi-static, contrary to the charge stripes in the cuprates. Thus, the charge order on the stripes in the cuprates is similar to usual CDWs.

In discussing the charge distribution of the $CuO_2$ planes in the previous subsection, we assumed that the vertical charge stripes run above the Cu sites. In principle, the stripes can alternatively be centered on the Cu–O bonds. Simulations of angle-resolved photoemission (ARPES) data obtained in Bi2212 show that the stripes in Bi2212 seem to be site-centered, not bond-centered.

Why do holes doped in the $CuO_2$ planes prefer to form stripes rather than to be distributed homogeneously? The combination of the electron-phonon, Coulomb and magnetic interactions results in the appearance of the charge stripes. The charge-stripe ordering in the cuprates, nickelates and manganites is evidence for a *strong, nonlinear electron-lattice coupling*. The holes in these compounds are self-trapped.

In the cuprates as in any system with strongly-correlated electrons, the electron-electron interactions are unscreened and, as a consequence, very strong. At low temperature, the magnetic order in the cuprates, frustrated by doped holes, does everything to expel them from the magnetic phase. In the presence of a strong and nonlinear electron-phonon interaction, doped holes become self-trapped, that is, in exchange of interaction with the lattice, a doped hole locally deforms the lattice in a way that it is attracted by the deformation. In its turn, this local deformation created by the first hole attracts another one, and so on. Of course, without the Coulomb repulsion, the holes attracted by the lattice deformation would rather gather in a cluster, not in a one-dimensional stripe. Thus, the charge stripes in the $CuO_2$ planes are a "product" of the electron-phonon, Coulomb and magnetic interactions. In addition, there are other factors favoring the charge-stripe formation; for example, gathering in one-dimensional stripes, the holes lower their kinetic energy in the transverse direction.

The parallel arrangement of the charge stripes minimizes the Coulomb repulsion between the neighboring stripes. What is the characteristic length of the charge-stripe order in the cuprates? Experimentally, the characteristic length of the charge-stripe order is about 80–110 Å, i.e. $\sim 21$–$28a$ [42, 43].

The striped phase in Fig. 6.2 which is nearly one-dimensional will cause the appearance of two incommensurate Bragg peaks symmetric about the fundamental lattice Bragg peaks for an ideal square $CuO_2$ plane. Experimentally however, upon doping, the fundamental lattice Bragg peaks are replaced by *four* new incommensurate peaks displaced by $\pm 2\delta$ from the fundamental lattice peaks (see Fig. 6.33a). It was initially proposed that the appearance of four incommensurate peaks is the consequence of the stripe orientation in the adjacent CuO$_2$ layers: the stripe orientation is alternately rotated by 90° layer by layer, as schematically shown in Fig. 6.10a. Later, another explanation for the appearance of four incommensurate Bragg peaks was proposed [19]. Since the charge-stripe ordering is not very long ($\sim 25a$), it is possible that, in each CuO$_2$ plane, the charge-stripe orientation in neighboring nanoscale clusters with vertical charge stripes may differ by 90°, as schematically depicted in Fig. 6.10b. Thus, the charge stripes with vertical and horizontal orientations may coexist in the same CuO$_2$ plane, causing the appearance of four incommensurate Bragg peaks. In this case it is possible that at high doping level thus in the overdoped region, the stripes in the neighboring clusters with perpendicular stripe orientations may merge, forming structures with the "L" and "⊔" shapes.

In the cuprates, nickelates and manganites, the charge stripes are insulating on the nanoscale scale. This means that there is a charge gap on the stripes. On a macroscopic scale, however, the charge stripes are conducting due to soliton-like excitations present on the stripes. Therefore, the in-plane charge transport in the cuprates is quasi-one-dimensional, as well as that in the nickelates and manganites.

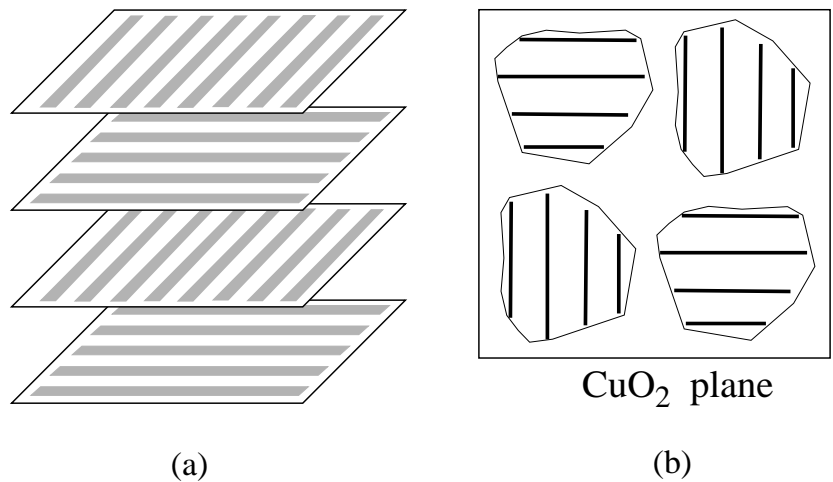

CuO$_2$ plane

(a)                                        (b)

*Figure 6.10.* (a) Stripe orientation in adjacent CuO$_2$ layers is assumed to be alternately rotated by 90°. (b) Sketch of charge-stripe clusters in a CuO$_2$ plane having two possible orientations in the same CuO$_2$ plane.

## 3.4  Phase diagram

Let us now discuss the phase diagram of hole-doped cuprates. The problem, however, is that the phase diagram for each superconducting cuprate, in a sense, is unique. Of course, the differences among all these phase diagrams are not drastic; nevertheless, the phase diagram for each cuprate has its specific features. As an example, comparing the phase diagrams for LSCO and YBCO, shown respectively in Figs. 3.8 and 3.10, one can notice that, in these two cuprates, the width of the antiferromagnetic phase at low doping level is different. In addition, there are some other features in each plot. In such a situation, we have no alternative than to consider in detail a phase diagram of *one* superconducting cuprate which reflects main features of the physics involved in *all* cuprates.

Figure 6.11 shows an idealized phase diagram of Bi2212. In the plot, the commensurate antiferromagnetic phase at low doping level is not shown because it was not yet observed in Bi2212: There is a technical problem to synthesize large-size good-quality *undoped* single crystals of Bi2212 which are necessary for INS and $\mu$SR measurements. Without the antiferromagnetic scale at low doping level, Figure 6.11 depicts six temperature/energy scales for Bi2212.

In Fig. 6.11, the spin-glass $T_g$ temperature scale was recently observed in Bi2212 by $\mu$SR measurements [45]. As was discussed above, this unconvention-

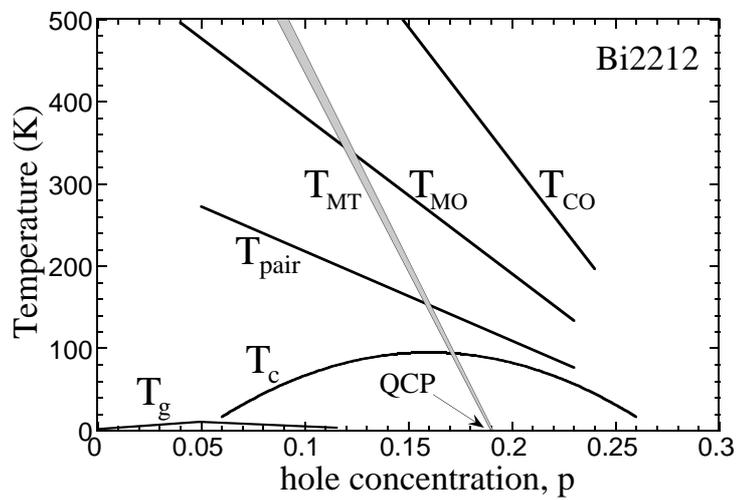

*Figure 6.11.* Phase diagram of Bi2212: $T_c$ is the critical temperature; $T_{pair}$ is the pairing temperature; $T_{MT}$ is the magnetic-transition temperature; $T_{MO}$ is the magnetic-ordering temperature; and $T_{CO}$ is the charge-ordering temperature (see text for more details) [19, 44]. The commensurate antiferromagnetic phase at low doping is not shown. The spin-glass temperature scale $T_g$ is shown schematically (QCP = quantum critical point).

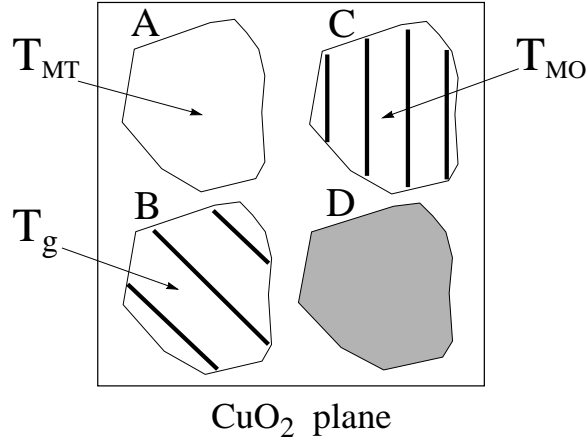

CuO$_2$ plane

*Figure 6.12.* Schematic representation of four types of clusters (phases) existing in the CuO$_2$ planes at different dopings. Cluster A includes the intact antiferromagnetic phase. Cluster B includes diagonal charge stripes and a magnetic ordering between the stripes. The stripes are schematically shown by the lines. Cluster C includes vertical charge stripes and a magnetic ordering between the stripes. In cluster D, holes are distributed homogeneously (Fermi liquid). The magnetic orderings in clusters A, B and C occur at $T_{MT}$, $T_g$ and $T_{MO}$, respectively.

al spin glass occurs in nanoscale clusters with diagonal charge stripes, shown schematically in Fig. 6.12. The $T_g$ temperature scale first rises approximately linearly as the doping level starts to increase from zero, reaches a maximum when it crosses the Néel temperature scale $T_N(p)$, and then falls to zero as the doping increases. In Bi2212, $T_g \simeq 8$ K at $p = 0.05$, and $T_g = 0$ somewhere at $p \simeq 0.16$ [45].

In Bi2212, the doping level of $p = 0.19$ is a quantum critical point which is located at absolute zero. In general, in a quantum critical point, a magnetic order is about to form or to disappear. At low temperature, superconductivity in the cuprates is the most robust at this doping level $p = 0.19$, and not at $p = 0.16$ where $T_c$ is a maximum.

In Fig. 6.11, the $T_{MT}$ temperature scale starts/ends in the quantum critical point (MT = Magnetic Transition). Then, it is more or less obvious that this temperature scale has the magnetic origin. The $T_{MT}$ temperature scale is analogous to a magnetic transition temperature of a long-range antiferromagnetic phase in heavy fermions [19]. The doping dependence $T_{MT}(p)$ can be expressed as

$$T_{MT}(p) \simeq T_{MT,0} \left[1 - \frac{p}{0.19}\right],\tag{6.1}$$

where $T_{MT,0} = 970$–990 K (see references in [19]). This temperature scale can be observed in resistivity, nuclear magnetic resonance (NMR) and specific-heat measurements, and will be discussed separately. Along the transition temperature $T_{MT}(p)$, magnetic fluctuations are very strong. The $T_{MT}$ temperature scale originates from a local antiferromagnetic ordering in hole-poor clusters (matrix) shown in Fig. 6.12. This type of clusters disappears completely at $p = 0.19$.

In Fig. 6.11, the three temperature scales, $T_{CO}$, $T_{MO}$ and $T_{pair}$, originate from nanoscale clusters with vertical charge stripes, shown schematically in Fig. 6.12. The $T_{CO}$ temperature scale corresponds to a charge-stripe ordering (CO = charge ordering). We already know that the striped phase in the cuprates is charge driven, and a structural phase transition precedes the charge ordering (see Fig. 6.5). The doping dependence $T_{CO}(p)$ can approximately be expressed as

$$T_{CO}(p) \simeq 980 \times \left[1 - \frac{p}{0.3}\right] \quad (\text{in K}). \tag{6.2}$$

The corresponding charge gap $\Delta_{cg}$, observed in tunneling and ARPES measurements, depends on hole concentration as

$$\Delta_{cg}(p) \simeq 251 \times \left[1 - \frac{p}{0.3}\right] \quad (\text{in meV}). \tag{6.3}$$

In the striped phase of cuprates shown in Fig. 6.2, the charge ordering is always followed by a magnetic ordering. In Fig. 6.11, the $T_{MO}$ temperature scale corresponds to a magnetic ordering (MO) occurring in insulating spin stripes located between the charge stripes. This is shown in Fig. 6.12. The doping dependence $T_{MO}(p)$ can be expressed as follows

$$T_{MO}(p) \simeq 566 \times \left[1 - \frac{p}{0.3}\right] \quad (\text{in K}). \tag{6.4}$$

Since the charge stripes in the $CuO_2$ planes fluctuate very rapidly, this magnetic ordering must also rearrange itself quickly. The dynamic quasi-two-dimensional spin stripes can be considered as a local memory effect of the commensurate antiferromagnetic phase located at low dopings. In the striped phase, the magnetic order occurring at $T_{MO}$ helps to stabilize the charge order.

In Fig. 6.11, the $T_{pair}$ temperature scale corresponds to the formation of Cooper pairs, the doping dependence of which can be expressed as

$$T_{pair}(p) \simeq \frac{T_{CO}(p)}{3} = \frac{980}{3} \times \left[1 - \frac{p}{0.3}\right] \simeq 326 \left[1 - \frac{p}{0.3}\right] \quad (\text{in K}). \tag{6.5}$$

The corresponding pairing energy scale $\Delta_p$ depends on the doping level as follows

$$\Delta_p(p) \simeq \frac{\Delta_{cg}(p)}{3} = \frac{251}{3} \times \left[1 - \frac{p}{0.3}\right] \simeq 83.6 \left[1 - \frac{p}{0.3}\right] \quad (\text{in meV}). \tag{6.6}$$

The pairing gap manifests itself in tunneling and ARPES measurements. It is important to emphasize that the ratio between $T_{pair}$ and $\Delta_p$, obtained experimentally, clearly indicates a strong-coupling regime of electron pairing in Bi2212:

$$2\,\Delta_p \simeq 6\,k_B\,T_{pair}. \tag{6.7}$$

The superconducting phase appears at a critical temperature $T_c$ shown in Fig. 6.11. In Bi2212 and in most of the cuprates, the doping dependence $T_c(p)$ can be expressed as

$$T_c(p) \simeq T_{c,max}[1 - 82.6(p - 0.16)^2]. \tag{6.8}$$

For Bi2212, $T_{c,max} = 95$ K. The superconducting phase is approximately located between $p = 0.05$ and 0.27, having the maximum critical temperature $T_{c,max}$ in the middle, thus at $p \simeq 0.16$. The corresponding phase-coherence energy scale $\Delta_c$ is proportional to $T_c$ as

$$2\Delta_c = \Lambda k_B T_c. \tag{6.9}$$

In different cuprates, the coefficient $\Lambda$ is slightly different: $\Lambda \simeq 5.45$ in Bi2212; $\Lambda \simeq 5.1$ in YBCO; and $\Lambda \simeq 5.9$ in Tl2201. The phase-coherence gap $\Delta_c$ manifests itself in Andreev-reflection, penetration-depth and tunneling measurements.

Let us consider a few issues directly related to the phase diagram in Fig. 6.11. First, in Bi2212, the extensions of three temperature scales $T_{CO}(p)$, $T_{MO}(p)$ and $T_{pair}(p)$ cut the concentration axis *approximately* in one point, at $p = 0.3$. Superconductivity however collapses at about $p \simeq 0.27$. Assuming that this collapse is due to a lack of long-range phase coherence, one can then understand why in Fig. 6.6, the distribution of doped holes in the $CuO_2$ plane at $0.27 < p < 0.3$ is shown inhomogeneous. Second, the two temperature scales $T_{CO}(p)$ and $T_{MT}(p)$ in Fig. 6.11 intersect the vertical axis approximately in one point, at $T \approx 980$ K. This seems logic. When, in the $CuO_2$ planes (at $p \to 0$), doped holes gather in nanoscale clusters at $T_{CO}$ to form charge stripes, the hole-poor matrix gets an opportunity to order itself magnetically, at least, locally. It is obvious that the latter process depends on the former one. At $p = 0$ they occur simultaneously (or almost simultaneously) at $T \sim 980$ K. In reality however, this point is not accessible because the melting point of Bi2212 is about 850 K. Third, in the phase diagram of Bi2212 shown in Fig. 6.11, one can see that three temperature scales have the magnetic origin, namely, $T_g(p)$, $T_{MT}(p)$ and $T_{MO}(p)$. This is not odd because these three temperature scales originate from magnetic orderings that occur in different clusters (or matrix) sketched in Fig. 6.12. The spin glass occurs at $T_g$ in the nanoscale clusters with diagonal charge stripes. The $T_{MT}$ temperature scale originates from local antiferromagnetic ordering in the hole-poor matrix which becomes divided into clusters as $p \to 0.19$. The magnetic ordering occurring in insulating spin stripes located between charge stripes sets in at $T_{MO}$. So, all these clusters and matrix are spatially separated. Four, the real doping dependences $T_{MT}(p)$, $T_{CO}(p)$, $T_{MO}(p)$ and $T_{pair}(p)$ are most likely not linear but *quasi*-linear. Five, in the literature, there is discrepancy among phase diagrams inferred from different sets of experimental data. This is because the phase diagram of every superconducting cuprate consists of several energy/temperature scales (at least, seven) which have different origins. Since different experimental techniques are sensitive to different types of correlations and have different resolutions and different characteristic times, one can then understand why there is a discrepancy among phase diagrams that can be found in the literature.

Finally, consider an interesting question: what type of clusters does the $T_c(p)$ temperature scale originate from? One may immediately suggest that the $T_c(p)$ temperature scale originates from the nanoscale clusters with vertical stripes. The answer is yes and no. As was already emphasized, the phase diagram in Fig. 6.11 is a superposition of different temperature scales originating from three types of clusters (or matrix) depicted in Fig. 6.12. The spin-glass temperature scale $T_g(p)$ has no direct relation with superconductivity; the other temperature scales in Fig. 6.11 are presented in two plots, according to their origin, as sketched in Fig. 6.13. One can see in Fig. 6.13a that the $T_c(p)$ temperature scale is attributed mainly, but not solely, to the matrix (clusters) with the local antiferromagnetic order. Why? The location of a quantum critical point at $p = 0.19$, where magnetic fluctuations are the strongest, is the reason for this. In the $CuO_2$ planes, the existence of the matrix (clusters) with local antiferromagnetic order as well as the $T_{MT}$ temperature scale is determined by the location of a quantum critical point. In a sense, the superconducting phase is "attracted" at low temperature by a quantum critical point. Of course, the charge-stripe fluctuations in clusters with vertical charge stripes generate spin excitations that mediate the long-range phase coherence. However, in order to propagate, the spin excitations must use a magnetic order, local or not. A quantum critical point provides the occurrence of magnetic order.

## 3.5    Pseudogap

The pseudogap is a depletion of the density of states above the critical temperature. Below a certain temperature, the cuprates do have a connected Fermi surface that appears to be consistent with conventional band theory. Above $T_c$, the pseudogap dominates the normal-state low-energy excitations. The pseudogap was observed for the first time in NMR measurements and therefore mistakenly interpreted as a spin gap [8]. Later, ARPES, tunneling, Raman, specific-heat and infrared measurements also provided evidence for a gap-like structure in electronic excitation spectra. Thus, it became clear that the pseudogap is not a spin gap but a gap to both spin and charge excitations; alternatively, there are two spatially separated pseudogaps: one is a spin gap, and the second is a charge gap. The magnitude of pseudogap(s) is large in the underdoped region and decreases as the doping level increases.

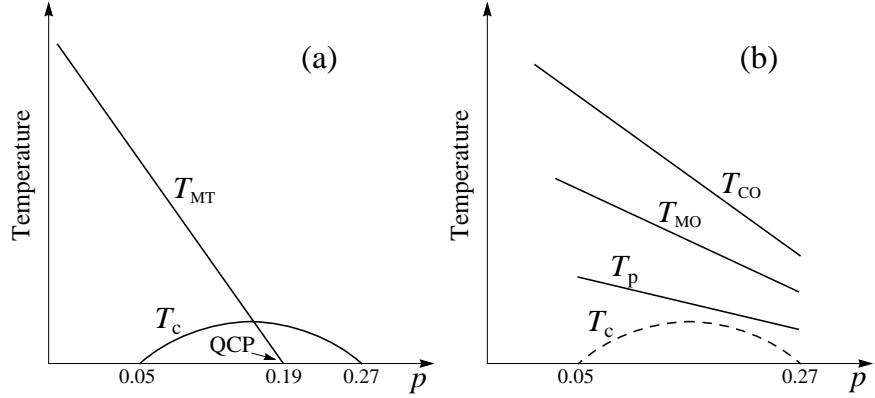

*Figure 6.13.* Temperature scales from Fig. 6.11: (a) the $T_{MT}(p)$ scale is determined by a quantum critical point (QCP) associated with antiferromagnetic phase, while (b) the $T_{CO}(p)$, $T_{MO}(p)$ and $T_{pair}(p)$ scales originate from the striped phase shown in Fig. 6.2. The $T_c(p)$ scale is a mutual temperature scale caused, first of all, by the quantum critical point and also by spin excitations present in striped-phase clusters.

So, what is the pseudogap in cuprates? There is no straight answer to this question because the answer depends upon the technique by which it was observed and upon the temperature at which it was seen. Above $T_c$, one may observe four different gaps in the cuprates: (i) below $T_{CO}$, a charge gap on the charge stripes, $\Delta_{cg}$, (ii) below $T_{MT}$, a gap having the magnetic origin, $\Delta_{NMR}$, (iii) it is possible that a spin gap $\Delta_{sg}$ sets in below $T_{MO}$ in hole-poor stripes in the striped phase shown in Fig. 6.2, and (iv) below $T_{pair}$, the pairing gap $\Delta_p$. So, generally speaking, the pseudogap in the cuprates, $\Delta_{pg}$, is

$$\Delta_{pg} = \sqrt{\Delta_{cg}^2 + \Delta_{NMR}^2 + \Delta_{sg}^2 + \Delta_p^2}. \qquad (6.10)$$

However, there is no experimental technique able to detect all of these gaps at once. The *magnitudes* of these four gaps are not affected much on cooling through $T_c$, but the pairing gap $\Delta_p$ becomes a part of the superconducting gap, $\Delta_{sc} = \sqrt{\Delta_p^2 + \Delta_c^2}$, predominant below $T_c$. For example, in the case of $\Delta_{cg}$, below $T_c$ there is a renormalization of low-energy excitations; thus, excitations inside the charge gap. Let us consider this renormalization.

### 3.5.1 Charge gap on the stripes

The pseudogap which can be observed in tunneling and ARPES measurements is a charge gap on charge stripes, manifesting itself in the form of wide humps present in the spectra at high bias. Figure 6.14a shows an ARPES spectrum obtained below $T_c$ in Bi2212. In the plot, the position of main quasiparticle peaks measures the magnitude of $\Delta_p$, while the hump at a bias of $V_{hump} \simeq 3\Delta_p/e$ is caused by the charge gap on the stripes. The ratio $V_{hump}/V_{peak} \simeq 3$, inferred from either ARPES or superconductor-insulator-normal metal (SIN) tunneling spectra, is independent of doping level. This ratio is determined by the ratio $\Delta_{cg}/\Delta_p \simeq 3$ in Eq. (6.6).

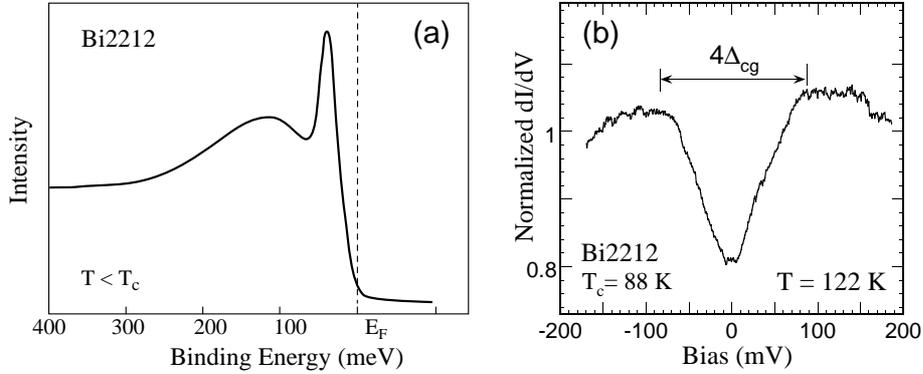

*Figure 6.14.* (a) ARPES spectrum obtained in a slightly overdoped Bi2212 single crystal with $T_c = 91$ K at $T \ll T_c$; and (b) Tunneling pseudogap in slightly overdoped Bi2212 obtained in a SIS junction at $T > T_c$ (see references in [19]).

In superconductor-insulator-superconductor (SIS) tunneling conductances, however, the bias ratio is about $V_{hump}/V_{peak} \simeq 2$ because the humps in SIS conductances do not correspond directly to $\Delta_{cg}$ but are a product of tunneling between genuine humps and quasiparticle peaks. Thus at $T \ll T_c$, tunneling $dI(V)/dV$ and $I(V)$ characteristics as well as ARPES spectra consist of two contributions caused by the superconducting condensate (bisolitons) and by the charge gap on the stripes. In SIN conductances, these two contributions are superimposed *linearly* while in SIS conductances, *nonlinearly*. This issue was discussed in detail in Chapter 12 of [19].

Figure 6.14b depicts the pseudogap in a tunneling conductance obtained by a break-junction technique in Bi2212 above $T_c$. In the measurements, the temperature of 122 K is above $T_{pair} \simeq 120$ K in slightly overdoped Bi2212 with $p \simeq 0.19$. The gap-like structure in tunneling conductances disappears upon increasing the temperature.

Consider now the charge-gap renormalization at $T_c$. Let us start with tunneling measurements. Figure 6.15a sketches the charge gap above and deep below $T_c$. Above the critical temperature, there are quasiparticle excitations at the Fermi level (at zero bias in Fig. 6.15a). As the temperature is lowered through $T_c$, quasiparticle excitations at the Fermi level, thus inside the gap, are renormalized: the charge gap deepens at low bias, as shown in Fig. 6.15a.

However, the maximum magnitude of the gap, in a first approximation, is unchanged. Analysis of ARPES data even suggests that a full gap opens up at the Fermi level, as shown in Fig. 6.15b. The magnitude of this gap is slightly smaller than the magnitude of $\Delta_p$, and its temperature dependence is reminiscent of the BCS temperature dependence (see references in [19]). If this is the case, then the charge gap at $T \ll T_c$ has an anisotropic s-wave symmetry with an anisotropy ratio of about 3 because $\Delta_{cg,min} \approx \Delta_p$ and $\Delta_{cg,max} \simeq 3\Delta_p$. Thus, the ratio is $\Delta_{cg,max}/\Delta_{cg,min} \approx 3\Delta_p/\Delta_p = 3$.

From Eqs. (6.5)–(6.7), the gap ratio $2\Delta_{cg}/(k_B T_{CO}) \simeq 6$ is sufficiently large, indicating a strong coupling between the lattice and holes. As noted above, the charge stripes represent self-trapped electronic states in the $CuO_2$ planes. In crystals, self-trapped states occur usually due to an interaction of quasiparticles with acoustic phonons. The electron-phonon interaction in the $CuO_2$ planes will be discussed separately.

Finally, it is worth to mention that in the ARPES spectrum shown in Fig. 6.14a, the dip that naturally occurs from a superposition of the quasiparticle peak and the hump **has no physical meaning**. In ARPES and SIN-tunneling spectra, this dip is situated at $e|V| \simeq 2\Delta_p(p)$, while in SIS tunneling conductances, it appears at $e|V| \approx 3\Delta_p(p)$.

### 3.5.2 $T_{MT}(p)$ temperature scale

The $T_{MT}$ temperature scale shown in Fig. 6.11 manifests itself in resistivity, NMR and specific-heat measurements. It originates from local antiferromagnetic ordering in hole-poor matrix (clusters) depicted in Fig. 6.12. Therefore, the energy scale related to the $T_{MT}$ temperature scale, $\Delta_{NMR}$, has also the magnetic origin. Figure 6.16 illustrates how the temperature $T_{MT}$ is determined in resistivity measurements: $T_{MT}$ is the temperature at which, upon cooling, the resistivity deviates from a linear temperature dependence.

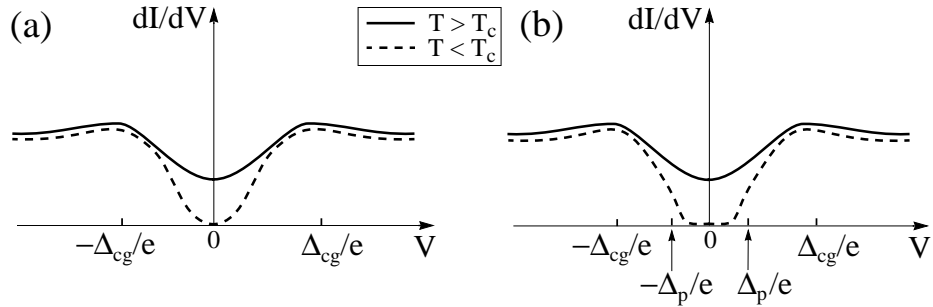

*Figure 6.15.* Gap on the charge stripes above and deep below $T_c$, inferred from (a) tunneling and (b) ARPES measurements (see references in [19]). In both plots, the solid lines are slightly shifted up for clarity.

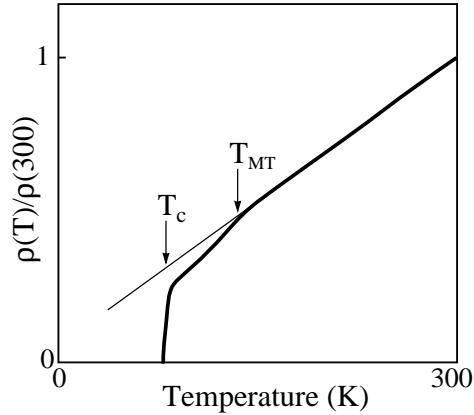

*Figure 6.16.* Pseudogap in resistivity measurements: $T_{MT}$ is the temperature at which the resistivity deviates on cooling from a linear dependence.

Why does the resistivity drop at $T_{MT}$? The temperature $T_{MT}$ is the onset temperature of *local* antiferromagnetic order in either the insulating matrix, if $p \ll 0.19$, or insulating clusters, if $p \rightarrow 0.19$. In nanoscale clusters with charge stripes, holes can move easily along the stripes. To sustain electrical current on a macroscopic scale, quasiparticles must tunnel through the insulating matrix (clusters). It is mistakenly believed that an antiferromagnetic ordering blocks electron transport. The opposite is true: electrons can tunnel through a thin layer of $NiO_2$ ordered antiferromagnetically about five times easier than through a thin layer of non-magnetic metal oxides [46]. This is why the temperature dependence of in-plane resistivity in the cuprates has a kink at $T_{MT}$: the resistivity simply drops at $T_{MT}$.

### 3.6    Soliton-like excitations on charge stripes

Depending on temperature, quasiparticles in superconducting cuprates are soliton-like excitations or pairs of these excitations, observed experimentally in tunneling measurements [16–19]. In the $CuO_2$ planes, the soliton-like excitations appear on the charge stripes immediately after the formation of the stripes, thus, somewhat below $T_{CO}$. They become paired at $T_{pair}$ shown in the phase diagram in Fig. 6.11. Consider first the elementary excitations; the pairs will be discussed in the following subsection.

In the cuprates the soliton-like excitations propagate in the middle of the charge gap present on the stripes, as depicted in Fig. 6.17. Since these excitations have the quantum numbers of an electron, they are fermions. Then, a question naturally rises: How can the fermions occupy the same energy level? There are two possible explanations to this fact: Either there is the spin-charge separation on the stripes or, between $T_{CO}$ and $T_{pair}$, there is no connected Fermi surface in the cuprates yet. Theoretically the spin-charge separation is indeed possible in one-dimensional systems, leading to the formation of the so-called Luttinger liquid. However, an experimental verification of this hypothesis, carried out in underdoped YBCO, showed that, within the resolution limits of the experiment, there is no spin-charge separation in YBCO [47]. Hence, it is most likely that, at $T_{pair} < T < T_{CO}$, there is no connected Fermi surface in the cuprates. This means that, at high temperatures, the charge stripes in the $CuO_2$ planes exist independently of each other. For example, in LSCO, upon cooling the chemical potential literally jumps at about 250 K. Thus, if this is the case, it is then possible that the soliton-like excitations become paired at $T_{pair}$ because they are required to. When the Fermi surface in the cuprates becomes connected somewhat above $T_{pair}$, the soliton-like excitations being at the same energy level must either disappear, except for two, or be paired. The paired quasiparticles are bosons, therefore, the Pauli exclusion principle is no longer applicable to them.

What type of excitations can occur on the charge stripes in the cuprates? Let us consider all possible excitations on a $2k_F$ charge stripe sketched in Fig. 6.9. In the case of hole-doped cuprates, we are exclusively interested in the stripe excitations with a charge of $+|e|$, where $e$ is the electron charge. Figure 6.18 shows schematically three types of excitations on the $2k_F$ charge stripes: a soliton, a kink-up and a kink-down. In Fig. 6.18, the stripe excitations are shown at rest; however, in reality, they are dynamic and propagate along the stripes which themselves fluctuate in the $CuO_2$ planes. As suggested by Za-anen and co-workers, the kink-up and the kink-down may have different spin orientations.

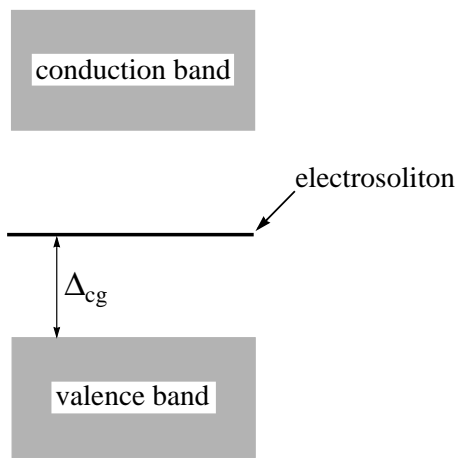

*Figure 6.17.* Energy level of soliton-like excitations that reside on charge stripes in the $CuO_2$ planes at high temperatures. $\Delta_{cg}$ is the magnitude of the charge gap on the stripes.

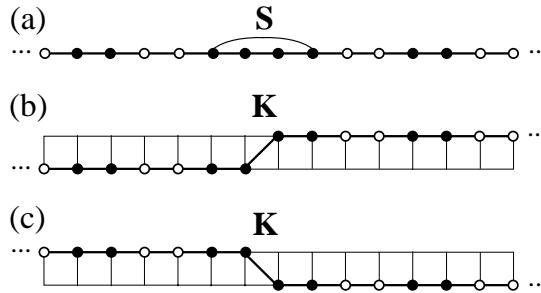

*Figure 6.18.* Sketch of three types of excitations on $2k_F$ charge stripes, shown at rest: (a) soliton; (b) kink-up, and (c) kink-down.

Such a description of soliton-like excitations in the cuprates is rather phenomenological. In a first approximation, they can also be described mathematically in the framework of the Davydov theory of electrosolitons and bisolitons for one-dimensional systems [3, 7, 9, 10]. Let us consider briefly the electrosoliton model. As discussed in Chapter 1, solitons are nonlinear excitations which are localized in space. In some systems, they represent *self-localized* states (i.e. self-trapped states). In solids, if the particle-field interaction (electron-phonon coupling) is strong, both the particle wavefunction and the lattice deformation will be localized. In the three-dimensional case, this localized entity is known as a (Holstein) polaron and, in one dimension, as a Davydov soliton (also known as an electrosoliton or polaronic soliton). Their integrity is maintained owing to a dynamical balance between dispersion (exchange inter-site interaction) and nonlinearity (electron-phonon coupling). Figure 6.19 schematically shows a self-trapped state of a particle (small polaron or Davydov soliton). In a self-trapped state, *both* the particle wavefunction and the lattice deformation are *localized*.

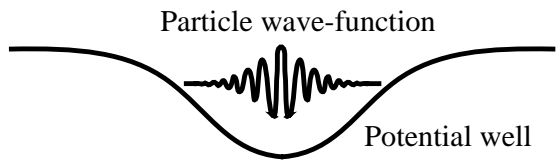

*Figure 6.19.* Sketch of a self-trapped state: a particle (electron or hole) with a given wavefunction deforms the lattice inducing a potential well which in turn traps the particle.

To describe the self-focusing phenomena, the nonlinear Schrödinger (NLS) equation is used,

$$\left[ i\hbar \frac{\partial}{\partial t} + \frac{\hbar^2}{2m} \frac{\partial^2}{\partial x^2} + G|\psi(x,t)|^2 \right] \psi(x,t) = 0. \qquad (6.11)$$

The equation is written in the long-wave approximation when the excitation wavelength $\lambda$ is much larger than the characteristic dimension of discreteness in the system, i.e. under the condition $ka = 2\pi a/\lambda \ll 1$. The equation describes the complex field $\psi(x,t)$ with self-interaction. The function $|\psi|^2$ determines the position of a quasiparticle of mass $m$. The second term in the NLS equation is responsible for dispersion, and the third one for nonlinearity. The coefficient $G$ characterizes the intensity of nonlinearity.

When the nonlinearity is absent ($G = 0$), the NLS equation has solutions in the form of plane waves,

$$\psi(x,t) = \Phi_0 \exp[i(ka - \omega(k)t)], \qquad (6.12)$$

with the square dispersion law $\omega(k) = \hbar k^2/2m$. With nonlinearity ($G \neq 0$) in the system having a translational invariance, the excited states move with constant velocity $v$. Therefore, it is convenient to study solutions of the NLS equation in the reference frame

$$\zeta = (x - vt)/a, \qquad (6.13)$$

moving with constant velocity. In this reference frame the NLS equation has solutions in the form of a complex function

$$\psi(x,t) = \Phi(\zeta) \exp[i(kx - \omega t)], \;\; k = mv/\hbar, \qquad (6.14)$$

where the function $\Phi(\zeta)$ is real.

The self-trapping effect occurs when two linear systems interact with one another. Consider an excess electron in a quasi-one-dimensional atomic (molecular) chain. If neutral atoms (molecules) are rigidly fixed in periodically arranged sites $na$ of a one-dimensional chain, then due to the translational invariance of the system, the lowest energy states of an excess electron are determined by the conduction band. The latter is caused by the electron collectivization. In the continuum approximation, the influence of a periodic potential is taken into account by replacing the electron mass $m_e$ by the effective mass $m = \hbar^2/(2a^2J)$ which is inversely proportional to the exchange interaction energy $J$ that characterizes the electron jump from one node site into another. In this approximation, the electron motion along an ideal chain corresponds to the free motion of a quasiparticle with an effective mass $m$ and the electron charge.

Taking into account small displacements of molecules of mass $M$ ($\gg m$) from their periodic equilibrium positions, there arises the short-range deformation interaction of quasiparticles with these displacements. When the deformation interaction is rather strong, the quasiparticle is self-localized. The local displacement caused by a quasiparticle manifests itself as a potential well that contains the particle, as schematically shown in Fig. 6.19. In turn, the quasiparticle deepens the well.

A self-trapped state can be described by two coupled differential equations for the field $\psi(x,t)$ that determines the position of a quasiparticle, and the field $\rho(x,t)$ that characterizes a local deformation of the chain and determines the decrease in the relative distance $a \rightarrow a - \rho(x,t)$ between molecules of the chain,

$$\left[ i\hbar \frac{\partial}{\partial t} + \frac{\hbar^2}{2m} \frac{\partial^2}{\partial x^2} + \sigma\rho(x,t) \right] \psi(x,t) = 0, \qquad (6.15)$$

$$\left( \frac{\partial^2}{\partial t^2} - c_0^2 \frac{\partial^2}{\partial x^2} \right) \rho(x,t) - \frac{a^2\sigma}{M} \frac{\partial^2}{\partial x^2} |\psi(x,t)|^2 = 0. \qquad (6.16)$$

The first equation characterizes the motion of a quasiparticle in the local deformation potential $U = -\sigma\rho(x,t)$. The second equation determines the field of a local deformation caused by a quasiparticle. The two equations are connected through the parameter $\sigma$ of the interaction between a quasiparticle and a local deformation. The velocity $c_0 = a\sqrt{k/M}$ is the longitudinal sound velocity in the chain with elasticity coefficient $k$. In the case of one quasiparticle in the chain, the function $\psi(x,t)$ is normalized by

$$\frac{1}{a} \int\limits_{-\infty}^{\infty} |\psi(x,t)|^2 dx = 1. \qquad (6.17)$$

In the reference frame $\zeta = (x - vt)/a$ moving with constant velocity $v$, the following equality $\partial\rho(x,t)/\partial t = -v/a \times \partial\rho/\partial\zeta$ holds. Then, the solution for $\rho(x,t)$ has the form

$$\rho(x,t) = \frac{\sigma}{k(1-s^2)} |\psi(x,t)|^2, \text{ if } s^2 = v^2/c_0^2 \ll 1. \qquad (6.18)$$

Substituting the expression for $\rho(x,t)$ into Eq. (6.15), we obtain the following nonlinear equation for the function $\psi(x,t)$,

$$\left[ i\hbar \frac{\partial}{\partial t} + \frac{\hbar^2}{2m} \frac{\partial^2}{\partial x^2} + 2\mathrm{g}J|\psi(x,t)|^2 \right] \psi(x,t) = 0. \qquad (6.19)$$

where

$$\mathrm{g} \equiv \frac{\sigma^2}{2k(1-s^2)J} \qquad (6.20)$$

is the dimensionless parameter of the interaction of a quasiparticle with local deformation. Substituting the function

$$\psi(x,t) = \Phi(\zeta)\exp[i(kx - \omega t)], \ \ k = mv/\hbar \qquad (6.21)$$

into Eq. (6.19) we get the equation

$$[\hbar\omega - \frac{1}{2}mv^2 - J\frac{\partial^2}{\partial\zeta^2} + 2\mathsf{g}J\Phi^2(\zeta)]\Phi(\zeta) = 0, \qquad (6.22)$$

for the amplitude function $\Phi(\zeta)$ normalized by $\int \Phi^2(\zeta)d\zeta = 1$. The solution of this equation is

$$\Phi(\zeta) = \frac{1}{2}\sqrt{\mathsf{g}} \times \mathrm{sech}(\mathsf{g}\zeta/2), \qquad (6.23)$$

with the dispersion law

$$\hbar\omega = \frac{1}{2}mv^2 - \frac{1}{8}\mathsf{g}^2 J. \qquad (6.24)$$

The last term in Eq. (6.24) determines the binding energy of the particle and the chain deformation produced by the particle itself. According to Eq. (6.23), the quasiparticle is localized in a moving reference frame

$$\Delta\zeta = 2\pi/\mathsf{g}, \qquad (6.25)$$

as shown in Fig. 6.20. In this region, the field localization is characterized by the function

$$\rho(\zeta) = \frac{\mathsf{g}\sigma}{4k(1 - s^2)}\,\mathrm{sech}^2(\mathsf{g}\zeta/2). \qquad (6.26)$$

The following energy is necessary for the deformation:

$$W = \frac{1}{2}k(1 + s^2)\int \rho^2(\zeta)d\zeta = \frac{1}{24}\mathsf{g}^2 J(1 + s^2). \qquad (6.27)$$

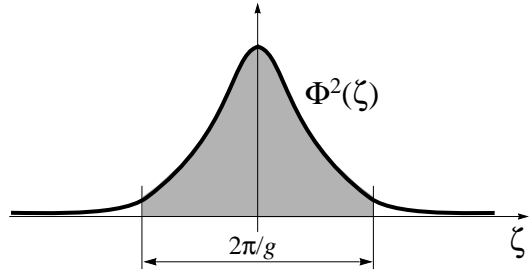

*Figure 6.20.* Size of a self-trapped soliton is about $2\pi/\mathsf{g}$ (compare with Fig. 2.4).

Measured from the bottom of the conduction band of a free quasiparticle, the total energy (including that of the deformation) transferred by a soliton moving with velocity $v$ is determined by the expression

$$E_s(V) = W + \hbar\omega = E_s(0) + \frac{1}{2}M_{sol}v^2 \tag{6.28}$$

in which the energy of a soliton at rest $E_s(0)$ and its effective mass $M_{sol}$ are determined, respectively, by the equalities

$$E_s(0) = \frac{1}{12}\mathrm{g}^2 J, \tag{6.29}$$

$$M_{sol} = m\,(1 + \frac{\mathrm{g}^2 J}{3a^2 k}). \tag{6.30}$$

The soliton mass $M_{sol}$ exceeds the effective mass of a quasiparticle $m$ if its motion is accompanied by the motion of local deformation.

The effective potential well where the quasiparticle is placed is determined in the reference frame $\zeta$ by the expression

$$U = -\sigma\rho(x,t) = -\mathrm{g}^2 J \operatorname{sech}^2(\mathrm{g}\zeta/2). \tag{6.31}$$

The self-trapped soliton is very stable. It moves with a velocity $v < c_0$; otherwise, the local deformation of the chain will be not able to follow the quasiparticle. Alternatively, the soliton can be stationary. It is worth to emphasize that the Davydov soliton is conceptually different from the small polaron which is three-dimensional and practically at rest because of its large mass.

The Davydov solitons belong to a large group of solitons the motion and transformations of which are described by the NLS equation. They are called the *envelope solitons*. As noted above, the NLS equation describes self-focusing phenomena, and the term $|\psi|^2$ in the NLS equation brings into the system the self-interaction. The second term of the NLS equation is responsible for dispersion, while the third term is responsible for nonlinearity [see, for example, Eq. (6.11)]. A solution of the NLS equation in real space is schematically shown in Fig. 6.21a. The shape of the enveloping curve (the dashed line in Fig. 6.21a) is the function $\Phi(x,t) = \Phi_0 \operatorname{sech}[(x-vt)/\ell]$, where $2\ell$ determines the width of the soliton. Its amplitude $\Phi_0$ depends on $\ell$ but *independent* of the soliton velocity $v$ [see, for example, Eq. (6.23)]. The envelope solitons can be regarded as particles but they are "mortal" (see Chapter 1). The interaction between two envelope solitons is similar to the interaction between two particles—they collide as tennis balls (see Chapter 5 in [19]).

In the envelope soliton in Fig. 6.21a, the stable groups have normally from 14 to 20 humps under the envelope, the central one being the highest one. The groups with more humps are unstable and break up into smaller ones. The waves inside the envelope move with a velocity that differs from the velocity of the soliton; thus, the envelope soliton has an *internal* dynamics. The relative motion of the envelope and the carrier wave is responsible for the internal dynamics of a NLS soliton. In momentum space, the envelope soliton is also enveloped by the "sech" hyperbolic function, as schematically shown in Fig. 6.21b. This is because the Fourier transform of the sech($\pi x$) function yields sech($\pi k$).

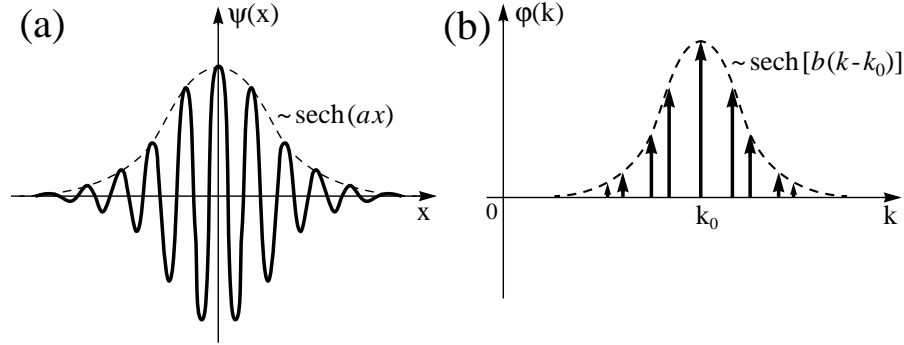

*Figure 6.21.* An envelope soliton (a) in real space, and (b) its spectrum, shown schematically.

Consider the last question related to the Davydov solitons. What can we expect from tunneling measurements carried out in a system with the Davydov solitons?

As discussed in Chapter 2, conductances $dI(V)/dV$ obtained in a SIN junction by tunneling measurements directly relate to the electron density of states per unit energy interval in a system (voltage $V$ multiplied by $e$ represents energy). The solitons present in a system will manifest themselves through the appearance of a peak in conductance. Since the internal dynamics in an envelope soliton is extremely fast, tunneling measurements can only provide information about the enveloping function. Let us present the enveloping curve of an envelope soliton in momentum space, sketched in Fig. 6.21b, as $\varphi(k) = C_0 \times$ sech $[d(k - k_0)]$, where $C_0$ and $d_0$ are constants [in fact, $C_0 = f(d_0)$]. Then, the function $|\varphi(k)|^2 = |\varphi(k - k_0)|^2 = |\varphi(\Delta k)|^2$ represents the density of states around $k_0$.

In the case of the cuprates, $k_0 = k_F$, where $k_F$ is the wave number at the Fermi level. Set zero energy level at the Fermi level, $E_F$. Since $E \propto k^2$, then $\varepsilon = E - E_F \propto k^2 - k_F^2 = (k - k_F)(k + k_F) \simeq 2k_F \Delta k$ for $\Delta k \ll k_F$. Thus, $\Delta k \propto \varepsilon/k_F$; substituting this expression into $|\varphi(\Delta k)|^2 \propto$ sech$^2(d_0 \Delta k)$, we obtain that $|\varphi(\varepsilon)|^2 \propto$ sech$^2(d \cdot \varepsilon)$, where $d \propto d_0/2k_F$. This means that, in tunneling measurements, the Davydov solitons will cause a peak centered at zero bias having the following form sech$^2(d'V)$, where in a first approximation, $d'$ is a constant. Since the amplitude $C_0$ is independent of the soliton velocity, this result is valid for any $v < c_0$, where $c_0$ is the longitudinal sound velocity.

To summarize, in SIN tunneling measurements performed in a system with the Davydov solitons, one should expect to observe a peak in conductance having the following shape

$$\frac{dI(V)}{dV} = A \times \text{sech}^2(V/V_0) \tag{6.32}$$

where $V$ is the applied bias, and $A$ and $V_0$ are constants. Bearing in mind that tunneling current is the sum under the conductance curve, $I(V) = \int \frac{dI(V)}{dV} dV + C$, where $C$ is a constant defined by the condition $I(V = 0) = 0$, the $I(V)$ characteristic centered at zero bias will have the following shape

$$I(V) = I_0 \times \tanh(V/V_0), \tag{6.33}$$

where $I_0$ is a constant. In these equations, $V_0$ determines the width of the conductance peak. One must however realize that the conductance peak corresponding to the solitonic states will appear in the background caused by other electronic states present in the system. As an example, Figure 6.22a depicts the conductance peak caused by the solitonic states; the corresponding $I(V)$ characteristic is sketched in Fig. 6.22b.

It is worth to mention that the soliton-like excitations are observed not only in the cuprates [19] but also in the manganites [19] and NbSe$_3$ [48].

## 3.7    Cooper pairs

Quasiparticles in superconducting cuprates below $T_{pair}$ are pairs of soliton-like excitations—bisolitons—observed experimentally in tunneling measure-ments [16–19]. The bisolitons are the Cooper pairs in the cuprates. The moderately strong and nonlinear electron-phonon interaction is responsible for coupling of the soliton-like excitations. In the $CuO_2$ planes, the bisolitons reside on the charge stripes. At temperatures somewhat above $T_c$, the bisolitons propagate at an energy level which is below the Fermi level by $\Delta_p$, as depicted in Fig. 6.23.

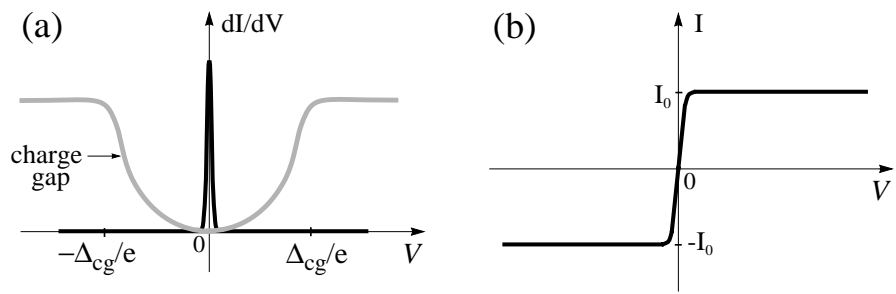

*Figure 6.22.*   (a) Sketch of a conductance peak caused by electrosolitons. The plot is adapted for the cuprates where the solitons propagate in the middle of a charge gap shown in grey. The height of the soliton peak depends on the density of added or removed electrons. (b) $I(V)$ characteristic corresponding to the soliton peak in plot (a). The $I(V)$ characteristic of the charge gap is not shown.

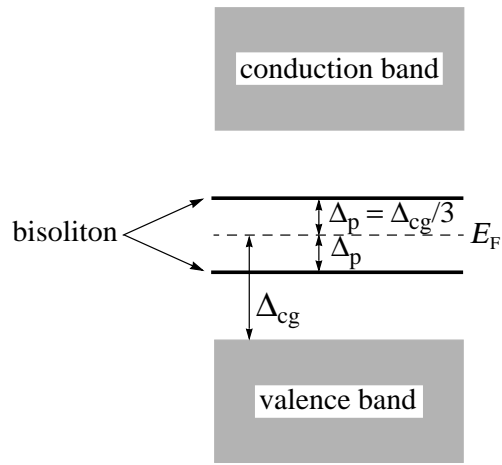

*Figure 6.23.* Energy levels of bisolitons at temperatures somewhat above $T_c$. The level below the Fermi level $E_F$ is the bonding bisoliton energy level, and that above $E_F$ is the antibonding one. The magnitude of the pairing gap is about one third of the magnitude of the charge gap, $\Delta_p \simeq \Delta_{cg}/3$.

What are the bisolitons in the cuprates? Figure 6.18 sketches three types of possible excitations on the $2k_F$ charge stripes present in the $CuO_2$ planes of the cuprates. For these three types of excitations, there are four *different* combinations of the pairing: Figure 6.24 schematically shows these four combinations. How can the stripe excitations shown in Fig. 6.18 form the pairs if they repel each other? Indeed, electrosolitons repel each other; however, in conventional superconductors, two electrons forming a Cooper pair also repel each other. The occurrence of an attractive potential between quasiparticles is central to the superconducting state. In the cuprates, a local lattice deformation is responsible for the pairing.

The combination (d) in Fig. 6.24 is asymmetrical and, therefore, seems less likely to be the case realized in the cuprates. For example, the only combination suitable for the chains in YBCO, as well as for quasi-one-dimensional organic superconductors, is the combination (a) in Fig. 6.24. Since superconductivity on the chains in YBCO is induced, it is most likely that superconductivity in the $CuO_2$ planes is caused by different stripe excitations; hence, either by (b) or (c) in Fig. 6.24. Indeed, the kink excitations have already been

<u>2k$_F$ stripes</u>

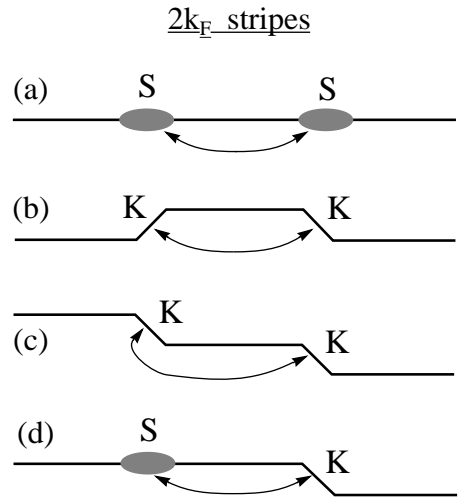

*Figure 6.24.*    Sketch of possible coupling between charge-stripe excitations on $2k_F$ stripes (see Fig. 6.18): (a) two solitons; (b) kink-up and kink-down; (c) two kinks (up or down); and (d) soliton and kink (up or down). The arrows schematically show the coupling.

observed by INS measurements in LSCO. Nevertheless, taking into account that our understanding of superconductivity in the cuprates on the *nanoscale* is still limited, none of these four combinations in Fig. 6.24 can be excluded.

Assuming that one of these four combinations in Fig. 6.24 can model the reality, the next two questions are: Do the stripe excitations couple in momentum space, and do they have the opposite momenta? In conventional superconductors, two electrons couple in momentum space, and the pairing is the most favorable when they have the opposite momenta. Is this the case for the cuprates too? It is difficult to say. If the pairing in the cuprates occurs in momentum space, then the stripe excitations on the same stripe, shown in Fig. 6.24, should have opposite momenta and thus move in the opposite directions. At the same time, as discussed in Chapter 4, the pairing in unconventional superconductors may occur not only in momentum space but also in real space. In this case, the stripe excitations on the same stripe in Fig. 6.24 can move in the same direction. Analysis of infrared measurements performed in Bi2212 indeed suggests that quasiparticles in the Cooper pairs seem to move in the same direction.

As discussed in Chapter 5, the electron pairing in conventional superconductors is local in space but non-local in time (see Fig. 5.5). Since the size of bisolitons in the cuprates can be as small as 15–20 Å, it is possible that the pairing in the cuprates is local in time but non-local in space, thus, opposite to that in conventional superconductors. It is worth noting that the pairing in real space is local in time and non-local in space.

Let us now discuss briefly the Davydov bisoliton model [3, 7, 9, 10]. Initially, the bisoliton model did not have any relation with superconductivity and was developed by Davydov and co-workers in order to explain electron transfer in living tissues. Later, the bisoliton model of electron transfer in a molecular chain was used to explain the phenomenon of superconductivity in quasi-one-dimensional organic compounds and cuprates. The bisoliton theory is based on the concept of bisolitons—electron pairs coupled in a singlet state due to a local deformation of the lattice.

If a quasi-one-dimensional soft chain is able to keep some excess electrons with charge $e$ and spin 1/2, they can be paired in a singlet state due to the interaction with a local chain deformation created by them. The potential well formed by a short-range deformation interaction of one electron attracts another electron which, in turn, deepens the well.

Assume that along a molecular chain, the elementary cells of mass $M$ are separated by a distance $a$ from one other. Within the continuum approach, we characterize their position by a continuous variable $x = na$. The equation of motion of two quasiparticles with effective mass $m$ in the potential field

$$U(x,t) = -\sigma\rho(x,t), \tag{6.34}$$

created by a local deformation $\rho(x,t)$ of the infinite chain, takes the form

$$\left[i\hbar\frac{\partial}{\partial t} + \frac{\hbar^2}{2m}\frac{\partial^2}{\partial x_i^2} + U(x_i,t)\right]\psi_j(x_i,t) = 0, \text{ where } i,j = 1,2, \tag{6.35}$$

and $\psi_j(x_i,t)$ is the coordinate function of quasiparticle $i$ in the spin state $j$.

The local deformation $\rho(x,t)$ is caused by two quasiparticles due to their interaction with displacements from equilibrium positions. The function $\rho(x,t)$ characterizing this local deformation of the chain is determined by the equation

$$\left(\frac{\partial^2}{\partial t^2} - c_0^2\frac{\partial^2}{\partial x^2}\right)\rho(x,t) + \frac{\sigma a^2}{M}\frac{\partial^2}{\partial x^2}\left(|\psi_1(x,t)|^2 + |\psi_2(x,t)|^2\right) = 0, \tag{6.36}$$

where $c_0 = a\sqrt{k/M}$ is the longitudinal sound velocity in the chain; $a\sigma$ is the energy of deformation interaction of quasiparticles with the chain, and $k$ is the coefficient of longitudinal elasticity.

Due to the translational symmetry of an infinite chain, it is possible to study the excitations propagating along the chain with a constant velocity $v < c_0$. In this case, the forced solution of Eq. (6.36) which satisfies the condition $\rho(x,t) \neq 0$ for all $x$ and $t$ under which $|\psi_j(x,t)|^2 \neq 0$ relative to the reference frame $\zeta = (x - vt)/a$ moving with velocity $v$, has the form

$$\rho(\zeta) = \frac{\sigma}{k(1-s^2)}\left[|\psi_1(\zeta)|^2 + |\psi_2(\zeta)|^2\right], \text{ where } s^2 \equiv v^2/c_0^2. \tag{6.37}$$

The total energy of local deformation of the chain is determined by

$$W = \frac{1}{2}k(1+s^2)\int \rho^2(\zeta)d\zeta. \qquad (6.38)$$

Substituting first $\rho(\zeta)$ into Eq. (6.35) and the latter into Eq. (6.36), one obtains the equation for the function $\Psi(x_1, x_2, t)$ which determines the motion of a pair of quasiparticles in the potential field given by Eq. (6.34), disregarding the Coulomb repulsion of electrons:

$$\left[ i\hbar\frac{\partial}{\partial t} + J\left(\frac{\partial^2}{\partial \zeta_1^2} + \frac{\partial^2}{\partial \zeta_2^2}\right) + G(|\psi_1(\zeta_1)|^2 + |\psi_1(\zeta_2)|^2 \right.$$
$$\left. + |\psi_2(\zeta_1)|^2 + |\psi_2(\zeta_2)|^2)\right]\Psi(x_1, x_2, t) = 0, \qquad (6.39)$$

where we introduced the following notations

$$J \equiv \frac{\hbar^2}{2ma^2} \text{ and } G \equiv \frac{\sigma^2}{k(1-s^2)}. \qquad (6.40)$$

We consider further the states with a small velocity of motion, $s^2 \ll 1$. In this case, the parameter $G$ can be replaced by the constant

$$G_0 \simeq \frac{\sigma^2}{k}. \qquad (6.41)$$

The coordinate function of a pair of quasiparticles in a singlet spin state is symmetric and can be written in the form

$$\Psi(x_1, x_2, t) = \frac{1}{\sqrt{2}}\left[\psi_1(\zeta_1)\psi_2(\zeta_2) + \psi_1(\zeta_2)\psi_2(\zeta_1)\right]e^{-iE_p t/\hbar}, \qquad (6.42)$$

where $E_p$ is the energy of two paired quasiparticles in the potential field $U(\zeta)$.

In a chain consisting of a large number $N$ of elementary cells and containing $N_1$ pairs of quasiparticles, the pairing is realized only from those states of free quasiparticles which have a wave number close to the wave number of the Fermi surface $k_F = \pi N_1/2aN$. Due to the conservation law of quasi-momentum, a pair moving with a velocity $v = \hbar k/m$ can be formed from two quasiparticles with the wave numbers

$$k_1 = 2k - k_F \text{ and } k_2 = k_F. \qquad (6.43)$$

If the wave functions of quasiparticles in the paired state are represented by the modulated plane waves

$$\psi_j(\zeta_i) = \Phi(\zeta_i)\exp(ik_j\zeta_i), \text{ where } i, j = 1, 2, \qquad (6.44)$$

the coordinate function of paired quasiparticles transforms into the following form

$$\Psi(x_1, x_2, t) = \sqrt{2}\Phi(\zeta_1)\Phi(\zeta_2)\cos[(k - k_F)(\zeta_1 - \zeta_2)] \times e^{i[k(\zeta_1 + \zeta_2) - E_p t/\hbar]}. \tag{6.45}$$

The appearance here of the cosine function results from the conditions of symmetry imposed.

Substituting the function $\Psi(x_1, x_2, t)$ into Eq. (6.39), one obtains the equation for the amplitude functions $\Phi(\zeta_i)$

$$\left[\frac{\partial^2}{\partial \zeta_i^2} + 4\mathbf{g}\Phi^2(\zeta_i) + \Lambda\right]\Phi(\zeta_i) = 0, \text{ where } i = 1, 2, \tag{6.46}$$

with the dimensionless parameter

$$\mathbf{g} \equiv \frac{G}{2J} = \frac{\sigma^2}{2kJ(1 - s^2)} \approx \frac{G_0}{2J} = \frac{\sigma^2}{2kJ} \text{ for } s^2 \ll 1, \tag{6.47}$$

that characterizes the coupling of a quasiparticle with the deformation field. The energy $E(v)$ of a pair of quasiparticles in this field is expressed in terms of the eigenvalue $\Lambda$ given by Eq. (6.46) via the relation

$$E(v) = E_p(0) + 2\frac{mv^2}{2} - \hbar v k_F, \tag{6.48}$$

where

$$E_p(0) = \Lambda J + E_F \tag{6.49}$$

characterizes the position of the energy level of a static pair of quasiparticles beneath their Fermi level

$$E_F = \frac{\hbar^2 k_F^2}{m}. \tag{6.50}$$

The deformation field $\rho(\zeta)$ is expressed in terms of the function $\Phi(\zeta)$ as follows

$$\rho(\zeta) = -\frac{2\sigma}{k(1 - s^2)}\Phi^2(\zeta). \tag{6.51}$$

Therefore, the energy of the local chain deformation in Eq. (6.38) is defined by

$$W = \frac{2G(1 + s^2)}{1 - s^2}\int\Phi^4(\zeta)d\zeta. \tag{6.52}$$

Equation (6.46) admits periodic solutions corresponding to the uniform distribution of a pair of quasiparticles over the chain. The real functions $\Phi(\zeta_i)$ of these solutions must satisfy the conditions of periodicity

$$\Phi(\zeta_i) = \Phi(\zeta_i + L), \text{ where } L = N/N_1, \tag{6.53}$$

and normalization

$$\int\limits_{0}^{L} \Phi^2(\zeta)d\zeta = 1. \tag{6.54}$$

The latter requires that each pair of quasiparticles should be within each period. The exact periodic solutions of Eq. (6.46) is expressed in terms of the Jacobian elliptic function $\mathrm{dn}(u, q)$ via the relation

$$\Phi_q(\zeta_i) = \sqrt{\frac{\mathsf{g}}{2}} E^{-1}(q) \times \mathrm{dn}(u, q), \ \ \text{where} \ \ u = \mathsf{g}\zeta/E(q). \tag{6.55}$$

An explicit form of the Jacobian function $\mathrm{dn}(u, q)$ depends on a specific value of the modulus $q$ taking continuous values in the interval $[0, 1]$. The eigenvalue $\Lambda$ of Eq. (6.46) is also given in terms of the modulus $q$ by the relation

$$\Lambda_q = -\mathsf{g}^2 q^2 / E(q). \tag{6.56}$$

The function $E(q)$ is a complete elliptic integral of the second kind which depends on the Jacobian function $\mathrm{dn}(u, q)$ and the complete elliptic integral of the first kind, $K(q)$.

We shall not follow further the exact calculations in the framework of the bisoliton theory. They can be found elsewhere [7, 9, 10, 19]. Instead, we consider the asymptotic case: the small density of quasiparticles, i.e. when the inequality $\mathsf{g}L \gg 1$ holds, where $\mathsf{g}$ is the dimensionless parameter which characterizes the coupling of a quasiparticle with the chain, and $L$ is the dimensionless distance between two bisolitons. In this case, the mass of a static bisoliton is

$$M_{bs} \simeq 2m + \frac{8\mathsf{g}^2 J}{3c_0^2}, \tag{6.57}$$

which exceeds two effective masses of quasiparticles, $2m$. In this expression, $J$ is the exchange interaction energy, and $c_0$ is the longitudinal sound velocity in the chain.

The energy gap in the quasiparticle spectrum resulting from a pairing is determined by

$$\Delta \simeq \frac{1}{3}\mathsf{g}^2 J. \tag{6.58}$$

The energy gap is half of the energy of formation of a static bisoliton. The bisoliton-formation energy includes not only the energy of quasiparticle pairing but also the energy of the formation of a local chain deformation. The magnitude of the energy gap decreases as the density of quasiparticles increases (when $\mathsf{g}L < 5$). The energy gap $\Delta$ and the bisoliton mass $M_{bs}$ do not depend on the mass $M$ of an elementary cell. This mass only appears in the kinetic energy of bisolitons. Therefore the isotope effect is very small, notwithstanding the fact that the basis of the pairing is the electron-phonon interaction.

The correlation length in a bisoliton (the size of a bisoliton) is given by

$$d = \frac{2\pi a}{\mathsf{g}}, \tag{6.59}$$

where $a$ is the lattice constant. The enveloping wave functions $\Phi_i(x)$ of quasiparticles within each period are approximated by the hyperbolic function

$$\Phi(x) = \sqrt{\frac{\mathsf{g}}{2}} \times \operatorname{sech}(\mathsf{g}x), \tag{6.60}$$

where $x$ is the axis along the chain. In this expression, one can see that the maximum amplitude of a bisoliton ($= \mathsf{g}/2$) is two times larger than that of an electrosoliton in Eq. (6.23), and its width is two times smaller than that of an electrosoliton.

When there is only one bisoliton in the chain, two quasiparticles in the bisoliton move in the combined effective potential well

$$U_{\uparrow\downarrow}(\zeta) = -2\mathsf{g}^2 J \times \operatorname{sech}^2(\mathsf{g}\zeta). \tag{6.61}$$

The radius of this well is a half of the radius of an isolated soliton, and its depth is twice larger than that of an electrosoliton [see Eq. (6.31)].

All these results were obtained without taking into account the Coulomb repulsion between quasiparticles. If we take into account the Coulomb repulsion as a perturbation, then, at small velocities, a pairing is still energetically profitable if the dimensionless coupling constant $\mathsf{g}$ is greater than some critical value,

$$\mathsf{g}_{cr} \approx \left[\frac{e_{eff}^2}{4a\pi^2 J}\right]^{1/2}, \tag{6.62}$$

where $e_{eff}$ is the effective screened charge. The critical value is estimated from the condition that the displacement of quasiparticles caused by the Coulomb repulsion is less than the bisoliton size $2\pi a/\mathsf{g}$.

The bisolitons are stable because they *do not* interact with acoustic phonons. This interaction is completely taken into account in the coupling of quasiparticles with a local deformation. Therefore, they do not radiate phonons. The velocity of bisoliton motion should not exceed the longitudinal sound velocity $c_0$ (in cuprates, $c_0 \sim 10^5$ cm/s). The bisolitons do not undergo a self-decay if their velocity is smaller than the critical one

$$v_{cr} = \frac{2\Delta}{\hbar k_F}, \tag{6.63}$$

where $k_F$ is the momentum at the Fermi surface.

When Davydov proposed the bisoliton model as the mechanism for high-$T_c$ superconductivity in the cuprates, he did not know about the existence of quasi-one-dimensional charge stripes in the $CuO_2$ planes. Hence, he needed to locate one dimensionality in the $CuO_2$ planes. Since the $CuO_2$ planes in the cuprates consist of quasi-infinite parallel chains of alternating ions of copper and oxygen, Davydov assumed that each -Cu-O-Cu-O- chain in a $CuO_2$ plane can be considered as a quasi-one-dimensional system. Therefore the current flows along these parallel chains. In the framework of the bisoliton model, he studied the charge migration in one of these chains and all the results obtained above were directly applied to the superconducting condensate in the cuprates. However, Davydov mistakenly assumed that the long-range phase coherence among the bisolitons sets in due to the overlap of their wavefunctions.

A comparison of the main characteristics of the bisoliton model and the data obtained in some cuprates, described in [19], shows that the bisoliton model is not a theory for high-$T_c$ superconductivity. Firstly, it lacks the mechanism of the onset of phase coherence. Secondly, the bisoliton model can describe *some* pairing characteristics but only *in a first approximation*. This is probably because in the bisoliton model the Coulomb repulsion between quasiparticles in a bisoliton is not taken into account. However, the main idea of the bisoliton model is correct: the moderately strong and nonlinear electron-phonon interaction mediates the pairing in the cuprates. The main result of the model is that, in the presence of a strong electron-phonon interaction, the BCS isotope effect can be absent or small. The bisoliton theory should serve as a starting point for the future theory of unconventional superconductivity.

As an example, consider the doping dependence of the distance between two holes in a bisoliton, derived in the framework of the Davydov model by using experimental data obtained in Bi2212. Figure 6.25 depicts this dependence, as well as the doping dependence $g(p)$. The dependence $d(p)$ in Fig. 6.25 is in good agreement with experimental data for Bi2212.

In Fig. 6.25, one can see that the values of the coupling parameter in Bi2212, g, is around 1. Such a result was in fact expected from the beginning. Why? As was mentioned in Chapter 1, the balance between nonlinearity and dispersion is responsible for the existence of solitons. The bisoliton model is based on the NLS equation. In the NLS equation, the second term is responsible for dispersion and the third one for nonlinearity [see, for example, Eq. (6.11)]. The coefficient in the second term, the energy of the exchange interaction, $2J$, characterizes the "strength" of dispersion, and the coefficient in the third term, the nonlinear coefficient of the electron-phonon interaction, $G$, characterizes the "strength" of nonlinearity. The parameter g represents the ratio between the two coefficients $G$ and $2J$ [see Eq. (6.47)]. Therefore, in a sense, the coupling parameter g reflects the balance between the nonlinear and dispersion forces. As a consequence, it cannot be very small g $\ll$ 1, or very large g $\gg$ 1. If g $\ll$ 1, dispersion will prevail, and the bisolitons will gradually diffuse, giving rise to "bare" quasiparticles. If g $\gg$ 1, nonlinearity effects prevail, and the bisolitons will become immobile and localized. Thus, g $\sim$ 1. In contrast, in superconductors described by the BCS theory the parameter g is of the order of $10^{-3}$–$10^{-4}$, meaning that, in conventional superconductors, the electron-phonon interaction is very weak and superconductivity is "linear."

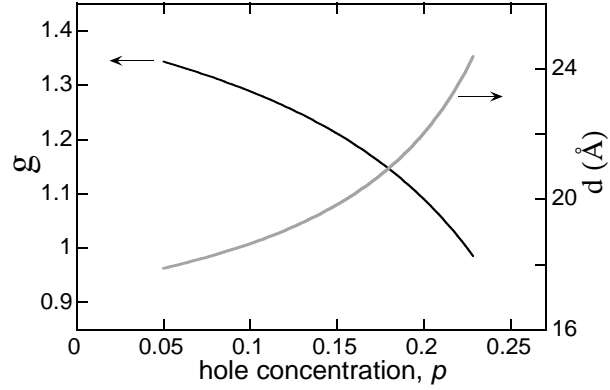

*Figure 6.25.* Doping dependence of the coupling parameter g (solid curve) and the corresponding average size of bisolitons (grey curve) in Bi2212 at low temperature, derived in the framework of the bisoliton model [19]. (Note that this plot slightly differs from that in [19] which is based on the point $d(0.16) = 15$ Å; this plot is founded on the point $d(0.16) = 20$ Å.)

In the previous subsection, we have derived SIN tunneling $dI(V)/dV$ and $I(V)$ characteristics for a system with electrosolitons. What can we expect from SIN tunneling measurements carried out in a system with bisolitons? From Fig. 6.23, it is obvious that the bisolitons will manifest themselves through the appearance of two peaks in conductance, situated symmetrically relative to zero bias. By using the same reasoning as that in the previous subsection, one can obtain that, in a SIN junction, tunneling conductance near these two peaks can be approximated by

$$\frac{dI(V)}{dV} = A \times \left[ \operatorname{sech}^2 \left( \frac{V + V_p}{V_0} \right) + \operatorname{sech}^2 \left( \frac{V - V_p}{V_0} \right) \right], \quad (6.64)$$

where $V$ is voltage (bias); $V_p$ is the peak bias, and $A$ and $V_0$ are constants. The corresponding $I(V)$ characteristic is then represented by

$$I(V) = I_0 \times \left[ \tanh \left( \frac{V + V_p}{V_0} \right) + \tanh \left( \frac{V - V_p}{V_0} \right) \right], \quad (6.65)$$

where $I_0$ is a constant. In the equations, $V_0$ determines the width of the conductance peaks. It is worth to recall that the bisoliton conductance peaks will appear in the background caused by other electronic states present in the system. Figure 6.26 visualizes Eqs. (6.64) and (6.65). The height of the bisoliton peaks relative to the the background depends on the density of added or removed electrons (holes): the height increases as the density increases. Figure 6.27 sketches this dependence for the cuprates, observed experimentally. In the case of the cuprates, one should however realize that there is a critical doping level ($p_{cr} \sim 0.3$) above which the bisolitons collapse.

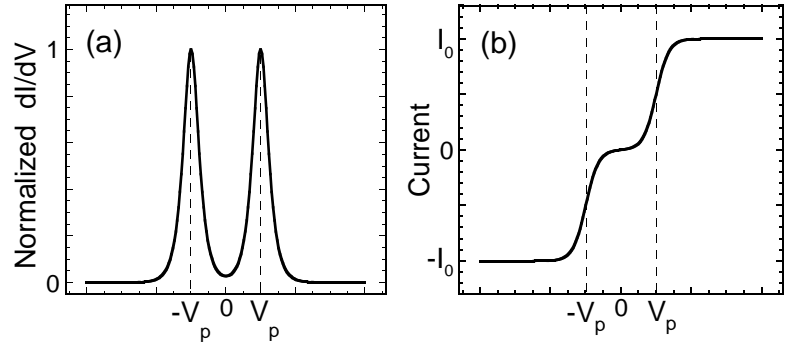

*Figure 6.26.* Visualization of (a) Eq. (6.64) and (b) Eq. (6.65) for a SIN junction. In plot (a), the conductance peaks are caused by bisolitons; the corresponding $I(V)$ characteristic is in plot (b). In plot (a), the conductance is normalized by its maximum (peak) value.

The bisolitons have experimentally been observed not only in the cuprates but also at low temperature in the manganate $La_{1.4}Sr_{1.6}Mn_2O_7$ [19] and charge-density-wave conductor $NbSe_3$ [48] which never exhibit superconductivity. Superconductivity requires not only the electron pairing but also the phase coherence. Since the size of bisolitons is small and their density is always low, the bisolitons cannot establish the long-range phase coherence. In this case, a question naturally arises: what happens with isolated bisolitons on lowering the temperature, $T \to 0$? Bisolitons propagate along charge-ordered one-dimensional structures such as charge stripes or charge-density-waves which are intrinsically insulating. Without having an opportunity to establish the long-range phase coherence, the bisolitons are "condemned" to condense into these charge-ordered one-dimensional structures and to become localized. In the manganites, for example, the bisolitons will join at low temperature holes on the charge stripes making them insulating not only on the nanoscale but also on a macroscopic scale. In Chapter 3, discussing charge-transfer organic salts, it was emphasized that the steep rise in resistivity at low temperature in $(TMTSF)_2PF_6$ and TmBCO occurs mainly due to a charge ordering (see Fig. 3.16). Therefore, even in the cuprates, without *dynamic* spin fluctuations, the bisolitons condense at low temperatures into the insulating charge stripes, becoming localized.

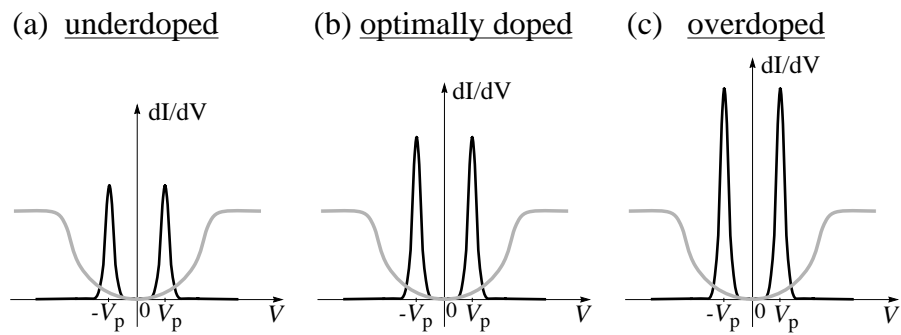

*Figure 6.27.* Sketch of the doping dependence of the height of quasiparticle peaks relative to the charge gap in the cuprates: (a) underdoped, (b) optimally doped, and (c) overdoped regions. The charge gap is schematically shown in grey. The absolute value of $V_p$ decreases as the doping level increases.

## 3.8  Phonons

Let us start with the isotope effect in the cuprates. The isotope effect is the first indicator of the BCS mechanism of superconductivity. The isotope effect was found to be extremely small in optimally doped cuprates. This fact was initially taken as evidence against the BCS mechanism of high-$T_c$ superconductivity and, mistakenly, against the phonon-pairing mechanism. If the pairing mechanism is different from the BCS mechanism, *this does not mean that phonons are irrelevant*.

In fact, there is a huge isotope effect in the cuprates. Figure 6.28 shows the oxygen ($^{16}$O vs $^{18}$O) isotope-effect coefficient $\alpha_O = d\ln(T_c)/d\ln(M)$, where $M$ is the isotope mass, as a function of doping level in LSCO, YBCO and Bi2212. In the plot, one can see that the oxygen-isotope effect in the cuprates is not universal: it is system- and doping-dependent. In the underdoped region, $\alpha_O$ can be much larger than the BCS value of 0.5 (according to the BCS theory, the isotope effect cannot be larger than 0.5). In the optimally doped region, the oxygen-isotope effect is indeed small. With exception of one point in LSCO, $p = 1/8$, the doping dependence of coefficient $\alpha_0$ is universal for these three cuprates. The coefficient $\alpha_0$ has a maximum at $p \to 0.05$. The copper ($^{63}$Cu vs $^{65}$Cu) isotope effect has also been studied in LSCO and YBCO. In LSCO, the copper-isotope effect is similar to the oxygen-isotope effect shown in Fig. 6.28. The copper-isotope effect in YBCO is small, even at low dopings, and can even be negative (as that in some charge-transfer organic superconductors and hydrides).

According to the bisoliton model, such a doping dependence of $\alpha_0$ in Fig. 6.28 indicates that the kinetic energy of charge carriers is large in the underdoped region and decreases as the doping level increases. ARPES measurements indeed show that the rate of band dispersion, i.e. the velocity of charge carriers, decreases as the doping level increases. Thus, by using the bisoliton model, one can explain the isotope effect in the cuprates.

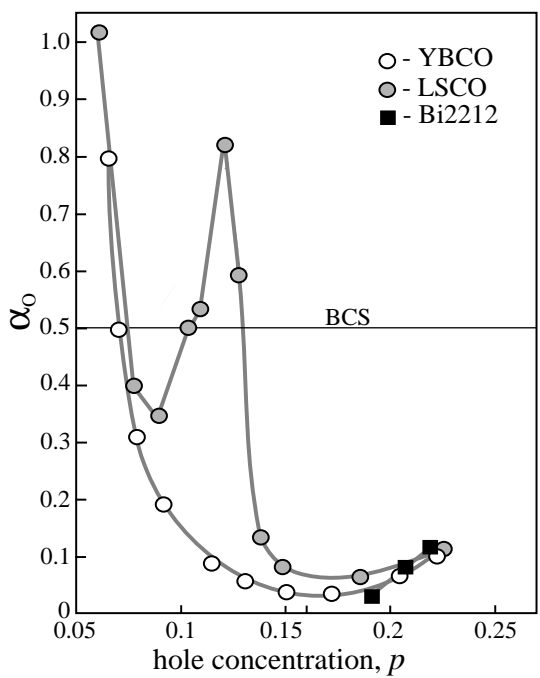

If phonons were not involved in the electron pairing in the cuprates, the isotope effect should be absent or very small. In fact, phonons are an essential part of the mechanism of high-$T_c$ superconductivity. Simply, phonons interact with charge carriers in conventional superconductors and in the cuprates in a different way. For example, in BCS-type superconductors, $T_c$ increases with lattice softening, while in the cuprates, $T_c$ increases with lattice stiffening, as shown in Fig. 6.29. So, the electron-phonon interaction is able to provide, at least, two different mechanisms of electron pairing: *linear* and *nonlinear*. In conventional BCS superconductors, the electron-phonon interaction is linear and weak, while, in the cuprates, it is moderately strong and nonlinear.

It is a paradox: the effect of isotope substitution on the transition temperature manifests itself when the electron-phonon interaction is weak, and can disappear when the electron-phonon interaction becomes stronger! The main result of the bisoliton model is that the potential energy of a static bisoliton, formed due to a local deformation of the lattice, does not depend on the mass of an elementary lattice cell. This mass appears only in the kinetic energy of the bisoliton.

It is difficult to underestimate the role of phonons in superconducting cuprates. Figure 6.30 shows the phonon spectrum $F(\omega)$ obtained in Bi2212 by

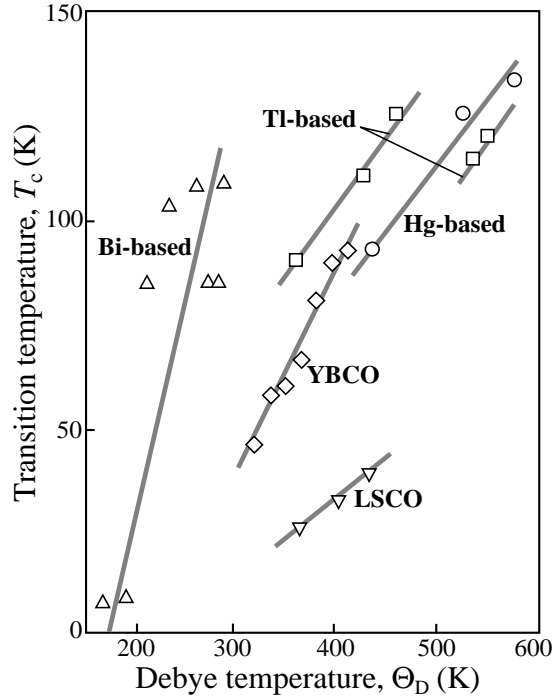

*Figure 6.29.* Critical temperature as a function of Debye temperature for different cuprates (see references in [19]).

INS measurements and two spectral functions $\alpha^2F(\omega)$ obtained in Bi2212 by two independent tunneling measurements. The spectral function $\alpha^2F(\omega)$ is the parameter of the electron-phonon interaction in the Eliashberg equations, which characterizes the coupling strength between charge carriers and phonon vibrations. In Fig. 6.30, one can see that charge carriers in Bi2212 are strongly coupled to the 20 meV acoustic mode and to the 73 meV optical mode. The 73 meV branch is associated with half-breathing-like oxygen phonon modes that propagate in the $CuO_2$ plane. The role of phonons at 50 meV in Bi2212 is controversial: one spectral function $\alpha^2F(\omega)$ shows a peak at 50 meV (dashed curve), while the other exhibits a dip (solid curve). The 50 meV branch is associated with either in-plane or out-of-plane Cu–O bond-bending vibrations. Leaving aside the question of the 50 meV phonons, it is clear that the optical phonons with $\omega = 73$ meV are coupled to charge carriers in Bi2212. Indeed, ARPES measurements performed in LSCO, YBCO, Bi2212 and Bi2201 show a kink in the dispersion at 55–75 meV, confirming the fact that optical phonons are coupled to charge carriers in the cuprates. Independently of the origin of these phonon modes, their energies are unusually high, indicating that the electron-phonon coupling is very strong. At this stage, the role of each phonon branch in the mechanism of unconventional superconductivity in the cuprates is still undetermined. Usually, in crystals, self-trapped states appear due to an interaction of quasiparticles with acoustic phonons.

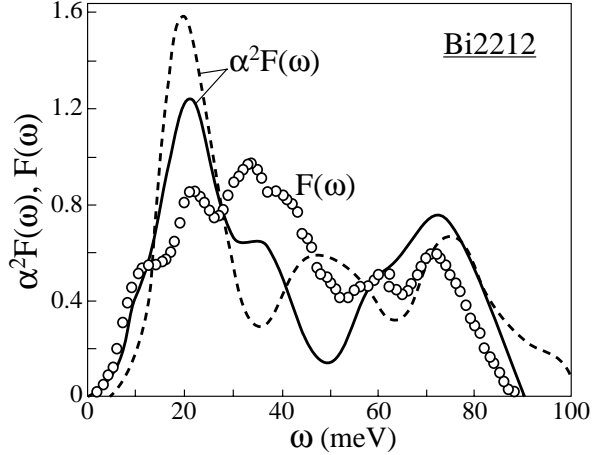

*Figure 6.30.* Phonon spectrum $F(\omega)$ (circles) obtained by INS measurements and two spectral functions $\alpha^2 F(\omega)$ (solid and dashed curves) obtained by tunneling measurements in slightly overdoped Bi2212 at low temperature, $T \ll T_c$ (see references in [19]).

## 3.9    Mechanism of phase coherence along the $c$ axis

Superconductivity requires not only the electron pairing but also the long-range phase coherence. Within a year after the discovery of high-$T_c$ superconductors, it was already established that superconductivity in the cuprates is quasi-two-dimensional, occurring in the $CuO_2$ planes. The mechanism of the interlayer coupling was never studied in detail because it was always assumed that the interlayer coupling originates from the Josephson coupling between superconducting $CuO_2$ layers. Later, analysis of some experimental data clearly indicated that the Josephson coupling between the $CuO_2$ planes cannot be responsible for the $c$-axis phase coherence in the cuprates.

The bisoliton wavefunctions in the cuprates lie in the $CuO_2$ planes, and they are quasi-two-dimensional. In the overdoped region, the average size of bisolitons is comparable with the mean distance between bisolitons. So, their wavefunctions can locally overlap in the $CuO_2$ planes, resulting in the onset of local in-plane phase coherence. Even, if we assume that the bisolitons are able to establish the phase coherence in every $CuO_2$ plane of the sample, they cannot do it perpendicular to the planes. The purpose of this subsection is to discuss the mechanism of phase coherence along the $c$ axis, i.e. the mechanism of the interlayer coupling. In the cuprates, this mechanism has the magnetic origin.

As was discussed in Chapter 3, many unconventional superconductors of the third group exhibit the coexistence of superconductivity and long-range antiferromagnetic order. For superconductors with a large-size coherence length, this may not seem too surprising since, over the scale of the coherence length, the exchange field of an antiferromagnet averages to zero. It was a surprise when in 2000 the coexistence of superconductivity and ferromagnetism was discovered in an alloy of uranium and germanium, $UGe_2$. Before 2000 superconductivity and ferromagnetism were always regarded as mutually exclusive phenomena. Soon after the discovery of superconductivity in itinerant ferromagnet $UGe_2$, two new itinerant ferromagnetic superconductors were discovered—zirconium zinc $ZrZn_2$ and uranium rhodium germanium $URhGe$. $ZrZn_2$ superconducts only when it is ferromagnetic (see, for example, Fig. 6.37b). The coexistence of superconductivity and weak ferromagnetism was found in the ruthenocuprate $RuSr_2RCu_2O_8$ (see Chapter 3). As discussed in Chapter 2, in the quasi-two-dimensional organic conductor $\lambda$-$(BETS)_2FeCl_4$, the superconducting phase is induced by a magnetic field exceeding 18 T. All these experimental facts clearly indicate that, in some cases, superconductivity needs spin fluctuations. Hence, they seem to mediate superconducting correlations.

### 3.9.1 Magnetic properties

Let us start first with the principal magnetic properties of the cuprates. The parent compounds of superconducting cuprates are antiferromagnetic Mott insulators. INS measurements show that the cuprates display a wide variety of magnetic properties. Because INS measurements require large-size homogeneous single crystals, INS studies have been performed only in a few cuprates: YBCO, LSCO, Bi2212 and Tl2201. Generally, all the cuprates exhibit common features of magnetic interactions. At the same time there are some particularities of magnetic correlations in each cuprate. YBCO is probably the most studied cuprate.

YBCO is a double-layer cuprate and, in a first approximation, it can be modeled as a set of weakly coupled $CuO_2$ bilayers. By neglecting the local anisotropy and other smaller interaction terms, the high-frequency spin dynamics can be described using the Heisenberg Hamiltonian for a single bilayer

$$H = \sum_{ij} J_\parallel S_i \cdot S_j + \sum_{ij'} J_\perp S_i \cdot S_{j'}, \qquad (6.66)$$

where $J_\parallel$ and $J_\perp$ are the intralayer and interlayer superexchange constants, respectively. The first term in the expression represents the nearest-neighbor coupling between Cu spins $S_i$ in the same plane and the second the nearest-neighbor coupling between Cu spins in different planes. In YBCO, $J_{\parallel} \simeq 120$ meV and $J_{\perp} \simeq 12$ meV. INS studies reveal that, in YBCO and other cuprates, spin-wave excitations present up to $2J_{\parallel}$.

By analogy with phonon-excitation spectra (see Fig. 5.3), spin-wave excitations in *bilayer* YBCO are also split into two channels: acoustic and optical. In the acoustic (odd) channel, pairs of neighboring spins in adjacent planes rotate in the same sense about their average direction. In the optical (even) channel, the spins in adjacent planes rotate in opposite directions, thus sensing the restoring force from the interplane coupling $J_{\perp}$.

Figure 6.31 schematically summarizes the temperature dependence of magnetic excitation spectra of underdoped YBCO. At any doping level, the low-energy spin excitations are absent in the optical (even) channel. The excitations in the even channel are almost unchanged across the superconducting transition, as shown in Fig. 6.31. The odd (acoustic) excitations undergo an abrupt sharpening on cooling through $T_c$. This sharp mode in the odd channel is called the *magnetic resonance peak*, which appears at the antiferromagnetic wave vector $Q = (\pi, \pi)$ exclusively below $T_c$. The resonance peak is caused by a collective spin excitation. The energy position of the resonance peak, $E_r$, as a function of doping level scales linearly with $T_c$. The antiferromagnetic correlations weaken in the overdoped region; however, the magnetic relaxation still remains predominant in the highly overdoped region.

Apart from the resonance peak which is commensurate with the lattice, INS measurements have also found four incommensurate peaks at some energy transfers, which appear in YBCO below $T_c$. Each spin scattering peak occurs at an incommensurate wave vector $Q \pm \delta$, as schematically shown in Fig. 6.32. In Fig. 6.32a, upon doping, commensurate antiferromagnetic Bragg peaks (grey stars) caused by antiferromagnetic ordering in an ideal Mott insulator disappear, and are replaced by four broadened incommensurate dynamic peaks (black circles and squares). This indicates that spin fluctuations are displaced from the commensurate peak by a small amount $\delta$, related in the underdoped region to the doping $p$ by $\delta = p$. The incommensurate peaks at $Q \pm \delta$ can either be caused by a sinusoidal spin-density-wave slowly fluctuating in space and time, or arise from the striped phase shown in Fig. 6.2. In the striped phase, a spin wave is in fact commensurate *locally*, but the phase jumps by $\pi$ at a periodic array of domain walls termed antiphase boundaries (charge stripes) can cause the appearance of the $Q \pm \delta$ peaks. Since the period of this magnetic structure is $8a$, where $a$ is the Cu–Cu distance in the $CuO_2$ planes, it is generally agreed that the incommensurate peaks originate from the striped phase.

At any doping level, the incommensurability and the commensurate resonance appear to be inseparable parts of the general features of the spin dynamics in YBCO, as shown in Fig. 6.32b. In optimally doped YBCO, the resonance peak appears at the maximum resonance energy $E_{r,max} = 41$ meV. Below 41 meV, there are well-separated incommensurate peaks. At low energies ($< 30$ meV), the intensity of the incommensurate peaks is strongly reduced. This can be caused by the opening of a spin gap below $T_c$. In Fig. 6.32b, above $E_{r,max}$, the peak is separated again, and the separation gradually increases with increasing energy up to $2J_\parallel$; although the peaks have a broader width and a much weaker intensity than those below $E_{r,max}$. In the underdoped region, the commensurate resonance peak is shifted to a lower energy $E_r < E_{r,max}$, scaling with $T_c$, and the incommensurate peaks appear below $E_r$. Upon heating through $T_c$, the incommensurate peaks are strongly renormalized upon approaching $T_c$ and disappear in the normal state.

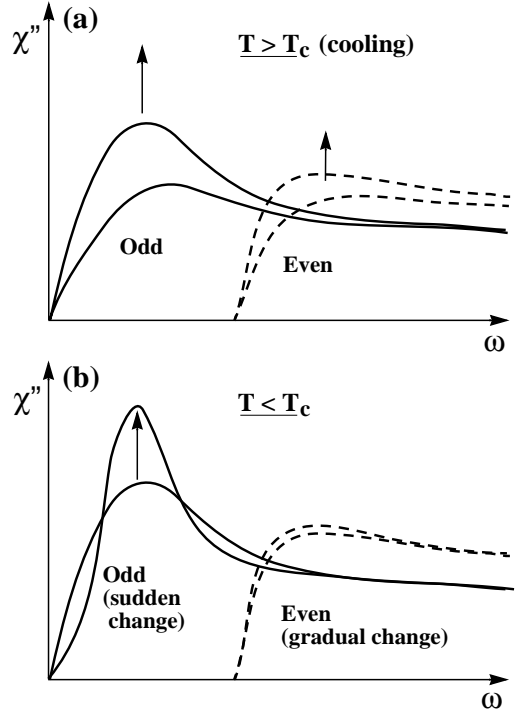

*Figure 6.31.* Schematic diagram summarizing the temperature evolution of magnetic excitation spectra of underdoped YBCO compound (a) in the normal state, and (b) across the superconducting transition. The excitations in the even channel evolve smoothly with temperature. The excitations in the odd channel undergo an abrupt sharpening at the resonance energy across $T_c$.

In LSCO, the low energy magnetic excitations have been extensively studied, and the observed spin fluctuations are characterized only by incommensurate peaks. The magnetic resonance peak has never been observed in LSCO.

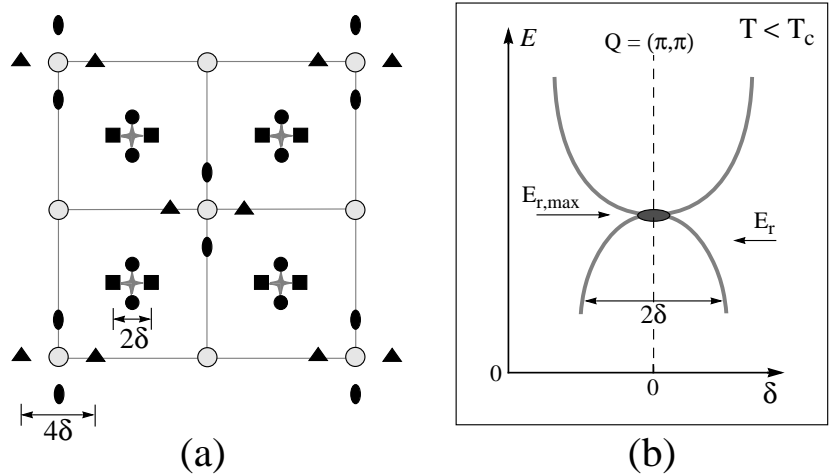

(a)                                    (b)

*Figure 6.32.* (a) Neutron scattering data in the reciprocal (momentum) space obtained in LSCO. The commensurate antiferromagnetic Bragg peaks, obtained in *undoped* LSCO at wave vector $Q = (\pi, \pi)$, are shown by grey stars. Upon doping, the commensurate peaks disappear and are replaced by four broadened incommensurate dynamic peaks (black squares and circles) with incommensurability $\delta$. The fundamental lattice Bragg peaks are shown by large grey dots. Upon doping, new peaks (black triangles and ovals) are observed, which are displaced by $2\delta$ from the fundamental lattice peaks. (b) Energy dependence of the peak position of incommensurate peaks versus incommensurability $\delta$ in YBCO. There are two legs below the maximum resonance energy $E_{r,max} = 41$ meV. Above the resonance, the signal is also split. In the underdoped region, the resonance peak appears at $E_r < E_{r,max}$.

Incommensurability in LSCO is consistent with that in YBCO with the same hole doping, but in LSCO, it persists in the normal state. In LSCO, the peak position is unchanged across $T_c$. Only around room temperature does the incommensurate structure begin to disappear. In LSCO, $\delta$ is energy-independent but depends strongly on the doping level. In the single-layer LSCO, spin excitations do not split into acoustic and optical channels as those in the double-layer YBCO. The similarity of spin dynamics in two different cuprates, LSCO and YBCO, demonstrate that the spin dynamics does not depend on the details of the Fermi surface, but have an analogous form to that for the striped phase. The resonance mode observed in YBCO has also been found in the double-layer Bi2212 and in the single-layer Tl2201. Incommensurability has not yet been seen in Bi2212 and Tl2201, but it is expected to be observed. As emphasized above, the incommensurate response and the resonance peak are inseparable parts of the same phenomenon. The case of LSCO, however, shows that the incommensurability can exist without the presence of the commensurate peak but not vice versa.

### 3.9.2 Correlations between magnetism and superconductivity

In YBCO, Bi2212 and Tl2201, the energy position of a magnetic resonance peak as a function of doping level scales linearly with $T_c$ and, as a consequence, with the phase-coherence energy gap. From Eq. (6.9), $2\Delta_c = \Lambda k_B T_c$, where $\Lambda \simeq 5.45$ in Bi2212; $\Lambda \simeq 5.1$ in YBCO; and $\Lambda \simeq 5.9$ in Tl2201. Figure 6.33 shows the two energy scales in Bi2212, $\Delta_p$ and $\Delta_c$, as a function of doping (compare with Fig. 6.11). Figure 6.33 also depicts the energy position of a magnetic resonance peak, $E_r$, at different doping levels in Bi2212, YBCO and Tl2201. In the plot, one can see that at different dopings $E_r \simeq 2\Delta_c$. This relation unambiguously shows that the magnetic resonance peak *intimately* relates to the onset of long-range phase coherence in the cuprates. This means that the resonance mode is either a consequence or the mediator of phase coherence. Indeed, for YBCO modest magnetic fields applied below $T_c$ suppresses significantly the intensity of a magnetic resonance peak.

In all layered magnetic compounds, including the undoped cuprates, the long-range antiferromagnetic (ferromagnetic) order develops at Néel temperature $T_N$ (Curie temperature $T_C$) along the $c$ axis. At the same time, in-plane magnetic correlations exist above $T_N$ ($T_C$). Thus, in quasi-two-dimensional magnetic materials the coupling along the $c$ axis represents the last step in establishing a long-range magnetic order. Interestingly, in all layered superconducting materials the phase coherence becomes long-ranged also due to the interlayer coupling occurring at $T_c$. For example, infrared reflectivity measurements performed in high-quality single crystals of LSCO show that the superconducting transition is accompanied by the onset of coherent charge transport along the $c$ axis, which was blocked above $T_c$. Therefore, in LSCO, the long-range phase coherence occurs at $T_c$ along the $c$ axis.

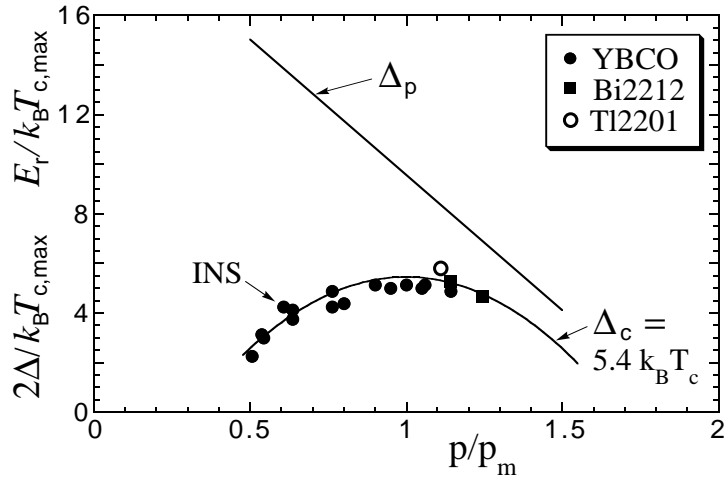

*Figure 6.33.* Phase diagram of the cuprates and the energy position of a magnetic resonance peak, $E_r$, in Bi2212 (squares), YBCO (dots) and Tl2201 (circle) at different hole concentrations ($p_m = 0.16$) [15, 19].

In LSCO, there exists direct evidence that superconductivity is intimately related to the establishment of antiferromagnetic order along the $c$ axis: $\mu$SR measurements performed in non-superconducting Eu-doped LSCO show that, at different dopings, the superconducting phase of pure LSCO is replaced in Eu-doped LSCO by a second antiferromagnetic phase, as depicted in Fig. 6.34. Thus in LSCO, it is possible to switch the entire doping-dependent phase diagram from superconducting to antiferromagnetic. Since LSCO is a layered compound, the main antiferromagnetic phase of Eu-doped LSCO and its second antiferromagnetic phase develop along the $c$ axis. Hence, the superconducting phase of pure LSCO is replaced in Eu-doped LSCO by an antiferromagnetic phase which arises along the $c$ axis. This clearly indicates that superconductivity in LSCO intimately relates to the onset of long-range antiferromagnetic order along the $c$ axis.

Elsewhere [19] it was shown that, if scaled, the phase diagram of the heavy fermion CePd$_2$Si$_2$ (see Fig. 6.37a) is almost identical to the two energy scales of Bi2212, namely, $T_{MT}$ and $T_c$ (see Fig. 6.11). It is generally agreed that superconductivity in CePd$_2$Si$_2$ is mediated by spin fluctuations. Therefore, this striking similarity suggests that the phase coherence in Bi2212, which sets in at $T_c$, is mediated by spin fluctuations.

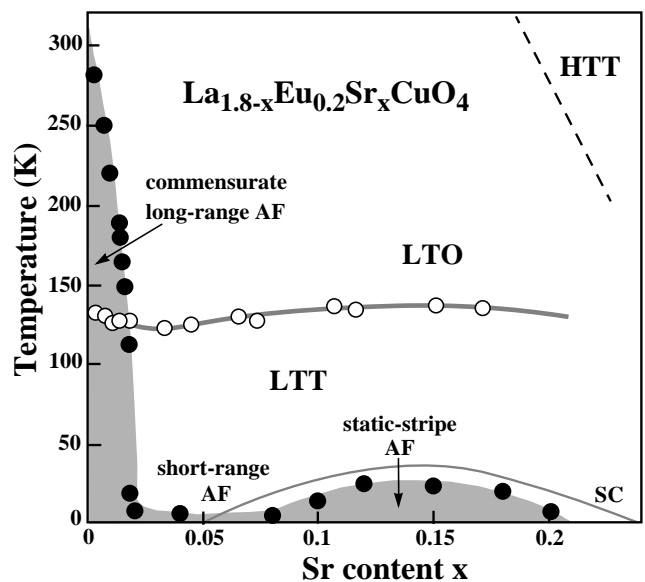

*Figure 6.34.* Phase diagram of La$_{1.8-x}$Eu$_{0.2}$Sr$_x$CuO$_4$ obtained by $\mu$SR measurements (see references in [19]). Full and open circles denote the magnetic and the structural transition temperatures, respectively. The superconducting phase in pure LSCO is marked by "SC." The structural transition from the high-temperature tetragonal (HTT) to the low-temperature orthorhombic (LTO) is indicated by the dashed line. The low-temperature tetragonal phase (LTT) appears below 120–130 K (AF = antiferromagnetic).

In antiferromagnetic superconductors, magnetic fluctuations which often exist above $T_c$ are enhanced on passing below $T_c$. For example, in superconducting YBCO, the antiferromagnetic ordering starts to develop above 300 K, as shown in Fig. 6.35. This antiferromagnetic commensurate ordering with a small moment was observed in underdoped and optimally doped YBCO. The magnetic-moment intensity *increases* in strength as the temperature is reduced below $T_c$, as depicted in Fig. 6.35. The magnetic-moment direction was found in one study to be along the $c$ axis and in-plane in the other (see references in [19]).

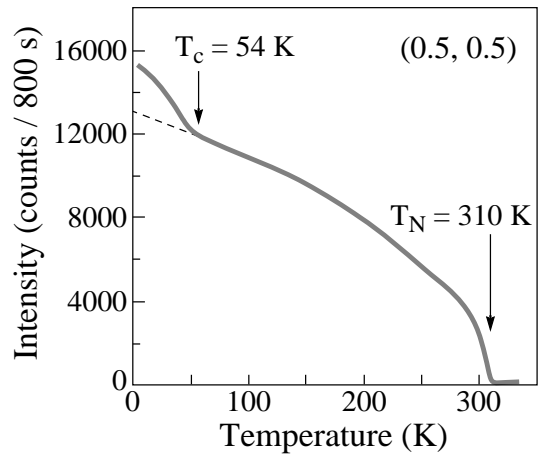

*Figure 6.35.* Temperature dependence of the magnetic intensity in underdoped YBCO ($T_c$ = 54–55 K), measured at the wave vector **Q** = (0.5, 0.5) by polarized and unpolarized neutron beams (see references in [19]. The antiferromagnetic order appears at a Néel temperature of $T_N \simeq 310$ K. The dashed line shows the background.

The energy positions of a magnetic resonance peak and incommensurate peaks are independent of temperature. However, the intensities of the commensurate peak and the incommensurate ones both exhibit a temperature dependence which is very similar to the temperature dependence of $\Delta_c$ (see Fig. 6.42). Figure 6.36 shows three temperature dependences of the peak intensities: the commensurate resonance peak in Bi2212 and YBCO, and the incommensurate peaks in LSCO (x = 0). The temperature dependences of $\Delta_c$ (see Fig. 6.42) and those in Fig. 6.36 exhibit below $T_c$ a striking similarity. To recall, the data in Fig. 6.36 reflect exclusively *magnetic* properties of the cuprates.

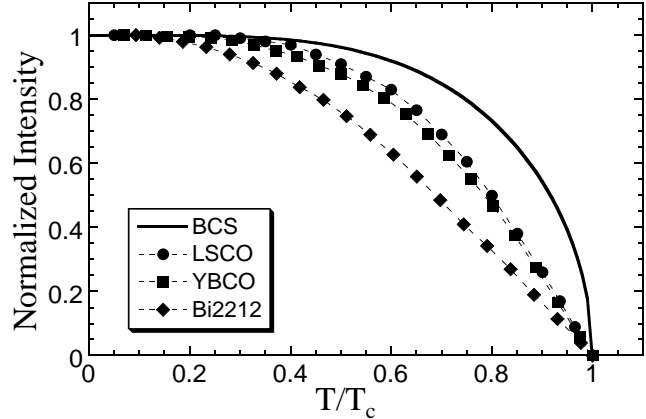

*Figure 6.36.*   Temperature dependences of the peak intensity of the incommensurate elastic scattering in LSCO (x = 0) ($T_c$ = 42 K) and the intensity of the magnetic resonance peak measured by INS in near optimally doped Bi2212 ($T_c$ = 91 K) and YBCO ($T_c$ = 92.5 K). The neutron-scattering data are averaged, the real data have a vertical error of the order of ±10–15%. The BCS temperature dependence is shown by the thick solid line [15, 19].

On the basis of tunneling data [17, 19], it was shown that the charge carriers in Bi2212 are strongly coupled to the spin excitation which causes the appearance of a magnetic resonance peak in INS spectra, and this magnetic excitation seems to mediate the phase coherence in Bi2212.

In the framework of a theoretical model developed for the cuprates, which takes into account a competition between interlayer direct hopping and hopping assisted by spin fluctuations, calculations show that, at least, in the underdoped region, the interlayer hopping assisted by spin fluctuations is predominant. Therefore, the interlayer direct hopping can be omitted. The model captures the main features of experimental data; for example, the anomalous behavior of the $c$-axis electronic conductivity in YBCO and thermoelectric power in LSCO (see references in [19]).

### 3.9.3    Magnetically-mediated superconductivity

The spin-fluctuation mechanism of superconductivity was first proposed as an explanation of superconductivity in heavy fermions [4]. This model is based on a short-range Coulomb interaction leading to an exchange coupling $J \times \mathbf{S}_i\mathbf{S}_j$ between near-neighbor copper spins $\mathbf{S}_i$ and $\mathbf{S}_j$ and strong magnetic spin fluctuations. The superexchange constant is denoted by $J$. In the cuprates, it has an extremely high magnitude, $J \sim 125$ meV $\approx 1500$ K.

The underlying microscopic physics can be described by the $t - J$ model defined by the Hamiltonian

$$H = H_t + H_J = -t \sum_{\langle nm \rangle \sigma} (d_{n\sigma}^\dagger d_{m\sigma} + H.c.) + J \sum_{\langle nm \rangle} S_n S_m, \qquad (6.67)$$

where $d_{n\sigma}^\dagger$ is the creation operator of a hole with spin $\sigma$ ($\sigma = \uparrow, \downarrow$) at site $n$ on a two-dimensional square lattice ($H.c.$ = Hermitian conjugated). The $d_{n\sigma}^\dagger$ operators act in the Hilbert space with no double electron occupancy. The spin operator is $S_n = \frac{1}{2} d_{n\sigma}^\dagger \sigma_{\alpha\beta} d_{n\beta}$, and $\langle nm \rangle$ are the nearest-neighbor sites on the lattice. At half-filling (one hole per site) the $t - J$ model is equivalent to the Heisenberg antiferromagnetic model, which has long-range Néel order in the ground state. Upon doping, the long-range antiferromagnetic order is destroyed; however, the local antiferromagnetic order is preserved. The magnetic coupling is not local both in space and in time. Magnetically-mediated superconductivity may exist only in samples in which the carrier mean free path exceeds the superconducting coherence length. In most cases this requires samples of very high purity. The spin-fluctuation mechanism of superconductivity results in the $d_{x^2 - y^2}$ symmetry of superconducting order parameter. The model gives the value of critical temperature in reasonable agreement with experimental data for high-$T_c$ superconductors.

In conventional superconductors, the Cooper pairs are formed via interactions between electrons and lattice vibrations (phonons). In superconducting heavy fermions, spin fluctuations are believed to mediate the electron pairing that leads to superconductivity. However, in reality, spin fluctuations seem to mediate only the long-range phase coherence in the heavy fermions, as well as in the cuprates.

Let us consider characteristic features of magnetically-mediated superconductivity. In conventional superconductors, superconductivity described by the BCS theory has its specific features, for example, the isotope effect, the s-wave symmetry of the order parameter, etc. What features are inherent to magnetically-mediated superconductivity?

*Quantum critical point.* As discussed above, magnetically-mediated superconductivity occurs near a quantum critical point. Figure 6.37 sketches two phase diagrams: the first diagram is typical for antiferromagnetic heavy-fermions, and the second is the phase diagram of the ferromagnetic heavy fermion UGe$_2$. In both phase diagrams, the density of charge carriers is changed by applying a pressure. Near a quantum critical density $n_c$, magnetic interactions become strong and long-ranged and overwhelm other channels. In Fig. 6.37, one sees that independently of the nature of magnetic interactions—antiferromagnetic or ferromagnetic—the superconducting phase occurs near a quantum critical point, where magnetic fluctuations are the strongest. In a sense, the superconducting phase is "attracted" by a quantum critical point.

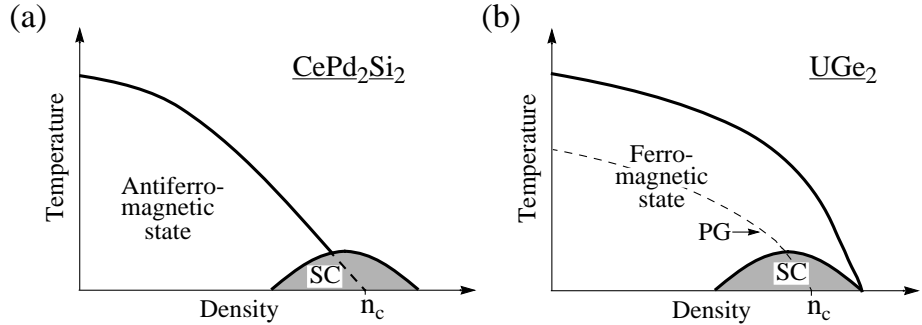

*Figure 6.37.* Phase diagrams of heavy fermions: (a) antiferromagnetic CePd$_2$Si$_2$, and (b) ferromagnetic UGe$_2$, shown schematically [19]. In both plots, the density is varied by pressure, and $n_c$ is the quantum critical point (SC = superconductivity). The phase diagram in plot (a) is typical for antiferromagnetic superconductors. In plot (b), the dashed line shows a pseudogap (PG) found in resistivity and magnetization measurements in UGe$_2$, indicating the presence of a quantum critical point.

*Superconducting dome.* One can see in Fig. 6.37 that the superconducting phase as a function of doping has a bell-like shape. Such a shape of the superconducting phase is typical for magnetically-mediated superconductivity.

*Symmetry of the order parameter.* Theoretically, the order parameter in antiferromagnetic superconducting compounds has a d-wave symmetry. The d-wave symmetry of the order parameter was indeed observed in a few magnetic compounds, including the cuprates. For ferromagnetic superconducting materials the situation is still not clear. Theoretically, the order parameter in superconductors with ferromagnetic correlations should have a p-wave (triplet) symmetry. However, there is no experimental confirmation of this conjecture.

*Temperature dependence.* In magnetic superconductors, the temperature dependence of magnetic (in)commensurate peak(s), shown in Fig. 6.36 for cuprates, is specific and lies below the s-wave BCS temperature dependence. At the same time, such a temperature dependence is similar to the temperature dependence of *coherence* superconducting characteristics; in other words, to the dependence $\Delta_c(T)$. In a first approximation, such a temperature dependence represents the squared BCS temperature dependence.

*Enhancement of spin fluctuations.* As discussed above, in antiferromagnetic superconductors, magnetic fluctuations are enhanced on passing below $T_c$. This is shown in Fig. 6.35.

*The magnetic resonance peak.* Below $T_c$ in antiferromagnetic superconductors, there often, but not always, occurs a specific magnetic excitation which causes the appearance of a magnetic resonance peak in INS spectra. The energy position of a magnetic resonance peak, $E_r$, is independent of temperature, and $E_r(p) = 2\Delta_c(p)$. However, the *intensity* of the resonance mode depends on temperature, and increases as the temperature decreases (see Fig. 6.36). It is assumed that the spin excitation causing the appearance of a magnetic resonance peak is a magnon-like. Generally speaking, magnons have a large degree of dispersion; if they were the cause of the appearance of a resonance peak, the peak should be quite wide. However, the width of a resonance peak in INS spectra is very narrow. Therefore, this excitation cannot be magnon-like. Alternatively, it was proposed that the resonance mode is a magnetic *exciton*. Magnons and excitons are non-interacting plane waves. To propagate, a magnon uses only the ground spin states—antiferromagnetic or ferromagnetic—while an exciton only excited spin states. Since excitons are also plane waves, they should have also a large degree of dispersion. On the other hand, analysis of experimental data shows that the spin excitation that causes the appearance of a resonance peak in INS spectra can be a magnetic soliton [19].

For example, in the heavy fermion UPd$_2$Al$_3$, the Néel temperature is about $T_N \simeq 14.3$ K, and $T_c \simeq 2$K. Upon cooling through $T_c$, an abrupt enhancement of magnetic fluctuations is observed in INS measurements, and a magnetic resonance peak appears at $E_r/k_B T_c \simeq 9.2$.

### 3.9.4    Interplay between the lattice and magnetism

Here we consider an important issue related directly to the mechanism of phase coherence in the cuprates—the interplay between the lattice and spin fluctuations. First of all, one should distinguish the onset of phase coherence in the CuO$_2$ planes and between the planes. The mechanism of interlayer phase coherence in the cuprates is magnetic, while the in-plane phase coherence occurs not only due to spin fluctuations but also due to the direct hopping of bisoliton wavefunctions. Furthermore, the lattice also plays an important role: as was discussed above, a structural phase transition always takes place somewhat above $T_{c,max}$ for each cuprate.

In the non-superconducting Gd- and Eu-doped LSCO, the frequency of spin fluctuations upon lowering the temperature monotonically decreases, having a kink at $T_{c,max} \simeq 38$ K [49]. *In the cuprates, superconductivity is associated with spin fluctuations which are rapid*. Below a certain frequency of spin fluctuations, $\omega_{min}$, the cuprates cannot superconduct [19]. Generally speaking, the frequency of spin fluctuations depends on fluctuations of charge stripes in the CuO$_2$ planes and, thus, on the underlying lattice. In the cuprates, if the charge stripes carrying bisolitons fluctuate in the CuO$_2$ planes not fast enough for exciting spin fluctuations capable of mediating phase coherence, superconductivity will never arise. Somewhat above $T_{c,max}$, there is a structural phase transition which flattens the CuO$_2$ planes and/or makes them more tetragonal. As a consequence, the charge stripes can now fluctuate quicker and induce spin excitations capable of mediating the phase coherence. Thus, even if the mechanism of interlayer coupling is magnetic, the lattice, in fact, determines the value of $T_{c,max}$ for each cuprate. In a sense, the mechanism of phase coherence in the cuprates is not pure magnetic but magneto-elastic.

In the cuprates, the underlying lattice determines the homogeneity of superconducting phase. As an example, Figure 6.38 depicts the surface of Bi2212, YBCO and Tl2201 at low temperature. This sketch is based on tunneling measurements, assuming that the surface is flat. In the sketch, the superconducting phase in YBCO and Tl2201 near the surface is more or less homogeneous, while in Bi2212, it forms nanoscale patches, as illustrated in Fig. 6.38. As a consequence, tunneling measurements performed in Bi2212 are able to provide information about the two energy gaps, $\Delta_c$ and $\Delta_p$. Above the superconducting nanoscale patches (shown in grey in Fig. 6.38), tunneling measurements can provide information about $\Delta_c$ (see the last subsection in this chapter), while between the patches where incoherent Cooper pairs are present, about $\Delta_p$. The homogeneity of superconducting phase deep inside Bi2212 is unknown.

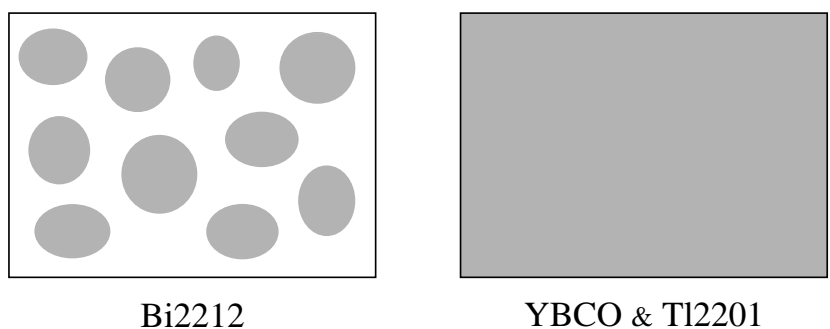

Bi2212                              YBCO & Tl2201

*Figure 6.38.* Superconducting phase on the flat surface of Bi2212, YBCO and Tl2201 at low temperature, shown in grey. In underdoped Bi2212, the size of the patches is about 30 Å. In the cuprates, the superconducting phase is associated with dynamic magnetic fluctuations.

In the absence of charge-stripe fluctuations which occur due to the underlying lattice, spin fluctuations become quasi-static, forming a spin wave with a periodicity of $8a$. So, one may say that the occurrence of superconductivity in the cuprates is really a "bypass product."

### 3.9.5    Origin of the resonance mode

At the moment of writing, the origin of spin excitation, which is referred to as the magnetic resonance peak, is unknown. It is clear that this spin excitation is a collective excitation, and has a magnitude of $S = 1$. It is also clear that, in LSCO, this excitation is absent because the Cu spins in this cuprate fluctuate not sufficiently quick to excite it. Here we discuss a possible origin of this spin excitation.

The magnetic interlayer coupling requires that the Cu spins in the $CuO_2$ planes must slightly be out-of-plane in order to have a small $c$-axis component, therefore they must be canted. It is possible that this may be the clue to the problem. As was shown elsewhere [19], every charge stripe carries spin excitations at its ends, as sketched in Fig. 6.39. In the striped phase shown in Fig. 6.2, below $T_{MO}$, insulating stripes between the charge stripes are antiferromagnetically ordered. The spin direction in the antiferromagnetic stripes rotates by $180°$ upon crossing a domain wall, as shown in Fig. 6.2. At the end of each charge stripe, two antiferromagnetic stripes separated by a charge stripe come into contact with one another. Because the spin direction in the two antiferromagnetic stripes rotates by $180°$ upon crossing the charge stripe, at the charge-stripe end the spin orientations in the two domains are opposite. Therefore, any spin orientation—up or down—at the charge-stipe end induces a local spin excitation, as shown in Fig. 6.39. In reality, the spin orientation at each end of charge stripes is the superposition of the two: $(\uparrow + \downarrow)/2$. It is then possible that the the most energetically favorable orientation of this spin is out-of-plane (in Fig. 6.39, perpendicular to the page). If the charge stripes fluctuate quickly, the spins at charge-stripe ends always give rise to an excitation which may be the one that is referred to as the magnetic resonance peak.

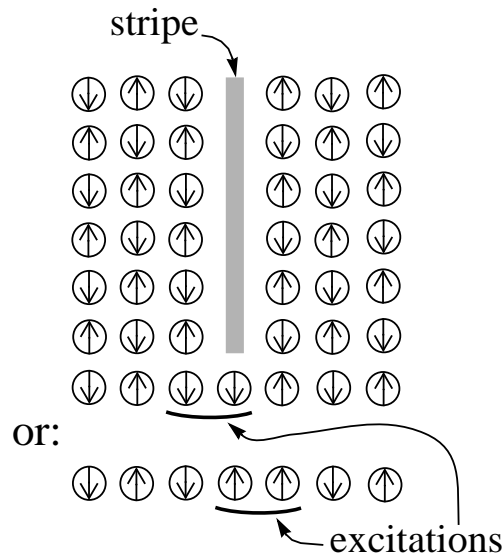

*Figure 6.39.* Sketch of a charge stripe in the antiferromagnetic environment. Independently of a spin orientation, each end of a charge stripe always carries a spin excitation.

### 3.10    Energy gaps $\Delta_p$ and $\Delta_c$

The energy gap is one of the most important characteristics of the superconducting state. Since the discovery of high-$T_c$ superconductivity in the cuprates, they have been extensively studied by different experimental techniques. Surprisingly in the cuprates, different techniques have initially provided different values of an energy gap. The discrepancy remained a mystery until it was realized that different experimental methods probe two different energy gaps. Moreover, the magnitudes of the two gaps strongly depend on the doping level.

The energy gap $\Delta_p$ is the bisoliton pairing gap shown in Fig. 6.23, and $2\Delta_p$ measures the strength of the binding of two quasiparticles. The phase-coherence energy gap $2\Delta_c$ is the condensation energy of a Cooper pair when the long-range phase coherence appears. Hence, quasiparticles in the cuprates undergo two condensations: the first when they become paired, and the second when the long-range phase coherence sets in. The total energy gain per quasiparticle is not $(\Delta_p + \Delta_c)$ but $\Delta_t = \sqrt{\Delta_p^2 + \Delta_c^2}$. Figure 6.40 schematically shows these three energy gaps relative to the Fermi level (F) in three dimensions. The bisoliton energy level (B) is below the Fermi level by $\Delta_p$, and it is separated by $\Delta_c$ from the energy level of the superconducting condensate (SC). In Fig. 6.40, each Cooper pair at the SC level has two possibilities: either to be excited or to be broken. The first possibility corresponds to the transition SC → B in Fig. 6.40, that requires a minimum energy of $2\Delta_c$. The second choice is the transition SC → F in Fig. 6.40 which requires a minimum energy of $2\Delta_t$. The bisolitons at the B energy level in Fig. 6.40 are uncondensed Cooper pairs; therefore, they are not in phase with the superconducting condensate. To break up a bisoliton which is at the B energy level, a minimum

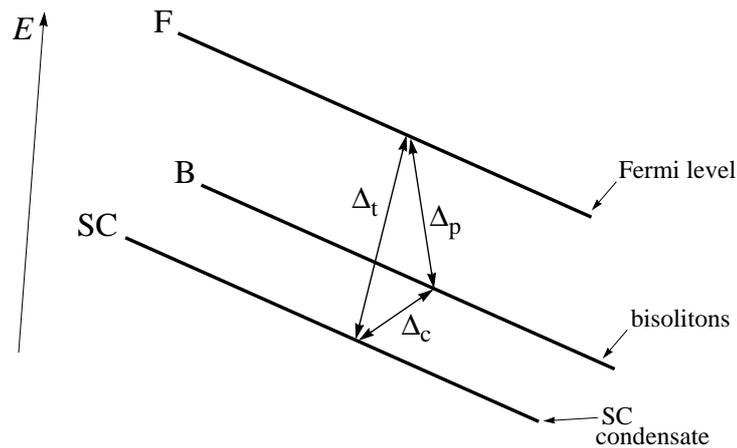

*Figure 6.40.*    Three energy levels of quasiparticles in unconventional superconductors in three dimensions (for more details, see text).

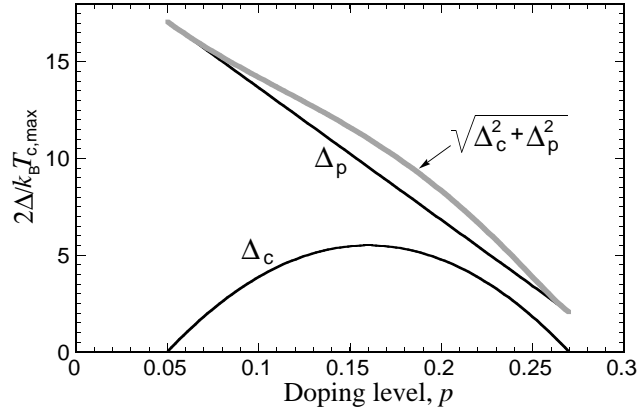

*Figure 6.41.* Low-temperature phase diagram of superconducting cuprates: the pairing energy scale $\Delta_p$ and the Cooper-pair condensation scales $\Delta_c$.

energy of $2\Delta_p$ must be supplied.

Figure 6.41 shows three energy scales $\Delta_c$, $\Delta_p$ and $\sqrt{\Delta_c^2 + \Delta_p^2}$ in Bi2212 as a function of doping level. $\Delta_c$ and $\Delta_p$ are given by Eqs. (6.9) and (6.7), respectively. Such a phase diagram is typical for hole-doped cuprates. From Fig. 6.41, one can see that the two energy scales $\Delta_p$ and $\sqrt{\Delta_c^2 + \Delta_p^2}$ have similar magnitudes. In the cuprates, depending on the type of an experiment (bulk or surface-layer sensitive; sensitive to single-quasiparticle excitations or to the coherence properties of the condensate), measurements *may* show one, two or three energy scales which are depicted in Fig. 6.41. It is worth noting that the pairing gap $\Delta_p$ is an in-plane energy scale, while the phase-coherence gap $\Delta_c$ is mainly a $c$-axis energy scale.

The temperature dependences of the two energy gaps, $\Delta_c$ and $\Delta_p$, are presented in Fig. 6.42. The $\Delta_p(T)$ dependence is similar to the BCS temperature dependence, while the $\Delta_c(T)$ dependence lies below the BCS temperature dependence, and is similar to the temperature dependences of (in)commensurate peak(s) depicted in Fig. 6.36.

In conventional superconductors, the order parameter $\Psi$ is proportional to the Cooper-pair wavefunction $\psi$, and has an s-wave symmetry (angular momentum $\ell = 0$). In the cuprates, the order parameter and the Cooper-pair wavefunction are different, and have different symmetries. Therefore, the symmetries of the energy gaps $\Delta_p$ and $\Delta_c$ are also different. In the cuprates, all phase-sensitive measurements show that the order parameter in hole- and electron-doped cuprates has the $d_{x^2-y^2}$ (d-wave) symmetry, shown in Fig. 6.43. Then, the energy gap $\Delta_c$ has also the d-wave symmetry. A key feature distinguishing the $d_{x^2-y^2}$ symmetry is that it has two positive and two negative lobes and four nodes between the lobes, as sketched in Fig. 6.43. As discussed above, spin fluctuations mediate superconductivity exclusively with a $d_{x^2-y^2}$ ground state.

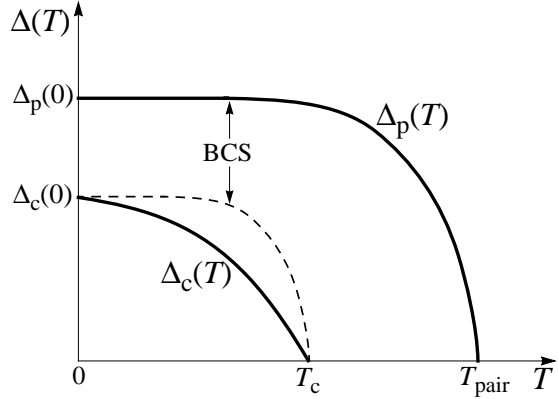

*Figure 6.42.* Temperature dependences of $\Delta_p$ and $\Delta_c$, shown schematically. The $\Delta_c(T)$ dependence lies below the BCS temperature dependence.

What is the symmetry of the pairing (bisoliton) wavefunction? The pairing wavefunction must have an s-wave symmetry. Many experiments mainly tunneling clearly show the presence of an s-wave component in the cuprates. To explain these experimental data, one must assume that the Cooper-pair wavefunctions have an s-wave symmetry. Fortunately, we know that phonons are mainly responsible for the pairing in the cuprates, and they indeed favor an electron pairing with an s-wave symmetry.

ARPES and tunneling measurements performed in Bi2212 show that the gap $\Delta_p$ is anisotropic and has a four-fold symmetry. In addition to a usual anisotropic s-wave symmetry, $\Delta_p$ can also have an extended s-wave symmetry schematically shown in Fig. 6.44. In the latter case, the gap is everywhere positive (negative) except four small lobes where the gap is negative (positive). In Fig. 6.44, the case of s-wave gap with nodes is an intermediate case between the two cases—anisotropic and extended.

In heavy fermions, the situation is very similar. For example, in $UBe_{13}$, Andreev-reflection measurements show the presence of a d-wave gap, while tunneling measurements show that $UBe_{13}$ is an s-wave superconductor (see references in [19]).

At the end of this subsection, let us consider a set of tunneling data to visualize the two gaps $\Delta_p$ and $\Delta_c$. In these tunneling data, the two gaps manifest themselves simultaneously. Figure 6.45 shows a temperature dependence of a conductance obtained in Ni-doped Bi2212 having $T_c \simeq 75$ K ($p \sim 0.2$). In the plot, one can see that the conductance peaks are composite, especially, those

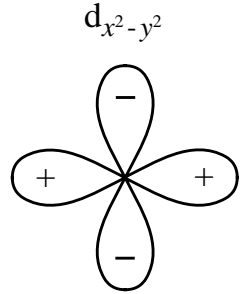

$d_{x^2-y^2}$

*Figure 6.43.* Superconducting order parameter in hole-doped cuprates, shown in real space: it has the $d_{x^2-y^2}$ shape.

at 41.8 K. As was discussed in [19], these data were measured by chance near an impurity, i.e. a Ni atom. Since Ni is a magnetic impurity in the $CuO_2$ planes, it is able to participate in the Cu-spin fluctuations. At the same time, Ni breaks up the Cooper pairs in the $CuO_2$ planes. In terms of the two gaps, this means that, in its vicinity, Ni destroys $\Delta_p$, whilst $\Delta_c$ remains practically unchanged. In Fig. 6.45, both gaps have practically the same magnitudes at 15 K, $\Delta_p \simeq \Delta_c \simeq 17.5$ meV (this can easily be seen in the corresponding $I(V)$ characteristics which are not shown here; for more details, see Chapter 6 in [19]). It is important to emphasize that such a situation is only possible **locally**. In Fig. 6.45, one can see that the quasiparticle peaks disappear between 65.3 K and 70.3 K. This means that, locally, $\Delta_p$ closes in this temperature interval. Taking the mean value $T_{\Delta_p} = (65.3 + 70.3)/2 \simeq 67.8$ K, we have $T_{\Delta_p}/T_c \simeq 0.91$. This enables us to determine the gap ratios for the two gaps

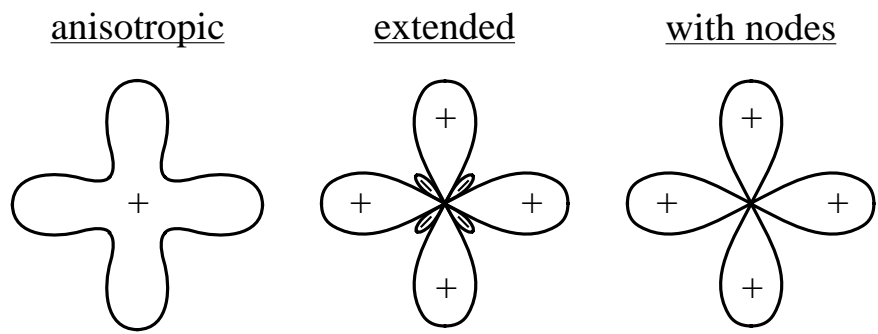

*Figure 6.44.* Possible types of s-wave symmetry of the pairing wavefunction in the cuprates, shown in real space: (a) anisotropic, (b) extended and (c) with nodes. The case (c) is intermediate between (a) and (b).

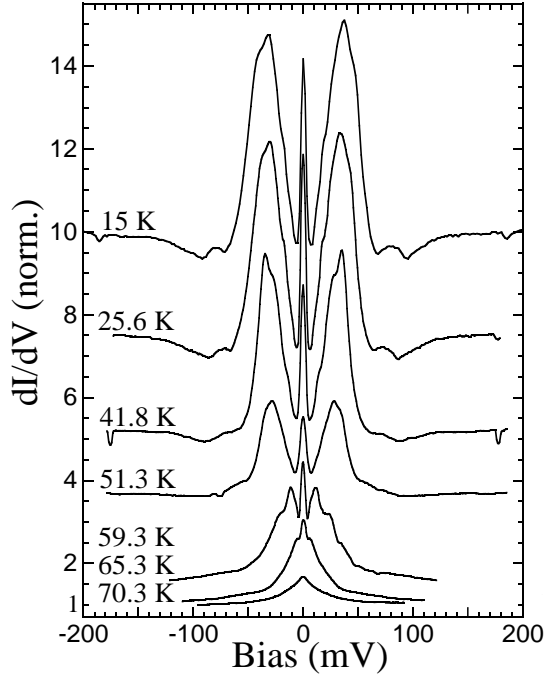

*Figure 6.45.* Temperature dependence of SIS-junction tunneling conductance obtained in a Ni-doped Bi2212 single crystal with $T_c = 75$ K [19]. The conductance scale corresponds to the 70.3 K spectrum, the other spectra are offset vertically for clarity. The peak at zero bias is due to the Josephson current in the junction.

in Bi2212:

$$\frac{2\Delta_p}{k_B T_{\Delta_p}} \simeq \frac{35\,\text{meV}}{5.83\,\text{meV}} \simeq 6\,(\pm\,0.2)\ \text{and} \qquad (6.68)$$

$$\frac{2\Delta_c}{k_B T_c} \simeq \frac{35\,\text{meV}}{6.45\,\text{meV}} \simeq 5.43. \qquad (6.69)$$

This example is also instructive because it helps to visualize gapless superconductivity in unconventional superconductors. In this example, the Cooper pairs do not exist between $T/T_c = 0.91$ and 1 near the Ni atom, while the Ni spin participates in local spin fluctuations. In Fig. 6.45, one can see that the 70.3 K conductance does not have quasiparticle peaks, but it exhibits a small zero-bias peak due to the Josephson current. This means that, in the Ni vicinity, between $T/T_c = 0.91$ and 1, the phase coherence is sustained. In this case, superconductivity is **locally** gapless. Thus, in the interval $T/T_c = 0.91$ and 1, all Cooper pairs (bisolitons) passing by the Ni atom are broken, while the fluctuations of Cu spins are not interrupted by the Ni atom. Thus, gapless superconductivity can arise in unconventional superconductors but exclusively locally. This is in contrast to conventional superconductors where gapless superconductivity can exist in a whole sample (see Chapter 2). From this example in Fig. 6.45, one can also have a feeling for the magnetic origin of the phase coherence mechanism in Bi2212.

### 3.11 Quantum critical point and the condensation energy

In cuprates, the doping level $p = 0.19$ is a quantum critical point where magnetic fluctuations are the strongest. Since spin fluctuations mediate the phase coherence in the cuprates, superconductivity at low temperature is the most robust at this doping level, $p = 0.19$, and not at $p = 0.16$. The superconducting condensation energy as a function of doping has a maximum at $p = 0.19$, as shown in Fig. 6.46. The maximum values of $U_0$ for YBCO and Bi2212 are $U_{0,max} = 2.6$ J/g atoms and $U_{0,max} = 2$ J/g atoms, respectively. From Fig. 6.46, superconductivity in the underdoped region is very weak. At $p = 0.19$, the superconducting-phase fraction is a maximum as well, as sketched in Fig. 6.8b. The temperature/energy scale $T_{MT}$ in Fig. 6.11 starts/ends in the quantum critical point. Hence, it has the magnetic origin.

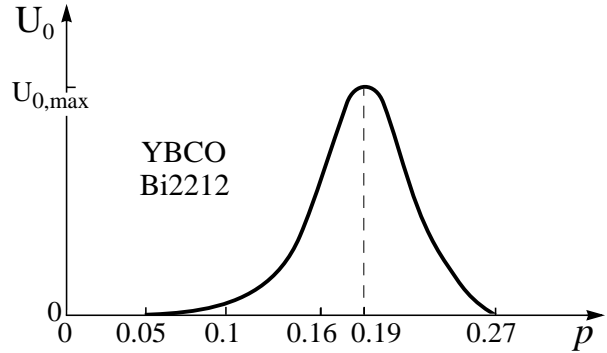

*Figure 6.46.* Superconducting condensation energy $U_0$ as a function of doping level, obtained in Bi2212 and YBCO by heat-capacity measurements (see references in [19]).

### 3.12 Effective mass anisotropy

Because of a layered structure of the cuprates, quasiparticles move much more easily in the $CuO_2$ planes than between the planes. Thus, anisotropy of the crystal structure of the cuprates affects transport properties. To account for the anisotropy, it is conventionally agreed that the effective mass changes with crystal direction. Instead of being a single-valued scalar $m$, the effective electron mass becomes a tensor. In the cuprates, to a good approximation, the effective electron mass is a diagonal tensor, and the in-plane effective masses have similar values, $m_a \approx m_b$. The value of the in-plane effective mass in the cuprates is slightly larger than the electron mass $m_e$ by a factor of between four and five: $m_{ab} \simeq (4\text{–}5)m_e$.

Anisotropy is defined by the ratio of the effective mass of quasiparticles in the various directions, $\gamma^2 = \frac{m_c}{m_a}$. In YBCO, the effective mass ratio is about $\gamma^2 \approx 30$. LSCO exhibits somewhat higher anisotropy, $\gamma^2 \approx 200$, while Bi- and Tl-based cuprates are much more anisotropic than YBCO: in Bi- and Tl-compounds, the ratio is about 50 000. Such a large anisotropy, which is totally foreign to conventional superconductors, means that electrons can barely move in the $c$-axis direction, and the cuprates are effectively two-dimensional.

## 3.13    Penetration depth

The penetration depth $\lambda$ is one of the most important parameters of the superconducting state because $\lambda$ directly relates to the superfluid density as $n_s \propto 1/\lambda^2$. Table 3.6 lists the penetration-depth data for some cuprates. In the table, one can see that, in the cuprates, $\lambda$ is very large, particularly, in the $c$-axis direction, meaning that $n_s$ is very low.

In the Uemura plot in Fig. 3.6, the $T_c$ value depends linearly on the superfluid density in underdoped cuprates, and this dependence is universal for all superconducting cuprates. As the doping level increases, $T_c$ first saturates and, in the overdoped region, decreases, making a "boomerang path", as shown in the lower inset of Fig. 3.6. Thus, in the overdoped region, the superfluid density falls with increased doping. The Uemura plot shows that the optimum doping level is different for different cuprates.

Figure 6.47 shows the doping dependence of the penetration depth in LSCO; thus, a sort of "inverted" Uemura plot. One notes from Figs 3.6 and 6.47, the penetration depth as a function of doping attains a minimum not in the optimally doped region but in the slightly overdoped region. In Bi2212, the in-plane penetration depth has the same trend.

It is worth noting that, in YBCO, due to the presence of chains, the penetration depth is smaller along the chains than that along the $a$ axis: in YBCO with $T_c = 93$ K, $\lambda_a = 1550\text{–}1600$ Å and $\lambda_b = 800\text{–}1000$ Å. The value $\lambda_{ab} = 1450$ Å listed in Table 3.6 for YBCO, represents the value of $\sqrt{\lambda_a \lambda_b}$.

## 3.14    Critical fields and current

All superconducting cuprates are type-II superconductors. Hence, they have two critical magnetic fields: $H_{c1}(0)$ and $H_{c2}(0)$. In Table 3.6, one observes that the magnitudes of the second critical magnetic field, $H_{c2}(0)$, are extremely high in the cuprates. At the moment of writing, the cuprates exhibit the largest values of $H_{c2}(0)$ amongst all unconventional superconductors.

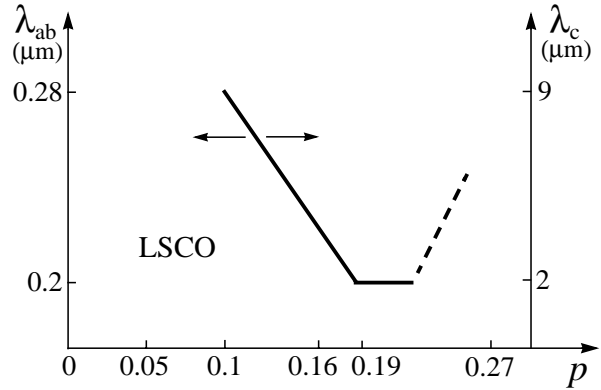

*Figure 6.47.* Absolute values of the in-plane $\lambda_{ab}$ and out-of-plane $\lambda_c$ magnetic penetration depth as a function of doping for LSCO (see references in [19]). The dashed line shows tendency of $\lambda_{ab}$ at high doping, expected from Fig. 3.6.

Furthermore, due to a highly anisotropic structure of the cuprates, there is a huge anisotropy of the critical magnetic fields applied parallel and perpendicular to the $CuO_2$ planes. Thus, in the cuprates, there are four different critical magnetic fields: $H_{c1,\parallel}(0)$, $H_{c1,\perp}(0)$, $H_{c2,\parallel}(0)$ and $H_{c2,\perp}(0)$. The symbols $\parallel$ and $\perp$ denote the critical value of $H$ applied parallel and perpendicular to the $CuO_2$ planes, respectively. In the cuprates, $H_{c2}$ is much larger when the field is applied parallel to the $CuO_2$ planes than that applied perpendicular to the planes. This is because most of the conduction is in the planes: a magnetic field applied parallel to the planes is not very effective in destroying superconductivity within the planes.

In the cuprates, the lower critical fields $H_{c1,\parallel}$ and $H_{c1,\perp}$ are very small. For example, in YBCO $H_{c1,\parallel} \sim 2 \times 10^{-2}$ T and $H_{c1,\perp} \sim 5 \times 10^{-2}$ T; in Hg1223, $H_{c1,\parallel} \approx 3 \times 10^{-2}$ T. It is interesting that the anisotropy in $H_{c1}$ has the sign opposite to that in $H_{c2}$: $H_{c2,\perp} < H_{c2,\parallel}$ but $H_{c1,\perp} > H_{c1,\parallel}$.

It is worth noting that the majority of $H_{c2}$ values in Table 3.6 are approximate because they are extrapolated from the resistivity data obtained near $T_c$. The magnetic fields accessible in the laboratory conditions are only of the order of 30 T. Secondly, in metallic superconductors described by the BCS theory, $H_{c2} \propto T_c^2$. In the cuprates, the relation is found to be different: $H_{c2} \propto T_c^{\sqrt{2}}$, determined in the cuprates with low $T_c$.

The critical current in layered cuprates is also very anisotropic. The highest values of the critical current $J_c$ were obtained in epitaxial thin films of YBCO. At liquid helium temperature, the critical current in the $ab$ plane is almost $10^8$ A/cm$^2$ and, along the $c$ axis, is of the order of $10^5$ A/cm$^2$.

## 3.15    Coherence length and the size of a Cooper pair

As defined in Chapter 2, the coherence length $\xi_{GL}$ is determined by variations of the order parameter $\Psi(\mathbf{r})$, whilst the Cooper-pair size $\xi$ is related to the wavefunction of a Cooper pair, $\psi(\mathbf{r})$. While the coherence length depends on temperature, $\xi_{GL}(T)$, the Cooper-pair size is temperature-independent. In contrast to conventional superconductors, the order parameter and the Cooper-pair wavefunction in all unconventional superconductors are independent of one another. Therefore, *generally speaking*, in unconventional superconductors, $\xi_{GL} \neq \xi$ at any temperature.

Let us consider first the in-plane $\xi_{GL,ab}$ and $\xi_{ab}$. Since the Cooper pairs in the cuprates reside into the $CuO_2$ planes, the size of a Cooper pair is, by definition, an in-plane characteristic and thus $\xi \equiv \xi_{ab}$. The magnetic field $H_{c2,\perp}$ directly relates to the in-plane coherence length $\xi_{GL,ab}$ through Eq. (2.67). As was analyzed in [19], *for the cuprates*, the field $H_{c2,\perp}$, in fact, yields the value of $\xi$ or, at least, a value which is very close to $\xi$. How is it possible? This fact may indeed look odd because, by applying a magnetic field to a system which is characterized by two coupling strengths, it is anticipated that the weaker "bond" will first be suppressed. In the cuprates, depending on the doping level, the strength of the electron-phonon interaction ($\sim 0.6$ eV) can be four times stronger than the strength of magnetic interaction ($J \sim 0.15$ eV). Experimentally, however, for cuprates the magnetic field $H_{c2,\perp}$ yields the value of $\xi$ which is mainly determined by the electron-phonon interaction. As discussed above, the mechanism of in-plane phase coherence in the cuprates is not purely magnetic: the direct wavefunction hopping largely contributes to the onset of in-plane phase coherence. Therefore, even if the in-plane spin fluctuations are suppressed, there will always be superconducting patches due to the direct wavefunction hopping. It is then obvious why, in the cuprates, the field $H_{c2,\perp}$ yields the value of $\xi$ and not $\xi_{GL,ab}$. The doping dependence of $\xi$ in Bi2212 is shown in Fig. 6.25.

To obtain $\xi_{GL,ab}$ in the cuprates, another method has been proposed. In LSCO, $\xi_{GL,ab}$ was determined by measuring the vortex-core size [50]. Figure 6.48 depicts the doping dependences of $\xi_{GL,ab}$ and $\xi$ in LSCO, obtained at low temperature. The dependence $\xi(p)$ was obtained through $H_{c2}$ as described in the previous paragraph. In Fig. 6.48, one can see that the dependence $\xi_{GL,ab}(p)$ has an inverted bell-like shape similar to the dependence $\lambda(p)$ in Fig. 6.47. Both these dependences directly follow from the fact that the dependence $\Delta_c(p)$ has a bell-like shape (see Fig. 10.14 in [19]). At $p \simeq 1/8$, $\xi_{GL,ab}(p)$ has a kink related to the $\frac{1}{8}$ anomaly (see Chapter 3). At $p \simeq 0.05$, the value of $\xi_{GL,ab}$ in LSCO is about 70 Å. In Fig. 6.48, one can see that, in the overdoped region, $\xi_{GL,ab} \simeq \xi$.

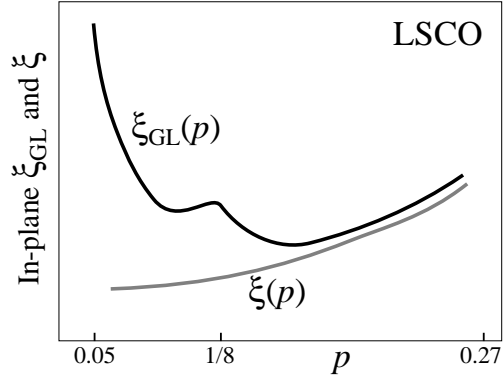

*Figure 6.48.* Sketch of doping dependences of in-plane $\xi_{GL}$ and $\xi$ in LSCO, obtained at low temperature [50].

In the cuprates, the in-plane magnetic correlation length is sufficiently large. For example, in LSCO, the value of in-plane magnetic correlation length is $\xi_m > 400$ Å [51]. If we substitute this value into Eq. (2.67), we obtain that $H_{c,m} < 0.2$ T. This value is of the same order of magnitude as $H_{c1,\perp} \sim 0.05$ T in YBCO (see above). So, it is possible that, in the cuprates, the in-plane magnetic correlation length determines $H_{c1,\perp}$. However, one must realize that $\xi_m$ has no relations with $\xi_{GL,ab}$.

The in-plane coherence length $\xi_{GL,ab}$ is determined by a length of dynamic magnetic characteristics, and not static ones as $\xi_m$. In the cuprates, $\xi_{GL,ab}$ relates most likely to the coherence length of incommensurate spin fluctuations which are in fact commensurate locally but manifest themselves in INS spectra as incommensurate. For example, in underdoped YBCO, the minimum coherence length of the magnetic resonance peak is practically doping-independent and is about 16 Å. At the same time, the minimum coherence length of the incommensurate spin fluctuations decreases from 35 to 24 Å as the doping level increases [52]. Thus, in underdoped YBCO, the doping dependence of minimum coherence length of incommensurate spin fluctuations is similar to the dependence $\xi_{GL,ab}(p)$ for LSCO, shown in Fig. 6.48, whilst the minimum coherence length of the magnetic resonance peak is doping-independent [52].

Consider now the out-of-plane $\xi_{GL,c}$ and $\xi_c$. The dependence $\xi_{GL,c}(p)$ must correlate with $\xi_{GL,ab}(p)$ shown in Fig. 6.48. In contrast, $\xi_c$ should increase as the doping level increases (see Fig. 10.15b in [19]) because, as the doping level increases, the cuprates become more three-dimensional. For cuprates, the values listed in Table 3.6 are the values obtained by using Eq. (2.68) and, formally, correspond to $\xi_{GL,c}$. However, it seems that these values are the $\xi_c$ values. Does this mean that, in the cuprates, the direct interlayer hopping also contributes to the onset of long-range phase coherence? This may be so, especially in the overdoped region.

## 3.16    Resistivity and the effect of the magnetic field

Consider temperature dependences of in-plane and out-of-plane resistivities in hole-doped cuprates at different doping levels. As schematically shown in Fig. 6.49, in the *undoped* region, the in-plane $\rho_{ab}$ and out-of-plane $\rho_c$ resistivities both are semiconducting; that is, the resistivities first fall with decreasing temperature, attaining their minimum values, and then sharply increase at low temperatures. As was emphasized above, the steep rise of the resistivities is due to a charge ordering along the charge stripes. The difference between the absolute values of $\rho_{ab}$ and $\rho_c$ is a few orders of magnitude. For example, in YBCO, $\rho_c/\rho_{ab} \sim 10^3$. It is important to note that $\rho_{ab}$ and $\rho_c$ attain their minimum values at different temperatures.

In the *underdoped* region, the in-plane and out-of-plane resistivities passing through their local minimum value both attain a maximum and then fall, vanishing below $T_c$, as shown in Fig. 6.49. The absolute values of the resistivities decrease in comparison with those in the undoped region, and the ratio $\rho_c/\rho_{ab}$ decreases as well. This means that, as the doping level increases, the two-dimensional cuprates become *quasi*-two-dimensional. For example, in LSCO ($x = 0.06$), the ratio is $\rho_c/\rho_{ab} = 4 \times 10^3$ and decreases to $\rho_c/\rho_{ab} \sim 10^2$ at $x = 0.28$. The sharp fall in $\rho_c$ and $\rho_{ab}$ occurring due to the transition into the superconducting state literally interrupts the rise corresponding to the insulating charge-ordering state.

In Fig. 6.49, near the *optimally* doped region, the in-plane resistivity above the critical temperature is now almost linear. However, the out-of-plane resistivity remains similar to that in the underdoped region, shifting to high temperatures and to low absolute values. Thus, in the optimally doped region, the in-plane resistivity is almost metallic, while the out-of-plane resistivity still exhibits the semiconducting behavior. In high-quality single crystals, the extension of the $\rho_{ab}(T)$ dependence passes through zero as that in Fig. 6.50a.

In the *overdoped* region, both $\rho_{ab}$ and $\rho_c$ become metallic, as shown in Fig. 6.49. Notably, $\rho_c$ in Bi2212 is probably the only exception from the general tendency. Figure 6.50 shows the in-plane and out-of-plane resistivities in an overdoped Bi2212 single crystal. As one can see in Fig. 6.50b, above $T_c \simeq 80$ K the out-of-plane resistivity remains semiconducting even in the overdoped region.

In the cuprates, there is also a weak in-plane anisotropy: $\rho_a$ and $\rho_b$ are not exactly the same. If, in YBCO, the in-plane anisotropy $\rho_a/\rho_b$ at 300 K varying from 1.23 in the underdoped region to 2.5 in the optimally doped region is principally due to the presence of the CuO chains, whilst in other cuprates, the weak in-plane anisotropy is due to self-organized charge stripes. For example, in undoped LSCO, the $\rho_b/\rho_a$ ratio increases with lowering the temperature, attaining 1.4 at 4.2 K.

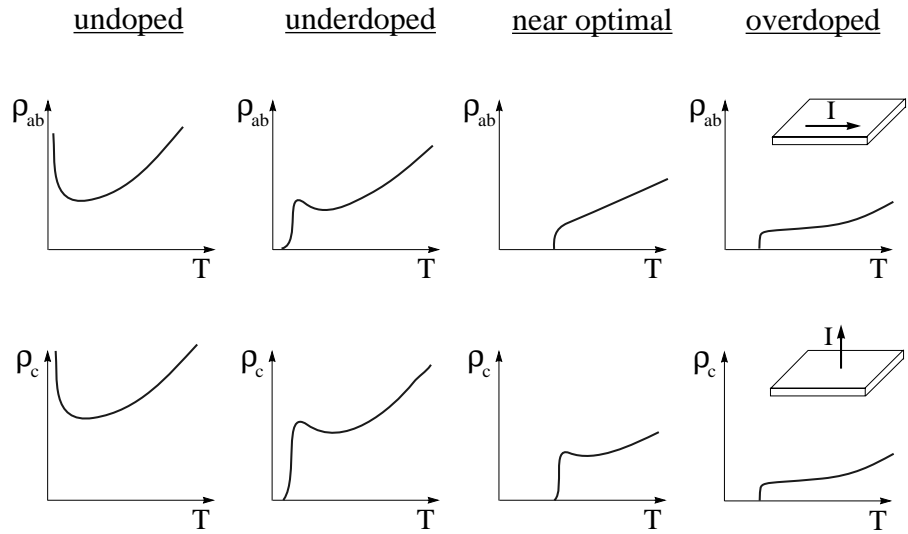

*Figure 6.49.* Schematic overview of transport properties of the cuprates at different dopings. In-plane resistivity is shown at the top, and out-of-plane resistivity at the bottom. Insets depict the direction of the current in each case.

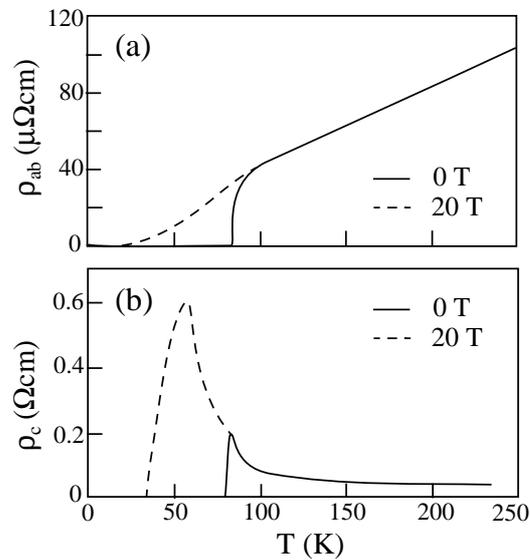

*Figure 6.50.* Temperature dependences of (a) in-plane and (b) out-of-plane resistivities in slightly overdoped Bi2212 with $T_c \simeq 80$ K (see references in [19]). The solid lines show the resistivities in zero magnetic field, and the dashed lines in a $dc$ magnetic field of 20 T.

The difference between conventional and unconventional superconductors can be illustrated by the following example. In conventional superconductors, by applying a sufficiently strong magnetic field, the part of the resistivity curve corresponding to the transition into the superconducting state remains steplike but is shifted to lower temperatures. The same takes place in half-conventional superconductors (see the next chapter). Contrary to this, the transition width in unconventional superconductors becomes broader with increasing magnetic field, implying that at temperatures just below $T_c(H = 0)$, there are large phase fluctuations. In layered unconventional superconductors, the transition widths in resistivity become broader in both directions, along and perpendicular to the layers.

The effect of an applied magnetic field on in-plane and out-of-plane resistivities in Bi2212 is shown in Fig. 6.50. An applied magnetic field smears the transition into the superconducting state, as shown in Fig. 6.50a. Such a behavior in resistivity is typical for the cuprates. If the magnitude of applied magnetic field is larger than $H_{c2}$, no transition into the superconducting state will be observed. Instead, the low-temperature parts of $\rho_{ab}$ and $\rho_c$, camouflaged by the onset of the superconducting phase in zero magnetic field, will be revealed. This trend can be seen in the behavior of $\rho_c$ shown in Fig. 6.50b.

It is worth noting that, in LSCO with $x = 1/8$, upon applying a magnetic field, the part of the resistivity curve corresponding to the transition into the superconducting state remains steplike, and is obviously shifted to lower temperatures [53]. Thus, the magnetic-field trend of resistivity in the cuprate LSCO ($x = 1/8$) is similar to that in conventional superconductors. This indicates that the magnitude of the energy gap $\Delta_c$ is small in LSCO with $x = 1/8$, meaning that spin fluctuations in this "anomalous" LSCO are much less dynamic that those, for example, in optimally doped LSCO.

## 3.17   Crystal structure and $T_c$

In conventional superconductors, there are no important structural effects. This, however, is not the case for the cuprates. Since superconductivity in the cuprates occurs in the $CuO_2$ planes planes, the structural parameters of these planes affect $T_c$ the most. The geometry of a $CuO_2$ plane is defined by the following factors: the length of the Cu–O bond; the degree of an orthorhombic distortion from square, and the degree of deviation from a flat plane (a buckling angle). The $T_c$ dependence on the Cu–O length has a bell-like shape [19]. And hence, for superconductivity, the length of the Cu–O bond in the $CuO_2$ planes has a certain optimum value.

The buckling angle of a $CuO_2$ plane is defined as the angle at which the plane oxygen atoms are out of the plane of the copper atoms. At fixed doping level, the highest maximum $T_c$ corresponds to the smallest maximum buckling angle. The highest critical temperature $T_c = 135$ K is observed in mercury compounds which have perfectly flat $CuO_2$ planes. The orthorhombic distortion is defined by the parameter $\frac{b-a}{b+a}$, where $a$ and $b$ are the lattice constants. All cuprates with the high critical temperatures ($> 100$ K) have tetragonal crystal structure. Therefore, for increasing $T_c$, the degree of orthorhombic distortion should be as small as possible. Thus, at fixed doping level, the highest $T_c$ will be observed in a cuprate with flat and square $CuO_2$ planes.

Consider now other parameters of the crystal structure outside the $CuO_2$ planes, which affect the critical temperature. Is there a correlation between the c-axis lattice constant and $T_c$? In the cuprates with two or more $CuO_2$ layers, there are two interlayer distances: the distance between $CuO_2$ layers in a bi-layer (three-layer, four-layer) block, $d_{in}$, and the distance between the bi-layer (three-layer, four-layer) blocks, $d_{ex}$. Usually $d_{in} + d_{ex} \simeq 15$ Å, $d_{in} \approx$ 3–6 Å and $d_{ex} \approx$ 9–12 Å. The intervening layers between the group of the $CuO_2$ planes are semiconducting or insulating. Transport measurements in the $c$-axis direction show that the $c$-axis resistivity depends exponentially on $d_{ex}$; however, there is no correlation between $T_c$ and $d_{ex}$. For example, in the infinite-layer cuprate (Sr, Ca)$CuO_2$, the distances $d_{in}$ and $d_{ex}$ are equal and short, $d_{in} = d_{ex} \simeq 3.5$ Å; however, $T_c \simeq 110$ K. Thus, the "optimal" region of the $d_{in}$ and $d_{ex}$ parameters is rather wide. Comparing three superconducting one-layer cuprates LSCO ($d_{ex} \simeq 6.6$ Å and $T_{c,max} = 38$ K), Hg1201 ($d_{ex} \simeq$ 4.75 Å and $T_{c,max} = 98$ K) and Tl2201 ($d_{ex} \simeq 11.6$ Å and $T_{c,max} = 95$ K), one can see that there is no correlation between $T_c$ and $d_{ex}$. The large difference in $T_c$, for example, between LSCO and Tl2201, is not due to the difference between the $c$-axis distances in these cuprates, but due to the difference in the structural parameters of the $CuO_2$ planes, which were discussed above.

The intervening layers can be divided into two categories: "structural" layers and charge reservoirs. The structural layers, like Y in YBCO, play a minor role in the variation of $T_c$. At the same time, the charge reservoirs make a large impact on $T_c$. Different charge reservoirs have different polarized abilities and different abilities to polarize other ions: the higher ones are the better. The distance between the charge reservoirs and the $CuO_2$ planes is also important: the shorter one is the better. In addition, the charge reservoirs also play the role of the structural layers. For example, in LSCO, the critical temperature is very sensitive to lattice strains induced by substituting Sr for different cations having different ionic radius. Thus, the intervening layers can affect the electronic structure of the $CuO_2$ planes drastically, especially in single-layer compounds.

It is important to note that an *isolated* $CuO_2$ layer will not superconduct. Even a $CuO_2$ layer situated on the surface of a crystal (this happens occa-sionally) will be semiconducting. This can easily be understood. In order to become superconducting, a $CuO_2$ layer must be structurally stabilized from both sides, above and below. Of course, this is a necessary but not sufficient condition.

## 3.18    Effect of impurities

One of the crucial tests for the superconducting state with a specific mechanism is how magnetic and non-magnetic impurities affect it. In conventional superconductors, non-magnetic impurities have a small effect on $T_c$, while magnetic impurities drastically affect it. In contrast, magnetic and non-magnetic impurities have the opposite effect on superconductivity mediated by magnetic fluctuations.

The coherence length of conventional superconductors is very large. Therefore, the effects of an impurity on superconductivity on a microscopic scale and a macroscopic scale are practically the same. This, however, is not the case for the superconducting cuprates, the coherence length of which is very short. Consequently, in the cuprates as well as in all unconventional superconductors, one must consider separately the effects of magnetic and non-magnetic impurities on a macroscopic and microscopic scale.

On a macroscopic scale, magnetic and non-magnetic impurities have a *similar* effect on $T_c$ in the cuprates. The partial substitution of Fe, Ni and Zn for Cu affects $T_c$ similarly with $dT_c/dx \approx -$ 4–5 K/at.%, independently of the substitutional element. An exception to this rule is Zn-doped YBCO, where Zn suppresses $T_c$ three times faster (–12 K/at.%) than Ni does for example. Experimentally, Zn atoms occupy not only Cu sites in the $CuO_2$ planes, but also Cu-chain sites. The effect of Zn-on-chain location is that Zn interrupts the phase coherence between nearest $CuO_2$ planes.

In the cuprates, such an effect on a macroscopic scale is because superconductivity in all unconventional superconductors occurs due to phonons *and* spin fluctuations. Thus, a magnetic impurity affects locally the pairing, but does not alter much local spin fluctuations. In contrast, a non-magnetic impurity affects locally spin fluctuations, but does not alter much the pairing. So, magnetic and non-magnetic doped atoms modify different "components" of unconventional superconductivity.

On a microscopic scale, magnetic and non-magnetic impurities cause very different effects on their local environment. Tunneling measurements performed above Zn and Ni impurities situated in the $CuO_2$ planes show that Zn creates voids around it, suppressing locally the superconducting state. The local Zn effect on superconductivity is reminiscent of the voids in swiss cheese. In contrast, a magnetic Ni atom has surprisingly little effect on its local environment in the $CuO_2$ planes: superconductivity is not interrupted at a Ni site.

The main conclusion from these remarkable results is that, in the $CuO_2$ planes, there is magnetically-mediated superconductivity.

Another surprising effect produced locally by a nonmagnetic Zn atom is that Zn induces a local magnetic moment of $0.8\mu_B$ either in hole-doped cuprates or electron-doped NCCO, where $\mu_B$ is the Bohr magneton. Magnetic moments on all Cu sites around a Zn atom have a staggered order; thus a nonmagnetic Zn atom does not destroy local antiferromagnetic correlations, but enhances them. Logically, magnetically mediated superconductivity should be enhanced around Zn atoms as well. However, this is not the case since Zn induces effective magnetic moments on neighboring Cu sites which are *quasi-static*, and they cannot participate in dynamic spin fluctuations. Magnetic Fe and Ni atoms locally induce an effective magnetic moment of $4.9\mu_B$ and $0.6\mu_B$ in hole-doped cuprates, and $2.2\mu_B$ and $2\mu_B$ in electron doped NCCO, respectively. Thus, in hole-doped cuprates, Ni located in the $CuO_2$ planes reduces slightly the effective magnetic moments on neighboring Cu spins.

The charge distribution in superconducting cuprates is inhomogeneous both on a microscopic and a macroscopic scale: charge-stripe domains always coexist either with insulating antiferromagnetic domains or conducting Fermi-sea domains, shown in Fig. 6.12. Therefore, the same foreign atom substituting Cu in the $CuO_2$ planes will eventually produce different effects on superconductivity, depending on its location—in an insulating, charge-stripe or conducting domain. However, it is reasonable to assume that any impurity or lattice defect will attract charge-containing domains—depending on the doping level—either charge-stripe or conducting domains. Since most studies of Cu substitution have been carried out in underdoped, optimally doped and slightly overdoped regions of the phase diagram, the conclusions made in these studies, first of all, reflect the effect on superconductivity by impurities located in charge-stripe domains.

## 3.19 Chains in YBCO

On the nanoscale, chains in YBCO are insulating at low temperatures, having a well-defined $2k_F$-modulated charge-density-wave order, where $k_F$ is the momentum at the Fermi surface. The CDW on the chains was clearly observed in tunneling and nuclear quadrupole resonance measurements. On a macroscopic scale, the chains conduct electrical current by solitons which were directly observed on the chains by tunneling measurements.

## 3.20 Superconductivity in electron-doped cuprates

The single-layer superconducting cuprates $Nd_{2-x}Ce_xCuO_4$, $Pr_{2-x}Ce_xCuO_4$ and $Sm_{2-x}Ce_xCuO_4$ are electron-doped. Thorough analysis of different types of measurements performed in the electron-doped cuprates suggests that the mechanism of superconductivity in these cuprates is similar to that in LSCO. However, there are specific features exclusive to the electron-doped compounds. Some superconducting characteristics of NCCO are listed in Table 3.6.

In the phase diagram of NCCO in Fig. 3.13, the antiferromagnetic and superconducting phases do not overlap. Recent $\mu$SR measurements performed in NCCO have demonstrated that the superconducting phase, in fact, enters into the antiferromagnetic phase [54]. Thus, in NCCO with low doping level, superconductivity and the antiferromagnetic ordering coexist. Even in optimally doped NCCO, superconductivity and antiferromagnetism coexist, as found by resent INS measurements [55]. Phase-sensitive measurements performed in the electron-doped cuprates detect a d-wave symmetry of the superconducting phase. At the same time, the presence of Cooper pairs with an s-wave wavefunction is also demonstrated by many measurements [56]. In NCCO with low electron concentration, two energy scales were clearly observed by tunneling measurements [56]. In addition, a pseudogap was also found in NCCO [19].

In these cuprates, doped electrons self-organize in a different way than holes that form charge stripes in the $CuO_2$ planes of hole-doped cuprates. It is suggested that doped electrons form charge stripes oriented along the diagonal direction relative to the –Cu–O–Cu–O– bonds in the $CuO_2$ planes. Such a charge ordering takes place in the nickelates. Experimentally, the strength of the electron-lattice coupling in the electron-doped cuprates is a few times weaker than the strength of the hole-lattice coupling [57]. This is, in fact, typical for all solids.

By applying a magnetic field in the electron-doped cuprates, the part of the resistivity curve, corresponding to the transition into the superconducting state remains steplike and is shifted to lower temperatures. As was discussed above, such a magnetic-field trend is similar to that in conventional superconductors. This fact indicates that spin fluctuations in these cuprates are less dynamic in comparison, for example, to those in hole-doped LSCO ($x = \frac{1}{8}$).

## 3.21    Superconductivity in alkali-doped $C_{60}$

As was discussed in Chapter 3, superconductivity in alkali-doped $C_{60}$ is unconventional. In principle, the mechanisms of superconductivity in the cuprates and the fullerides are similar: Phonons are responsible for the electron pairing, while spin fluctuations mediate the long-range phase coherence. Since the crystal structure of the fullerides, shown in Fig. 3.18, is simpler than that of the cuprates, it is much easier to visualize the processes of electron pairing and the onset of phase coherence in the fullerides than those in the cuprates. Hence, let us consider briefly the mechanism of superconductivity in the fullerides. Furthermore, we shall need this information in Chapters 8–10.

All the fullerides are electron-doped superconductors. In the Uemura plot of Fig. 3.6, one of the fullerides, $K_3C_{60}$, is situated amongst other unconventional superconductors. Thus, the density of charge carriers in the fullerides is very low. Superconductivity occurs when each $C_{60}$ molecule is doped, *on average*, by approximately three electrons. The coherence length in alkali-doped $C_{60}$ is short, $\sim 30$ Å, while the penetration depth is very large, $\sim 4000$ Å. The values of $H_{c2}$ in the fullerides are sufficiently large for electron-doped superconductors, $\sim 30$–55 T. Some superconducting characteristics of the fullerides are listed in Tables 2.2 and 3.7.

For $C_{60}$, there is evidence that phonons take part in the electron pairing. At the same time, all superconducting alkali-doped $C_{60}$ exhibit strong antiferromagnetic correlations due to alkali spins. Phonon effects in the fullerides are often masked by spin-fluctuation effects. It is important to note that single crystals of $C_{60}$ are structurally unstable.

Each $C_{60}$ ball is a complex organic molecule with an even number of conjugate bonds. As discussed in Chapter 1, in such molecules, there exists a superconducting-like state. In contrast to $\sigma$ electrons which are located close to the atomic nuclei, the $\pi$ electrons in such molecules are not localized near any particular atom, and they can travel throughout the entire molecular frame. Hence, such complex organic molecules are very similar to a metal: the framework of atoms plays the role of a crystal lattice, while the $\pi$ electrons that of the conduction electrons. In molecules with an even number of carbon atoms, the $\pi$ electrons form bound pairs analogous to the Cooper pairs in a superconductor. The pair correlation mechanism is principally due to two effects: (i) the polarization of the $\sigma$ core, and (ii) $\sigma - \pi$ virtual electron transitions. When such organic complex molecules are doped, added electrons create structural instabilities which travel throughout the entire molecular frame like dislocations. In doped organic complex molecules, the doped electrons tend also to be paired in order to minimize the free energy of the system. This reasoning is also valid for the $C_{60}$ molecules.

In a $C_{60}$ crystal, the $C_{60}$ balls are closely packed; thus, the doped electrons and the electron pairs can easily jump from one $C_{60}$ molecule to another.

In the fullerides, below a certain temperature, the alkali spins become locally ordered; they prefer an antiferromagnetic order. Below this temperature, the $C_{60}$ balls are "immersed" locally in the antiferromagnetic environment. The occurrence of *quasi-static* magnetic order, local or not, does not automatically lead to the onset of long-range phase coherence for electron pairs residing on $C_{60}$ molecules. Only sufficiently quick spin fluctuations are able to mediate the superconducting phase coherence. In the fullerides, it is most likely that dynamical spin fluctuations occur due to doped electrons and/or electron pairs *circulating* around $C_{60}$ molecules and/or *jumping* between $C_{60}$ molecules. Locally, they frustrate the magnetic environment, creating spin excitations. If in the cuprates, the charge stripes excite dynamical spin fluctuations, in the fullerides, the electron pairs themselves seem to be responsible for this. In some $C_{60}$ compounds, the $C_{60}$ balls are not static but rotate around their center of gravity.

## 3.22    Future theory

The future theory of unconventional superconductivity in the cuprates must deal with the following processes: the formation of fluctuating charge stripes and antiferromagnetic stripes between them; charge-stripe excitations and their pairing; spin excitations induced by fluctuating charge stripes, and the onset of phase coherence mediated by these spin excitations. The bisoliton theory and the spin-fluctuation theory can be used as a starting point, but both of them must be modified. In the framework of the bisoliton model, it is necessary to take into account the Coulomb repulsion. The spin-fluctuation model must deal with spin excitations different from magnons: excitons and/or solitons.

As discussed above, the bisoliton theory is a "one-dimensional" theory. Even the BCS theory intrinsically contains one dimensionality: in the framework of the BCS theory, the condition $\mathbf{k}_1 = -\mathbf{k}_2$ for the electron pairing reflects the presence of one dimensionality in the theory.

## 3.23    Two remarks

At the end of this chapter, it is worth touching upon two issues directly related to the mechanism of unconventional superconductivity in the cuprates. The first one concerns the presence of a magnetic resonance peak in LSCO. In INS studies performed in highly underdoped YBCO with a $T_c$ near 35 K, the magnetic resonance peak was barely visible in INS spectra [58]. Then, it is possible that the spin excitation manifesting itself as a magnetic resonance peak in INS spectra exists also in the LSCO cuprates, but it is extremely weak.

The second issue deals with a scenario of unconventional superconductivity in all members of the third group, which is slightly different from that presented in this chapter. In the following chapter, we shall discuss the mechanism of superconductivity in superconductors of the second group. The superconducting state in these materials is characterized by the presence of two interacting superconducting subsystems. One of them is one-dimensional and exhibits genuine superconductivity of unconventional type (i.e bisolitons), while superconductivity in the second subsystem being three-dimensional is induced by the first one and of the BCS type.

The evidence presented in [19] unambiguously indicates that (i) the Cooper pairs in the cuprates are soliton-like excitations and (ii) the energy gap $\Delta_c$ occurs due to spin fluctuations which are responsible for the long-range phase coherence. The scenario of unconventional superconductivity described in this chapter postulates that, in the cuprates, there is only one set of Cooper pairs—bisolitons, and the order parameter $\Psi$ is, in a sense, external for them, result-ing in the occurrence of an additional energy gap $\Delta_c$. In fact, the evidence presented in [19] also admits the existence of another scenario of superconductivity which assumes the presence of two interacting subsystems as those in superconductors of the second group. The first superconducting subsystem represents bisolitons, as described in this chapter. The second subsystem responsible for the long-range phase coherence is the electron pairs bound by spin fluctuations. In this case, the two energy gaps $\Delta_p$ and $\Delta_c$ coexist in "parallel" like those in superconductors of the second group.

The scenario of unconventional superconductivity described in [19] and in this chapter was chosen because a large number of experimental facts are in favor of this scenario. Since the second scenario of unconventional superconductivity can *in principle* be realized, it would be unfair categorically to deny such a possibility. There is perhaps a 1% chance that superconductivity with such a mechanism can exist (and was discussed in [59]).

The main purpose of this book is to discuss room-temperature superconductivity and how to synthesize a room-temperature superconductor. At this stage, it should already be obvious to the reader that a room-temperature superconductor must be a member of the third group of superconductors. Therefore, we need to know the actual mechanism of unconventional superconductivity. What is interesting is that all discussions and calculations presented in Chapters 8–10 are, in fact, valid for both these scenarios of unconventional superconductivity. So, for the practical realization, this remark is unimportant; however, this issue is very important from an academic point of view.

## 3.24  Tunneling in unconventional superconductors

Tunneling spectra of conventional superconductors contain information about one energy gap. Tunneling spectra of unconventional superconductors, for example, of the cuprates, may contain information about three energy gaps, namely, $\Delta_p$, $\Delta_c$ and a charge gap $\Delta_{cg}$. So, tunneling spectra of unconventional superconductors are much more complicated than those of conventional superconductors. An explanation of several features of tunneling spectra obtained in the cuprates was presented in the last chapter of [19]. Here we consider an explanation of another set of tunneling data representing a "three-piece puzzle."

As was discussed above, the Cooper pairs in unconventional superconductors can be excited; therefore, they can leave the superconducting condensate being still paired. This is in contrast to conventional superconductors in which the Copper pairs cannot be excited, having only two alternatives shown in Fig. 5.8: either to be a part of the condensate or to be broken. In unconventional superconductors, the third option for the Cooper pairs corresponds to the transition SC $\rightarrow$ B in Fig. 6.40, which requires a minimum energy of $2\Delta_c$. This $2\Delta_c$ amount of energy can be supplied by two tunneling electrons each having an energy of $\Delta_c$. In a sense, this case corresponds to tunneling of Cooper pairs, and was considered in detail in [19]. However there is another option: the minimum required energy of $2\Delta_c$ can be supplied by one tunneling electron. Here we discuss this case.

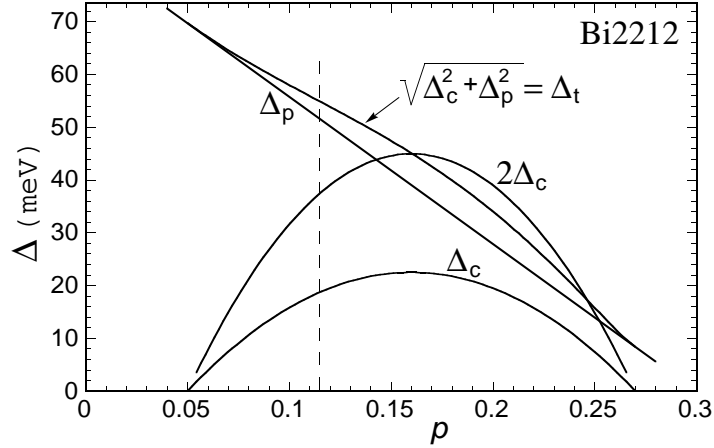

*Figure 6.51.* Phase diagram from Fig. 6.41 with an additional energy scale $2\Delta_c$, adapted to Bi2212 with $T_{c,max} = 95$ K. The dashed line marks a doping level of $p = 0.115$.

Figure 6.51 shows the phase diagram from Fig. 6.41 with an additional energy scale $2\Delta_c$. For this plot, the vertical axis is adapted to Bi2212 with $T_{c,max} = 95$ K; so one can grasp directly the gap values in Bi2212. Figure 6.52a depicts the variation of tunneling gap obtained in a SIN junction on the surface of underdoped Bi2212 with $T_c \simeq 79$ K along a line. In Fig. 6.52a, one can see that along a 140 Å line the gap value varies rapidly. Furthermore, some conductances exhibited double-gap structures. The main question is why does there exist such a variation of gap values? What do they correspond to?

The doping level of this Bi2212 sample is about $p = 0.115$ and denoted in Fig. 6.51 by the dashed line. If we use the gap values $2\Delta_c(0.115)$, $\Delta_p(0.115)$ and $\Delta_t(0.115)$ in Fig. 6.52a, shown by the dashed horizontal lines, then, one can see that the minimum values of the tunneling gap correspond exactly to $2\Delta_c(0.115)$. The upper values are slightly above $\Delta_t(0.115)$. Hence, the triangles in Fig. 6.52a seem to reflect the excitation of Cooper pairs (bisolitons); the circles correspond to the break of uncondensed bisolitons present always on the surface, and the squares reflect the break of condensed bisolitons. Such a rapid spatial variation of gap value in Bi2212 can be understood by referring to Fig. 6.38: spin fluctuations are not homogeneous on the surface of Bi2212. This means that the triangles in Fig. 6.52a indicate the patches with the phase coherence on the Bi2212 surface. Thus, at this stage, we have already made good progress in understanding the data; however, there is another piece of the puzzle.

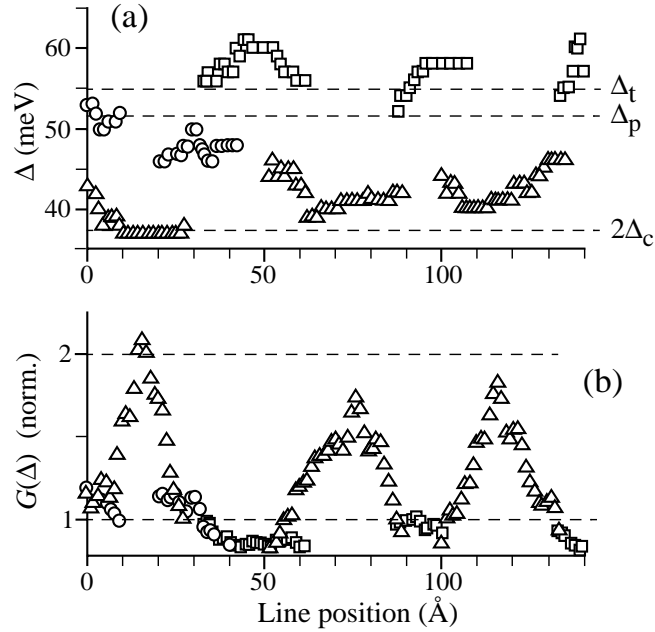

*Figure 6.52.* (a) Variation of tunneling gap values obtained in a SIN junction on the surface of underdoped Bi2212 with $T_c \simeq 79$ K ($p \simeq 0.115$) along a line. The dashed lines mark the values of $2\Delta_c$, $\Delta_p$ and $\Delta_t$ at $p \simeq 0.115$. (b) Relative height of main conductance peak at negative bias related to the data in plot (a). The data are taken from [60].

Figure 6.52b represents the relative height of the main conductance peak at negative bias, related to the data in Fig. 6.52a. In the two plots in Fig. 6.52, one can see that there is a correlation between the gap value and the height of the peak. In patches with the phase coherence, the height of the conductance peak is about twice higher than that in patches without the phase coherence. Why? Does it mean that, in the different patches, the hole concentration is different? The answer is no. It would be difficult to explain this correlation without seeing the corresponding conductances. Figure 6.53 depicts two conductances taken in the different patches. In this plot, one can see that the quasiparticle peaks in the conductance measured in a patch with phase coherence are indeed higher than those in the other conductance. However, one must also notice that these peaks are narrower than the other ones. If we subtract in both conductances a contribution made by a pseudogap (charge gap) and calculate amounts of quasiparticle excitations under the obtained curves, we would find that these amounts are approximately the same. This means that, in a first approximation, the hole concentration in the different patches is more or less the same.

Equation (5.71) contains the clue to the last piece of the puzzle. In tunneling measurements, the width of conductance peaks is inversely proportional to the lifetime of quasiparticles. For Bi2212, this means that quasiparticles related to the Cooper-pair excitations live about three time longer than quasiparticles created by the break of bisolitons. Therefore, one may conclude that, in the cuprates, spin fluctuations stabilize the bisolitons and extend their lifetime.

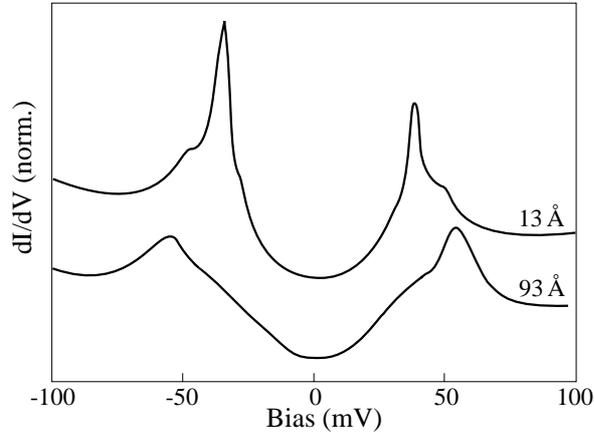

*Figure 6.53.* Two "extreme" conductances obtained in different locations along the line in Fig. 6.52, indicated by the numbers in plot [60]. The conductances are shifted vertically for clarity.

It is worth noting that this explanation of the data is fully in agreement with another set of tunneling data. In SIS-junction measurements performed in slightly overdoped Bi2212 single crystals, the value of the Josephson product $I_c R_n$ is the highest when a tunneling gap is $\sim \Delta_c$, and decreases as the tunneling gap increases [61, 19].

To conclude, the tunneling data presented in Figs. 6.52 and 6.53 are now understood; in addition, we have obtained useful information concerning the lifetime of quasiparticles in Bi2212.

Chapter 7

# SECOND GROUP OF SUPERCONDUCTORS: MECHANISM OF SUPERCONDUCTIVITY

This chapter is the shortest in the book because a potential content of this chapter is already presented in Chapters 5 and 6: one half in Chapter 5 and the other half in Chapter 6. The content of the present chapter is just a "bridge" between the two halves.

The second group of superconductors incorporates superconducting compounds which are *low-dimensional* and *non-magnetic*. The superconducting state in these materials is characterized by the presence of two interacting superconducting subsystems. One of them is low-dimensional and exhibits genuine superconductivity of unconventional type (i.e bisolitons), while superconductivity in the second subsystem which is three-dimensional is induced by the first one and of the BCS type. So, superconductivity in this group of materials can be called *half-conventional* or, alternatively, *half-unconventional*. Charge carriers in these superconductors are electrons. As a consequence, their critical temperature is limited by $\sim 40$ K and, in some of them, $T_c$ can be tuned. All materials of this group are type-II superconductors with a upper critical magnetic field usually exceeding 10 T.

## 1.    General description of the mechanism

Figure 7.1 presents a short summary of the mechanism of superconductivity in compounds of the second group. As was emphasized above, these materials are characterized by the presence of two subsystems. Below a certain temperature, the "nonlinear" Cooper pairs (bisolitons) are formed in a low-dimensional subsystem. Then, the electron pairing is induced into the second subsystem which is three-dimensional. In the second subsystem, the Cooper pairs are "linear", i.e. of the BCS type. So, the Cooper pairs are "unconventional" in the low-dimensional subsystem, while they are "conventional" in the three-dimensional subsystem. The electron-phonon interaction is responsible for electron pairing in both subsystems; however, the strength of this interaction is different in the two subsystems: it is moderately strong in the first subsystem, while it is weak in the second.

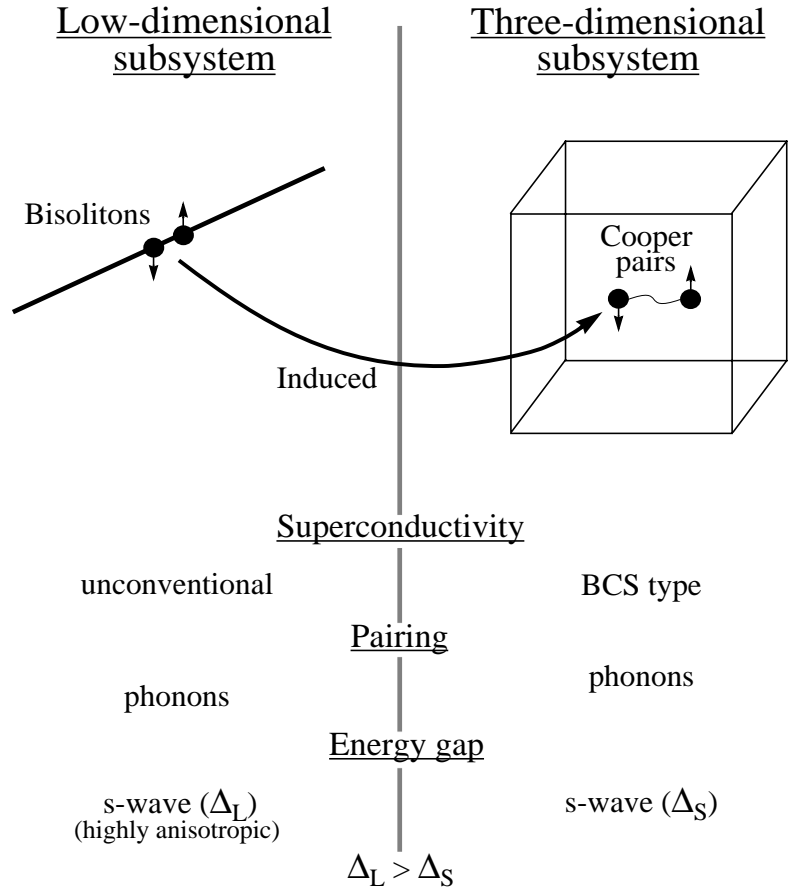

*Figure 7.1.* Short summary of the mechanism of superconductivity in compounds of the second group. In these materials, there are two interacting subsystems: one of them is low-dimensional, while the second is three-dimensional. The first subsystem exhibits genuine superconductivity of unconventional type (i.e bisolitons), whilst superconductivity in the second subsystem is induced by the first one and of the BCS type. For more details, see text.

The Cooper pairs of the first subsystem along are not able to establish the long-range phase coherence because their size is small and their density is low. In superconductors of the second group, the long-range phase coherence occurs due to the onset of phase coherence among the Cooper pairs of the second subsystem. Therefore, the order parameter of the superconducting state is the wavefunction of Cooper pairs of the second subsystem, as that in conventional superconductors.

The energy gaps in the two subsystems have both an s-wave symmetry typical for electron pairing due to phonons. They are both anisotropic; however, while the energy gap in the second subsystem is only slightly anisotropic, it is highly anisotropic in the first one and, probably even with nodes (like that in the cuprates). The maximum magnitude of the first gap is a few times larger than that of the second gap. This is typical for an induced superconductivity (see Chapter 4).

In terms of the band theory, the Fermi surface of superconductors of the second group consists of, at least, two disconnected sections. Each section corresponds to one of the two subsystems.

The process of the formation of low-dimensional Cooper pairs (bisolitons) is described in Chapter 6, and the process of the formation of conventional (three-dimensional) Cooper pairs is presented in Chapter 5. Since the onset of long-range phase coherence in superconductors of the second group occurs automatically due to the overlap of Cooper-pair wavefunctions in the second subsystem as that in conventional superconductors, this process does not require a separate description.

## 1.1   Effect of isotope substitution on $T_c$

In superconductors of this group, the critical temperature must be sensitive to the mass of some elements, similar to the effect of isotope substitution in conventional superconductors. However, the critical temperature can be insensitive to the mass of elements that form the low-dimensional subsystem. As discussed in Chapter 6, in a system with bisolitons, the isotope effect can be very small.

## 1.2   Effect of impurities on $T_c$

Since the electron-phonon interaction is responsible for electron pairing in superconductors of the second group, the effect of magnetic and non-magnetic impurities on the critical temperature is similar to that in conventional superconductors (see Chapter 5). Thus, magnetic impurities drastically suppress the superconducting transition temperature, whereas non-magnetic impurities do not alter $T_c$ much, if their concentration is relatively small.

## 1.3   Magnetic-field effect on resistivity

In conventional superconductors, by applying a sufficiently strong magnetic field, the part of resistivity curve corresponding to the transition into the superconducting state (see, for example, Fig. 2.1) remains steplike but is shifted to lower temperatures. The same effect takes place in superconductors of the second group because the onset of long-phase coherence in half-conventional superconductors is identical to that in conventional superconductors.

## 2.    MgB$_2$

Among superconductors of the second group, MgB$_2$ is the most studied one. Let us briefly discuss some of its superconducting properties. Some characteristics of MgB$_2$ have already been discussed in Chapter 3: Table 3.2 lists some of them.

In MgB$_2$, superconductivity occurs in the boron layers (see Fig. 3.3). So, the two subsystems in MgB$_2$ are both located into the boron layers. Band-structure calculations of MgB$_2$ show that there are at least two types of bands at the Fermi surface. The first one is a narrow band, built up of boron $\sigma$ orbitals, whilst the second one is a broader band with a smaller effective mass, built up mainly of $\pi$ boron orbitals.

The presence of two energy gaps in MgB$_2$ is a well documented experimental fact. The larger energy gap $\Delta_\sigma$ occurs in the $\sigma$-orbital band, the smaller gap $\Delta_\pi$ in the $\pi$-orbital band. The gap ratio $2\Delta/(k_B T_c)$ for $\Delta_\sigma$ is about 4.5. For $\Delta_\pi$, this ratio is around 1.7, so that, $\Delta_\sigma/\Delta_\pi \simeq 2.7$. Both energy gaps have an s-wave symmetry. The larger gap is highly anisotropic, while the smaller one is either isotropic or slightly anisotropic. The induced character of $\Delta_\pi$ manifests itself in its temperature dependence. Figure 7.2 depicts the temperature dependences of the two energy gaps, obtained in tunneling and Andreev-reflection measurements. In the plot, one can see that $\Delta_\sigma$ follows the temperature dependence derived in the framework of the BCS theory. At the same time, the temperature dependence of $\Delta_\pi$ lies below the BCS dependence at $T \to T_c$. For phonon-mediated superconductivity, this fact indicates that $\Delta_\pi$ is induced. Superconductivity in the $\pi$-band is induced either by interband scattering or Cooper-pair tunneling.

Superconductivity in MgB$_2$ is mediated by phonons. The boron isotope effect is sufficiently large, $\alpha \simeq 0.3$, while the Mg isotope effect is very small. Due to its layered structure, below $T_c$, MgB$_2$ has a highly anisotropic critical magnetic field: $H_{c2,\parallel}/H_{c2,\perp} \sim 7$. The muon relaxation rate in MgB$_2$ is about 8–10 $\mu s^{-1}$. In the Uemura plot (see Fig. 3.6), MgB$_2$ is literally situated between the large group of unconventional superconductors and the conventional superconductor Nb.

In MgB$_2$, the effect of B substitution on $T_c$ is well studied. The boron partial substitution by non-magnetic Al, C and Be leads to a decrease in $T_c$. The results of these experiments show that this decrease is mainly due to a structural transformation in the boron layers: the B–B distance decreases.

Finally, it is worth noting that MgB$_2$ is very similar to graphite both crystallographically and electronically. In MgB$_2$, each atom Mg donates two electrons to the boron subsystem. So, each boron acquires one electron and the electron configuration of a carbon atom: B$^-(2s^22p^2) \equiv$ C$(2s^22p^2)$. Thus, the B$^-$ sheets are electronically like graphite sheets. The main difference between MgB$_2$ and graphite is in the $c$-axis layer separation: it is about 15% shorter in MgB$_2$.

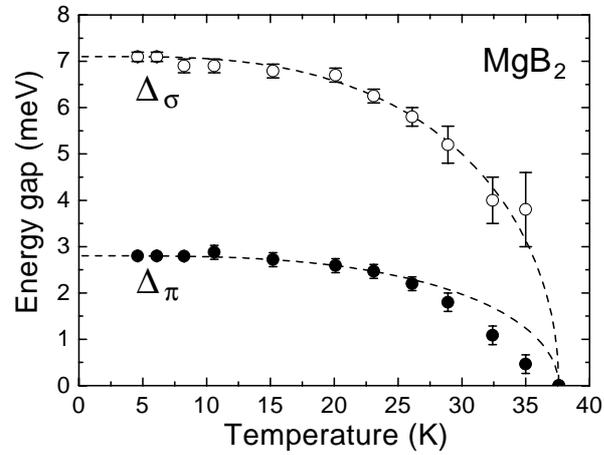

*Figure 7.2.* Temperature dependence of the two energy gaps in MgB$_2$, obtained in tunneling and Andreev-reflection measurements [62]. The dashed lines are the BCS temperature dependences.

# Chapter 8

# COOPER PAIRS AT ROOM TEMPERATURE

*What is now proved was once only imagined.*

—William Blake (1751 - 1827)

The ultimate goal of the last three chapters of the book is to discuss materials that superconduct at room temperature. However, before we discuss the materials, it is first necessary to discuss, from the physics point of view, the possibility of the occurrence of superconductivity at room temperature. Thus, in this chapter, we discuss the electron pairing at room temperature and, in the following chapter, the onset of long-phase coherence at room temperature. The materials will be discussed in Chapter 10.

As was mentioned in the Preface, the last three chapters of the book are mainly addressed to specialists interested in synthesizing a room-temperature superconductor and to researchers in the field of superconductivity.

According to the first principle of superconductivity (see Chapter 4), superconductivity requires electron pairing. Indeed, electron pairing is the keystone of superconductivity. Therefore, the first question that we must deal with is: can the Cooper pairs exist at room temperature? So, the purpose of this chapter is to show that the Cooper pairs can exist at room temperature. In fact, they do exist at temperatures much higher than room temperature.

As was discussed in Chapter 1, the expression "a room-temperature superconductor" is used here implying a superconductor having a critical temperature of $T_c \simeq 350$ K. From a practical point of view, it is much better, however, to have a superconductor with $T_c \approx 450$ K.

## 1.    Mechanism of electron pairing at room temperature

As described in the previous three chapters, only one mechanism of electron pairing is known at the moment of writing—only the electron-phonon interaction is capable of mediating the electron pairing in solids. There are two types of the electron-phonon interaction—linear and nonlinear. The linear interaction is weak, whilst the nonlinear one is moderately strong. It is obvious that, at high temperatures, only the nonlinear electron-phonon interaction can bind two electrons in a pair. This means that, in a room-temperature superconductor, the Cooper pairs will be represented by bisolitons having a small size $\xi$. For our ultimate goal, it is not important whether they are bound in real or momentum space (see Chapters 4 and 6).

### 1.1    Electrons versus holes

Generally speaking, for room-temperature superconductivity it is not important whether quasiparticles in the material are electron- or hole-like. However, it is an experimental fact that, in solids, the strength of electron-phonon interaction is a few times weaker than the strength of hole-phonon interaction. Therefore it is most likely that in a room-temperature superconductor the quasiparticles will be hole-like.

## 2.    Selection process by Nature

The matter was created by Nature with some order in mind. Physicists are interested in understanding the laws of this order. There are many ways to unravel Nature's secrets—by an observation, measurement, modelling, etc. One widely-used method is by applying knowledge accumulated in one domain to another one. This is exactly what we are going to do in this chapter.

The living matter, not by accident, is organic and water-based. "In order to create the living matter, Nature needed billions of years." This quotation is the first part of the prologue to this book. Indeed, we must realize that, during billions of years, Nature has selected materials to make us and other living creatures function. She has done a wonderful job. "This experience is unique, and we must learn from it." The last quotation is the second part of the prologue and one of the main points of this book.

During evolution, Nature has tried many materials to create the living matter. One of the main criteria for the material was that it must support an effective signal transfer. Nowadays, we know that, in the living matter, the signal transfer occurs due to charge (electron) transfer. In *some* cases, the electron transfer occurs in pairs with opposite spins (see Chapter 1). The quasiparticle pairing simplifies their propagation because, for quasiparticles, it is more profitable *energetically* to propagate together than separately, one by one (see Chapter 6). So, the electron pairing occurs in living tissues first of all because of an energy gain; the electron spin is a secondary reason for the pairing. However, we are more interested in the quasiparticle pairing because of spin.

This means that electron pairs in a singlet state exist at room temperature in the living matter. Therefore, we can use these materials for synthesizing a room-temperature superconductor. One should however realize that superconductivity does not occur in living tissues because it requires not only the electron pairing but also the onset of long-range phase coherence. We shall discuss the latter issue in the following chapter.

The idea to use organic compounds as superconducting materials is not new. In 1964, Little proposed that long organic chains can exhibit superconductivity with a mechanism different from the BCS scenario [2]. Later, Kresin and co-workers [23] showed that the superconducting-like state exists *locally* in complex organic molecules with conjugate bonds (see Fig. 1.3). Even, before Little's paper [2], Pullman and Pullman have already emphasized that "the essential fluidity of life agrees with the fluidity of the electronic cloud in conjugated molecules. Such systems may thus be considered as both the cradle and the main backbone of life" [24]. Davydov has devoted his life to studying the electron and energy transfer in organic chains [7, 10]. Even before the discovery of superconductivity in organic salts, Davydov has already known that, in some biological processes, the quasiparticles in living tissues are paired. Soon after the discovery of superconductivity in organic compounds in 1979, Davydov and Brizhik proposed the bisoliton model of superconductivity in organic materials [3]. Later, Davydov has used the bisoliton model to explain the phenomenon of high-$T_c$ superconductivity occurring in cuprates [7, 9, 10]. Thus, the superconducting state and *some* biological processes have, at least, one thing in common—in a sense they both do not like the electron spin created by Nature at the Big Bang.

## 2.1 Solitons and bisolitons in the living matter

Solitons are encountered in biological systems in which the nonlinear effects are often the predominant ones. For example, many chemical reactions in biological systems would not occur without large conformational changes which cannot be described, even approximately, as a superposition of the normal modes of the linear theory.

The shape of a nerve pulse was determined more than 100 years ago. The nerve pulse has a bell-like shape and propagates with the velocity of about 100 km/h. The diameter of nerves in mammals is less than 20 microns and, in a first approximation, can be considered as one-dimensional. For almost a century, nobody has realized that the nerve pulse is a soliton. Thus, all living creatures including humans are literally stuffed by solitons. Living organisms are mainly organic and, as a consequence, are bad conductors of electrical current—solitons are what keeps us alive.

The last statement is true in every sense: the blood-pressure pulse is some kind of solitary wave; the muscle contractions are stimulated by solitons, etc.

In organic materials, the charge transfer by bi-solitons is energetically profitable (see Chapter 6). In a bi-soliton, moving or static, two quasiparticles are held together by a local lattice deformation. The formation of tri-solitons is forbidden by the Pauli exclusion principle.

## 3.    Cooper pairs above room temperature

The presence of bisolitons in the living matter signifies that the Cooper pairs exist in some organic materials at 37 C $\simeq$ 310 K. Recently, living organisms have been found in extreme conditions: some survive without sunlight, some survive in water near the boiling point. Probably the most extreme case is the discovery of so-called black smokers or chimneys on the ocean bottom on a depth of about 2 km, and shrimps dwelling next to these chimneys. The measurement of water temperature at which the shrimps reside yielded $T \sim 270$ C. This means that the Cooper pairs exist in certain organic materials at a temperature of about 550 K.

Polythiophene is a one-dimensional conjugated polymer. Figure 8.1a shows its structure. It has been known already for some time that, in polythiophene, the dominant nonlinear excitations are positively-charged electrosolitons and bisolitons [63]. This means that the Cooper pairs with a charge of $+2|e|$ exist at room temperature in polythiophene. Figure 8.1b depicts a schematic structural diagram of a bisoliton on a polythiophene chain. Bisolitons have also been observed in other one-dimensional conjugated organic polymers such as polyparaphenylene and polypyrrole [63].

From these facts one can make a very important conclusion, namely that, *for occurrence of superconductivity at room temperature, the quasiparticle pairing will* **not** *be the bottleneck but the onset of long-range phase coherence will.* This question is discussed in the following chapter.

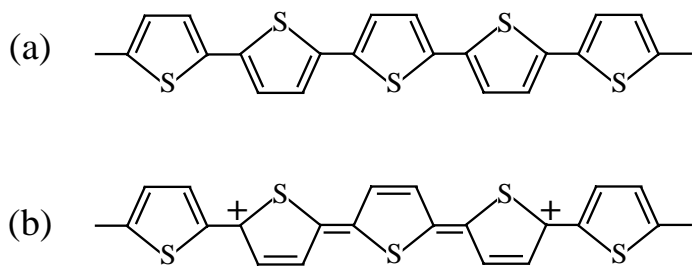

*Figure 8.1.*    (a) Chemical structure of polythiophene. (b) Schematic structural diagram of a positively-charged bisoliton on a polythiophene chain [63].

In the literature, one can find a few papers reporting superconductivity above room temperature. The results of most of these papers represent USOs (see Chapter 1). However, at least two works seem to report genuine results or, it is better to say, *almost* genuine results (see below why). The first report presents results of resistivity and *dc* magnetic susceptibility measurements performed in a thin surface layer of the complex material $Ag_xPb_6CO_9$ ($0.7 < x < 1$) [64]. The data indicate that in $Ag_xPb_6CO_9$ at 240–340 K there is a transition reminiscent of a superconducting transition. The authors suggest that the crystal structure of $Ag_xPb_6CO_9$ is quasi-one-dimensional.

The second work reports evidence for superconductivity above 600 K in single-walled carbon nanotubes, based on transport, magnetoresistance, tunneling and Raman measurements [38]. The Raman measurements have been performed on single-walled carbon nanotubes containing small amounts of the magnetic impurity $Ni : Co$ ($\leq 1.3$ %). In single-walled carbon nanotubes, the energy gap obtained in tunneling measurements is about $\Delta \simeq 100$ meV [38]. As described in Chapter 3, bulk superconductivity was already observed in single-walled carbon nanotubes at 15 K [36].

As was discussed a few moments earlier, the electron pairing occurs in some organic materials at $\sim 550$ K. Since both these materials, $Ag_xPb_6CO_9$ and the nanotubes, contain carbon, it is most likely that these reports present evidence for electron pairing above room temperature, not for bulk superconductivity. Of course, fluctuations of phase coherence may always exist locally. On the basis of these results, the reader can conclude once more that, for room-temperature superconductivity, *the onset of long-range phase coherence will be the bottleneck, not the quasiparticle pairing*.

## 3.1 Pairing energy in a room-temperature superconductor

Let us estimate the value of pairing energy in a room-temperature superconductor at $T = 0$. First of all, it is worth to recall that, in a superconductor, the pairing energy (gap) $\Delta_p(0)$, *generally speaking*, has no relation with a critical temperature $T_c$. The pairing energy $\Delta_p(0)$ is proportional to $T_{pair}$, the pairing temperature. In conventional superconductors, $\Delta_p(0) \propto T_c$ because the onset of long-range phase coherence in the BCS-type superconductors occurs due to the overlap of Cooper-pair wavefunctions and, therefore, $T_{pair} \simeq T_c$. However, in a general case, $T_c \leq T_{pair}$. For example, in the cuprate Bi2212 at any doping level, $1.3\, T_c < T_{pair}$, as shown in Fig. 6.11.

According to the fourth principle of superconductivity presented in Chapter 4, the pairing energy gap must be $\Delta_p(0) > \frac{3}{4} k_B T_c$. For the case $T_c = 350$ K, this condition yields $\Delta_p > 23$ meV. Figure 8.2 shows the pairing gap $\Delta_p(0)$ as a function of $T_{pair}$. In the plot, the energy scale $\frac{3}{4} k_B T$ marks the lowest allowed values of $\Delta_p(0)$ at a given temperature. It is worth noting that this condition is universal and applies to any superconductor including conventional ones.

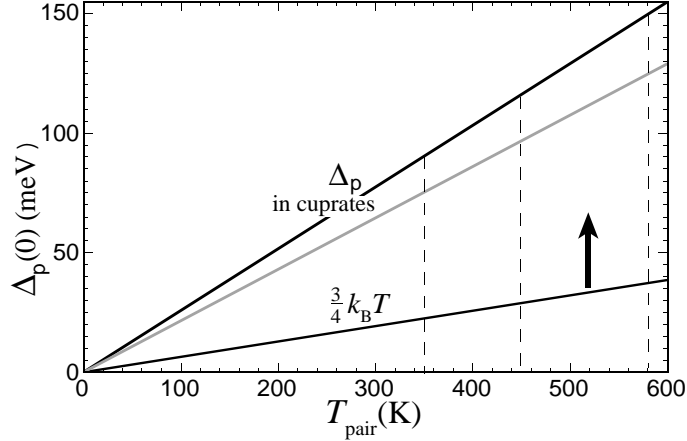

*Figure 8.2.* Pairing energy of quasiparticles $\Delta_p(0)$ as a function of pairing temperature $T_{pair}$. In plot, the thick vertical arrow indicates the allowed values of $\Delta_p(0)$ which lie above the energy scale $\frac{3}{4} k_B T$. The grey line represents the case of strong electron pairing $2\Delta_p = 5\, k_B T_{pair}$. In the cuprates, $2\Delta_p = 6\, k_B T_{pair}$, as illustrated in plot. The dashed lines indicate the temperatures 350, 450 and 580 K (see text).

As discussed above, the electron-phonon (hole-phonon) interaction in a room-temperature superconductor is most likely moderately strong and nonlinear. This means that, in a room-temperature superconductor, the pairing-gap ratio $2\Delta_p/(k_B T_{pair})$ must be at least 5. For example, in the cuprates $2\Delta_p/(k_B T_{pair}) = 6$ as specified in Eq. (6.7) and shown in Fig. 8.2. The grey line in Fig. 8.2 represents the case $2\Delta_p = 5\, k_B T_{pair}$. Then assuming that $T_{pair} \sim 350$ K, we obtain that, in a room-temperature superconductor, the *minimum* value of the pairing gap is about $\Delta_{p,min} \simeq 75$ meV. Is it realistic to anticipate in a superconductor such a value of $\Delta_p$? The answer is yes. In the copper oxide Bi2212, the pairing energy at a doping level of 0.05 equals 70 meV, as shown in Fig. 6.51. This experimental fact is obtained in tunneling and angle-resolved photoemission (ARPES) measurements. If, in a room-temperature superconductor $2\Delta_p = 6\, k_B T_{pair}$ as in the cuprates, then $\Delta_{p,min} \simeq 90$ meV.

In Eq. (6.58), one can see that the magnitude of the pairing energy in a bisoliton depends on the coupling parameter g and the exchange interaction energy $J$. Since the value of g in most cases is $\sim 1$ (see the discussion in Chapter 6), the exchange interaction energy $J$ mainly determines the magnitude of $\Delta_p$. This means that the pairing energy is large in materials with strongly-correlated electrons. It is worth noting, however, that the expression for $\Delta_p$ in Eq. (6.58) is obtained in the framework of the bisoliton model without taking into account the Coulomb interaction between electrons.

Assume that, in a room-temperature superconductor, $T_{pair} = 1.3\,T_c$ ($\simeq 450$ K). This is more realistic. Then, for the gap ratio $2\Delta_p/(k_B T_{pair}) = 5$, this means that $\Delta_p(0) \simeq 97$ meV and, for the gap ratio $2\Delta_p/(k_B T_{pair}) = 6$, $\Delta_p(0) \simeq 116$ meV. For clarity, Table 8.1 lists all these values of $\Delta_p(0)$. Thus, in a room-temperature superconductor, the pairing energy should be about $\Delta_p(0) = 90$ meV. As discussed above, in single-walled carbon nanotubes the energy gap obtained in tunneling measurements is around $\Delta \simeq 100$ meV [38].

*Table 8.1.* Pairing energy of quasiparticles $\Delta_p(0)$. $T_{pair}$ is the pairing temperature and $k_B$ is the Boltzmann constant

| $T_{pair}$ (K) | $\Delta_p(0)$ (meV) if $\frac{2\Delta_p}{k_B T_{pair}} = 5$ | $\Delta_p(0)$ (meV) if $\frac{2\Delta_p}{k_B T_{pair}} = 6$ |
|---|---|---|
| 350 | 75 | 90 |
| 450 | 97 | 116 |
| 580 | 125 | 150 |

## 3.2 Pairing energy in the case $T_c \simeq 450$ K

For *large-scale* applications, it is necessary to have a superconductor with $T_c \simeq 450$ K (see Chapter 1). Assuming that in such a superconductor $T_{pair} \approx T_c$ (= 450 K), the values of the pairing energies for the two cases, $2\Delta_p/(k_B T_{pair}) = 5$ and 6, are listed in Table 8.1. The table also presents the values of the pairing gap for the case $T_{pair} = 1.3\,T_c$ ($\simeq 580$ K). This case is also shown in Fig. 8.2. So, in a superconductor with $T_c \simeq 450$ K, the pairing energy must be about 120 meV.

## 4. Summary

The Cooper pairs in the form of bisolitons do exist at room temperature and, even, at much higher temperatures in organic materials forming the living matter. Every human being is a "carrier" of an enormous amount of bisolitons. Such Cooper pairs are most likely formed in real space. However, this question is not important from a standpoint of practical application.

For the occurrence of superconductivity at room temperature, it is necessary to solve the problem of the onset of long-range phase coherence. This issue is the main topic of the following chapter.

# Chapter 9

# PHASE COHERENCE AT ROOM TEMPERATURE

*Imagination is more important than knowledge.*

—Albert Einstein

The superconducting state is a quantum state occurring on a macroscopic scale. The electron pairing is the keystone of superconductivity. However this is only a necessary condition for the occurrence of superconductivity but not a sufficient one. Superconductivity requires also the condensation of the electron pairs in momentum space, i.e. the formation of a quantum condensate which is similar to the Bose-Einstein condensate (the second principle of superconductivity discussed in Chapter 4). The process of the Cooper-pair condensation taking place at $T_c$ is also known as the onset of long-range phase coherence, implying that below $T_c$ the Cooper-pair wavefunctions are in phase (see Fig. 4.1).

The main purpose of this chapter is to discuss the onset of long-phase coherence in a room-temperature superconductor. Here we shall mainly deal with three questions. First, what mechanism (interaction) can be responsible for the onset of phase coherence in a room-temperature superconductor? Second, is it realistic to anticipate the onset of long-range phase coherence at $T_c \simeq 350$ K? Third, what magnitude of the coherence energy gap, $\Delta_c$, should there be in a room-temperature superconductor?

## 1. Mechanisms of phase coherence

As described in Chapters 5–7, only two mechanisms of phase coherence are known at the moment of writing: the overlap of Cooper-pair wavefunctions (the Josephson coupling) and spin fluctuations (magnetic). The first mechanism is responsible for the onset of long-range phase coherence in supercon-ductors of the first and the second groups, whilst the magnetic mechanism is characteristic of unconventional superconductors of the third group. Can these two mechanisms of phase coherence be responsible for room-temperature superconductivity?

## 1.1    The Josephson coupling

The mechanism of the Josephson coupling is effective when the distance between Cooper pairs is smaller than the average size of Cooper pairs. In this case, the Cooper-pair wavefunctions become overlapped resulting in a Bose-Einstein-like condensation of the Cooper pairs. This mechanism of the Cooper-pair condensation is the simplest and does not lead to the appearance of a "new" order parameter. Simply, the Cooper-pair wavefunctions "magnified" multiply become the order parameter of the condensate. A characteristic feature of this mechanism of phase coherence is that the Cooper pairs condense immediately after their formation. Thus, the two processes—the electron pairing and the onset of long-range phase coherence—occur almost simultaneously. In other words, $T_c \simeq T_{pair}$.

The Josephson coupling is *very* robust and effective at *any* temperature as long as the Cooper-pair wavefunctions exist and remain overlapped. It is however unlikely that, in a room-temperature superconductor, the Josephson coupling can lead to the onset of long-range phase coherence. There are at least two reasons against the involvement of this mechanism in room-temperature superconductivity.

First, from the previous chapter we know that, in a room-temperature superconductor, the Cooper pairs will be represented by bisolitons. The size of bisolitons is usually small, say, a few lattice constants. In addition, the density of bisolitons, by definition, is always small—much smaller than the density of free electrons in metals. Taken together, this means that, in a room-temperature superconductor, the effective overlap of bisoliton wavefunctions can hardly be realized. In fact, this is exactly what happens in some living tissues and organic polymers in which the bisolitons exist but the long-range phase coherence does not occur.

Second, as will be discussed in the following chapter, the structure of room-temperature superconductors must be low-dimensional, for example, like that in the cuprates. This means that in such superconductors the Cooper-pair wavefunctions are also low-dimensional. Even if the overlap of Cooper-pair wavefunctions can partly occur in the conducting planes or chains, this process is absolutely ineffective between the planes or chains (depending upon the dimensions). For example, in the cuprates somewhat above $T_c$, in the $CuO_2$ planes locally there are fluctuations of phase coherence due to the overlap of bisoliton wavefunctions, but it does not lead to the onset of phase coherence between the $CuO_2$ planes (see Chapter 6).

To conclude, it is unlikely that, in a room-temperature superconductor, the overlap of bisoliton wavefunctions can lead to the onset of long-range phase coherence. Nevertheless, in Chapter 10 we shall discuss a possibility of artificial formation of a bisoliton condensate occurring due to the overlap of bisoliton wavefunctions.

## 1.2 Spin fluctuations

Can spin fluctuations mediate the long-range phase coherence at room temperature? Undoubtedly, yes. Spin fluctuations mediate the phase coherence in the cuprates. The highest critical temperature observed in the cuprates is 164 K (under pressure). This temperature is *only twice* smaller than the room temperature. Hence, it is logical to anticipate that this mechanism can be responsible for the onset of long-phase coherence at room temperature. In fact, for the magnetic mechanism, formally, there is no temperature-limit if the third and the forth principles of superconductivity are satisfied (see Chapter 4). Then, one can conclude that a room-temperature superconductor must most likely be a member of the third group of superconductors.

The characteristic features of the magnetic mechanism of phase coherence were discussed in Chapter 6. It is worth to emphasize that, in a superconductor in which spin fluctuations mediate the phase coherence, there are always two energy gaps, $\Delta_p$ and $\Delta_c$ (see Fig. 6.40). This also means that the order parameter of the superconducting state $\Psi$ is different from the Cooper-pair wavefunction $\psi$. Generally speaking, in superconductors of the third group, the order parameter has either a d-wave symmetry (in antiferromagnetic materials) or a p-wave symmetry (in ferromagnetic materials).

## 1.3 Other mechanisms of phase coherence

Theoretically, phonons can also mediate the phase coherence in a superconductor. However, in reality, it is impossible. Why? According to the third principle of superconductivity, the mechanism of phase coherence must be different from the mechanism of quasiparticle pairing. Since only the electron-phonon interaction is able to bind electrons in pairs, this means that the same mechanism cannot mediate the long-range phase coherence.

## 2. The magnetic mechanism

In this section we shall discuss the magnetic mechanism of phase coherence in a room-temperature superconductor and requirements to the material and to the coherence energy gap. We begin with the simplest question: Must the spin correlations in a room-temperature superconductor be antiferromagnetic or ferromagnetic?

## 2.1    Antiferromagnetic or ferromagnetic?

*Theoretically*, in a superconductor in which *the electron pairing* is mediated by magnons (i.e. spin fluctuations), the critical temperature is higher in antiferromagnetic materials than in ferromagnetic ones. We are however discussing a slightly different case—the onset of *phase coherence* due to spin fluctuations. These two cases are not the same. On the other hand, the strength of spin fluctuations in both these cases is important.

Experimentally, the critical temperature indeed is on average higher in antiferromagnetic materials than in ferromagnetic compounds. However, the exploration of ferromagnetic materials was started only in 2000 (see Chapter 6). Thus, it is too early to make a final conclusion. There is even at least one factor in favor of ferromagnetic compounds: *generally speaking*, the Curie temperature $T_C$ in ferromagnetic materials is on average higher than the Néel temperature $T_N$ in antiferromagnetic ones. Therefore, in the framework of our project, ferromagnetic materials should not be excluded fully. Nevertheless, on the basis of experimental data accumulated at the time of writing, it is better to start with antiferromagnetic compounds.

## 2.2    Requirements to magnetic materials

In order to impose requirements on the material, it is necessary, first, to understand the most important features of the magnetic mechanism. As was noted above, the characteristic features of the magnetic mechanism of phase coherence are presented in Chapter 6. Basically, in order to mediate superconducting correlations, there are three major requirements to be met by the material.

First, the localized states in the *undoped* material should have spin-1/2 ground states. At the time of writing, there is no magnetic superconductor having the localized states with a spin other than 1/2. This is an experimental fact.

Second, the material must be in a state near a quantum critical point. This will ensure the presence of local spin fluctuations and a high value of $T_c$ (see Figs. 6.37 and 6.13a). In a quantum critical point where a magnetic order is about to form or to disappear, the spin fluctuations are the strongest. At the moment of writing, we do not know yet how to determine from a single measurement the presence/absence of a quantum critical point in a certain compound. So, this should be the first intermediate goal: how to determine quickly the presence/absence of a quantum critical point in a given compound.

In superconductors of the third group, besides the first and the second requirements, superconductivity also demands the presence of *dynamic* spin fluctuations, not quasi-static ones. These dynamic spin fluctuations mediate the long-range phase coherence. For example, in the cuprates the spin excita-tions mediating the phase coherence are induced by fluctuating charge stripes. Therefore, in a given magnetic material, the fast spin excitations must be induced either by charge fluctuations, as those in the cuprates, or *artificially*. The presence of fast spin fluctuations can be observed, for example, by inelastic neutron scattering (INS) measurements. Working with cuprates, nickelates and manganites, the experience accumulated by the INS community for the last ten years in this domain is very useful. However, an attempt to induce artificially a spin excitation able to mediate the superconducting correlations has never been made before. This is a challenge for the near future, and we shall discuss this case in Chapter 10. By analogy with the cuprates, the frequency of spin fluctuations must be of the order of $\omega_{sf} \sim 10^{12}$–$10^{13}$ Hz = 1–10 THz [19, 49].

It is necessary to note that these spin excitations must be coupled to quasiparticles; otherwise, they are useless.

## 2.3    Coherence energy gap

Let us estimate the value of condensation energy of a Cooper pair in a room-temperature superconductor at $T = 0$, i.e. $2\Delta_c(0)$. The quantity $\Delta_c$ is also known as the phase-coherence gap, which is proportional to $T_c$ (in conventional superconductors, the coherence energy gap is absent). In superconductors of the third group, the magnitude of $\Delta_c$ depends on the value of magnetic (super)exchange energy $J$ between the neighboring spins appearing in Eqs. (6.66) and (6.67).

According to the fourth principle of superconductivity presented in Chapter 4, the coherence energy gap must be $\Delta_c(0) > \frac{3}{4}k_B T_c$. For the case $T_c = 350$ K, this condition means that $\Delta_c > 23$ meV. Figure 9.1 shows the phase-coherence gap $\Delta_c(0)$ as a function of $T_c$. In the plot the energy scale $\frac{3}{4}k_B T_c$ marks the lowest allowed values of $\Delta_c(0)$ at a given temperature. On the other hand, the magnitude of $\Delta_c$ must be smaller than $\Delta_p$. Thus, in the case $T_c \simeq 350$ K, the magnitude of $\Delta_c$ must be smaller than the corresponding values of $\Delta_p$ listed in Table 8.1.

What gap ratio should we expect for $\Delta_c$? There is no straight answer to this question. For example, for the pairing energy $\Delta_p$, there is a reference value—the BCS gap ratio $2\Delta_p/(k_B T_c) = 3.52$. For the value of the coherence gap $\Delta_c$, the only condition known at the time of writing is the forth principle of superconductivity presented in Chapter 4. Then, the gap ratio $2\Delta_c/(k_B T_c)$ can formally take any value between $\frac{3}{4}k_B T_c$ and $2\Delta_p/(k_B T_c)$. As discussed in the previous chapter, the minimum value of the latter ratio is around 5. In this case, the maximum allowed value of $\Delta_c$ is about 75 meV. This result is obtained by assuming that $T_{pair} \simeq T_c$ (= 350 K). However, one must realize that the maximum allowed value of $\Delta_c(0)$ can be much larger than 75 meV. As an example, if we assume that $T_{pair} \simeq 1.5\,T_c$ and $2\Delta_p/(k_B T_{pair}) = 6$, then the maximum allowed value of $\Delta_c(0)$ equals 135 meV. For instance, in the cuprate Bi2212, the value of the ratio $2\Delta_c/(k_BT_c)$ is 5.45, as shown in Fig. 9.1. Then, by using this gap ratio, one can easily obtain that, in the case $T_c = 350$ K, $\Delta_c(0) \simeq 82$ meV.

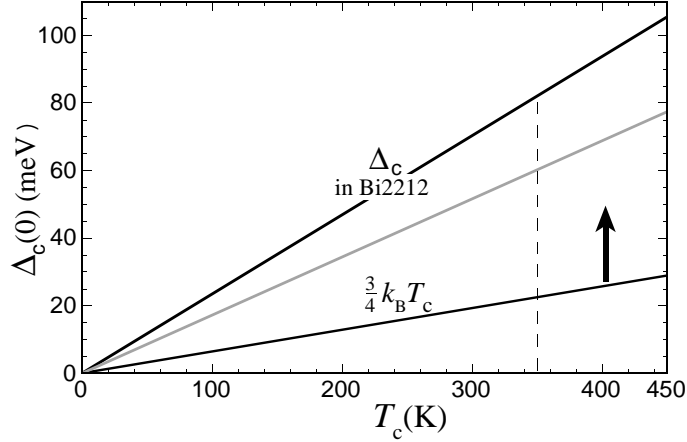

*Figure 9.1.* Phase-coherence energy gap $\Delta_c(0)$ as a function of critical temperature $T_c$. In plot, the thick vertical arrow indicates the allowed values of $\Delta_c(0)$ which lie above the energy scale $\frac{3}{4}k_BT_c$. The grey line represents the case $2\Delta_c = 4\,k_BT_c$. In the cuprate Bi2212, $2\Delta_c = 5.45\,k_BT_c$, as illustrated in plot. The dashed line indicates the critical temperature of $T_c = 350$ K.

If we assume that, in a room-temperature superconductor, $2\Delta_c/(k_BT_c) \sim 4$, then $\Delta_c(0) \approx 60$ meV. The grey line in Fig. 9.1 illustrates the gap ratio $2\Delta_c/(k_BT_c) = 4$. Is it realistic to anticipate that in a superconductor $\Delta_c \simeq 60$ meV? The answer is yes.

Consider the values of $\Delta_c$ in the cuprates: in optimally doped Tl2201 with $T_c \simeq 95$ K, the magnitude of the coherence energy gap is about $\Delta_c(0) \simeq 24$ meV. In Hg1223 which has the highest $T_c$ value of 135 K, $\Delta_c(0) \geq 30$ meV (at the time of writing, the exact value of the gap ratio for Hg1223 is unknown). Under pressure, the critical temperature of Hg1223 rises to $T_c \approx 164$ K. This means that the magnitude of the phase-coherence gap rises as well, becoming $\Delta_c(0) \geq 36$ meV.

As was noted above, the magnitude of $\Delta_c$ depends on the value of magnetic (super)exchange energy $J$. What value of $J$ should we expect in a room-temperature superconductor? Since the relation between $\Delta_c$ and $J$ is unknown, we are only able to estimate $J$. In optimally doped Bi2212 with $T_c \simeq 95$ K, $3\Delta_c(0) \simeq J$ (see Fig. 8.4 in [19]). By using the same ratio between $\Delta_c$ and $J$ for the case $T_c = 350$ K, we have $J \sim 3\Delta_c = 180$ meV. For comparison, in the undoped cuprates $J = 110$–150 meV (the highest $J \simeq 150$ meV is in NCCO).

Thus, the value of $J \sim 180$ meV is large but looks realistic if compared with those in the cuprates.

In the case $T_c \simeq 450$ K and $2\Delta_c/(k_B T_c) \sim 4$, the magnitude of the phase-coherence gap, $\Delta_c(0) \approx 78$ meV, is very large. Realizing $T_c \simeq 450$ K may be a real challenge for the future.

## 3. $T_c$ and the density of charge carriers

Superconductivity requires the presence of electron paring and the onset of long-phase phase coherence. In general, these two phenomena are independent of one another. However, it is not exactly the case for superconductors of the third group because these superconductors are systems with strongly correlated electrons (holes). This means that, in these compounds, the electronic, magnetic and crystal structures are strongly coupled, and the changes in one subsystem influences the other two. As an example, let us consider the so-called Uemura plot (relation) depicted in Fig. 3.6.

In Fig. 3.6, one can see that the critical temperature of unconventional superconductors is connected with the density of charge carriers, $n_s$, divided by the effective mass of the carriers $m^*$. At low temperature, the muon-spin relaxation rate $\sigma$ in Fig. 3.6 is proportional to the ratio $n_s/m^*$. Since in superconductors of the third group, the onset of phase coherence occurs mainly due to spin fluctuations, the critical temperature of these superconductors should be independent of the ratio $n_s/m^*$ because the density $n_s$ and the effective mass $m^*$ are the characteristics of charge carriers. Therefore, they should in principle have an effect only upon the pairing characteristics. Experimentally however, in unconventional superconductors at low doping level, $T_c \propto n_s/m^*$. As explained in the previous paragraph, this is because in third-group superconductors, the electronic, magnetic and crystal structures are coupled.

The Uemura relation found in unconventional superconductors can be understood in the following way. As discussed above, in superconductors of the third group, the occurrence of superconductivity and, therefore $T_c$, depends on the frequency of spin fluctuations $\omega_{sf}$. The spin fluctuations are induced by charge (stripe) fluctuations. It is obvious that the frequency of charge fluctuations depends on the effective mass of charge carriers: the charge carriers having a light mass can fluctuate faster than those with a heavy mass. Independently of this, a large number of charge carriers can induce spin fluctuations easier than a small number of them. In this way, these two characteristics of charge carriers, $n_s$ and $m^*$, affect the critical temperature in unconventional superconductors.

Since the relation $T_c \propto n_s/m^*$ is universal for superconductors of the third group, one may expect that the Uemura relation will manifest itself also in room-temperature superconductors, if the spin fluctuations in these room-temperature superconductors are induced by charge (stripe) fluctuations. Fig-ure 9.2 shows the Uemura relation at critical temperatures $T_c \sim 350$ K and $T_c \sim 450$ K. In the plot, one can see that, for the case $T_c \sim 350$ K, $\sigma \approx 9$–11 $\mu s^{-1}$. What is interesting is that the muon-spin-relaxation rate in $MgB_2$ has a similar value, $\sigma = 8$–10 $\mu s^{-1}$. As was discussed in Chapter 3, $MgB_2$ is similar to graphite both electronically and crystallographically (see Figs. 3.3 and 3.19). However, $MgB_2$ is an electron-doped superconductor.

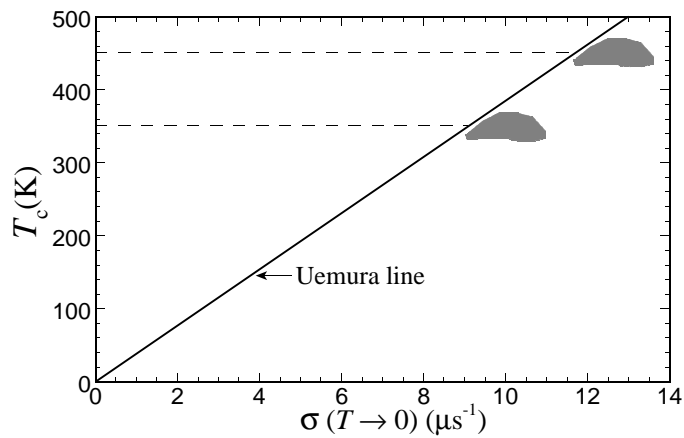

*Figure 9.2.* The Uemura relation $T_c \propto \sigma(T \to 0)$ [29] in room-temperature superconductors. $\sigma$ is the muon-spin-relaxation rate, and $\sigma \propto 1/\lambda^2 \propto n_s/m^*$, where $\lambda$ is the magnetic penetration depth; $n_s$ is the density of charge carriers, and $m^*$ is the effective mass of charge carriers. The Uemura line is taken from Fig. 3.6. The dashed lines indicate the temperatures of 350 K and 450 K. The grey areas reflect the Uemura relation $T_c \propto \sigma$ for these two cases, $T_c \sim 350$ K and $T_c \sim 450$ K.

What conditions does the Uemura relation impose on the material of room-temperature superconductors? The Uemura relation means that, in room-temperature superconductors, the density of charge carriers must be as large as possible, and their effective mass must be as small as possible. From Chapter 8 we know that the Cooper pairs in room-temperature superconductors will be represented by bisolitons. By definition, the density of bisolitons is low, much less than that of free electrons in metals. In every compound where the bisolitons exist, their density has an upper limit above which the compound becomes quasi-metallic. This means that, in room-temperature superconductors, the effective mass of charge carriers *must* be very small: $m^* \sim m_e$ or, even, $m^* < m_e$, where $m_e$ is the electron mass. For example, in single-walled nanotubes having a small diameter, the theoretical value of $m^*$ is $m^* = 0.36\,m_e$ (see Chapter 3). In general, a small value of the bisoliton effective mass signifies that the size of a bisoliton must be large [63].

It is worth noting that, in the case when the spin excitations that mediate the phase coherence are induced *artificially*, the Uemura relation $T_c \propto n_s/m^*$ is no longer applicable.

## 4. Transition temperature interval

For every superconductor, the absolute value of $T_c$ is important; however, one must also bear in mind that the high $T_c$ can be useless if the transition temperature interval $\Delta T_c$ is very large. This matter relates directly to the issue of homogeneity of the superconducting phase. It would be a surprise if one could synthesize a room-temperature superconductor having immediately a transition temperature interval of, say, a few degrees. This is practically impossible. It is most likely that the first samples will have a very large $\Delta T_c$. The problem of large $\Delta T_c$ must be the next goal to solve. As an example, the first test specimens of superconducting cuprates have had a very large $\Delta T_c$. It took more than a year of synthesizing samples of the superconducting cuprates which showed a sharp superconducting transition.

# Chapter 10

# ROOM-TEMPERATURE SUPERCONDUCTORS

*If you do not expect the unexpected, you will not find it.*
—Heraclitus [of Ephesus] (ca 550–475 BC)

The main purpose of this chapter is to discuss materials that superconduct above room temperature. In the context of practical application, this chapter is the most important in the book. In the two previous chapters we learned that, from the physics point of view, the occurrence of superconductivity is not limited by a certain temperature and, under suitable conditions, superconductivity can occur above room temperature. After all, why should the occurrence of superconductivity at *room temperature* be a special event? Of course, this is important for humans but not for Nature. At the Big Bang, Nature did not plan to set the Earth temperature near 300 K and to limit the occurrence of superconductivity by this temperature. In fact, even the occurrence of superconductivity on a *macroscopic scale* was not planned by Nature (see Chapter 1). Therefore, from the physics point of view, one must not emphasize that the onset of superconductivity above room temperature is an extraordinary event. This is just an event.

I do not want to say that it is easy to synthesize a room-temperature superconductor, not at all. I want to stress that it can be done. The second important point of this chapter is that, in order to realize this project, we may use some of Nature's experience. This will save a lot of time.

The main ideas of this chapter are based on experimental facts. At the same time, some ideas presented here are based exclusively on intuition. As a result, some text in this chapter will be presented in the first person, contrary to the tradition. As a matter of fact, I have already started doing this in the previous paragraph.

A few words about new ideas. Reading this chapter, some readers may think that this is science fiction. One should however understand that the border between reality and fiction depends on time. At present, some things look fictional but can be real tomorrow. Who could have imagined 100 years ago that soon the sky would be full of planes, everyone would carry a mobile phone, some even a TV. One hundred years ago, the word "a TV" was simply meaningless. What about computers? Video games? Brain surgery? Artificial insemination? Etc. The view on superconductivity evolves as well. *Some* ideas described in this chapter are indeed not "ready" for the present time but can be tomorrow's reality. Read the prologue to Chapter 8.

Let us consider one example. Charge inhomogeneity in the cuprates was first discussed by Gor'kov and Sokol in 1987 [6]. However, to the best of my knowledge, the existence of charge inhomogeneity was in general considered for the first time by Krumhansl and Schrieffer in 1975 [65]. At the end of a paper in which they discuss the motion of domain walls (solitons) in materials with a Peierls transition, they wrote: "Finally, we record a few speculative ideas, which may be worth further development. First, if these domain walls are present in the low-temperature phase of pseudo-one-dimensional crystals which have undergone Peierls transition, the Peierls energy gap in those walls could go to zero, the material becoming locally metallic. One could then have a distribution of conducting sheets (walls) in an insulating matrix. ..." [65]. So, in 1975 this idea was speculative; however, it is obvious to every solid-state physicist today (see Fig. 6.2).

A last remark before we discuss the plan of this chapter. I truly believe that, **at least**, one idea presented in this chapter leads to room-temperature superconductivity; maybe, not immediately, but surely in the near future. Only the experiment is the final judge for these ideas. As was mentioned in the Preface, I anticipate that in 2011 superconductivity will celebrate its $100^{th}$ jubilee having a transition temperature above 300 K.

This chapter is organized as follows. First, we shall analyze the properties of superconducting materials described in Chapter 3. On the basis of this analysis and the experimental facts presented in Chapters 8 and 9, we shall then discuss requirements for characteristics and the structure of room-temperature superconductors. Next, we shall view a plan of our main project and, finally, each item of the plan will be discussed in detail in the following subsections.

## 1. Superconducting materials: Analysis

In order to "create" new superconductors, one must first understand the common features of existing ones and the trend in the development of new materials. In other words, in order to predict the future, one should know the past. Hence, it is worthwhile to analyze the properties of superconducting materials presented in Chapter 3.

In Chapter 3 the superconducting materials are classified into three groups according to the mechanism of superconductivity in each compound. The first group consists of conventional superconductors; the second comprises half-conventional ones, and unconventional superconductors form the third group. Comparing these three groups of superconductors, one can conclude that

- the third group is the largest and has the highest rate of growth in the last twenty years, and

- superconductors of the third group exhibit the highest critical temperature.

Indeed, the critical temperatures of superconductors of the first and the second groups do not exceed 10 K and 40 K, respectively. This is because $T_c$ in these superconductors is limited by the strength of the **linear** electron-phonon interaction. In superconductors of the third group, the critical temperature depends on the strength of *dynamic* spin fluctuations. At the time of writing, the cuprates show the highest $T_c$. If the rate of the growth of the third group will remain in the future at the same level, then, one will soon need to make an internal classification of this group. On the basis of these observations, it is obvious that, in the framework of our project, we should discuss further exclusively superconductors of the third group.

All superconductors of the third group are

- magnetic or, at least, have strong magnetic correlations,

- low-dimensional,

- with strongly correlated electrons (holes),

- near a metal-insulator transition,

- probably, near a quantum critical point (impossible to check), and

- type-II superconductors.

Superconductors of the third group have

- small-size Cooper pairs (represented by bisolitons),

- a low density of charge carriers $n_s$,

- a universal $T_c(n_s/m^*)$ dependence (see Fig. 3.6), where $m^*$ is the effective mass of charge carriers,

- large values of $H_{c2}, T_c, \lambda$ (magnetic penetration depth) and a large gap ratio $2\Delta_p/(k_B T_c)$,

- anisotropic transport and magnetic properties,

■ a complex phase diagram,

■ the moderately strong and nonlinear electron-phonon interaction,

■ an unstable lattice,

■ charge-donor or charge-acceptor sites (charge-reservoirs), and

■ a complex structure (with the exception of hydrides, deuterides and a few heavy fermions).

In the third group:

■ the $T_c$ value of hole-doped superconductors is on average a few times higher than that of electron-doped superconductors.

■ superconductors with $T_c > 20$ K have no metal-metal bonds (only heavy fermions have metal-metal bonds).

■ oxides and organic superconductors represent an absolute majority of this group.

Considering the common features of superconductors of the third group, one must however realize that some of these features are direct consequences of the other. For example, the anisotropic character of transport and magnetic properties of superconductors of the third group is a direct consequence of a low-dimensional structure of these superconductors. The strong and nonlinear electron-phonon interaction results in a large value of the pairing energy gap and, therefore, in a large value of the gap ratio $2\Delta_p/(k_B T_c)$. Since in superconductors of the third group, the Cooper pairs are represented by bisolitons having a small size (a consequence of the strong and nonlinear electron-phonon interaction) and a low density, this leads to the penetration depth and, consequently, the ratio $\lambda/\xi$ being large. Therefore, all superconductors of the third group are type-II. The presence of strongly correlated electrons in these superconductors results in a complex phase diagram, and so on. Hence, some of these common features of superconductors of the third group are more important than others.

## 2.    Requirements for high-$T_c$ materials

We are now in a position to discuss requirements for materials that superconduct near room temperature. In this section, we shall first consider the characteristics of room-temperature superconductors that are important for (i) electron pairing and (ii) phase coherence. Then, we shall discuss (iii) the crystal structure and (iv) materials of room-temperature superconductors.

## 2.1 Electron pairing

From Chapter 8, we know that, in room-temperature superconductors, the Cooper pairs should be represented by positively-charged bisolitons. On the basis of the analysis of the properties of superconducting materials presented in the previous section and Chapter 8, for the presence of positively-charged bisolitons, room-temperature superconductors must be

- hole-doped,

- low-dimensional,

- with strongly correlated holes,

- near a metal-insulator transition,

- with the moderately strong and nonlinear hole-phonon interaction,

- with an unstable lattice, and

- most likely, organic.

The electrosolitons and bisolitons appear in low-dimensional systems having strongly correlated electrons and the moderately strong, nonlinear electron-phonon interaction. These systems are in a state near a metal-insulator transition and have an unstable lattice. (In fact, the expression "systems with strongly correlated electrons" partially assumes that the electron-phonon interaction in these systems is strong and nonlinear.) Experimentally, the bisolitons exist above room temperature in organic compounds.

## 2.2 Phase coherence

From Chapter 9, we know that, in room-temperature superconductors, the mechanism of phase coherence should most likely be magnetic. On the basis of the analysis of the properties of third-group superconductors presented in the previous section and Chapter 9, for the onset of long-range phase coherence due to spin fluctuations, room-temperature superconductors must

- be magnetic or, at least, have strong magnetic correlations,

- have the localized states with a spin of 1/2,

- have **dynamic** spin fluctuations,

- be in a state near a quantum critical point,

- most likely follow the Uemura relation $T_c \propto n_s/m^*$ (see Fig. 9.2), and

- most likely be antiferromagnetic.

The last constraint is based on experimental facts accumulated at the time of writing. However, ferromagnetic materials should also be examined in the near future (see the discussion in Chapter 9). The other requirements have already been discussed in Chapter 9.

## 2.3    Structure

On the basis of the analysis of the properties of superconducting materials presented in the previous section, the structure of room-temperature superconductors must

- be low-dimensional,

- be complex (with more than two sites per unit cell),

- have an unstable lattice,

- have electron-acceptor sites, and

- have no metal-metal bonds.

In materials able to superconduct at room temperature, the unit cell must have at least two *interacting* subsystems: one subsystem is quasi-metallic, and the other is magnetic. The electron pairing takes place in the first subsystem, whilst the onset of long-range phase coherence occurs with the participation of the second subsystem. For example, in the cuprates, the unit cell has three subsystems. In addition to the quasi-metallic and magnetic subsystems mentioned above (see Fig. 6.2), the third subsystem represents charge reservoirs. The charge reservoirs in the cuprates are the layers that intercalate the $CuO_2$ layers. They are usually insulating or semiconducting. In contrast, in organic superconductors the second and the third subsystems coincide: the charge reservoirs, after donating/accepting electrons to/from organic molecules, become magnetic. Thus, in organic superconductors, one subsystem performs two functions: to donate/accept electrons and to mediate the phase coherence. To conclude, a room-temperature superconductor must have:
1) **a subsystem with bisolitons**,
2) **charge reservoirs**, and
3) **magnetic atoms/molecules**.
Experimentally, the second and the third subsystems can be represented by the same atoms/molecules.

## 2.4    Materials

From Chapter 8 and the analysis of the properties of superconducting compounds presented in the previous section, the material of room-temperature superconductors must be

- multicomponent,

- most likely, organic, and

- probably, oxidic.

It is an experimental fact that bisolitons exist in some organic materials at temperatures much higher than room temperature.

In Chapter 1, we have discussed the requirements for materials that superconduct at high temperatures, presented by Geballe in 1993 [22]. One can now compare these requirements with those itemized in this section. The requirements summarized in this section include the Geballe constraints.

## 3. Three basic approaches to the problem

In this section, we discuss the basic approaches to the problem of room-temperature superconductivity. In general, one may suggest three approaches to the problem.

**The first approach**: synthesis. The basic idea for synthesizing a compound able to superconduct above room temperature is straightforward. One should take a material containing bisolitons above room temperature and dope (intercalate) it with atoms/molecules able to accept electrons and having an unpaired electron after the intercalation. In this case, the intercalant atoms/molecules will be magnetic. For instance, the structure of a Bechgaard salt shown in Fig. 3.15 is a good example. Alternatively, a material containing bisolitons can be doped by atoms/molecules of two types. The atoms/molecules of one type stands duty exclusively as charge reservoirs. This basic idea is more or less obvious; the main question is what materials to use and how to achieve a right intercalation. That is all.

**The second approach** consists in improving the performance of known superconducting materials, for example, the cuprates. This approach is also obvious.

**The third approach**. We already know that, by definition, the density of bisolitons in a system cannot be large; otherwise, the system will become metallic. As a result, the bisolitons cannot condense due to the overlap of their wavefunctions because, in general, the average distance between bisolitons is larger than the size of a bisoliton. Theoretically, this obstacle can be overcome by creating a train of bisolitons in a specially-prepared polymer. Let us call it the *bisoliton overlap* approach.

It is worth noting once more that some ideas presented in the following sections cannot be realized at present because of technological difficulties but can surely be realized in the near future.

## 4.  The first approach

In order to synthesize a room-temperature superconductor, one needs to know what materials to use. As was discussed above, the structure of every superconductor of the third group has at least two interacting subsystems: the electron pairing takes place in one subsystem, whilst the onset of long-range phase coherence occurs with the participation of the second subsystem. In spite of the fact that these two subsystems are interacting, we shall consider materials of one subsystem independently of materials of the other. This is because a number of various combinations between the materials of one subsystem and the materials of the other is very large, and an attempt to analyze all these combinations is simply impractical. On the other hand, discussing these two groups of materials independently of one another, one should take into account that, in practice, certain materials of the two subsystems can be incompatible, i.e. cannot be used together.

We shall discuss first "pairing" materials and then "magnetic" materials which can be used to intercalate the "pairing" ones.

## 4.1    Materials for electron pairing

In this subsection, we shall consider materials containing bisolitons above room temperature *intrinsically*. For simplicity, let us classify the materials for electron pairing as

- organics,

- living tissues, and

- oxides.

Such a sorting is conventional because these three groups, in fact, overlap. For example, the living tissues are organic and contain often oxygen.

### 4.1.1    Organic materials

Let us start this "journey" by trial and error with organic materials which are well studied and commercially available.

**Polythiophene**. As was already discussed in Chapter 8, polythiophene is a one-dimensional conjugated polymer having the structure shown in Fig. 8.1a. The dominant nonlinear excitations in polythiophene are positively-charged electrosolitons and bisolitons [63]. In a thiophene ring, the four carbon $p$ electrons and the two sulfur $p$ electrons provide the six $p$ electrons that satisfy the $(4n + 2)$ condition necessary for aromatic stabilization. From a theoretical point of view, the bisolitons exist in polythiophene because the energy levels of its two degenerate ground states are not equal [63]. In other words, the two valence-bond configurations shown in Fig. 10.1a are not equivalent. Poly-thiophene has a few derivatives and one of them shown in Fig. 10.1b is called poly(3-alkylthienylenes) or P3AT for short. In contrast to polythiophene, P3AT is soluble.

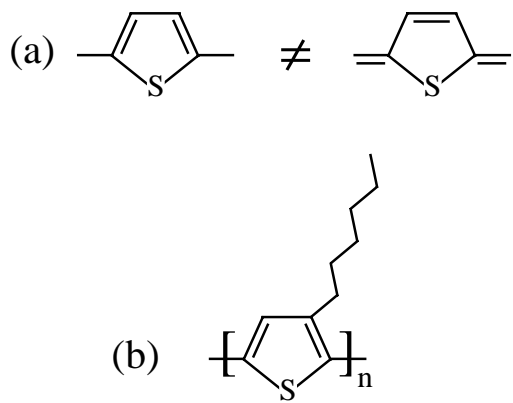

*Figure 10.1.* (a) Two valence-bond configurations of polythiophene shown in Fig. 8.1 are not equivalent. (b) Chemical structure of soluble poly(3-alkylthienylenes) [63].

Polythiophene, its derivatives and other organic conjugated polymers are usually doped by using the so-called electrochemical method [63]. The reaction is carried out at room temperature in an electrochemical cell with the polymer as one electrode. To remove electrons from organic polymers, oxidation is usually used. Through doping, one can control the Fermi level or the chemical potential. In the framework of our project, we are interested in doping polythiophene by magnetic atoms/molecules. These "magnetic" materials will be discussed below. In practice, it is impossible to foresee the structure of a doped organic compound, even knowing materials before the beginning of a doping procedure. Depending on their origin, concentration and size, the dopant species after the diffusion can take different positions relative to the polythiophene chains.

Figure 10.2 shows several examples of possible positions of dopant species relative to polythiophene chains. It is less likely that the dopant species will occupy positions in the planes of polythiophene chains, as sketched in Fig. 10.2a. They will most likely intercalate the polythiophene planes, as illustrated in Fig. 10.2b. In fact, the dopant atoms/molecules are even able to induce a reversible structural transition [63]. Upon doping, the polythiophene chains and dopant species can for example form a checker-board pattern shown schematically in Fig. 10.2c. For instance, in Na-doped polyacetylene, the $Na^+$ ions and polyacetylene chains form a modulated lattice with a "triangular" pattern depicted in Fig. 10.2d. In Na-doped polyacetylene, such a lattice appears exclusively at moderate doping levels.

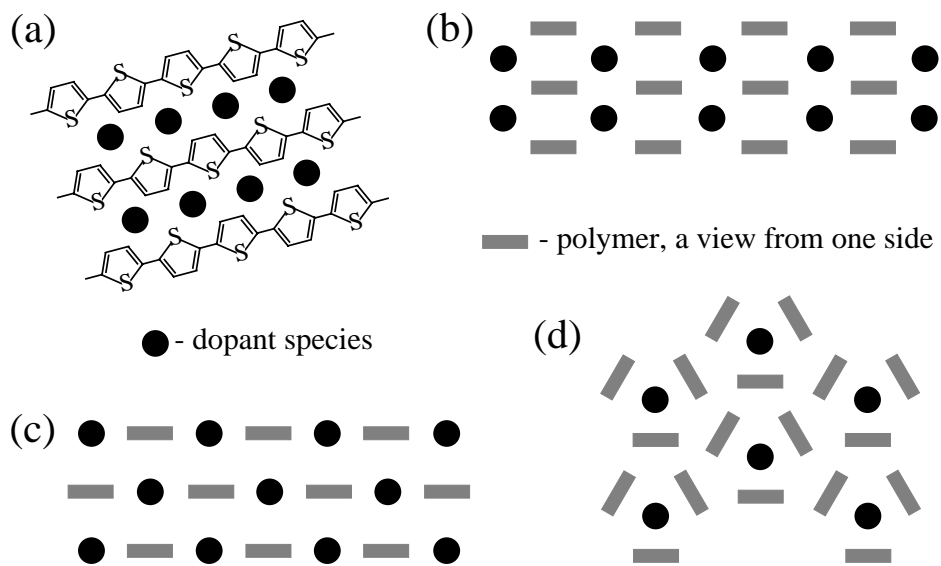

*Figure 10.2.* Possible positions of dopant species relative to infinite polythiophene chains: (a) in the plane of polythiophene chains; (b) between the planes; (c) a checker-border pattern, and (d) a "triangular" pattern realized in Na-doped polyacetylene [63].

Until now we have considered polythiophene chains having the infinite length. By analogy with the structure of organic superconductors (see, for example, Fig. 3.15), one should also try to use polythiophene chains having a finite length. Taking into account that the width of a bisoliton is a few lattice constants [10, 63], then, the length of pieces of polythiophene chains, $\ell$, should be then 2–3 times larger; thus $\ell \sim 15a$, where $a$ is the lattice constant. Since some atoms/molecules must be attached to the free ends of polymer pieces, the two ends of a polymer piece can be closed by one another resulting in the formation of a ring. Upon doping these rings, the dopant species can occupy the centers of the rings, as schematically shown in Fig.10.3. It is worth noting that the structure shown in Fig. 10.3 is similar to that of the fullerides depicted in Fig. 3.18.

Using various magnetic atoms/molecules, one should not forget to control the doping level of the organic polymers. It can be done, for example, by adding a small amount of atoms/molecules of another type, which may or may not be magnetic after the diffusion. Undoubtedly, some of these doped polythiophene-chain materials will superconduct. The main question is what maximum value of $T_c$ can be attained in these organic compounds. This mainly depends on the ability of "magnetic" materials to mediate the long-range phase coherence.

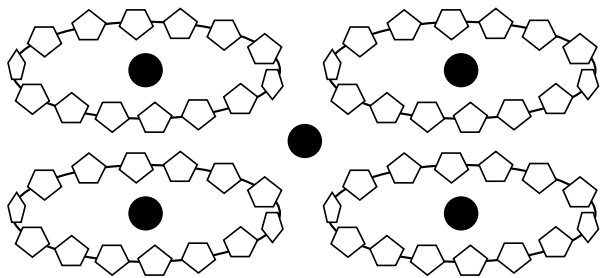

*Figure 10.3.* Possible arrangement of polythiophene rings and diffused magnetic atoms/molecules.

**Other conjugated polymers**. Other conjugated polymers containing bisolitons can be used instead of polythiophene. It is known that positively-charged bisolitons exist, for example, in polyparaphenylene, polypyrrole and poly(2,5-diheptyl-1,4-phenylene-alt-2,5-thienylene) (PDHPT) [63]. The structure of polyparaphenylene is depicted in Fig. 10.4a. A bisoliton on a polyparaphenylene chain is schematically shown in Fig. 10.4b. For example, a derivative of polyparaphenylene, *p*-sexiphenyl depicted in Fig. 10.4c, is widely used in organic light-emitting diodes. Why not use it as a basic material for a room-temperature superconductor? The structure of PDHPT is illustrated in Fig. 10.4d.

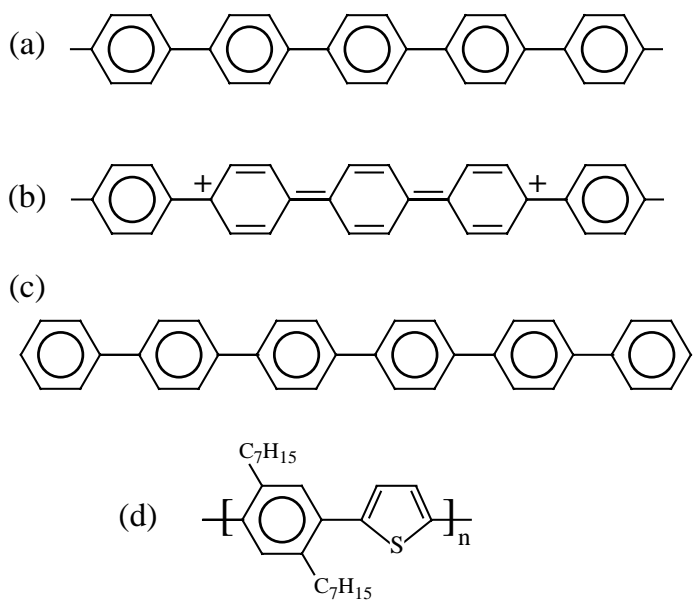

*Figure 10.4.* (a) Chemical structure of polyparaphenylene. (b) Schematic structural diagram of a positively-charged bisoliton on a polyparaphenylene chain [63]. (c) Molecular structure of *p*-sexiphenyl, and (d) the chemical structure of soluble poly(2,5-diheptyl-1,4-phenylene-alt-2,5-thienylene) (PDHPT).

As established experimentally, the physical properties of polymers strongly depend upon preparation conditions. For instance, the same polymer prepared by different techniques has different conductivities [63]. Unfortunately for experimentator, this fact adds one more degree of freedom for achieving the goal.

**Graphite**. For the last thirty years, graphite is one of the most studied materials. Several books are dedicated to a description of the physical properties of graphite (see, for example, [34]). It is also one of the most promising superconducting materials. Graphite intercalation compounds (GICs) able to superconduct were discussed in Chapter 3. Depending on their structure and the preparation technique, there are stage 1 and stage 2 GICs. All superconducting GICs are alkali-doped and, therefore, magnetic due to alkali spins ordered antiferromagnetically. In the superconducting GICs, the charge carriers are however electrons, not holes. Graphite-sulphur (CS) composites exhibit superconductivity at $T_c = 35$ K [35]. It is assumed that, in the CS composites, superconductivity occurs in a small fraction of the samples. The resistance in the CS composites remains finite down to the lowest measured temperature, indicating that superconducting clusters are isolated from each other [66].

The physical properties of graphite as well as other organic polymers depend on the preparation method. In most experiments, highly oriented pyrolytic graphite (HOPG) is used. In practice, all large-size single crystals of graphite, including commercially available ones, never have an ideal structure: they always have intrinsic carbon defects. The last statement is also valid for large-size single crystals of superconducting cuprates. Both graphite and the cuprates have the layered structure.

There exist both theoretical predictions and experimental evidence that electronic instabilities in pure graphite can lead to the occurrence of superconductivity and ferromagnetism, even at room temperature ([66, 67] and references therein). Some experiments indeed show that the superconducting and ferromagnetic correlations in graphite coexist [35, 66]. In graphite, an intrinsic origin of high-temperature superconductivity relates to a topological disorder in graphene layers [66]. (A single layer of three-dimensional graphite is called graphene.) This disorder enhances the density of states at the Fermi level. For example, four hexagons in graphene (see Figs. 3.19 and 10.5) can in principle be replaced by two pentagons and two heptagons [67]. Such a defect in graphene modifies its band structure. The disorder in graphene transforms an ideal two-dimensional layer into a network of quasi-one-dimensional channels preferable for bisolitons. In high magnetic fields ($> 20$ T), the in-plane resis-

Armchair edge

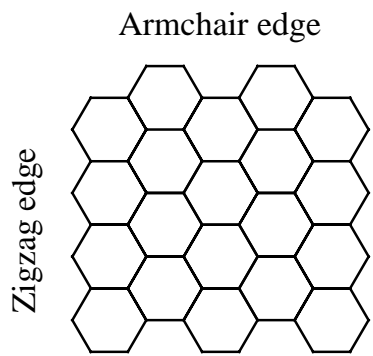

*Figure 10.5.* Two basic types of graphite edges [67].

tance of graphite exhibits an anomalous behavior attributed to the formation of charge-density-wave (CDW) [66]. This means that, in a high magnetic field, mobile bisolitons condense into the localized CDW states.

Interestingly, the magnetization of HOPG samples shows ferromagnetic hysteresis loops up to 800 K [66–68]. The details of this hysteresis depend on the sample, sample heat treatment and the direction of the applied field. The ferromagnetic signal in graphite is weak; however, as shown experimentally, ferromagnetic impurities cannot be responsible for this ferromagnetic ordering [68]. Most experimental results suggest that this ferromagnetism is intrinsic. Its origin is attributed partly to topological defects and in part to strong electron correlations in graphite [66–68]. In practice, the graphene sheets are always finite. Their electronic properties are drastically different from those of bulk graphite. It is experimentally established that the electronic properties of nanometer-scale graphite are strongly affected by the structure of its edges [66–69]. The graphene edges induce electronic states near the Fermi level. Any graphene edge can be presented by a linear combination of the two basic edges: zigzag and armchair, shown in Fig. 10.5. The free energy of an armchair edge is lower than that of a zigzag edge [69]. It is assumed that the zigzag edges are partly responsible for the ferromagnetic ordering [67]. Finally, it is worth noting that the magnitude of a ferromagnetic moment in graphite is age-dependent. The storage of graphite samples at ambient conditions results in a drastic decrease of the magnetization. As an example, a one-year storage brings the samples to a diamagnetic state without noticeable changes in their composition and lattice parameters [67].

Are graphite-based compounds able to superconduct above room temperature? Undoubtedly, yes. From experimental data, the bisolitons seem to exist in graphite above $T \sim 600$ K. The appearance of bisolitons at high temperatures in graphene depends on the graphene structure: graphene sheets must be topo-logically disordered. This can be achieved in a few ways. As discussed above, the hexagons in graphene can be replaced by pentagons and heptagons. Some carbon atoms in graphene can be substituted by B, N or Al [67]. In the framework of our project, B and Al are probably better than N because each N adds an additional electron to graphene, while room-temperature superconductivity requires holes. Instead of graphene sheets, one can for instance use nanostripes of graphene [70]. Alternatively, one may use the molecules of hexabenzocoronene shown in Fig. 1.3c or other large molecules of conjugated hydrocarbons. To stabilize structurally graphene nanostripes, hexabenzocoronene or other large molecules of conjugated hydrocarbons, one can prepare a single crystal consisting of two alternating layers: graphene and a layer of one of these large molecules or graphene nanostripes. Then, these single crystals must be doped by magnetic atoms/molecules responsible for long-range phase coherence.

In principle, it should be not a problem for bisolitons to occur in graphite and graphene-based compounds above room temperature. The main problem for the occurrence of bulk superconductivity in graphite and other organic compounds above room temperature is the question of the onset of long-range phase coherence. One can dope graphite by atoms/molecules spins of which are ordered antiferromagnetically after the diffusion. Alternatively, one may try to enhance intrinsic weak ferromagnetism of graphite/graphene.

Since the bonds between adjacent layers in graphite are weak, an artificial "sandwich" doping can be used for graphite. By using the electrochemical method to dope organic compounds [63], one cannot control the positions of dopant species in the crystal structure after the diffusion. Furthermore, the doping occurs only in a thin surface layer. For graphite however, one can consciously control the positions of intercalant species. The idea is as follows. By using a scanning tunneling microscope (STM), one can manipulate single atoms/molecules putting them into the proper positions on a clean graphite surface [71]. After such a delicate doping, one can cover the dopant species by one or two graphene sheets. The STM doping is then repeated again and so on. In the future, this procedure can in principle be computerized.

**Fullerenes**. Closed-cage molecules consisting of only carbon atoms are called fullerenes. A molecule of the fullerene $C_{60}$ is schematically shown in Fig. 3.17. There are other fullerenes such as $C_{20}$, $C_{28}$, $C_{70}$, $C_{72}$, $C_{100}$ etc. The alkali-doped fullerenes (fullerides) able to superconduct were discussed in Chapter 3. The unit cell of superconducting fullerides $M_3C_{60}$ is depicted in Fig. 3.18, where M is an alkali atom. The superconducting fullerides are electron-doped. To exhibit room-temperature superconductivity, the single crystals of $C_{60}$ must be doped by holes. Thus, one should find suitable dopant species for this purpose.

Theoretical calculations show that fullerenes having a diameter smaller than that of $C_{60}$, such as $C_{28}$ [72] and $C_{20}$ [73], are able to exhibit a higher value of $T_c$ relative to that of $C_{60}$. Hence, in addition to buckyballs $C_{60}$, the fullerenes $C_{28}$ and $C_{20}$ are also promising candidates with which to form a room-temperature superconductor.

One can use fullerenes not only in pure but also in polymerized form. As an example, Figure 10.6 shows various one- and two-dimensional polymeric solids formed from $C_{60}$. Similarly to graphite, polymerized rhombohedral $C_{60}$ displays weak ferromagnetism above room temperature [67–69].

In addition to experiments on single crystals of pure or polymerized fullerenes, the fullerenes can also be used in a combination with graphite or/and nanotubes. One can intercalate the graphene sheets in graphite by fullerenes, as sketched in Fig. 10.7. Such an intercalation can for example be achieved by using the STM doping discussed above. In this case, the intercalation of these graphene–fullerene compounds by "magnetic" species can simultaneously be done by STM.

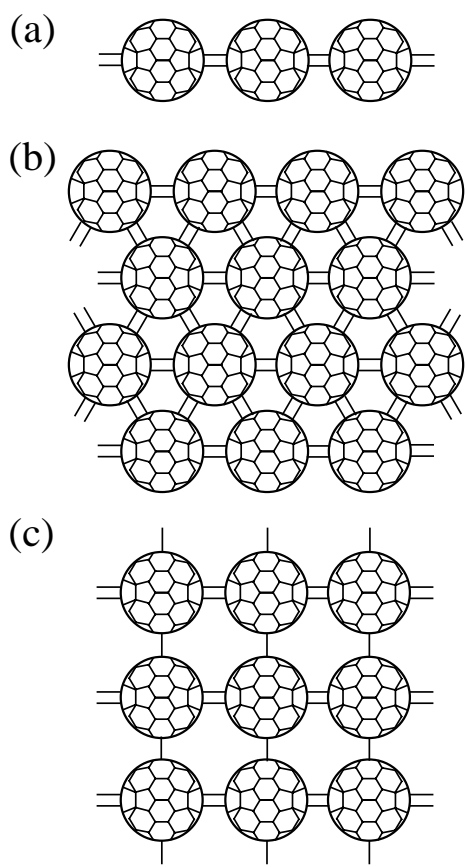

*Figure 10.6.* Various one- and two-dimensional polymeric solids formed from $C_{60}$ [74]. The $C_{60}$ balls are shown schematically.

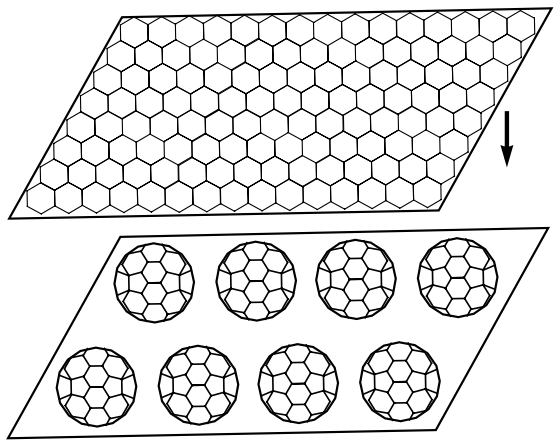

*Figure 10.7.* Intercalation of graphene sheets by $C_{60}$ molecules.

Unlike graphene and other long organic polymers, the fullerenes have an advantage to be packed into any form. Using insulating or semiconducting nanotubes, one can form a one-dimensional "wire" from fullerenes by packing them into the interior of one of these nanotubes, as schematically depicted in Fig. 10.8a. Inorganic single-walled nanotubes which are either insulating or semiconducting have recently been reported in the literature, such as $MoS_2$ [75], $TiO_2$ [76] and BN [77]. Theoretically, $B_2O$ and $BeB_2$ nanotubes may exist as well [78]. In practice, the boron nitride nanotubes have already been filled successfully by $C_{60}$ molecules, forming quasi-one-dimensional insulating wires [77]. The utilization of BN nanotubes having various diameters results in different stacking configurations of $C_{60}$ molecules [77]. In the framework of our project, one should fill the nanotubes not only with fullerene molecules but also with "magnetic" species, sticking to a certain order in the filling. This filling order should be a subject for a separate investigation. Depending on the diameter of a nanotube, one may use nanoscale pistons for applying a pressure, as sketched in Fig. 10.8b. By varying the pressure, one can change the filling factor and, therefore, the $T_c$ value. These nanoscale pistons in Fig. 10.8b can also be used as electrical contacts. Since carbon nanotubes which will be discussed next can also be insulating or semiconducting, one may use them instead of inorganic nanotubes.

**Carbon nanotubes**. In addition to spherical fullerenes, tubular carbon-based structures are also called fullerenes. Here we shall call them carbon nanotubes or just nanotubes for short. Carbon nanotubes can be multi- and single-walled. For simplicity, we shall discuss the single-walled nanotubes. They show metallic, insulating and semiconducting properties depending on the helicity with which a graphene sheet is wrapped to form the tubule. The armchair nanotubes are usually metallic, while the zigzag ones are semiconducting. Figure 3.20 shows a piece of armchair nanotube. Similarly to graphite, the carbon nanotubes may also exhibit weak ferromagnetism [67–69]. Two nanotubes can be joined by electron beam welding, forming a molecular junction [79].

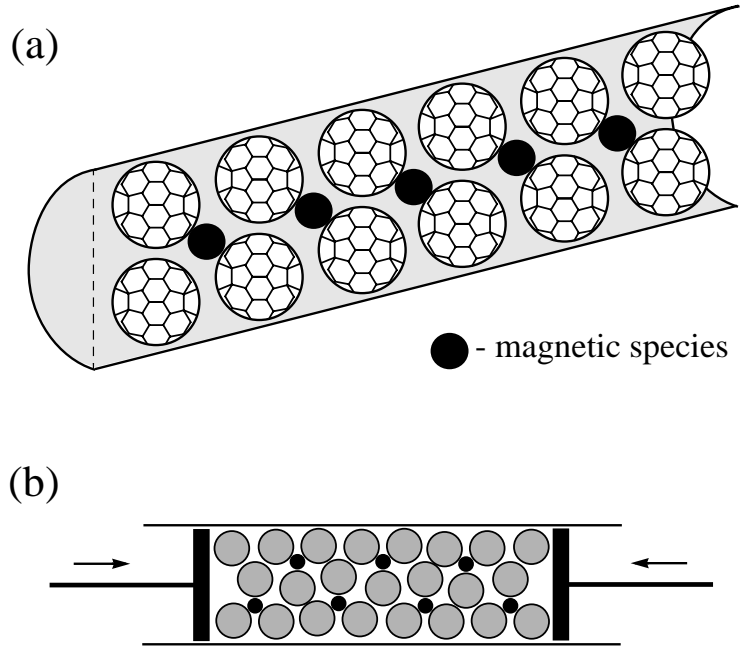

*Figure 10.8.* (a) $C_{60}$ molecules inside an insulating nanotube [77], intercalated by magnetic atoms/molecules. (b) Applying a pressure to $C_{60}$ molecules shown in plot (a). The nanoscale pistons can be used as electrical contacts.

Due to their remarkable electronic and mechanical properties, a great future, in the context of practical application, awaits the carbon nanotubes. The nanotubes are also a promising candidate with which to form a room-temperature superconductor. As discussed in Chapter 3, the single-walled carbon nanotubes with a diameter of $4.2 \pm 0.2$ Å exhibit *bulk* superconductivity below $T_c \simeq 15$ K [36]. The nanotubes with a smaller diameter may display a higher $T_c$. The onset of *local* superconductivity was observed in single-walled carbon nanotubes containing a small amount of the magnetic impurities Ni and Co at 645 K [38]. By embedding these nanotubes into a *dynamic* magnetic medium, one can witness bulk superconductivity above 450 K.

Since the electronic properties of the nanotubes depend on the wrapping angle of a graphene sheet, one can invent a technique of twisting a nanotube along its main axe. In such a way, one can vary the pairing temperature $T_p$ and, as a result, $T_c$. This suggestion is obviously for the distant future.

The carbon nanotubes can in principle be used in a combination with graphene sheets. They will structurally support the nanotubes, as sketched in Fig. 10.9. The "magnetic" species can for example be situated between the nanotubes. Alternatively, instead of graphene sheets, one can use layers of magnetic materials ordered antiferromagnetically. The nanotubes in the adjacent layers can for instance be oriented perpendicular to each other. This will make the superconductor two-dimensional.

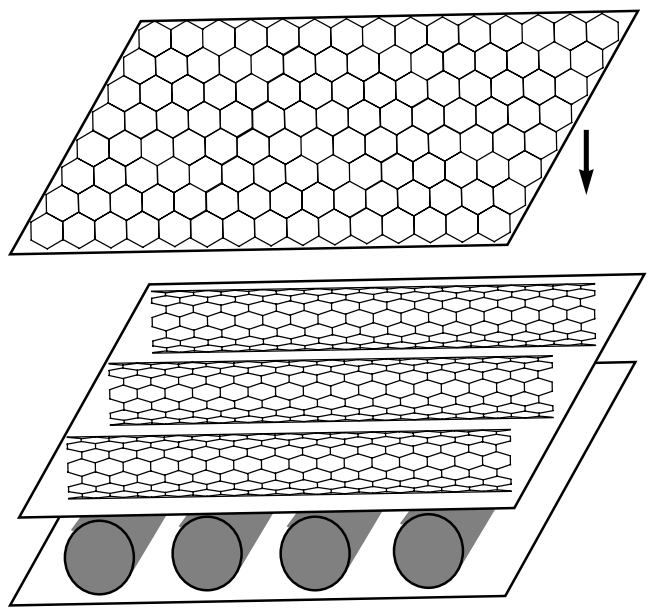

*Figure 10.9.* Intercalation of graphene sheets by carbon nanotubes. In the adjacent layers, the main axes of nanotubes may be mutually orthogonal.

Single-walled carbon nanotubes filled with $C_{60}$ molecules are called *peapods*. In peapods, the inner diameter of nanotubes is slightly larger than the outer diameter of $C_{60}$ molecules. In order to discuss the following idea, it is worth to recall that, in a solid, only *dynamic* spin fluctuations are able to mediate the long-range phase coherence. In the cuprates, for example, charge fluctuations in the $CuO_2$ planes induce these dynamic spin fluctuations. By analogy with the cuprates, $C_{60}$ or other fullerene molecules moving inside a nanotube, as shown in Fig. 10.10a, can induce dynamic spin fluctuations in the magnetic surroundings. The more practical cases are depicted in Figs. 10.10b and 10.10c: the fullerene molecules remain always inside the same nanotube. However, in order to realize this idea, one must invent a technique allowing to initiate the fullerene-molecule movement relative to a nanotube. Instead of carbon nanotubes, one can in principle use the inorganic nanotubes discussed above. Ideally, these inorganic nanotubes can be magnetic themselves; then, there is no need for an additional doping procedure.

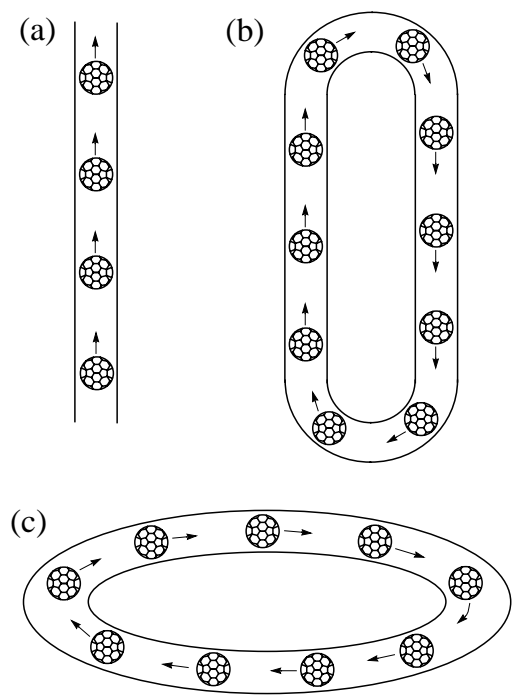

*Figure 10.10.* Moving $C_{60}$ molecules inside nanotubes: (a) a straight nanotube; (b) two parallel nanotubes closed at each end by rounded pieces of nanotubes, and (c) a round nanotube. The purpose of such a dynamics is explained in the text.

### 4.1.2 Living tissues

It is worth to recall that the main idea of the whole book is that, in order to synthesize a room-temperature superconductor, we may use some of Nature's experience accumulated during billions of years, even if, this experience in principle has nothing to do with superconductivity (see the discussions in Chapters 1 and 8). After all, Nature is smarter than humans.

In principle, one can use living tissues to form a room-temperature superconductor. All the descriptions presented in the previous subsection can equally be applied to living tissues which are usually one- or two-dimensional. One should use living tissues containing bisolitons. The main question is what living tissues contain the electron pairs. From Chapter 1, we know that (i) in redox reactions, electrons are transferred from one molecule to another in pairs with opposite spins, and (ii) electron transport in the synthesis process of ATP (adenosine triphosphate) molecules in conjugate membranes of mitochondria and chloroplasts is realized by pairs.

Mitochondria and chloroplasts are integral parts of almost every living cell. In principle, one can easily use their membranes to form a room-temperature superconductor as described in the previous subsection. The redox reactions occur practically in every cell. One should find out what parts of the cells are responsible for the redox reactions, and then use these tissues to form a room-temperature superconductor.

In addition to these tissues, DNA (deoxyribonucleic acid) is also a good choice: proximity-induced superconductivity was already observed in DNA below 1 K [37]. It is also assumed that DNA will exhibit genuine superconductivity if one can find a technique to dope it. The double helix of DNA has a diameter of 20 Å, and it can be a few microns long. So, DNA is truly a one-dimensional system. Charge transport through DNA is crucial for its biological functions such as the repair mechanism after radiation and biosynthesis. So far, the findings concerning conductivity of DNA are controversial [80]. Some measurements indicate that DNA behaves as a well conducting one-dimensional molecular wire. In contrast, other measurements show that DNA is insulating. The available data are more or less consistent with a suggestion that DNA is a wide-bandgap semiconductor [80]. This proposal is also in good agreement with the idea that electron transport in DNA occurs due to electrosolitons and/or bisolitons, accompanied by molecular distortion (local deformation of the lattice).

One of the main advantages of DNA is that it can easily be attached to electrical contacts, as shown in Fig. 10.11. A large single molecule of double-stranded DNA can be manipulated by attaching short, single-stranded DNA molecules to each of its ends. Then, chemical labels on the ends of these single-stranded molecules are used to attach the DNA molecule to electrical leads or microscopic golden beads, as depicted in Fig. 10.11.

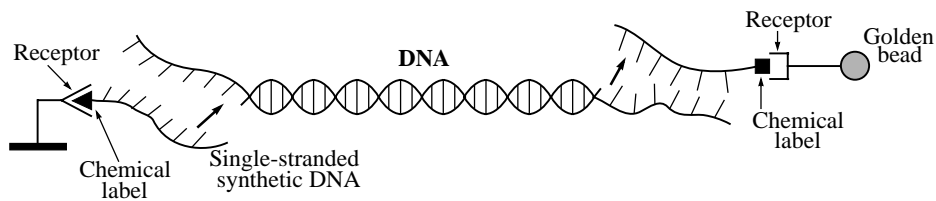

*Figure 10.11.* A single molecule of double-stranded DNA with single-stranded DNA molecules attached to each of its ends [81]. Chemical labels on the ends of single-stranded molecules are used to attach the DNA to an electrical lead or to a bead which can then be used as an electrical contact. This configuration is in fact used for manipulation of DNA molecules [81], but can stand duty as an electrical circuit.

It is worth noting that, in living cells, DNA molecules are in a solution, and they actively interact with the solution. Therefore, one should consider to experiment with DNA not only in a dry environment but also in the proper solution.

The experiments should not be restricted by the living tissues discussed above, one can use other tissues as well. The main requirement for them is that they must contain bisolitons.

### 4.1.3 Oxides

Oxides, as a class of materials, deserve to be considered as materials able potentially to superconduct at room temperature because:

- at present, oxides are the largest group of superconductors having the highest growth rate;

- at present, oxides exhibit the highest value of $T_c$ (cuprates);

- at least, one oxide ($Ag_xPb_6CO_9$) exhibits some signs of *local* superconductivity above room temperature [64], and

- oxygen, as a chemical element, is a unique acceptor of electrons and plays a crucial role in the living matter.

All the living matter *cannot* function without oxygen: the redox reactions are essential part of biochemical processes occurring in living organisms, in some of which electrons are transferred from one molecule to another in pairs with opposite spins (see the previous subsection). Oxygen is also a constituent element of DNA.

Contrary to the widely-used expression "the living matter is organic" (including humans), oxygen, in fact, represents the heaviest part of a human body (61 %), while C and N contribute only 23 % and 2.6 % to the total weight of the body, respectively [82]. Of course, it is mainly water that makes us heavy (materials containing water will be discussed below).

As was considered in Chapter 8, in a thin surface layer of the complex oxide $Ag_xPb_6CO_9$ ($0.7 < x < 1$) at 240–340 K, there is a transition reminiscent of a superconducting transition [64]. Taking together, all these facts indicate that oxides should seriously be considered as potential candidates for a room-temperature superconductor. For example, the layers of $CoO_2$ doped by holes will undoubtedly superconduct (see the respective subsection below). The question is what $T_c$ value will they exhibit? Other layered oxides with the spin $S = \frac{1}{2}$ ground state should also be on the list of potential candidates. As a pre-selection procedure, one must use the common requirements for materials discussed above.

In addition to layered and quasi-one-dimensional oxides, one may try to synthesize zero-dimensional oxides similar to fullerenes. An attempt to synthesize inorganic fullerene-like molecules has been successful [83]. Thus, one may in principle synthesize various fullerene-like oxides. After the doping, some of them may superconduct.

## 4.2    Materials for phase coherence

A superconductor of the third group must be magnetic or, at least, have strong magnetic correlations. While oxides can be magnetic naturally, like the cuprates for example, organic and living-tissue-based compounds must be doped by magnetic species which will be responsible for long-range phase coherence.

Unfortunately, during evolution, Nature did not need to develop such magnetic materials. Hence, we should only rely on accumulated scientific experience and work by trial and error. The general requirements for the magnetic properties of room-temperature superconductors were discussed above and in Chapter 9. In addition, a few hints can be suggested.

By doping organic materials or living tissues, one should take into account that, after the diffusion, the dopant species must not be situated too close to the organic molecules/tissues. Otherwise, they will have a strong influence on bisoliton wavefunctions and may even break up the bisolitons. On the other hand, the dopant species cannot be situated too far from the organic molecules/tissues because bisolitons must be coupled to spin fluctuations.

Since in superconductors of the third group, spin fluctuations must be coupled to quasiparticles, the dopant atoms/molecules (at least, the majority of them) should donate/accept electrons to/from molecules (or complex structures) responsible for electron pairing. In the framework of our project, they must accept electrons, creating holes in a material responsible for electron pairing. In all known cases, the dopant species donate/accept either 1 or 3 electrons. For achieving a high $T_c$, the dopant species should accept 2 electrons. In this case, the electron pairs can wander around much more easily. In reality, however, this is impractical because, after accepting/donating *two* electrons, the dopant species will remain non-magnetic.

As was estimated in Chapter 9, in a room-temperature superconductor of the third group, the value of magnetic (super)exchange energy between the adjacent spins, $J$, should be of the order of 150–200 meV. This value is large but realistic. In my opinion, the most difficult task to be resolved is to create *dynamic* spin fluctuations with $\omega_{sf} \sim 10^{12}$–$10^{13}$ Hz. In the cuprates for example, a structural phase transition precedes the transition into the superconducting state. This structural transition allows the charge stripes to fluctuate quicker, provoking a transition into the superconducting state. Therefore, synthesizing a room-temperature superconductor, one must pay attention to its structure: the "distance" between failure and success can be as small as 0.01 Å in the lattice constant.

In addition to "magnetic" species which will accept electrons, one may try to dope a material for electron pairing also by a small amount of molecular magnets to vary the strength of magnetic correlations. Molecular magnets are the nanoscale clusters containing a transitional metal, and they are promising components for the design of new magnetic materials [84].

As discussed in Chapter 9, in the framework of our project, one should start with antiferromagnetic compounds. Nevertheless, ferromagnetic materials should not be excluded from the project. At present, one can find several publications/preprints reporting *room-temperature* ferromagnetism in various compounds [85–91].

Finally, what materials to use. We are interested in materials which accept electrons after being intercalated. For example, all the compounds listed in Table 3.7 donate electrons. They are used to intercalate fullerenes. From Chapter 3, the materials able to accept electrons from organic molecules are the following atoms and molecules: Cs, I, Br (atoms) and $PF_6$, $ClO_4$, $FeCl_4$, $Cu(NCS)_2$, $Cu[N(CN)_2]Br$ and $Cu[N(CN)_2]Cl$ (molecules). So, one should most likely start with these materials.

### 4.2.1 Artificially-induced spin excitations

From the classical standpoint, magnetic field is detrimental to superconductivity. Experimentally, however, this is not always the case: magnetic field can not only destroy superconductivity but also induce it. As discussed in Chapter 2, the superconducting phase in the quasi-two-dimensional organic conductor $\lambda$-(BETS)$_2$FeCl$_4$ is induced by magnetic field [26, 27]. The superconducting phase as a function of magnetic field has a bell-like shape, occurring between 18 and 41 Tesla with a maximum $T_c \simeq 4.2$ K in the middle [27]. The field is applied parallel to the conducting layers. This experimental fact demonstrates that what *should* be detrimental to superconductivity can in fact induce it. The idea which we are going to discuss now is based on this fact.

Let us consider an idea of inducing the superconducting phase by *ac* electromagnetic field, applicable to magnetic compounds containing bisolitons. Assume that we have a thin film of an organic compound containing bisolitons, which is doped by magnetic species. The thickness of this thin film is of the order of, or less than, a typical penetration depth in superconductors of the third group, thus, $\sim 1000$–$2000$ Å. We know that in order to become superconducting, this complex compound should have dynamic spin fluctuations with a frequency of $\omega_{sf} \sim 10^{12}$–$10^{13}$ Hz = 1–10 THz. If such spin fluctuations are absent, one may in principle induce them artificially. Let us place the thin film in a weak *ac* electromagnetic field with $\omega \sim 1$–$10$ THz. The quasiparticles in the film will follow the field, inducing spin excitations with a similar fre-quency. If the bisolitons in the thin film are not broken by the electromagnetic field, they will be coupled to spin fluctuations which, in turn, *may* mediate the long-range phase coherence for bisolitons. In this case, the thin film will become superconducting, at least, partially.

In the above example, when the thin film will become superconducting, the *ac* field will in part be expelled from the film but not fully because the thickness of the film is of the order of the penetration depth. In such a situation, one may expect the presence of nonlinear oscillations of the order parameter and other self-interaction effects. Depending on an application, these effects may not be important. In this case, one will have a room-temperature superconductor for use in certain devices.

Let us estimate the upper limit for the frequency of *ac* field. The energy $\hbar\omega$ must be smaller than the coherence gap $\Delta_c(0)$. It is worth to recall that $\Delta_c(0) < \Delta_p(0)$, always. For $\Delta_c(0) \approx 60$ meV, the condition $\hbar\omega < \Delta_c(0)$ yields $\omega < 14$ THz.

One may wonder why the frequency of spin excitations able to mediate superconductivity should be $\omega_{sf} \sim 10^{12}$–$10^{13}$ Hz. In the absence of any complete theory of unconventional superconductivity, there is no definite answer to this question. However, if we assume that the Cooper pairs in unconventional superconductors, i.e. bisolitons, are formed in real space, then, it is possible to give a quantitative answer. Independently of the issue of real-momentum space pairing, spin excitations which mediate the phase coherence must be coupled to bisolitons. Let us denote the average lifetime of bisolitons by $\tau$. If the frequency of spin excitations is $1/(2\pi\omega_{sf}) \gg \tau$, the spin excitations will not notice the presence of a *given* bisoliton. Thus, $\omega_{sf}$ must be $1/(2\pi\omega_{sf}) \sim \tau$. Since in organic polymers, $\tau \sim 10^{-13}$ s [63], then we obtain that $\omega_{sf} \sim 1/(2\pi\tau) \sim 2 \times 10^{12}$ Hz.

## 4.3    Materials containing water

During the writing of this book a new unconventional superconductor was discovered: the layered cobalt oxyhydrate $Na_xCoO_2 \cdot yH_2O$ ($\frac{1}{4} < x < \frac{1}{3}$ and $y = 1.3$–1.4) exhibits superconductivity [92]. The structure of the parent compound $Na_xCoO_2$ consists of alternating layers of $CoO_2$ and Na. In the hydrated $Na_xCoO_2$, the water molecules form additional layers, intercalating all $CoO_2$ and Na layers. After the hydration of $Na_xCoO_2$, the *c*-axis lattice parameter increases from 11.16 Å to 19.5 Å [93]. Thus, the elementary cell of $Na_xCoO_2 \cdot yH_2O$ consists of three layers of $CoO_2$, two layer of $Na^+$ ions and four layers of $H_2O$. The $Na^+$ ions are found to occupy a different configuration from the parent compound. The displacement of the $Na^+$ ions is required in order to accommodate the water molecules which form the structure that replicates the structure of ice. The oxygen positions are fixed, while the positions of hydrogens are randomized as they are in ice. The Na and $H_2O$ sites are only partially occupied, while the $CoO_2$ layers in these structures are robust and consist of edge-sharing tilted octahedra. Each octahedron is made up of a Co ion surrounded by six O atoms at the vertices. Within each $CoO_2$ layer, the Co ions occupy the sites of a triangular lattice. The $1 - x$ fraction of Co ions is in the low spin $S = \frac{1}{2}$ $Co^{4+}$ state, while the $x$ fraction is in the $S = 0$ $Co^{3+}$ state. In the triangular lattice, the spins of $Co^{4+}$ ions are ordered antiferromagnetically.

Superconductivity in $Na_x CoO_2 \cdot yH_2O$ occurs in the $CoO_2$ layers. The superconducting phase as a function of $x$ has a bell-like shape, situated between 0.25 and 0.33 with a maximum $T_c \simeq 4.5$ K near $x = 0.3$ [94]. $Na_x CoO_2 \cdot yH_2O$ is a strongly anisotropic type-II superconductor with $k \sim 10^2$ [93] and, even, $10^4$ [95] at low temperatures. In $Na_x CoO_2 \cdot yH_2O$, $H_{c2} \sim 4$–5 T which yields a coherence length of 100 Å [95]. There is a strong inverse correlation between the $CoO_2$ layer thickness and $T_c$: the critical temperature increases as the thickness decreases [93]. The substitution of deuterium for hydrogen in water molecules has no apparent effect on $T_c$ [96].

With relevance to the phonon density of states, several features in the acoustic (0–40 meV) and optical (50–100 meV) phonon branches have been observed by neutron scattering in $Na_x CoO_2$ [93]. Similar features are also found in the acoustic channel of hydrated $Na_x CoO_2$, indicating that the phonons associated with the $CoO_2$ and Na layers are similar in the two materials. However, in the optical phonon branch of superconducting $Na_x CoO_2 \cdot yH_2O$, the neutron scattering in the energy range 50–120 meV is much stronger than that in the parent compound. This additional scattering is assumed to be caused by hydrogen [93].

All experimental facts indicate that the presence of water is crucial to superconductivity [91–97]. There is a marked resemblance in superconducting properties between $Na_x CoO_2 \cdot yH_2O$ and the cuprates [94].

We are now in a position to discuss the significance of the discovery of superconductivity in $Na_x CoO_2 \cdot yH_2O$. In spite of the fact that the critical temperature in $Na_x CoO_2 \cdot yH_2O$ is below 5 K, this discovery is probably the most important since 1986 when superconductivity in cuprates was found [5]. Let us consider why. The presence of superconductivity in the $CoO_2$ layers indicates that, *under suitable conditions*, all oxide layers with the spin $S = \frac{1}{2}$ ground state are most likely able to superconduct when they are slightly doped by charge carriers. This ability is not only the privilege of the $CuO_2$ layers ($RuO_2$ layers also superconduct). Secondly, after the discovery of superconductivity in $Na_x CoO_2 \cdot yH_2O$, the cuprates are not the only Mott insulators able to superconduct (see p. 94): sodium cobalt oxide $Na_x CoO_2$ is also a Mott insulator. Thirdly, $Na_x CoO_2 \cdot yH_2O$ is the first superconductor containing water (ice) the presence of which is crucial for the occurrence of superconductivity. This fact runs counter to common sense. A similar feeling among scientists was after the discovery of superconductivity in cuprates in 1986. From the experimental data, it is clear that the presence of ice in $Na_xCoO_2 \cdot yH_2O$ is important for superconductivity, at least, for two reasons. The $H_2O$ layers enlarge the $c$-axis lattice parameter and, apparently, the formation of hydrogen bonds between H and O situated in the $CoO_2$ layers [97] plays an significant role in the occurrence of superconductivity in $Na_xCoO_2 \cdot yH_2O$. It is important to note that, in DNA, it is exactly the hydrogen bonds that hold complementary base pairs together [81]. In other words, life on Earth is based on hydrogen bonds. It is not by accident that we continue to find parallels between superconductivity and the living matter. It is most likely that, in $Na_xCoO_2 \cdot yH_2O$, the *randomized* hydrogen bonds make the $CoO_2$ layers structurally unstable, resulting in the occurrence of superconductivity. In addition to the two aforementioned reasons, the water in $Na_xCoO_2$ may also screen Co atoms from strong Coulomb force of the Na atoms, assisting the occurrence of superconductivity in the $CoO_2$ layers.

The main point of this subsection is that, in the framework of our project, one should try to experiment with water using it as an intercalant. In addition, as was already discussed earlier, one must attempt to dope the $CoO_2$ layers by holes.

## 5.    The second approach

In this section, we consider the second approach to the problem of room-temperature superconductivity. To recall, this approach consists in improving the performance of known superconducting materials. This approach is not new and was numerously used in the past. Human imagination has no limits, and one should use it in the framework of this approach.

Before attempting to improve the critical temperature of a superconductor, one must first consciously choose this superconductor. Let us briefly discuss my ideas concerning what system to choose and how to improve its performance. These ideas should not be considered as an action plan; they just represent my personal thoughts and can be ignored by the reader. In this case, one can continue the reading with the following section.

At present, the cuprates exhibit the highest value of $T_c$. Hence, it is logical to experiment with the cuprates attempting to improve their critical temperature. In Fig. 6.51, one can see that, in underdoped cuprates, the pairing temperature exceeds the room temperature. Therefore, the cuprates may in principle exhibit superconductivity at room temperature.

Superconductivity in the cuprates occurs in the $CuO_2$ layers. In a crystal, the layers that intercalate the $CuO_2$ layers basically perform two functions: they structurally support the $CuO_2$ layers and play a role of charge reservoirs. From Chapter 6, we know that, in a superconductor of the third group, the more the lattice is unstable, the higher the critical temperature is. Therefore, one should find suitable intercalant layers which make the $CuO_2$ layers structurally unstable in a greater degree than those in Hg1223 having the highest critical temperature of 135 K (at ambient pressure). As an example, by applying a pressure $P$ to a single crystal of Eu-doped LSCO, it was shown that the derivative $\frac{dT_c}{dP}$ strongly depends on the direction of applied pressure, taking *negative* as well as *positive* values [98]. In the cuprates and other superconductors, the lattice holds the key to a higher $T_c$. As was discussed in the previous section, one may also try to intercalate the $CuO_2$ layers by organic molecules/layers.

The second suggestion represents an idea which was also discussed in the previous section, namely, to induce in a cuprate spin excitations able to mediate the phase coherence above its "normal" $T_c$.

## 6.    The third approach

In the framework of the first approach to the problem of room-temperature superconductivity, materials containing bisolitons must be doped by "magnetic" species which are responsible for the onset of long-range phase coherence. To find a right combination of two materials and arrange them properly in a crystal structure is not easy and needs a lot of time. In principle, the presence of magnetic materials is not necessary for the occurrence of superconductivity if one can organize the direct overlap of bisoliton wavefunctions leading to the onset of long-range phase coherence. To solve this task is also not easy because the density of bisolitons cannot be large and their size is small. Therefore, one should be ingenious in tackling this problem. As was mentioned above, this approach can be called the *bisoliton overlap* approach.

Before discussing a concrete suggestion, it is first necessary to introduce a few notions. Figure 10.12a shows schematically a chain polymer, for example polyparaphenylene depicted in Fig. 10.4a, and two bisolitons on the polymer. Let us denote the size of a bisoliton by $d$ (see also Fig. 6.20), and the minimum permissible distance between two bisolitons on the same chain by $L_{min}$. Denote also the distance between two parallel polymer chains by $\ell$, as shown in Fig. 10.12b. Two bisolitons on the neighboring chains can overlap [63]. This can occur if the distance between the adjacent chains is not large, i.e. $\ell \sim d$. In general, $d$ is a few lattice constants, and $L_{min}$ is of the order of several $d$.

Figure 10.12c shows a few parallel polymer chains and the bisolitons moving with a velocity $v$ along these chains. The number of the chains is $n \geq L_{min}/d$. Assume that the distance between the adjacent chains is of the order of $d$ and, in one train, the bisolitons on the neighboring chains travel with a small delay in time $\Delta t \sim d/v$ relative to one another. In this case, the bisolitons A and B shown in Fig. 10.12c will have the same phase due to the direct overlap of bisoliton wavefunctions. However, this series of bisolitons will not be in phase with the following one. In order to establish the phase coherence between the two trains of bisolitons, the wavefunctions of the bisolitons B and

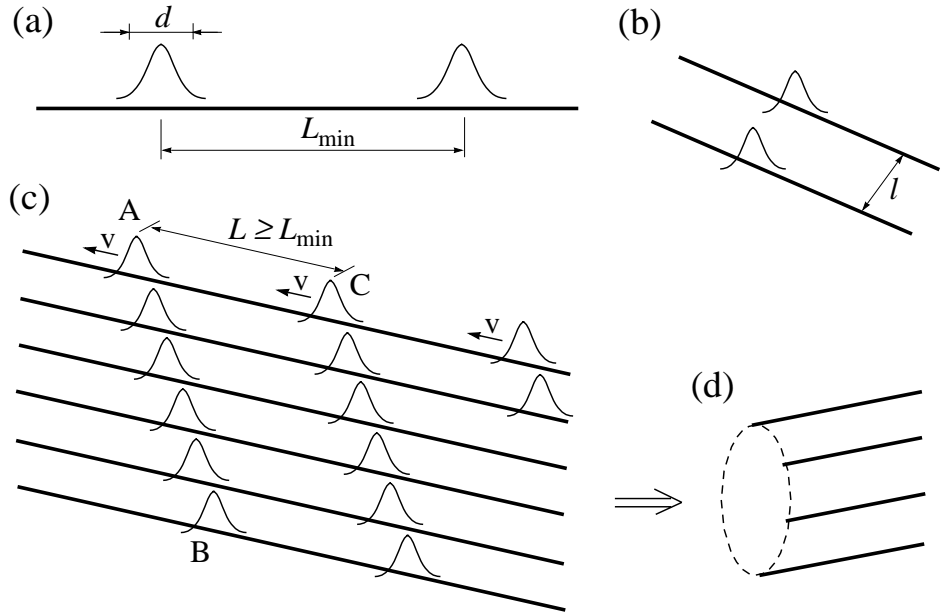

*Figure 10.12.* Bisolitons on polymer chains shown schematically by thick straight lines. (a) $d$ is the size of a bisoliton (see Fig. 6.20), and $L_{min}$ is the minimum permissible distance between two bisolitons on the same chain. (b) $\ell$ is the distance between two parallel polymer chains. (c) Bisoliton trains on parallel polymer chains, moving with a velocity $v$. The number of chains is $n \geq L_{min}/d$, and $L \geq L_{min}$ is the distance between bisolitons on the same chain. In one train, bisolitons on the adjacent chains travel with a delay in time $\Delta t \sim d/v$ relative to one another. (d) The polymer chains from plot (c) wrapped to form a tube (for more details, see text).

C in Fig. 10.12c must overlap. This can be achieved by wrapping these chains to form a tube (a sort of a nanotube) shown schematically in Fig. 10.12d. The radius of this nanotube is $\sim d\,n/2\pi$. In this case, all the bisolitons, from the first to the last one, will be in phase. However, the system as whole will not be superconducting because the ends of all the chains are not connected together (otherwise, it is impossible to arrange the trains of bisolitons).

We are now in a position to discuss a suggestion which can be realized in practice and can be successful. Instead of using several polymer chains, one should take one chain and twist it in shape of a helix, as shown in Fig. 10.13a. The helix step must be $h \sim d$ and its radius should equal, or be larger than, $R_{min} \simeq \frac{1}{2\pi}\sqrt{L_{min}^2 - h^2}$. By sending a bisoliton series along this spiral polymer with a repetition time of $\sqrt{(2\pi R)^2 + h^2}/v$, one will observe a bisoliton condensate traveling along the polymer. This can be achieved not only at 350 K but even at 500 K. Instead of a chain polymer, one may use a carbon nanotube. Alternatively, a nanostripe of graphene [70] having a width of a few carbon bonds can also be used.

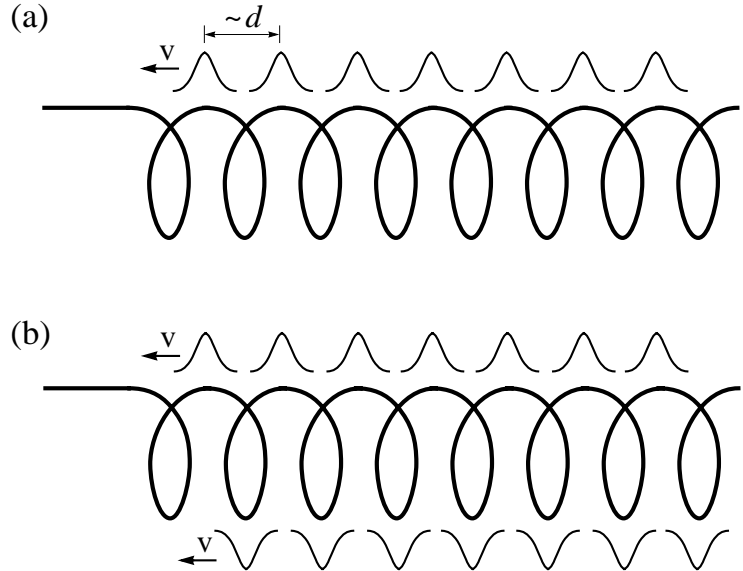

*Figure 10.13.* Bisoliton trains on a helix polymer having a step of the order of the bisoliton size *d*: (a) one train representing a bisoliton condensate moving along the polymer, and (b) two trains traveling on the opposite sides of the helix. These two bisolitons trains are not in phase (for more details, see text). The bisoliton velocity *v* is relative to the polymer.

To realize this idea in practice, one will however face, at least, three problems. First, technologically, it is not easy to twist a polymer in a helix with a fixed radius and a fixed step. Second, one should invent a device producing trains of bisolitons with a frequency of $v/\sqrt{(2\pi R) + h^2}$. Finally, certain molecules upon receiving a charge isomerize, i.e. change their shape upon receiving a charge [80]. This change is small [80] but in a long polymer containing a large number of bisolitons the total change can be noticeable. This can be a real problem for some polymers.

In the case shown in Fig. 10.13a, the polymer may slightly bend following the rotation of bisolitons. To avoid this unwanted bend of the helix, one may use two trains of bisolitons, as shown in Fig. 10.13b. In this case, the two trains of bisolitons are not in phase with one another but, being on the opposite sides of the helix, their actions on the helix structure are mutually compensated. In the general case when $N$ bisolitons travel *within a single helix step*, the helix should have the following dimensions $h \sim d$ and $R \geq \frac{1}{2\pi}\sqrt{(NL_{min})^2 - h^2}$. The repetition time in the general case equals $\sqrt{(2\pi R)^2 + h^2}/(vN)$.

In Fig. 10.13, one can notice that the structure of the polymer is similar to that of DNA. It is amazing that we continue to find additional parallels between superconductivity and the living matter. In fact, one may even adopt the DNA structure for avoiding a bend of the helix discussed above. To realize this, two spirals twisted in the opposite directions should be put together, resulting in a structure similar to that of DNA. The ends of these two spirals must be connected. If in a single helix, a bisoliton train circles in one direction, in the double helix, two bisoliton trains circle in the opposite directions symmetrically along the main axis of the polymer.

What about DNA itself: can it exhibit superconductivity when it is charged by a bisoliton train? The double helix of DNA has a radius of 10 Å, and its step is 34 Å. Such a structure is suitable for bisolitons having the following characteristics $d \sim 34$ Å and $L_{min} \sim \sqrt{(2\pi R)^2 + d^2} \approx 71$ Å. In practice, the ratio $L_{min}/d$ should be at least 5. This is however not the case for DNA. Thus, the structure of DNA is not suitable to support a bisoliton condensate: its radius is too small. However, this does not mean that DNA cannot support a few independent bisolitons. Therefore, DNA can be used as material for electron pairing in the framework of the first approach discussed above.

It is worth noting that the structure of the inorganic polymer $(SN)_x$ has the form of a helix. As discussed in Chapter 3, $(SN)_x$ becomes superconducting below $T_c = 0.3$ K when doped with bromine. Its unit cell contains two parallel spirals of $(SN)_x$ twisted in the opposite directions [99]. The onset of long-phase coherence in $(SN)_x$ occurs due to, however, not the overlap of bisoliton wavefunctions but spin fluctuations of unpaired electrons on $Br_3^-$ and $Br_5^-$ clusters situated between the $(SN)_x$ spirals.

At the end of this section, let us discuss the issue of practical application for such a type of superconductors. It is obvious that such room-temperature superconductors will have a limited number of applications. For example, they cannot be used for large-scale applications. On the other hand, they may be perfect for use in microchips for example. Then, the next question which needs a solution is the matter of good-quality electrical contacts. This subject is the topic of the following section.

## 7.    Electrical contacts

Without doubt, a room-temperature superconductor will be available in the near future. Most likely, it will be a superconductor containing organics. One of the main problems for use of organic materials is the issue of electrical contacts. Organic materials cannot be soldered onto metal leads in the conventional sense of this expression because metals do not wet organics. Therefore, the quality of electrical contacts for a room-temperature superconductor will be the next problem needed a solution. In an ideal contact, none of the electrons entering or leaving a piece of material under the test will be scattered back by the contact. In early experiments with carbon nanotubes, the quality of microfabricated contacts was bad. In these experiments, the transport therefore appeared to be diffusive rather than ballistic. Of course, in a contact between a superconductor and a normal metal, independently of the quality of the contact there will always be electrons scattered back because of the Andreev reflection.

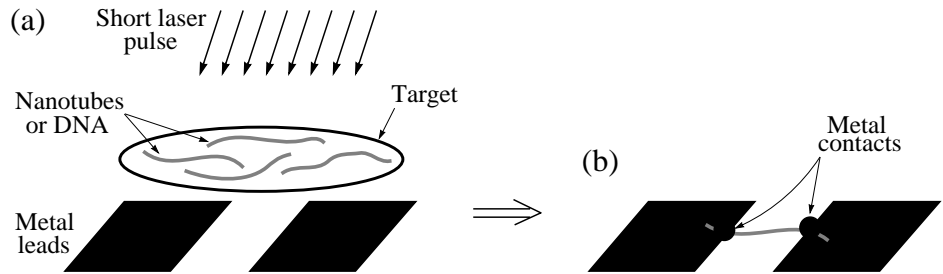

*Figure 10.14.* Laser-based technique used to solder DNA molecules [37] and carbon nanotubes [100] onto metal leads. (a) A target covered with DNA molecules or nanotubes is placed above metal leads. A short pulse ($\sim$ 10 ns) of a focused laser beam is then fired at the target, and the target is evaporated. The detached molecules fall down. One molecule may be connected to the leads due to locally molten metal on each side of the slit as shown in plot (b). For more details, see text.

As discussed above, a DNA molecule can be connected onto metal pads or beads "chemically." Alternatively, DNA as well as carbon nanotubes can be soldered onto the leads by using a laser-based technique mentioned in Chapter 3. Let us consider briefly this nano-soldering technique [37, 100]. A target covered with DNA molecules or nanotubes is placed above metal leads (a golden membrane with a slit of $\sim$ 300 nm), as shown in Fig. 10.14a. A short pulse ($\sim$ 10 ns) of a focused laser beam (power $\sim$ 10 kW) is fired at the target to detach the molecules from the target. It is anticipated that, at least, one molecule will fall and connect the edges of the slit below. Since the metal leads on each side of the slit are locally molten, the molecule gets soldered into the leads, and is suspended, as shown in Fig. 10.14b. Approximately, one out of ten attempts is successful. The attempts to solder a nanotube or a DNA molecule lying immediately on the leads were not successful. Thus, the originality of this technique lies in the suspended character of organic giant molecules [100].

Of course, for industrial production of microchips based on a room-temperature superconductor, this technique must be improved.

# Index